# Performance of the CMS Tracker Optical Links and Future Upgrade Using Bandwidth Efficient Digital Modulation

Stefanos Dris

High Energy Physics
Imperial College London
Prince Consort Road
London SW7 2BW

Thesis submitted to the University of London for the degree of Doctor of Philosophy

September 2006

# Abstract

The Compact Muon Solenoid (CMS) experiment at the Large Hadron Collider (LHC) particle accelerator will begin operation in 2007. The innermost CMS sub-detector, the Tracker, comprises ~10 million detector channels read out by ~40 000 analog optical links.

The optoelectronic components have been designed to meet the stringent requirements of a high energy physics (HEP) experiment in terms of radiation hardness, low mass and low power. Extensive testing has been performed on the components and on complete optical links in test systems. Their functionality and performance in terms of gain, noise, linearity, bandwidth and radiation hardness is detailed.

Particular emphasis is placed on the gain, which directly affects the dynamic range of the detector data. It has been possible to accurately predict the variation in gain that will be observed throughout the system. A simulation based on production test data showed that the average gain would be ~38% higher than the design target at the Tracker operating temperature of -10°C. Corrective action was taken to reduce the gains and recover the lost dynamic range by lowering the optical receiver's load resistor value from 100Ω to 62Ω. All links will have gains between 0.64 and 0.96V/V.

The future iteration of CMS will be operated in an upgraded LHC requiring faster data readout. In order to preserve the large investments made for the current readout system, an upgrade path that involves reusing the existing optoelectronic components is considered. The applicability of Quadrature Amplitude Modulation (QAM) in a HEP readout system is examined. The method for calculating the data rate is presented, along with laboratory tests where QAM signals were transmitted over a Tracker optical link. The results show that 3-4Gbit/s would be possible if such a design can be implemented (over 10 times the equivalent data rate of the current analog links, 320Mbits/s).

*To my family*

# Acknowledgements

First and foremost I would like to thank my two supervisors, Dr Costas Foudas (Imperial College) and Dr Jan Troska (CERN). They have both contributed to this work in very different, but equally important ways.

I am also grateful to Francois Vasey for giving me the opportunity to work in the optical links section of the CMS Electronics (CME) group at CERN (now part of the Microelectronics group), and Geoff Hall for giving me a place in the HEP silicon group at Imperial College. I would not have been able to work at CERN without the funding provided by CME, headed initially by Peter Sharp and later by Jordan Nash.

I consider myself extremely fortunate to have had the immense support and unconditional love of my family and of Valentina.

# Contents

## Chapter 4: Predicting the Gain Spread of the CMS Tracker Readout Optical Links

## Chapter 5: Upgrading the Tracker Optical Links: Digital Modulation

**Chapter 6: Upgrading the Tracker Optical Links: Laboratory tests**

**Chapter 7: Conclusions and Future Work**

# Chapter 1

# Introduction to the LHC and CMS

*The Large Hadron Collider (LHC) accelerator is introduced, along with the motivation for building it. The Compact Muon Solenoid (CMS) experiment is one of the large, general-purpose experiments that will benefit from the LHC's unprecedented performance to explore new physics and validate existing theory. The design criteria of CMS are outlined, with particular focus on its innermost tracking sub-detector.*

## 1.1 The Large Hadron Collider

The Large Hadron Collider (LHC) at the European Organization for Nuclear Research (CERN) [1] is a proton-proton collider being installed in the 27km ring that formerly housed the Large Electron Positron (LEP) collider. In circular accelerators, particles are constrained in a vacuum pipe with the help of electromagnets. The relation between the proton momentum $p$ in GeV/c and the magnetic field $B$ in Tesla is:

$$p = 0.3Br \qquad (1.1)$$

where $r$ is the ring radius in metres. Since the ring radius of the LHC is that of the LEP tunnel, the only parameter that can be varied to in order to increase the maximum beam energy is $B$. Superconducting magnets with a maximum field of 8.33T will therefore be used to accelerate each proton beam to achieve collisions with centre-of-mass energies of 14TeV.

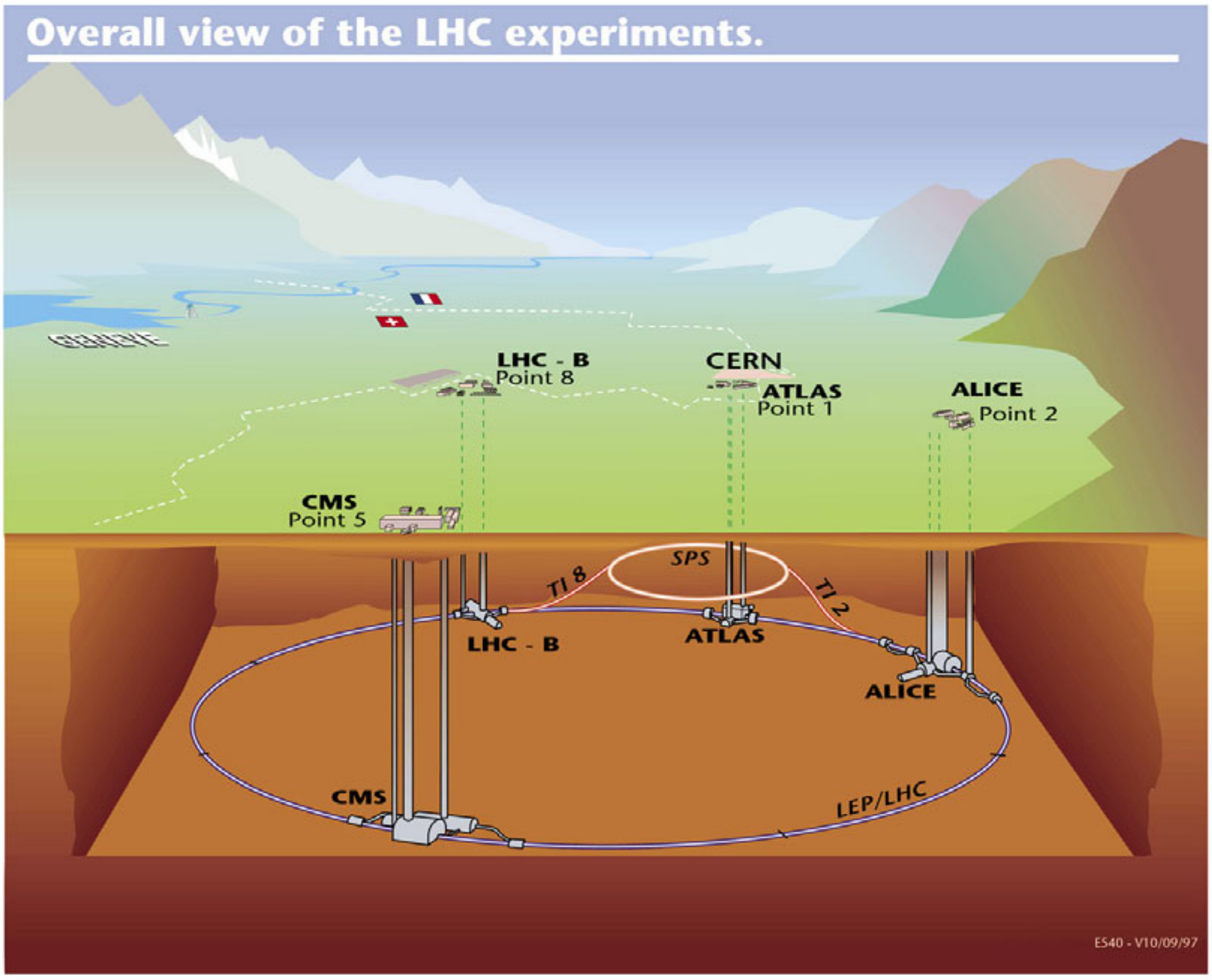

**Figure 1.1:** The LHC and experiments at CERN [2].

The LHC's two proton beams rotate in opposite directions and consist of proton bunches that are separated by 25ns and intersect at four points in the ring where experiments are placed (Figure 1.1), providing a peak luminosity of $10^{34}cm^{-2}s^{-1}$ with a projected lifetime of 10 hours and an expected integrated luminosity of $100 fb^{-1}$ per year.

The four experiments built to exploit LHC physics are optimised for various searches. Two general purpose detectors, CMS (Compact Muon Solenoid) [3] and ATLAS (A Toroidal Lhc ApparatuS) [4] will identify and analyse a wide range of particles.

### PHYSICS AT THE LHC

The Standard Model (SM) is currently the best theory for particle interactions, having passed all experimental tests of its predictions in recent years. In the framework of the SM, particles acquire their mass by interaction with the Higgs field. This implies the existence of the Higgs Boson, a particle which is yet to be discovered. Previous and present accelerators and their experiments (at LEP, SLAC and Fermilab) have not been able to detect this particle, but have excluded the range of possible masses up to ~114.4GeV [5]. While the theory gives no explicit value for the mass of the new particle, it can predict decay modes of the Higgs for a given choice of Higgs mass. This allows the design of accelerators and experiments to be optimized for the search of the elusive particle. The LHC will be able to explore the energy range above that accessible with current machines.

There are a number of decay modes for the SM Higgs, depending on the mass of the particle. For a low-mass Higgs ($M_H$~120GeV), the most promising decay mode will be into two photons (Figure 1.2(a), $H^0 \rightarrow \gamma\gamma$). For a heavy Higgs, the most promising channel for discovery of the particle is its decay into four leptons (Figure 1.2(b), $H^0 \rightarrow ZZ \rightarrow 2l^+ 2l^-$), which will be clearly identifiable in the detectors.

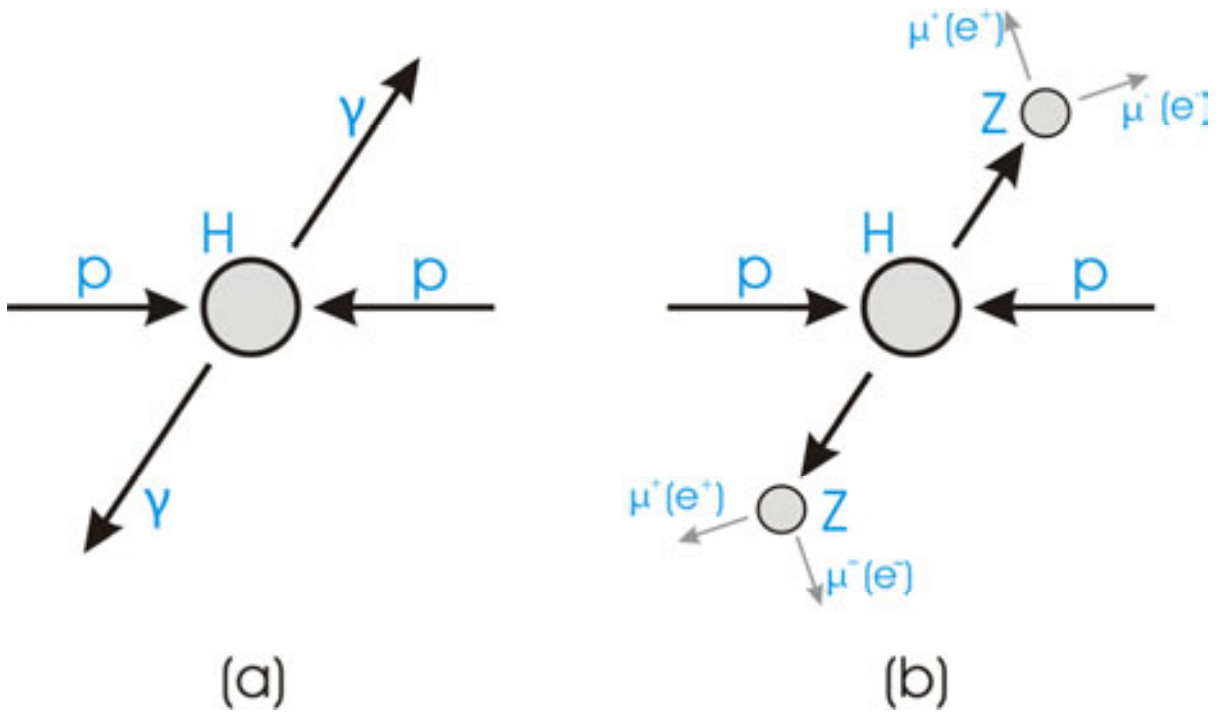

**Figure 1.2:** Promising decay channels for an SM Higgs of different masses.

In addition to the Higgs, the LHC also offers discovery potential for various other new particles. In an extension of the SM called supersymmetry, supersymmetric partners (sparticles) of the currently known SM particles are postulated to exist. The sparticles have masses in excess of their 'normal' partners, and are expected to be within the energy reach of the LHC. In addition, CP violation may be observed at the LHC, which could provide the answer to the question of why the universe appears to be dominated by matter, rather than anti-matter.

It is clear that a general purpose experiment must be excellent in all types of detection in order to take advantage of the wide range of new physics expected to be accessible at the LHC. This not only places stringent requirements on the detector elements, but also on the electronics needed for readout and transfer of the information for analysis of interesting events. The work described in this thesis concerns the readout system of the tracking detectors of CMS. The next section describes this experiment in more detail.

## 1.2 CMS Design Criteria

### 1.2.1 Overview of Sub-Detectors

CMS is comprised of a series of detectors in concentric cylindrical layers around an axial beampipe (Figure 1.3). Hermeticity is achieved through two endcaps on each end. Protons enter from both sides, with collisions occurring at the center of the structure. Resulting particle trajectories pass radially through the following layers (Figure 1.4): The silicon tracker, the crystal electromagnetic calorimeter (ECAL), the hadronic calorimeter (HCAL), the 4T solenoid magnet, and a four station muon detection system made up of tracking and dedicated trigger chambers. Detailed descriptions of each sub-detector can be found in the Technical Design Reports [6-10], and only a brief overview of each will be given here.

Detecting muons is key for the discovery of the Higgs Boson. The detector technology chosen for the CMS muon system allows muon identification and momentum measurement, as well as triggering. Muon identification is achieved with precision positional information coming from the drift tubes and cathode strip chamber detectors, while precise momentum measurement is carried out in conjunction with information from the Tracker. These are complemented by fast resistive plate chambers to provide fast triggering capability.

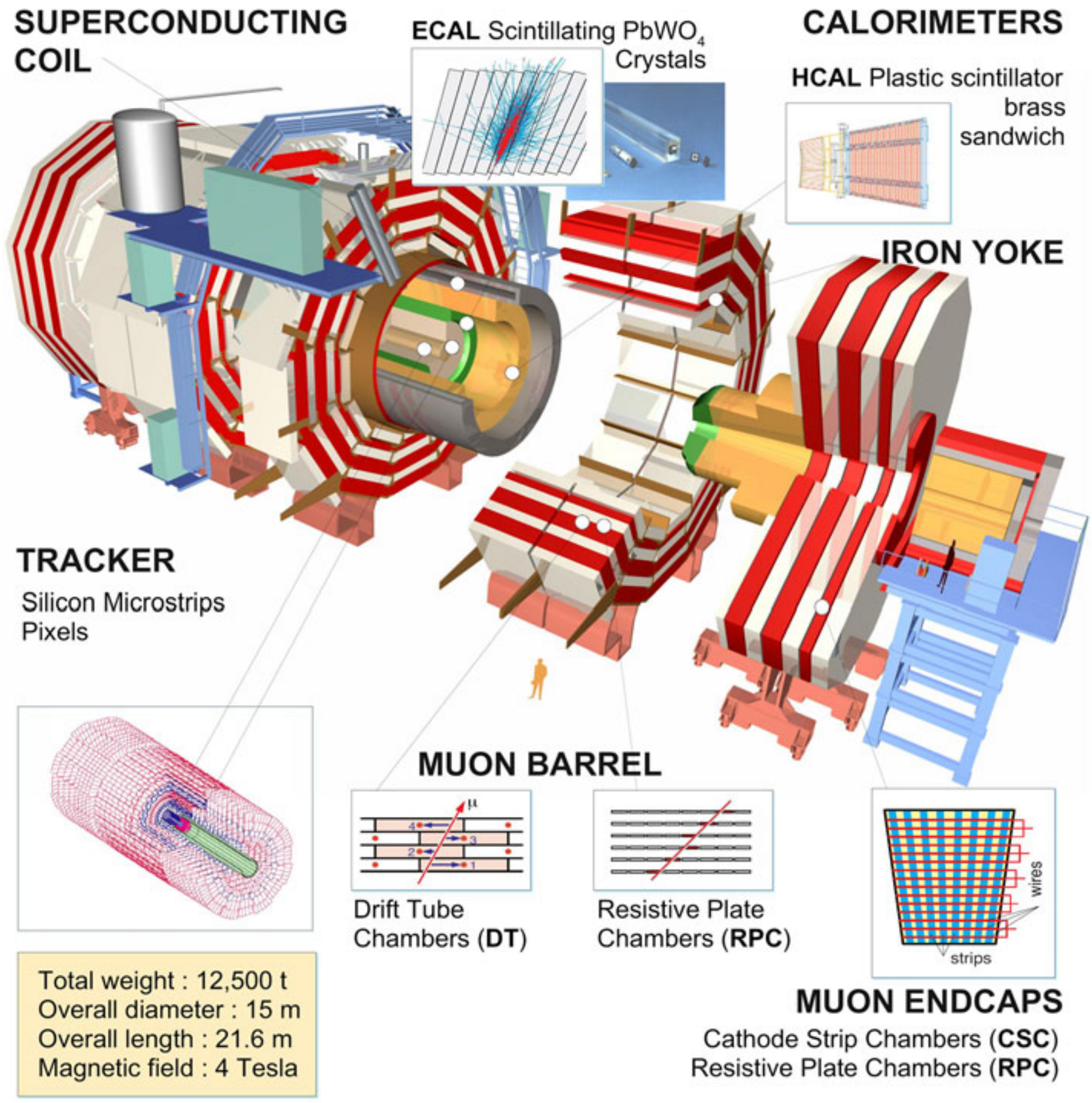

**Figure 1.3:** 'Exploded' 3-D view of the CMS experiment [11].

The choice of magnetic field in CMS directly impacts the muon momentum resolution. A superconducting solenoidal magnet has been chosen to bend particles in the transverse plane of the detector, thus allowing precision vertex determination due to the small transverse dimension of the beam. For similar bending power, a solenoid is smaller than a toroid. The high magnetic field chosen (4T) has been chosen to benefit the momentum resolution of the Tracker. Moreover, high precision in the ECAL is achieved by comparison of electron momentum in the Tracker with the energy deposition in the ECAL. Finally, the high magnetic field ensures that low momentum charged particles are confined in the central region of CMS, hence limiting the radiation damage in the bulk of the detector.

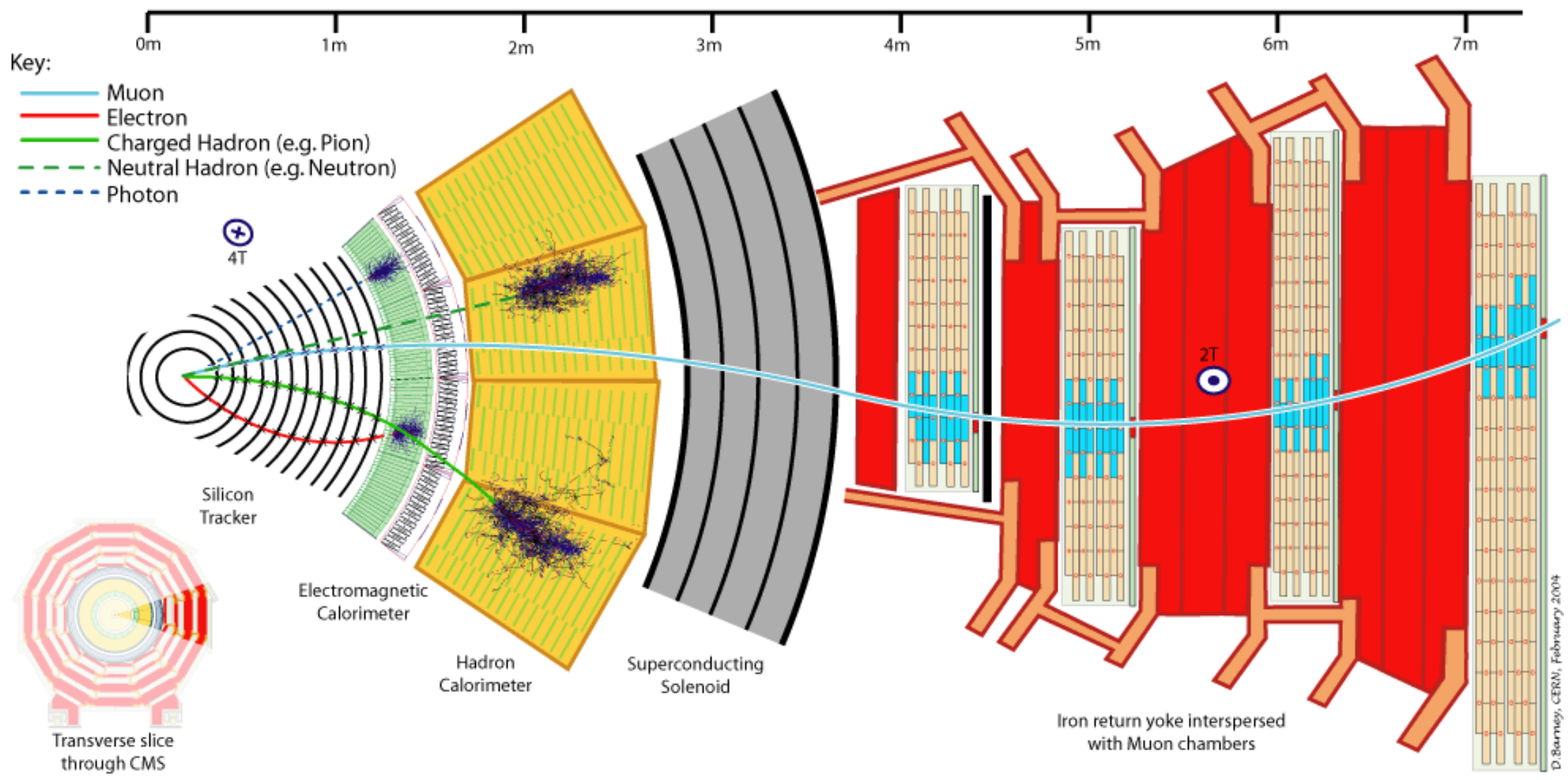

**Figure 1.4:** Cross-sectional view of the CMS detector [11].

The HCAL contains and measures the energy and direction of all hadronic showers, while also aiding in the identification of photons, electrons and muons. It consists of copper absorber and plastic scintillator tiles. The HCAL is fully hermetic, a feature that is important in the search for new particles where the signatures contain missing transverse energy ($E_T$) in the form of neutrinos and their supersymmetric partners. The non-instrumented material in front of the HCAL has been minimized to avoid unidentified conversions occurring before the particles reach the sensors. This is crucial to the determination of missing $E_T$, since it allows the measurement of the energy of all detectable particles.

The ECAL has been designed for high precision energy measurement of photons, which is crucial in the discovery of a low-mass SM Higgs (~120GeV). CMS has opted for a homogeneous scintillating crystal ($PbWO_4$) calorimeter. These high-density crystals ensure the system is highly compact. The amount of material in front of the ECAL has been minimized to reduce electromagnetic interactions occurring outside the sub-detector's volume which degrade its energy resolution.

The innermost part of the CMS experiment is occupied by the silicon microstrip Tracker. Its function is to identify and precisely measure the momentum of muons, electrons and photons, over a large energy range. The Tracker is also used to identify leptons and photons. The latter is crucial for suppressing backgrounds to the decay of the Higgs to two photons. The performance requirements of the Tracker have been met while ensuring that there is no excess material in front of the calorimeters to avoid degradation in their resolution and maintain efficiency.

Moreover, the material budget also affects the Tracker's resolution and efficiency for isolated electrons, due to Bremsstrahlung effects. A more detailed description of the Tracker is given next.

### 1.2.2 The Tracker

The design of the Tracker is constrained not only by the physics requirements, but also by the budget available to the sub-detector itself (~20% of the entire CMS budget [10]). Hence, commercially available components must be used as much as possible to minimize costs associated with the development of customized components. In addition, the physics performance requirement dictates that the Tracker material budget (detectors and readout systems) must be kept at a minimum. Given that it is important to instrument as much of the Tracker volume as possible, availability of space for systems inside the detector (the 'front-end') is severely restricted.

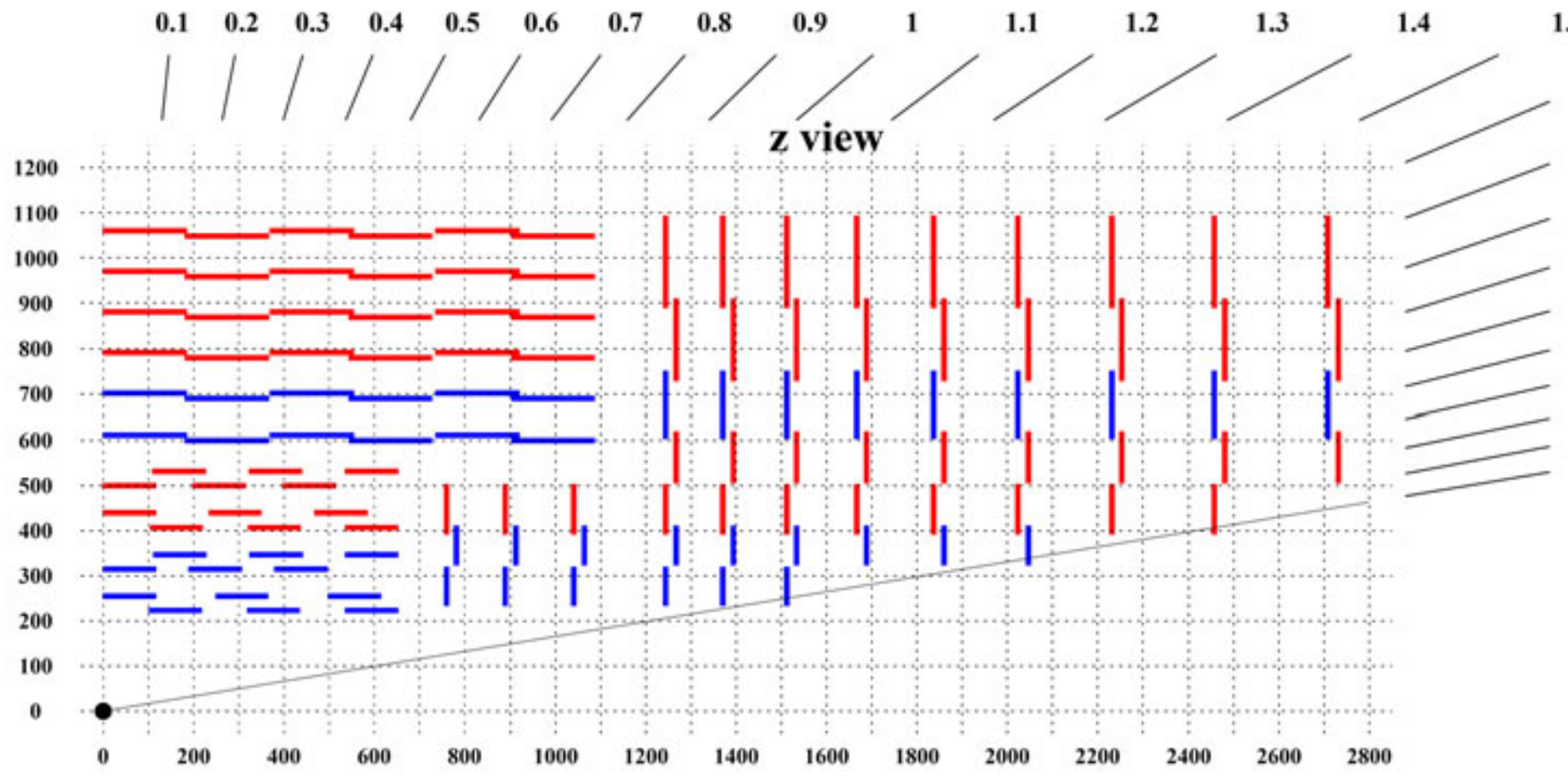

**Figure 1.5:** Layout of the CMS Tracker, showing the ¼-view of the z-plane [12]. Collisions occur at coordinate (0,0), with the two beams traveling from the left and right of the collision point. The red lines represent single-sided modules, while the blue lines are double-sided. Pixel layers that occupy the space closest to the collision point are not shown.

The physics goals, in conjunction with the constraints outlined above, have led to the design of the layout of the Tracker shown in Figure 1.5 [10]. The Tracker comprises silicon pixel and silicon microstrip layers in a cylindrical volume ~5.4m long and ~2.2m in diameter. The design allows the measurement of particle momentum in conjunction with the magnetic field, and the reconstruction of vertices of both the primary proton-proton interaction points and the secondary vertices due to particle decays. At peak luminosity, approximately 1000 charged particle tracks in the pseudorapidity range $|\eta|<3$ will be recorded every 25ns. High channel density is required for good spatial resolution, while the detector is made of many low mass layers so that particles are not deflected from their paths. Tracks are formed by recording hits in the microstrips as particles pass through each layer.

The pixel detector is placed close to the interaction region in order to measure vertices accurately as well as "seed" tracks. The pixels are arranged in a 3-layer barrel, with two 2-layer endcaps on each side. The silicon microstrips cover a very large volume, ensuring that significant bending occurs within the sub-detector, and therefore accurate momentum measurement of very high energy charged particles can be made. There are approximately 10 million silicon microstrips arranged in nested layers around the beam pipe, with 10 cylindrical layers (the barrel region) parallel to the pipe and two endcaps each consisting of 9 disks.

The Tracker barrel region contains of four inner barrel layers (TIB) and six outer barrel layers (TOB). The first two layers of both the TIB and TOB are double-sided. These consist of two single sided sensors mounted back to back, tilted by an angle of 100mrad with respect to each other, effectively giving two-dimensional hit positions. The endcap system consists of three inner endcap disks (TID) and nine outer endcap disks (TEC), placed on either side of the barrel region.

The distinction between inner and outer regions is not merely topological. The corresponding sensors are different, due to the need to optimize three parameters: Depletion voltage, signal-to-noise ratio and radiation tolerance. The depletion voltage is proportional to the square of the thickness and the inverse of the resistivity. However, low-resistivity sensors are more radiation tolerant. In the inner regions where radiation levels are higher, the detectors are low-resistivity sensors, with a thickness of 320μm. In the outer regions the sensors are 500μm thick, to compensate for the higher noise due to higher capacitance in their longer strips. The Tracker will be operated at a temperature of -10°C to limit the extent of radiation damage [10].

The silicon microstrips are essentially p-n diode structures, consisting of an n-type substrate with a $p^+$-type layer facing the interaction point and an $n^+$-type backing plate. During operation, they are sufficiently reverse-biased so as to fully deplete them. When a charged particle passes through a microstrip, electron-hole pairs generated in the depletion region are swept out and a current signal is produced.

### 1.2.3 Tracker Readout Requirements

The performance of the readout system is critical to the overall performance of the Tracker as the detectors themselves. With over 10 million microstrip channels in the Tracker, the requirements on the readout system are formidable. A highly parallel data transmission system, unprecedented in any other application, is needed to read out the information from each detector channel and transfer the data out of CMS. The large number of channels, as well as the data rate required to keep up with the 40MHz LHC bunch crossing rate, makes the use of traditional (for physics experiments) twisted pair copper wire prohibitive.

The amount of material used in the innermost regions of CMS must be minimized, as explained earlier. Taking into account that the Tracker must be as hermetic as possible to maintain uniform performance, it is highly desirable to minimize the volume occupied by cabling of the readout system. In addition, the readout system must be sufficiently radiation resistant, since parts of it are in the high radiation region of the detector.

To ensure that the readout system remains within budget, CMS has opted to use Commecial Off The Shelf (COTS) components wherever possible to avoid the development costs of customized components. However, this means that the COTS components must be qualified for use in the hostile environment of the LHC.

Optical data transmission has been selected for the Tracker readout system. This is especially attractive for the environment of the LHC since transmission speeds are potentially much higher than in copper-based systems. Moreover, due to space constraints, the small size of optical transmitters, receivers and fibers make this solution ideal for use inside the Tracker volume. Optical systems are also immune to ElectroMagnetic Interference (EMI), which would be detrimental to a highly parallel copper-based system of this scale. Finally, electrical isolation between the detector front-end and the remote readout electronics ensures that no grounding problems can occur.

## 1.3 Super LHC Upgrade

The LHC is expected to start operation (i.e. circulating beam) in the Spring of 2007, with the initial physics program starting later that year. With the initial luminosity expected to be ~$2 \cdot 10^{33}$ $cm^{-2}s^{-1}$, the integrated luminosity by 2008 should be ~$10fb^{-1}$. Once the luminosity reaches the design target of $10^{34}$ $cm^{-2}s^{-1}$, it will take around 5-6 years to collect data corresponding to $200\text{-}300fb^{-1}$. The enormity of this number can be comprehended by considering that Run II of the Tevatron at Fermilab (the most powerful accelerator in operation) has provided an integrated luminosity ~$1.7fb^{-1}$ in five years, since mid-2001.

With the LHC having collected enough data, and the quadrupole focusing magnets having an expected lifetime of less than 10 years, an upgraded accelerator could materialize before 2015. Plans are currently being considered for an upgraded collider using the existing tunnel. From a machine point of view, the primary objective for the so-called Super LHC (SLHC) [13] is to increase the luminosity an order of magnitude, to $10^{35}$ $cm^{-2}s^{-1}$.

At SLHC luminosity, the current tracking detectors will not work [14]. A replacement of inner tracking will be necessary in CMS, with new detector technologies required for the region near the collision point (radius<20cm). It is therefore very difficult to estimate the requirements on the readout system imposed by a re-design of the Tracker before it materializes. It is clear, however, that the number of readout channels will undoubtedly increase and this will almost certainly translate into a need for higher data rate readout. It will therefore be necessary to switch to digital readout in a future implementation of the optical links, though the required rate is currently unknown.

There are two options currently being considered for the future digital readout link: A completely new system that fully replaces of the current link, or a system based on the current link components, with added electronics to realize the digital conversion. The former has the advantage that it will rely on commercial technologies, and possibly on radiation-hard variants of COTS components, which implies less effort in research and development, and possibly quicker completion of the project. On the other hand, the cost of the current optoelectronic components represents a large fraction of the CMS Tracker electronics budget. A lot of time and effort has gone into the development of the low-power components capable of withstanding the harsh operational environment. Hence, a digital system reusing the existing components while delivering sufficient performance for SLHC operation could potentially be a cost-effective alternative to a full replacement of the installed links. A bandwidth-efficient modulation scheme similar to what is used in Asymmetric Digital Subscriber Line (ADSL) would be necessary for such an upgrade. The feasibility of the conversion must be explored in terms of performance that can be achieved and implementation complexity.

## 1.4 Summary

An overview of the LHC and CMS was presented in this chapter. The performance requirements on each CMS sub-detector has been outlined, with emphasis on the Tracker and its readout system, in preparation for the more detailed examination of functionality and performance that follow in the ensuing chapters. Finally, the need for an upgrade of the links was pointed out, in the context of the SLHC upgrade that will take place in about 10 years.

## 1.5 Contribution to the Work Presented in this Thesis

I have participated in the production tests (and analysis of the results) for the laser transmitter (Chapter 2, section 2.2.1) and Analog Optoelectronic Receiver (Chapter 2, section 2.3). The production test setups for these two components pre-date the beginning of my PhD work and were built by the CMS optical links group, headed by Francois Vasey at CERN. I have designed and carried out the tests described in section 2.4. Matthew Noy and Jonathan Fulcher (both of Imperial College) wrote the proprietary software required for reading out data from the CMS Tracker Front-End Driver (FED) in this particular configuration.

In Chapter 3, section 3.1, I describe in detail the optical link's functionality. While I have not participated in the original system design, this section is a result of extensive experience working with deployed optical links in the laboratory and in test beams. I have developed the setup algorithms presented in section 3.2.

The work described in Chapters 4, 5 and 6 is entirely my own.

# Chapter 2

## The CMS Tracker Readout Link Components

*The aim of this chapter is to introduce the components making up the readout optical links of the CMS Tracker. A brief overview of their functionality is given and typical performance of each device is presented through production test results. The optical links transmit data from the CMS cavern to the counting room, where the optical receivers are hosted on Front End Driver boards (FEDs). The FED includes other components which complete the analog part of the readout chain. The FED's analog components' performance is evaluated and checked against the specification for the entire readout system.*

## 2.1 Introduction

A description of the principle optical link components will be given in this chapter. Where appropriate, results from production test data is presented to demonstrate the performance of each component. More details on the production test setups and procedures can be found in [1].

### 2.1.1 The Tracker Readout System

The CMS Tracker has opted for an analog readout system, driven by the reduction of the front-end chip complexity, power dissipation and potentially better position resolution through charge sharing between strips [2]. The readout link is illustrated diagrammatically in Figure 2.1, with photographs of the optical link components also shown.

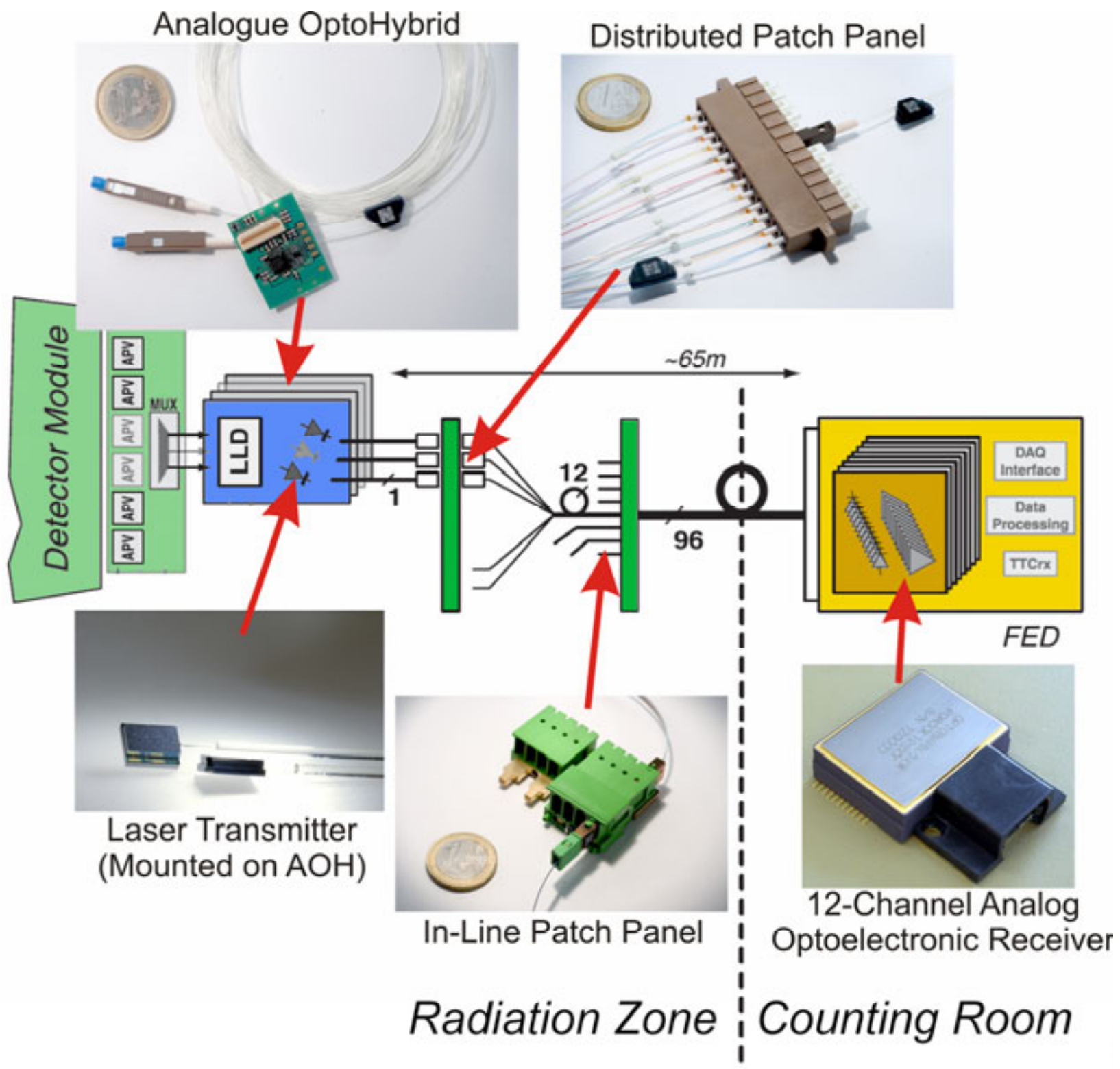

Figure 2.1: The CMS Tracker analog readout link.

Analog Pipeline Voltage (APV) ASICs [3] at the detector hybrids read out groups of 128 microstrips. Each APV channel (Figure 2.2) comprises a preamplifier coupled to a shaping amplifier producing a 50ns CR-RC pulse. Each channel's shaper output is sampled at the LHC frequency of 40MHz into a 192 cell deep pipeline. 32 cells are reserved for buffering events awaiting readout; hence the pipeline depth allows a programmable latency of up to 4μs. On arrival of a level 1 trigger, the appropriate pipeline cell columns are marked for readout, and not overwritten until this is completed. Each channel of the pipeline is read out by the Analog Pulse Shape Processor (APSP) which can operate in one of three modes [3]:

- In *peak* mode, one sample per channel is read from the pipeline (timed to be at the peak of the analog pulse).
- In *deconvolution* mode, three samples are read, the output being a weighted sum of all three.
- In *multi* mode, one trigger takes three samples at a time, enabling the observation of the time evolution of charge deposition on the silicon strips.

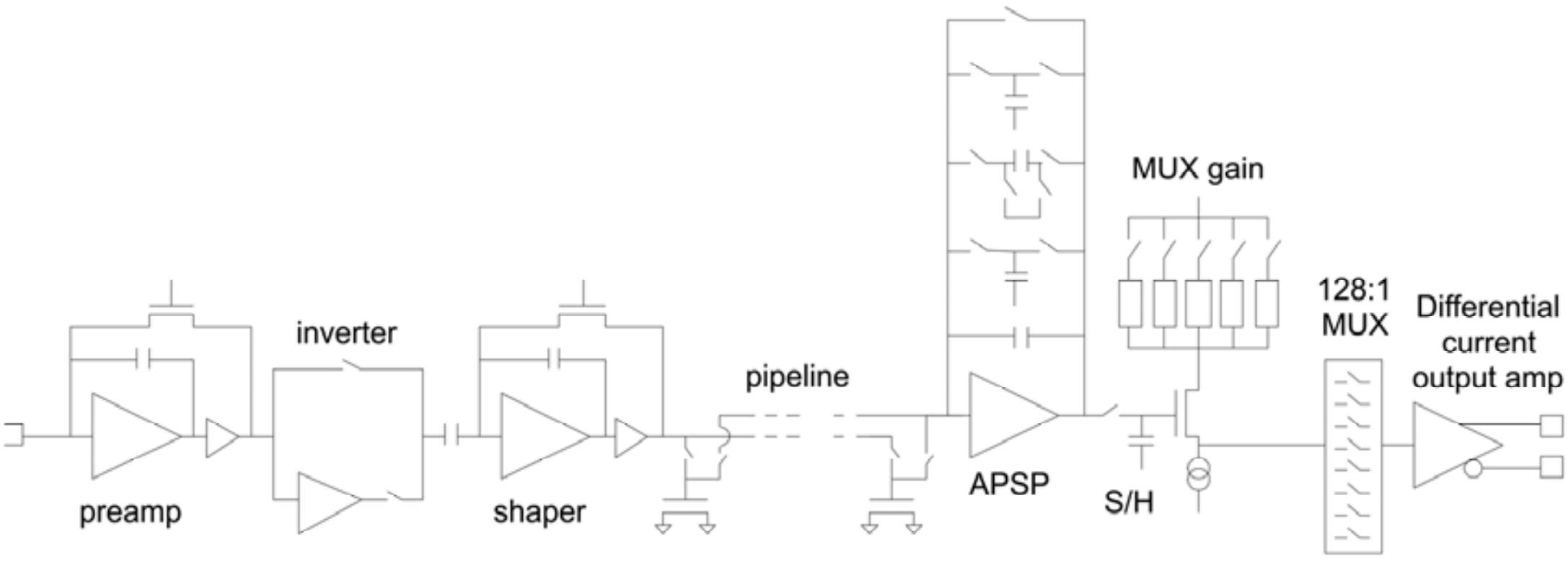

**Figure 2.2:** Block diagram of the APV analog chain [4]. There is one chain (from preamp to MUX stage) corresponding to each one of the 128 APV channels.

The deconvolution operation results in a re-shaping of the analog pulse shape to one that peaks at 25ns and returns rapidly to the baseline. The output is then sampled/held and sent to the 128:1 multiplexer. The multiplexer operates at 20MHz, resulting in a non-consecutive channel order for the analog samples.

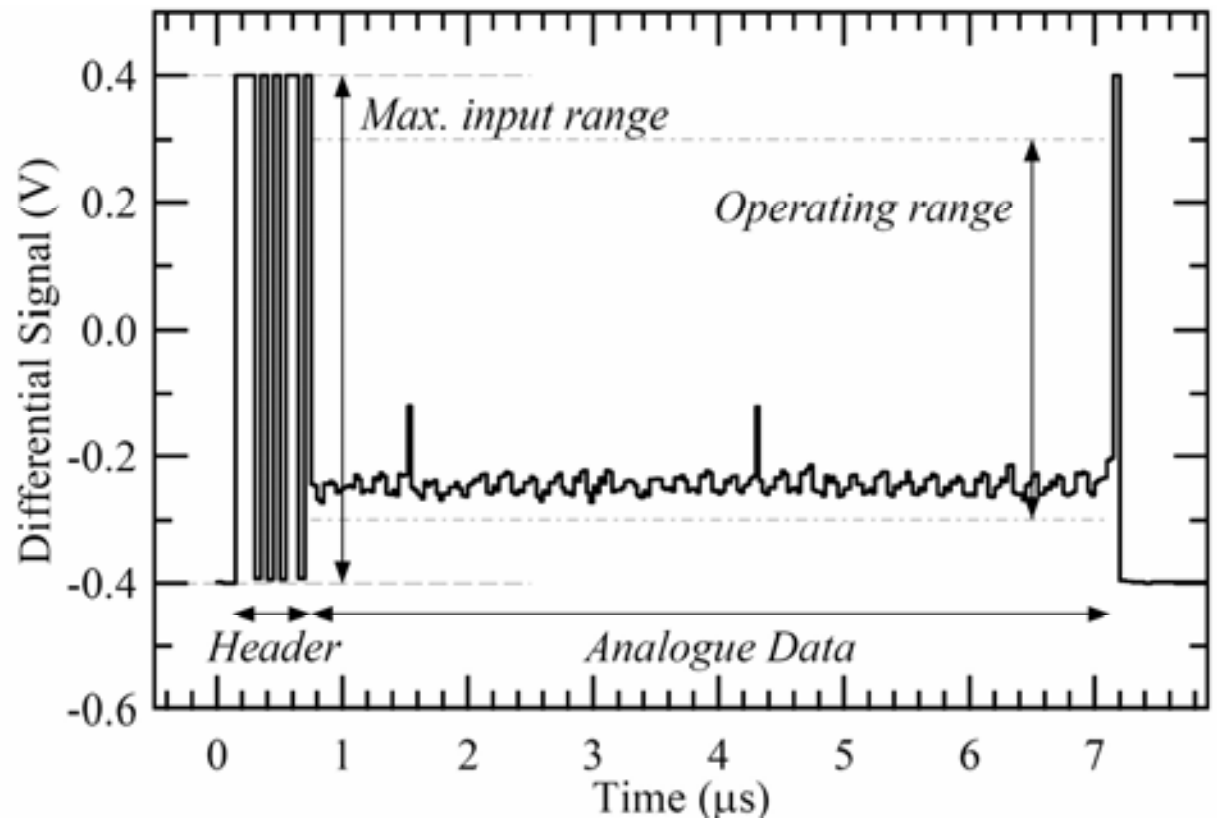

**Figure 2.3:** Typical data frame at the input of the readout link [5]. The analog data is time multiplexed at a ratio of 256:1 with a sample-width of 25ns. A digital header –which includes the address of the transmitting APV– and a stop bit is added on either side of the data.

Pulse height data from pairs of APVs are multiplexed by a 2:1 multiplexer (the APVMux) onto a differential line which provides the input to the Analog OptoHybrid (AOH) [6]. Figure 2.3 is an illustration of a typical data frame. The analog detector signals are encapsulated in a frame also containing digital data; a digital header (3 bits, all logic '1') followed by the APV pipeline address used to store the analog data, and an error bit. A stop bit is added after the analog data. At the output of the APV, the height of the digital header is nominally +/-4mA. This translates to +/-400mV (or 800mV differential) at the input to the AOH. At the nominal APV gain of 1MIP/mA for thin (320µm) detectors, the digital header is equivalent to an 8MIP signal.

The AOH performs the task of electrical to optical conversion for transmission over a ~65m fiber optic cable to the counting room, where the 9U VME Front End Driver (FED) board [7] processes the detector event data before they are subjected to the High Level Trigger system. The optical data are converted back to electrical signals by 12-channel analog optoelectronic receivers (ARx12) [8] mounted on FEDs. Each FED receives optical data from 192 APVs (96 optical fiber channels, each carrying multiplexed signals from 2 APV chips). The FED is shown in Figure 2.4.

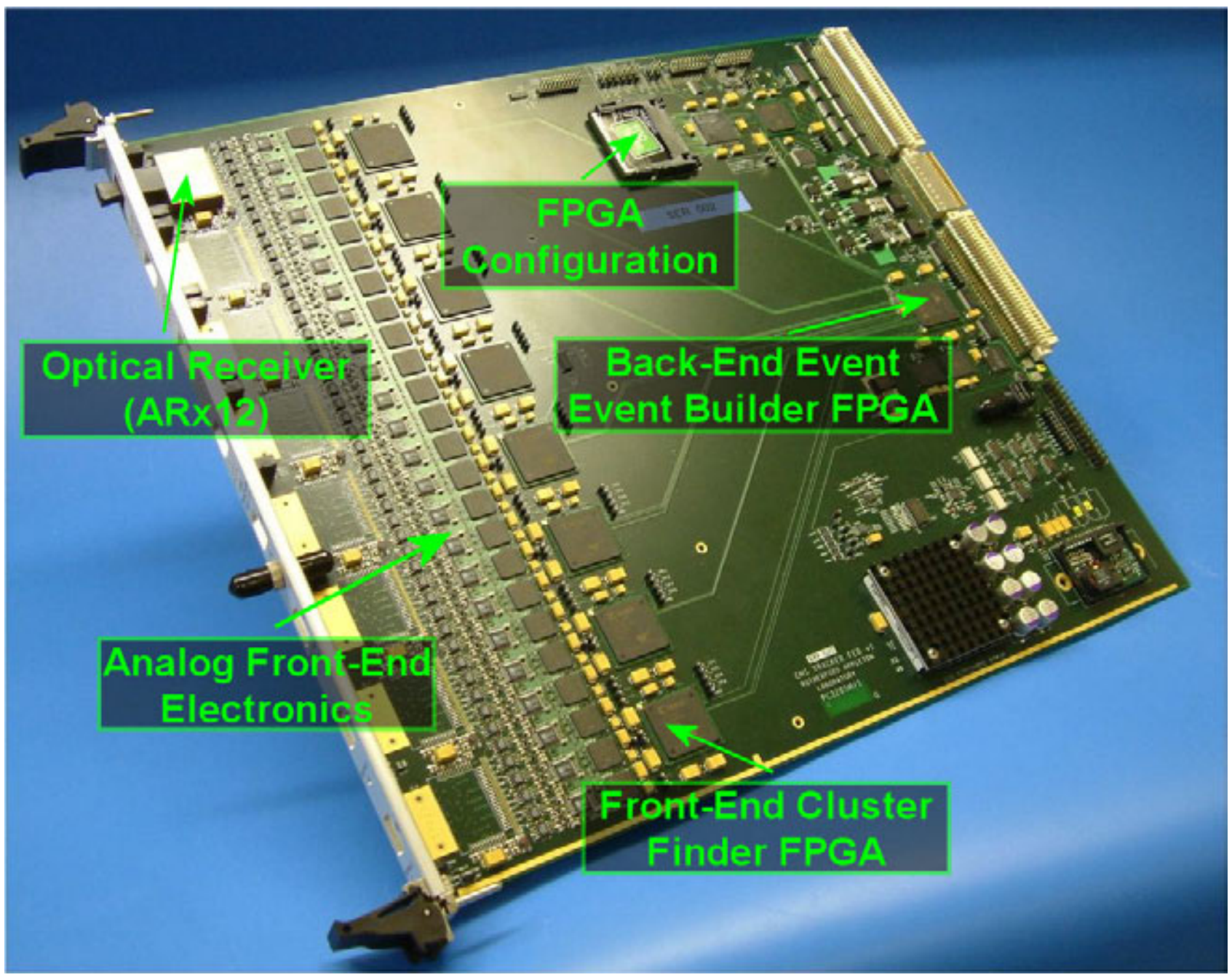

**Figure 2.4:** The VME FED (only one out of eight ARx12s is attached on the board shown) [9].

Optical data received from the detector are converted to single-ended electrical signals by the ARx12s on the FED. The electrical signals are matched to the input of a 10-bit, differential input Analog to Digital Converter (ADC) by the analog front-end electronics. The digitized data is then processed by a Field Programmable Gate Array (FPGA) in one of the eight front end modules. In normal operation, the front end FPGA performs zero suppression, in order to reduce the amount of data sent to the DAQ system. Its operation includes pedestal subtraction, channel re-ordering, common-mode noise rejection, and clustering. The front end FPGA only retains information regarding particle signals. The data from all eight front end modules is collected at the back end FPGA. A header containing bunch crossing and trigger numbers is added to the data, which are then sent out to the Data Acquisition (DAQ) system.

### 2.1.2 The Optical Links

Optical fiber is an ideal medium for such a transmission system, due to its immunity to electromagnetic interference (EMI), potential for low power dissipation and low mass. The sheer number of channels required to be read out in the CMS Tracker make the use of copper transmission prohibitive.

Analog links were preferred in order to eliminate the need for digitization on the detector hybrids (i.e. inside CMS), thereby simplifying the front-end electronics and lowering powers consumption inside the Tracker volume. The motivation for a low-power design is provided by the potential increase in overall performance. As stated in Chapter 1, the amount of material in the Tracker must be kept at a minimum for best performance. The material budget is strongly influenced by the front-end electronic functions since the topology of the detector hybrid is less dependent on chip design but more by the number of interconnections and cooling requirements. Since all power sources must be cooled, it follows that lower power in the front-end results in less material in services (e.g. cooling pipes), and hence increased physics performance.

In addition, charge sharing between silicon microstrips will be frequent due to non-normal incidence of particles, magnetic field effects and energy loss variations. An analog system can therefore achieve superior positional resolution compared to binary digital readout.

Another advantage of analog readout is the expected robustness of the system in the LHC environment. Systems are easier to debug and operate if pulse height information is available. Analog data will give greater ability to identify problems and to apply online corrections. Degradation due to radiation damage both to detector signals and electronics, as well as minor damage such as connector failures may occur. Pulse shapes, amplitudes, occupancies and radiation damage effects can be readily monitored. Electronics such as those required for digitization and pedestal subtraction will be located off-detector, far from the radiation zone. They will then be easily accessible and this should provide greater reliability than locating them within the Tracker volume. In addition, commercial components can be used directly, without the need for customization for radiation hardness.

The on-detector elements (lasers and photodiodes) of the ~40 000 readout and ~2 500 control links will be distributed throughout the detector volume in close proximity to the silicon detector elements. This places strict requirements on small package size, mass, power dissipation, immunity to electromagnetic interference and radiation hardness. These requirements have been met with the extensive use of commercially available components with a minimum of customization.

The transmission scheme employed is analog Pulse Amplitude Modulation (PAM) at a sample rate of 40MHz. The Signal to Noise Ratio (SNR) of the link is specified over a bandwidth of 100MHz to have an equivalent digital resolution of at least 8 bits. Hence, the analog modulation scheme can be thought of being comparable to a digital baseband PAM system using 256 distinct levels (8 bits) at a symbol rate of 40MSymbols/s. The equivalent digital data rate is therefore 320Mbit/s (=8×40MHz).

The analog optical readout links operate single-mode at 1310nm wavelength. The Linear Laser Driver (LLD) ASIC [10] on the AOH directly modulates the edge-emitting laser diode drive current to achieve light amplitude modulation. Single fibers from the pigtailed lasers are connected via small form-factor MU-type single-way connectors to a fan-in, which merges single fibers into a 12-fiber ribbon (Distributed Patch Panel in Figure 2.1). At a second break-point within the CMS Detector, the transition to a rugged multi-ribbon cable (8×12-fiber ribbons/cable) is made via 12-channel MFS-type array connectors (In-Line Patch

Panel). In the counting room each ribbon connects directly to a 12-channel analog optical receiver (ARx12) module on the FED, using MPO12 connectors. The total length of each readout link is ~65m, of which ~10m is within the high radiation environment.

### 2.1.3 Main Link Performance Specifications

As described earlier, the signals produced by the detectors are sampled and pulse-shaped at 40MHz by the APVs and the data is then sent in analog form over the optical links. Hence, at the output of the link, the analog waveform essentially consists of 25ns pulses whose height is proportional to the signal coming from the corresponding detector microstrips. In the counting room, the readout link's output is sampled at 40Mhz and then digitized to produce discrete values. A synchronous system based on sampling pulses has been preferred over reading out continuous data values from the detectors, since it eases implementation complexity, only requiring a simple CR-RC shaper in the front-end.

It follows that synchronization is an important factor in this system, since the sampling of the signal at the output must be made as close as possible to the final settled value of each analog pulse[1]. In addition, the rise times and settling times of the readout links must be small enough to allow sampling at the final value. These are roughly related to the bandwidth of the links, which is required by the specifications [11] to be a minimum of 70MHz for all optical links. Moreover, the maximum settling time specification (to within 1% of the final value) is 20ns. This ensures that there is (at least) a 5ns window for which each pulse can be sampled at its final value and hence minimize the amplitude error. This is possible since the sampling point at the back-end (i.e. on the FED) can be time-shifted in ~1ns steps, providing sufficient granularity for synchronization within the 5ns window.

Pulse Amplitude Modulation (PAM) is used in the readout system, and the required equivalent digital dynamic range is 8bits. The system must be able to differentiate between 256 levels, and the peak Signal to Noise Ratio (SNR) has therefore been specified at 48dB. Since the link's input range is 600mV differential (or ±300mV) the required SNR translates to a maximum noise specification of 2.4mV, referred to the input. The amplitude resolution of the system has been chosen to be easily sufficient for the range of signals expected and the performance goals outlined previously (i.e. excellent positional resolution by exploitation of charge sharing).

[1] The 'final value' is at the flattop of each (square) pulse.

Gain is an important parameter in an analog readout system, since it determines both the signal sizes that can be viewed at the output, as well as the resolution of the system. Since the optical link consists of analog components whose individual gains will exhibit a spread in values, the overall link gain will also vary throughout the 40 000 deployed CMS Tracker links. To equalize the link gains, the AOH was designed with four different gain settings (see section 2.2.1). The target gain during normal operation is 0.8V/V, which corresponds to being able to view signals as high as 3.2MIPs [2] in 256 FED ADC counts (8 bits). The specified range of gains is 0.3-1.3V/V, though it will be shown in Chapter 4 that the real spread in the system will be much smaller (0.64-0.96V/V).

Finally, the readout links are required to have a linear transfer characteristic over the operating range of 600mV differential. To this end, the Integral Non-Linearity (INL) is specified to be (typically) 1%. Linearity is crucial to the operation of the readout system, given that analog PAM is employed. In order to distinguish between 256 levels at the output, signals should be linearly amplified across the entire operating range of the link. Non-linearities cause distortion which cannot be compensated for. For example, considering that pedestal and common-mode subtraction is performed off-detector (and therefore *after* transmission through the readout link), the analog baseline of each APV frame (see Figure 2.3) will vary throughout the Tracker detectors modules. Hence a non-linear transfer characteristic would mean that a 1MIP signal would produce a different size output, based on which module the signal was coming from.

### 2.1.4 Environmental Effects on Optical Link Components

The radiation level in the Tracker is expected to reach $3.4 \cdot 10^{14}$ $cm^{-2}$ particles and 150kGy dose, over the nominal 10-year lifetime [2]. This operating environment is not only extreme by comparison to any standard communications application, but also unprecedented for high energy physics experiments, due to the LHC having a much larger energy and luminosity than any particle accelerator ever built. Component reliability is therefore an issue, since many components inside the Tracker will not be accessible for repair and maintenance once installed in CMS.

[2] For thin (320μm) silicon detectors.

In addition to the radiation effects, the Tracker will operate inside a 4T magnetic field, as well as a temperature of -10°C. This operating temperature has been chosen [2] to combat radiation-induced bulk damage in the silicon detectors. The behavior of silicon detectors after irradiation is a major concern. 'Reverse annealing' effects dominate the bulk damage, increasing both the leakage current and depletion voltage after irradiation has stopped. These effects can be greatly reduced by operating and maintaining the detector at low temperature throughout its lifetime.

The cold temperature should not be a problem as far as the optical links are concerned, since it is within the standard telecoms operating range. It will, however, generally cause an increase in gain for both electronic and optical components. The presence of the magnetic field means that all front-end components must be non-magnetic. No degradation in performance from the 4T field is expected [12].

The AOH is placed inside the Tracker volume and hence must endure the harsh radiation environment. The same is true of the connectors, and length of fiber inside the radiation zone (see Figure 2.1 for positions of components). Commercial Off-The-Shelf (COTS) optical fiber and connectors have been irradiated as part of the quality assurance program during system development [13]. The tests showed that there was no radiation induced degradation in the return loss of connectors, while a slight increase in attenuation was observed in the fibers. Single-mode fibers showed induced losses of only 0.04-0.06dB/m after irradiation to 100kGy at a dose rate of ~720Gy/hr. With only ~10m of fiber in the high-radiation zone, the degradation in attenuation of ~0.4-0.6dB. This is comparable to the connectors' re-mating uncertainty, and is therefore acceptably low.

The InGaAsP laser transmitters used in the AOH have also been comprehensively tested for radiation hardness [14]. Accounting for annealing and the expected particle fluences which depend on the performance of the LHC over a 10-year time period, the laser threshold is expected to increase by ~5-6mA in addition to a 5% drop in efficiency. Hence the AOH's Linear Laser Driver (LLD) ASIC has been designed with a programmable DC bias current in order to track laser threshold changes during operation, and thus provide protection against this failure mode.

## 2.2 The Analog OptoHybrid

The AOH converts the electrical signals it receives from the APVMUX chip to optical for transmission over fiber to the counting room. The AOH consists of an LLD ASIC and 2 or 3 laser diodes pigtailed with single-mode fiber that is terminated using MU connectors. Tracker modules consisting of double-sided sensors employ 3-laser AOHs, while those consisting of single-sided sensors contain 2-laser AOHs. Figure 2.5 is a block diagram of the device.

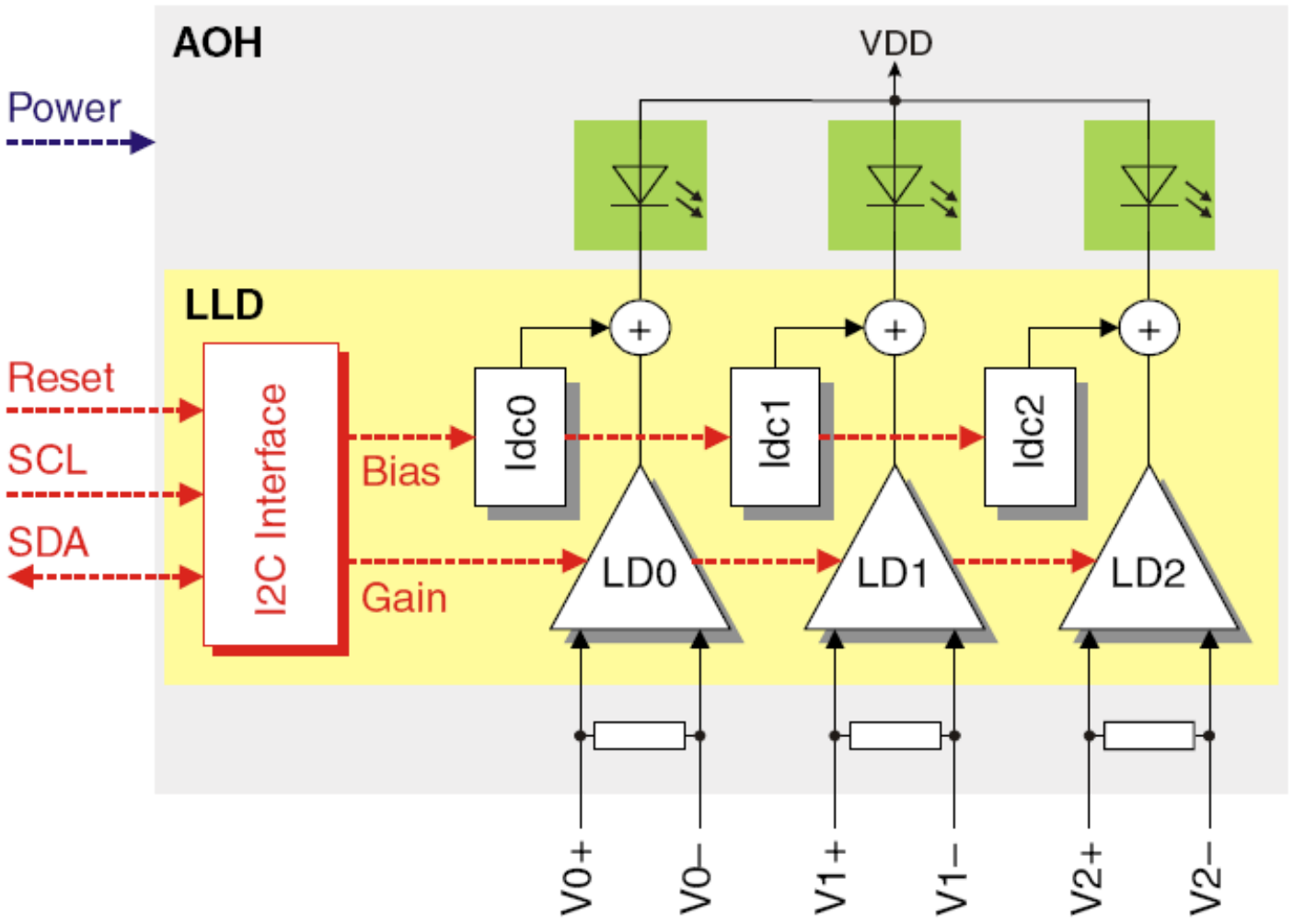

**Figure 2.5:** Block diagram of the AOH from [15], also detailing the LLD chip. In 2-laser AOHs, the middle channel of the LLD is not used.

The laser driver chip provides currents to each laser (via three independent channels) in order to bias them at their quiescent operating points. Direct modulation of each laser is achieved by supplying a current proportional to the input current of the corresponding LLD channel (analog Pulse Amplitude Modulation). The input voltage signals are differential, in response to which the LLD adds and subtracts current from the output signal, which is centered on the programmable laser bias current.

The lasers generate the required optical power for transmission of the signals to the optoelectronic receiver. In order to compensate for irradiation induced damage and component performance spread, the AOH includes $I^2C$ control links for adjustment of the laser bias current and the gain of the laser driver.

### 2.2.1 Linear Laser Driver

The LLD ASIC was developed using radiation-tolerant 0.25μm CMOS technology, in accordance to the requirements of the CMS Tracker environment. Radiation and accelerated ageing tests have shown that the device will operate within its specifications for the expected lifetime of CMS (10 years) [10]. A relatively small increase in the LLD bias current is expected (less than 15%), as shown in [16].

The LLD chip contains three independent channels, each consisting of a linear driver and a laser diode bias generator. A differential input voltage is converted into a single-ended, linearly amplified output current that is added to a programmable DC bias current. The bias current ranges from 0 to ~55mA, and is user-selectable in 128 linear steps through an $I^2C$ control interface.

The laser transmitters typically have a threshold of ~3mA at the nominal Tracker operating temperature (-10°C) [17]. However, the detector signals that are input to the AOH are differential, and modulate the laser by subtracting and adding current to the bias current supplied by the LLD. It follows that the actual operating point is chosen to be higher than the laser threshold current, to accommodate the full negative swing of the input signals (see Chapter 3, section 3.1.3 for more details on the functionality). The actual operating bias points will therefore range between ~4.5 and 10mA (equivalent to ~10-21 $I^2C$ steps for the LLD chip). As stated in section 2.1.4, the expected laser threshold increase during the experiment's lifetime will be ~5-6mA. Hence, at most, the operating bias point of any given link will be under 20mA, even after irradiation. This is well within reach of the LLD bias current range (up to ~55mA), and radiation-induced threshold increase can be compensated for.

Adjustment of the gain of each channel is also possible, with four gain settings available (nominally 5.0, 7.5, 10.0, 12.5mS)[3]. This allows for equalization of the gains of the ~40 000 Tracker readout links, close to the specified gain of the optical links. In addition it allows for compensation against the variation in the gain of a given link, caused by temperature changes and radiation damage. From section 2.1.4, the expected laser efficiency drop due to radiation damage over the lifetime of the experiment will be ~5%. Since the LLD gain settings allow increasing the gain up to 250% (referred to the lowest setting), there is clearly sufficient range for compensation of radiation damage. Immunity of the control interface to Single Event Upsets (SEUs) is achieved through the use of triple redundant digital logic and a majority voting decision scheme.

[3] These are also referred to in this thesis as Gain 0, Gain 1, Gain 2 and Gain 3 respectively.

### LLD PERFORMANCE

The device achieves linear operation over an 8-bit dynamic range, in accordance to the specification of the whole readout chain. The LLD's Integral Non-Linearity (INL) is better than 0.5% over an input range of ±300mV. Power dissipation is 10mW/channel at the lowest bias and gain settings, and can reach up to a maximum 110mW/channel.

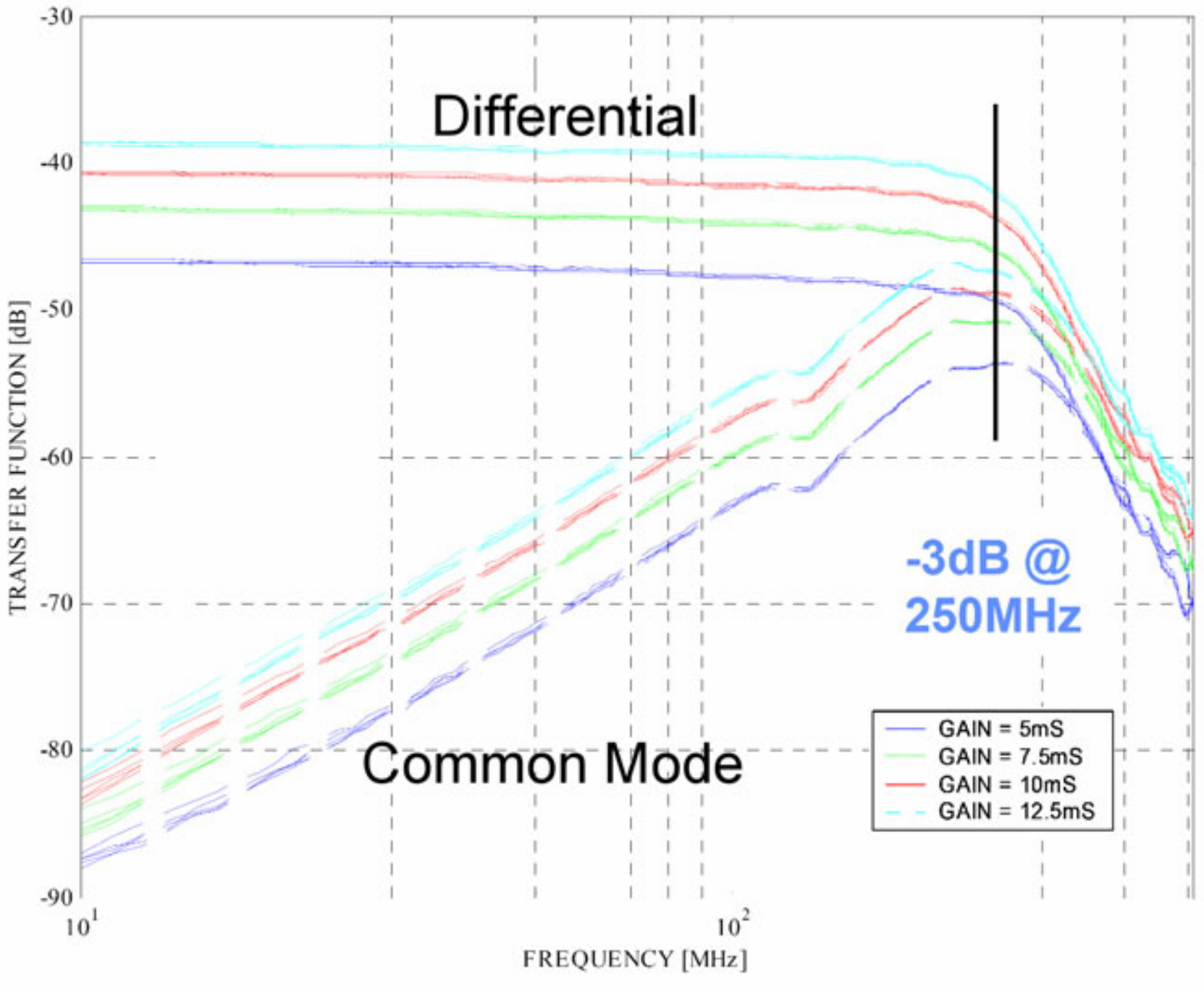

**Figure 2.6:** Frequency responses (differential and common-mode) of several LLDs for all four gain settings. Plot taken from [16].

Dynamic performance tests have also been performed on the LLD [10]. Pulse response measurements have shown the rise and fall times to be below 2.5ns, with settling times[4] around 12-15ns. This is in concert with the requirements of the readout link, allowing 13-15ns for correctly sampling the output (since the width of each signal sample is 25ns). This is in excess of the 20ns settling time specification for the entire readout link. The frequency response of the LLD is illustrated in Figure 2.6, showing a 3-dB bandwidth of ~250MHz. The equivalent input noise into this bandwidth is gain and laser bias dependent, but has been measured to be less than 1mV rms in all cases [10], corresponding to a peak SNR of over ~55dB. Hence the LLD performs comfortably within the 48dB SNR requirement for the optical link.

[4] Unless otherwise stated, in this document the settling time is the time it takes the output signal to settle to within 1% of its final value.

### 2.2.2 Laser Transmitter

The electrical signals produced by the LLD are converted to optical by the laser transmitter. It consists of an edge-emitting, Multi Quantum Well (MQW), InGaAsP laser diode [18] that emits light into a single-mode optical fiber pigtail. Edge emitting lasers were chosen for the analog readout optical links due to their good thermal management, highly linear transfer characteristics, high output powers and low threshold currents [19]. Low threshold current implies lower power is dissipated in the front-end for biasing the lasers at their operating point. This is crucial to the low-power design of the Tracker which, as explained earlier in this chapter, can potentially increase performance. The lasers type used is an established technology with proven reliability and laser-fiber coupling techniques.

Commercial packaged lasers are too bulky for use inside the detector volume, and hence a semi-custom package was developed for the lasers in the Tracker optical links. The semi-custom package consists of the laser die attached to a silicon optical sub-assembly and fiber-pigtail, with a simple cover. This bare assembly minimizes the links' contribution to the material budget [2] of the CMS Tracker, where it is critical to have as little dead material as possible, as this degrades the performance of the detector.

The laser diodes and fiber operate single-mode at 1310nm wavelength, hence avoiding modal noise associated with multimode fiber systems. While coupling light into a single-mode fiber can be much more difficult than into a multimode fiber, it was estimated in [19] that this would only incur a 10% cost penalty, given the system's modest power requirements. Moreover, edge-emitting lasers pigtailed with single-mode fiber are a standard in the telecoms industry, thus ensuring development and purchasing costs were kept to a minimum.

### LASER TRANSMITTER PERFORMANCE

Typical transfer characteristics of the laser transmitters in use by the CMS Tracker are shown in Figure 2.7 [20]. The linearity is extracted from the transfer characteristic by fitting a straight line over the input range and plotting the residual referred to the input (this is the Integral Non-Linearity (INL) shown in Figure 2.7). In all cases shown, INL is better than 1% for a range of ~20mA. To put this in perspective, at the typical LLD gain setting (5.0mS or Gain 1), the link's specified linear input range of 600mV corresponds to ~4.5mA at the input of the laser transmitter. Even at the highest gain setting (12.5mS or Gain 3), the linear range required of the link corresponds to ~7.5mA. Clearly, the linearity of the laser transmitters far exceeds the specification of the link.

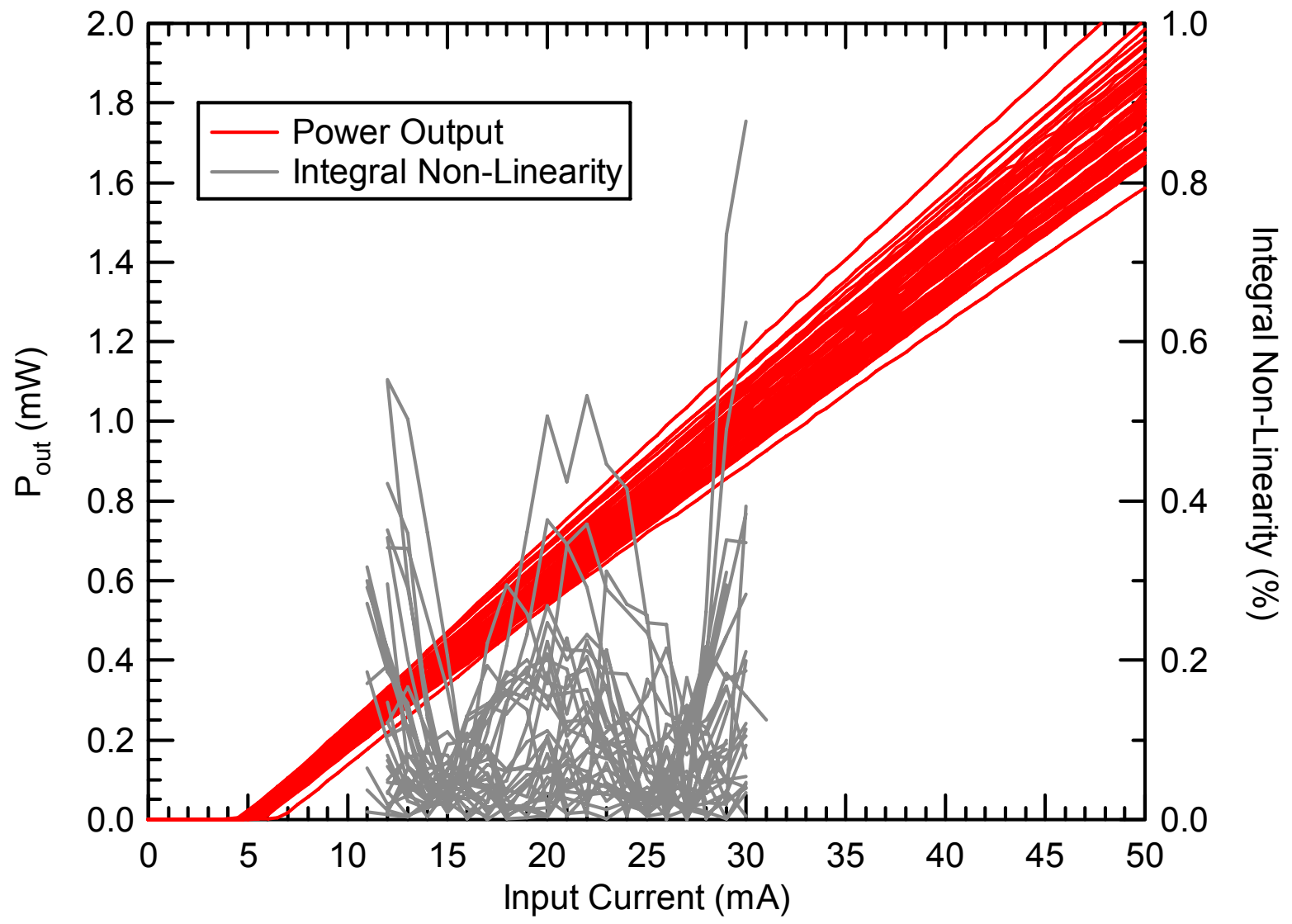

**Figure 2.7:** Showing typical transfer characteristics and Integral Non-Linearity, obtained from production testing of laser transmitters.

The bandwidth of the lasers is in the GHz range. This leads to very low rise times, with all laser transmitters below 0.5ns. This is illustrated in Figure 2.8, which shows a typical spread in rise time values obtained during production testing [20].

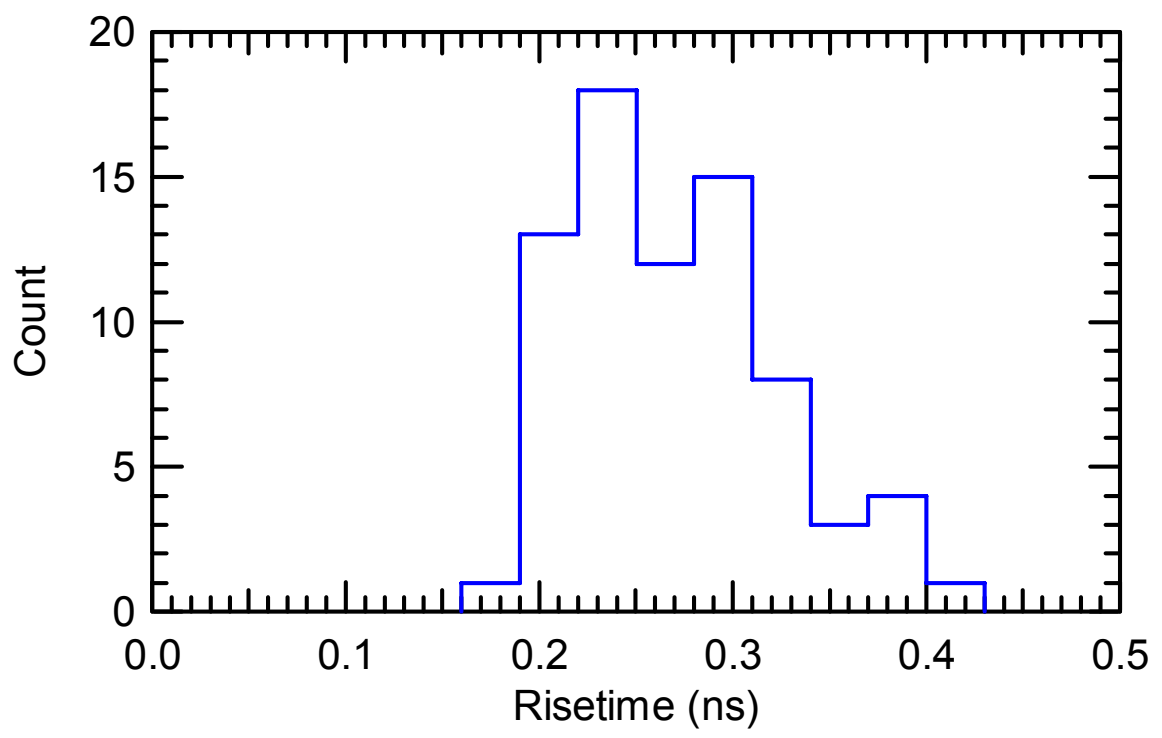

**Figure 2.8:** Typical histogram of laser transmitter rise time from production tests.

During production testing, the laser Relative Intensity Noise (RIN) is extracted from noise measurements made with a CMS-type optical receiver at the output. The distribution of RIN for several lasers belonging to the same production batch is shown in Figure 2.9, where the noise of the optical receiver has been accounted for. In this calculation, with a conservative estimate of the effective bandwidth of the laser and receiver (80MHz), most devices perform better than the typical specification of -130dB/Hz. The typical specification corresponds to 1.6mV equivalent link input noise, for this bandwidth.

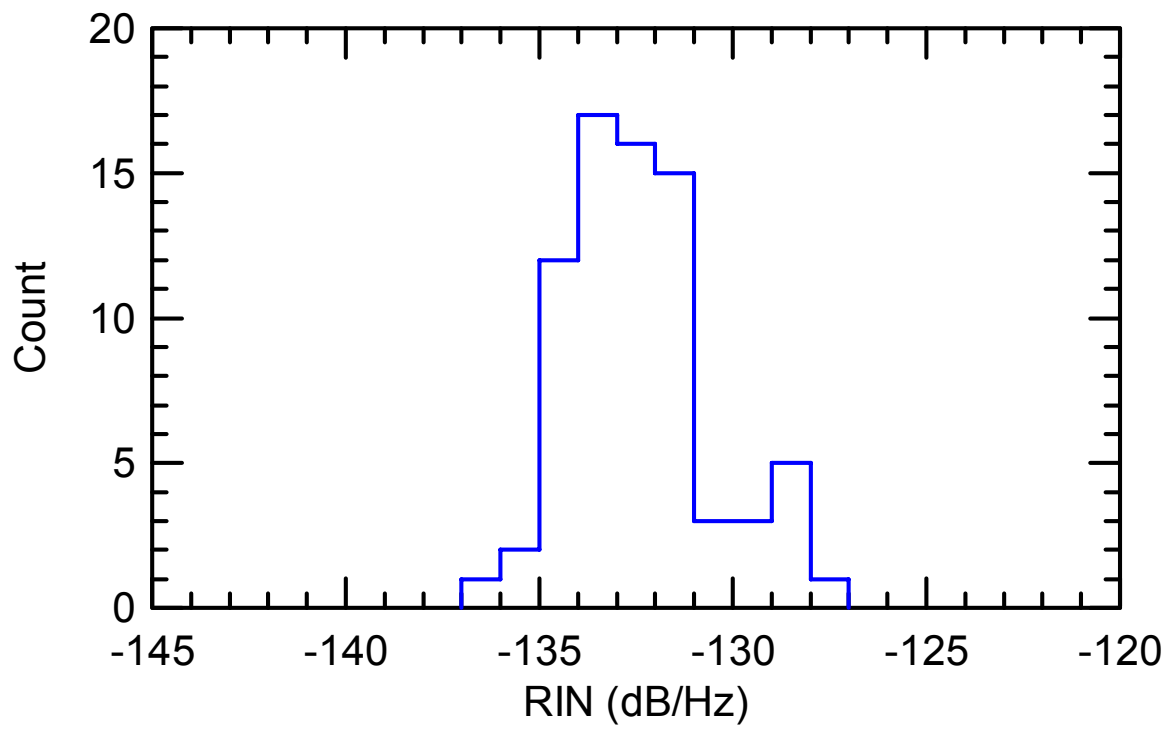

**Figure 2.9:** Typical histogram of laser RIN from production tests.

## 2.3 The 12-Channel Analog Optoelectronic Receiver

The 12-Channel Analog Optoelectronic Receiver (ARx12) is based on a modified commercial 12-channel module [8], where the digital amplifier ASIC has been replaced by a custom-designed analog variant, designed by Helix [21]. In addition to the amplifier ASIC, the analog receivers contain 12-channel p-i-n photodiode (PD) arrays housed in compact modules. Module production is carried out by NGK [22]. The ARx12 conforms to the overall CMS Tracker readout link specification, featuring an integral linearity deviation of 1% over an 8-bit dynamic range. The 3-dB bandwidth is typically 100MHz. The settling time specification for the ARx12 is 15ns, to allow some margin from the 20ns specification of the overall link.

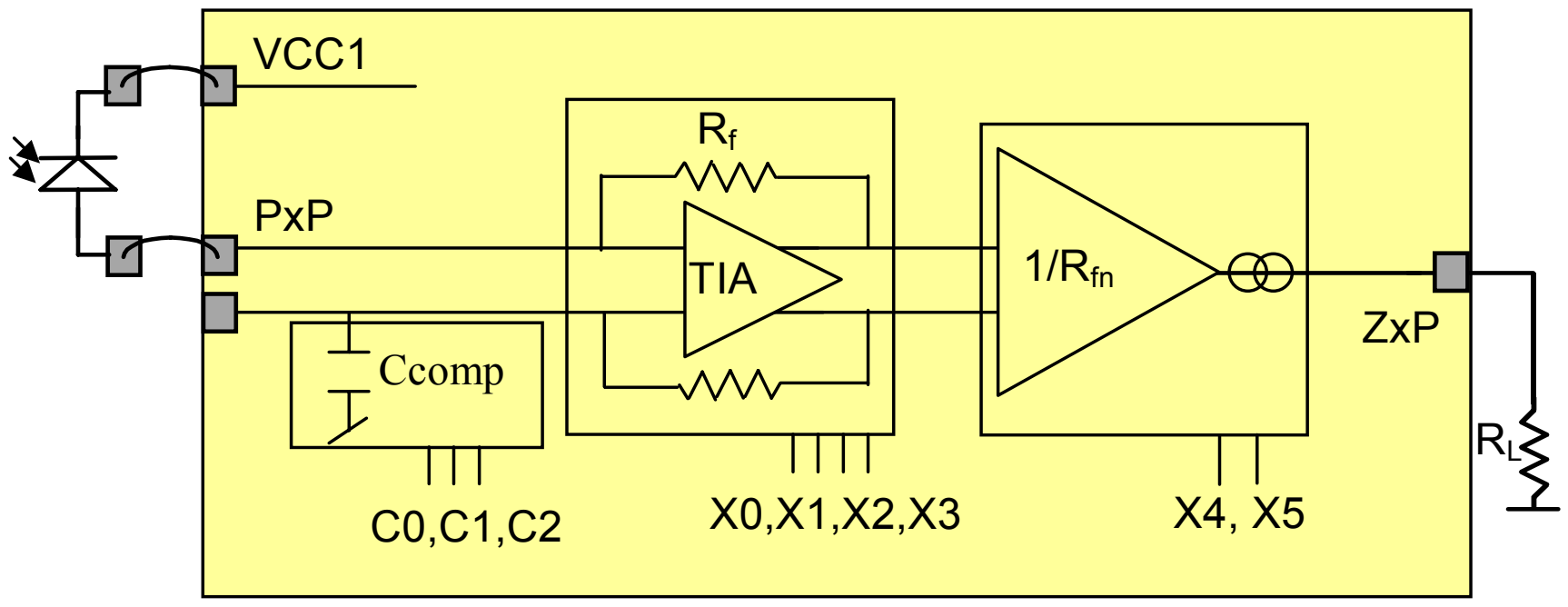

**Figure 2.10:** Block diagram of a single channel of the analog receiver ASIC. Diagram taken from [8].

The receiver ASIC is a low-noise 12-channel photoreceiver. The analog photoreceiver channel consists of a transimpedance stage followed by a transconductance stage (Figure 2.10). The photocurrent is first converted to voltage by the transimpedance amplifier. The feedback resistor ($R_f$) is subject to process variations of approximately ±25%, which causes the gain of the signal to also vary by 25%. The transconductance stage that follows converts the voltage back to a current. It also compensates for resistor process variations, having a transfer function $1/R_{fn}$. Since both $R_f$ and $R_{fn}$ are subject to the same relative process fluctuations, the net effect on the overall gain is cancelled out. The output current is then converted to the desired output voltage with an external load resistor $R_L$. A precision resistor (1% tolerance) is used, with the specified range of values being 50-200Ω. The simplified transfer function is as follows:

$$\Delta V_{ZxP} = R_L \cdot \frac{R_f}{R_{fn}} \cdot \Delta I_{ph} \qquad (2.1)$$

*Where $\Delta I_{ph}$ is the photo current injected by the photodiode into the ASIC.*

As can be seen in Figure 2.10, the transimpedance amplifier is fully differential, with one input connected to a photodiode and the other to an on-chip capacitance, $C_{ARx}$. The receiver channel has been designed to fulfil the settling time specification of 15 ns. Typically, this corresponds to a bandwidth of approximately 100MHz. Even if the bandwidth specification has been met, a slight mismatch between the photodiode capacitance ($C_{pd}$) and $C_{ARx}$ can result in extremely slow settling times [23]. To compensate for the slight difference between the two capacitances, the ASIC contains programmable capacitances with the pins $C_0$, $C_1$, and $C_2$. The compensation capacitance $C_{ARx}$ consists of fixed and programmable portions. The compensation-capacitance value is given by[5]:

$$C_{ARx} = 800fF + C_0 \cdot 100fF + C_1 \cdot 200fF + C_2 \cdot 400fF \qquad (2.2)$$

The transimpedance stage has been designed to be linear for input signals of up to 300μA. This requires the full range of the differential input stage to be used, which in turn implies that the DC level of the photocurrent must be centered on the input. A programmable photocurrent offset $I_{phoff}$ (settings X0 to X3) is used to optimize the DC level and ensure full-range linear amplification. The output level of the transconductance stage (and therefore of the ASIC) can be adjusted using a programmable offset current. Pins X4 and X5 are used for this purpose.

### 2.3.1 ARx12 Performance

STATIC PERFORMANCE

The devices are tested in a system that includes 12 laser transmitters, one for each channel of the optoelectronic receiver. Typical transfer characteristics of optoelectronic receiver channels, obtained from production tests [24], are shown in Figure 2.11. The Equivalent Input Non-Linearity (EINL) is also shown, in mV. The 1% specification, corresponding to an EINL of 6mV, is indicated by the dotted line. All devices exhibit an input linear range exceeding the 600mV required by the specifications.

The overall optical link noise specification is 2.4mV, referred to the input (which corresponds to a peak SNR of 48dB). Figure 2.12 shows the noise measurement made during ARx12 production tests, including the noise of the transmitter in the test setup, which varies as a function of input laser bias current (or, equivalently, link input voltage as shown in Figure 2.12).

[5] In the final production module, only $C_1$ and $C_2$ are connected to output pins while $C_0$ is set to 0.

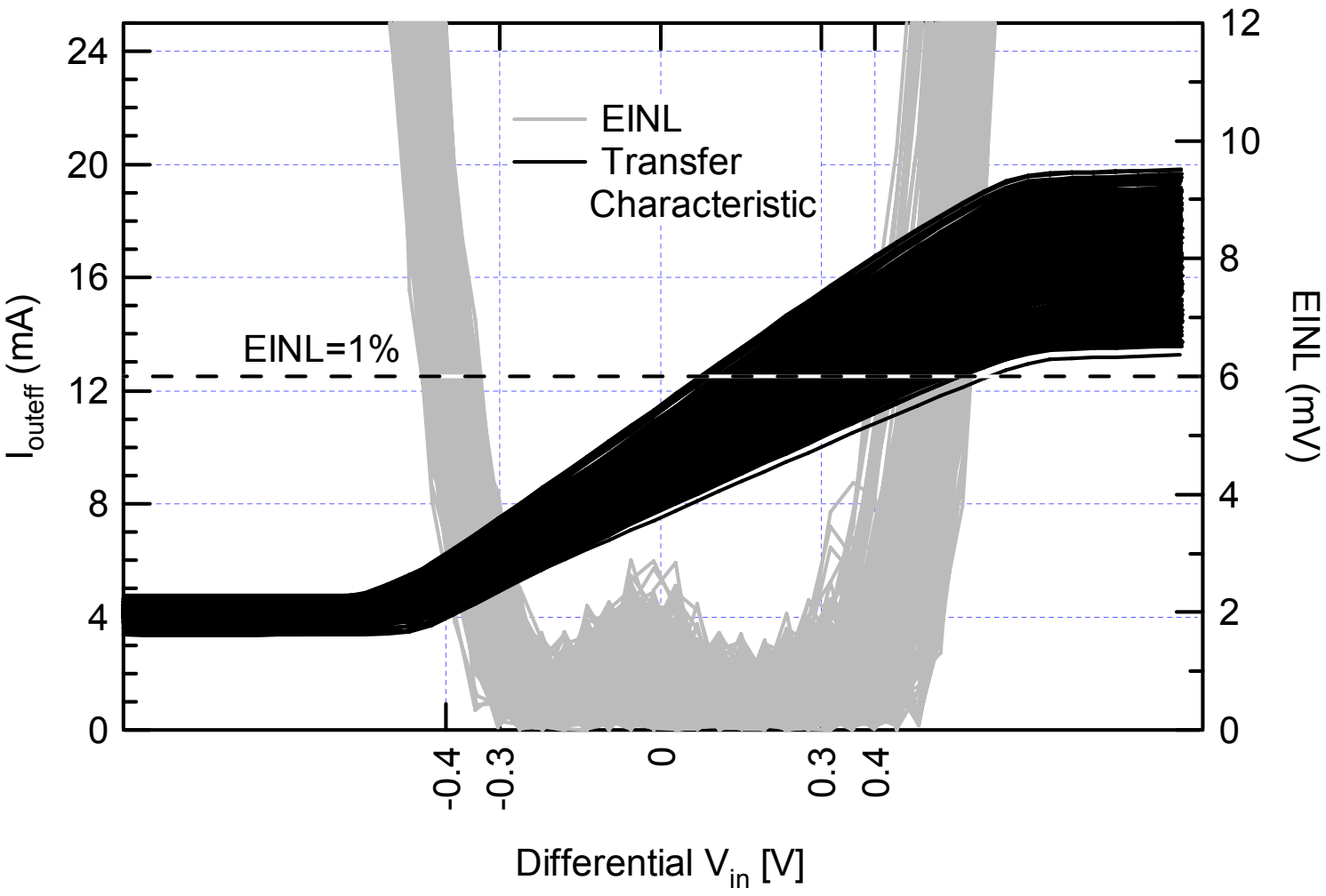

**Figure 2.11:** Typical transfer characteristics of ARx12 devices from the same production test batch. Corresponding Equivalent Input Non-Linearity (EINL) is also shown (in gray). The 1% EINL is indicated by the dotted line.

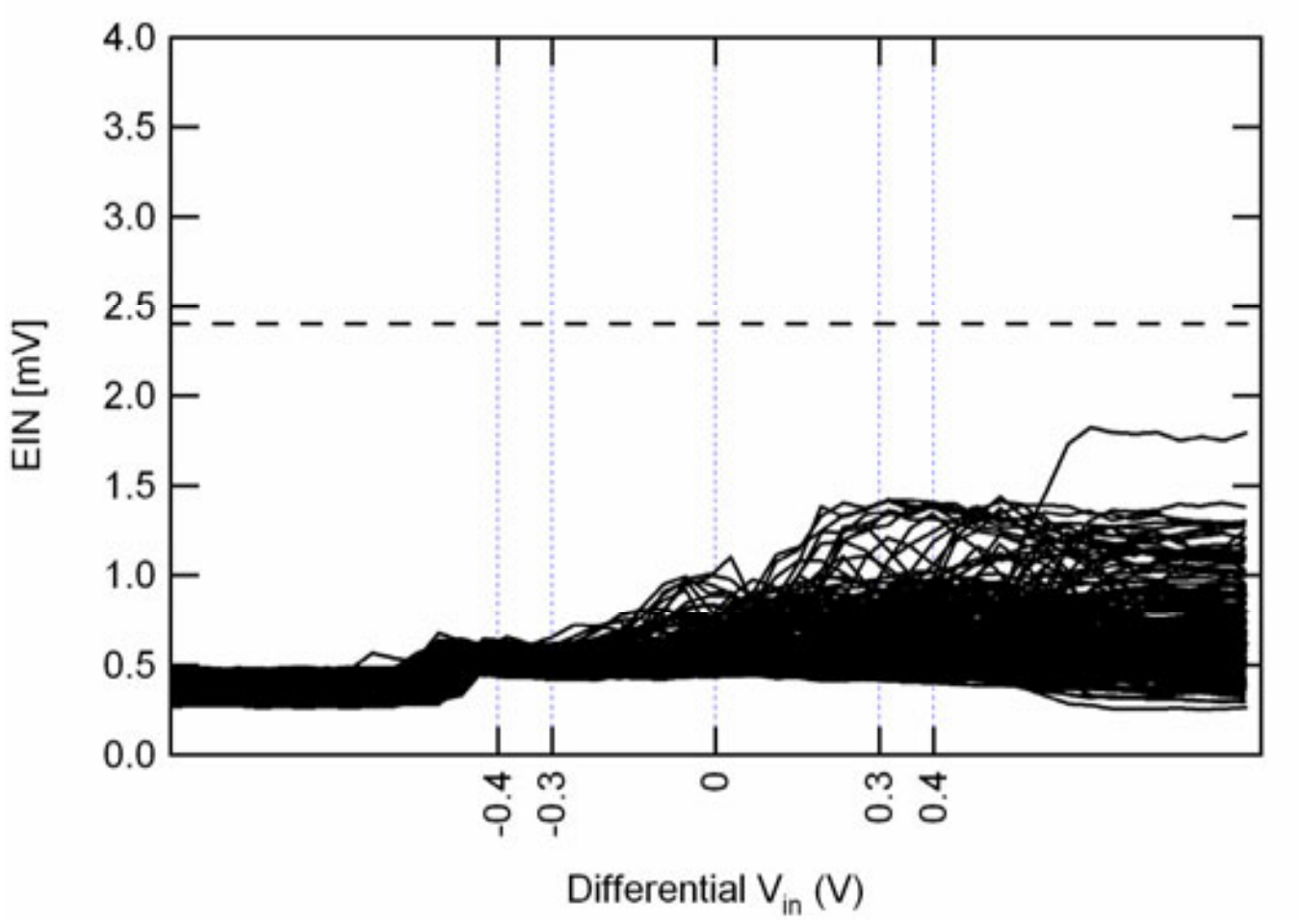

**Figure 2.12:** Typical equivalent input noise plots for ARx12 devices belonging to the same production test batch. The dotted line indicates the limit of the specification.

DYNAMIC PERFORMANCE

Figure 2.13 shows typical pulse responses obtained from production tests, using the different capacitance settings (see section 0). All channels of the 20 modules belonging to the same example production batch are displayed. From the pulses, the rise times and settling times have been extracted (Figure 2.14). All channels of all tested modules are plotted against the different capacitance values that can be selected using the control settings $C_1$ and $C_2$ (see Table 2.1 for the relationship between switches and capacitor value).

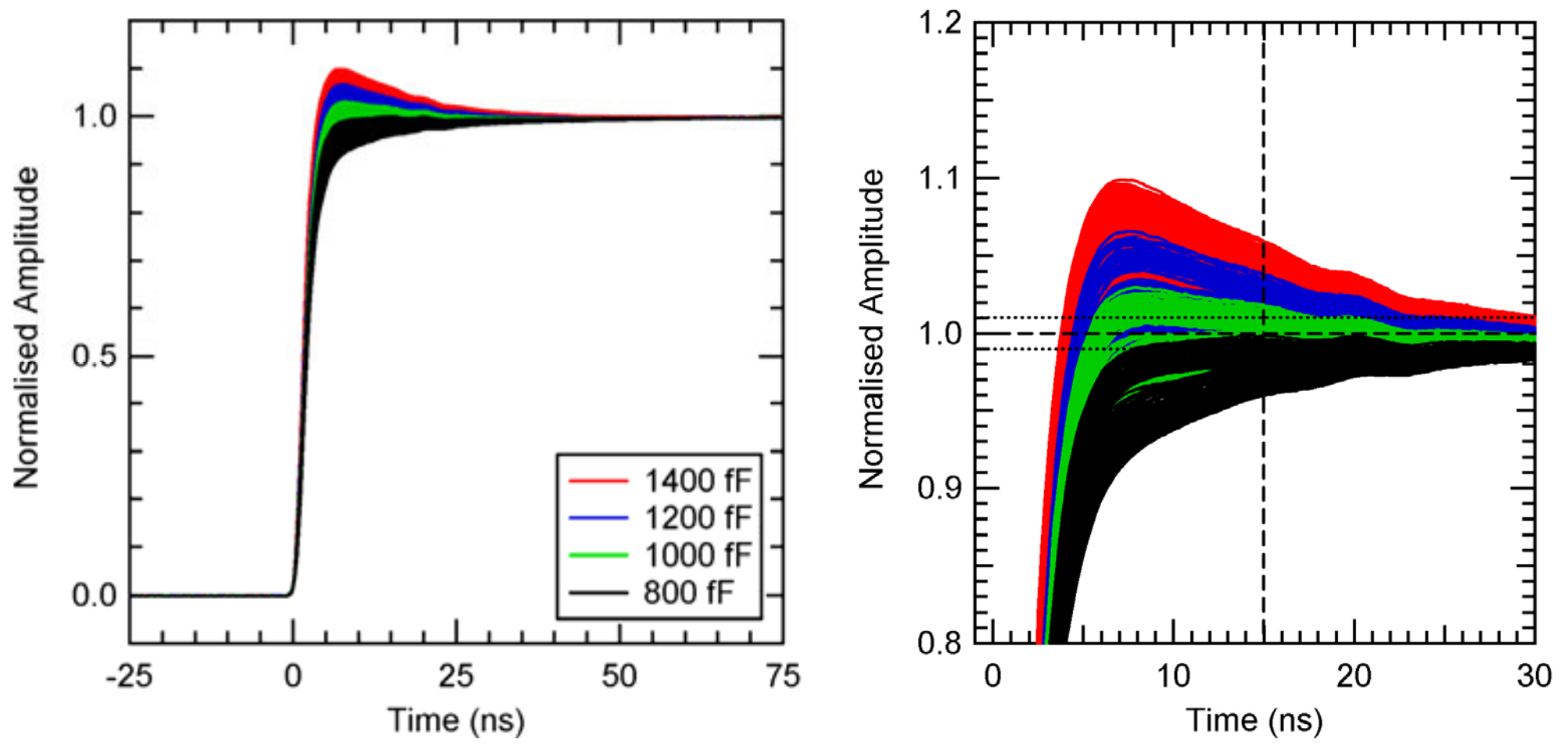

**Figure 2.13:** Typical pulse responses for all measured channels of ARx12 devices belonging to the same production test batch. The results are shown for all different capacitance settings.

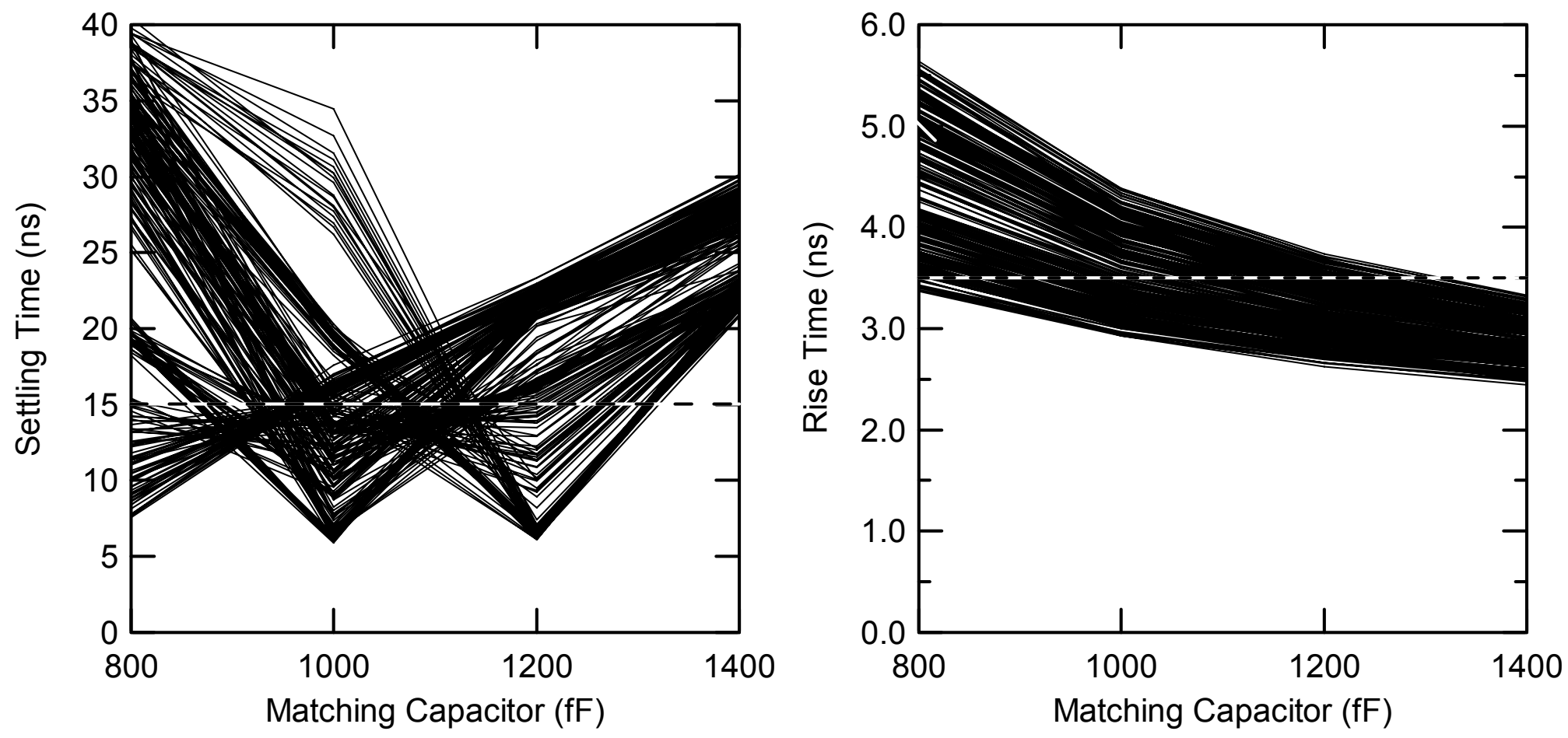

**Figure 2.14:** Settling times (left) and rise times (right) for the pulses of Figure 2.13.

**Table 2.1:** Relationship between *C*-switch settings and matching capacitor value.

| Switch $C_1$ | Switch $C_2$ | Matching Capacitor value (fF) |
|---|---|---|
| OFF | OFF | 1400 |
| ON | OFF | 1200 |
| OFF | ON | 1000 |
| ON | ON | 800 |

The settling time plot in Figure 2.14 exhibits three dominant troughs, indicating that there is no single optimum capacitance setting for all channels. It has been shown [24] that in a given ARx12 module, the channels exhibit similar (but not identical) pulse responses for each of the *C*-settings. Hence, one can select the module's 'optimum' *C*-setting[6] with the algorithm described next.

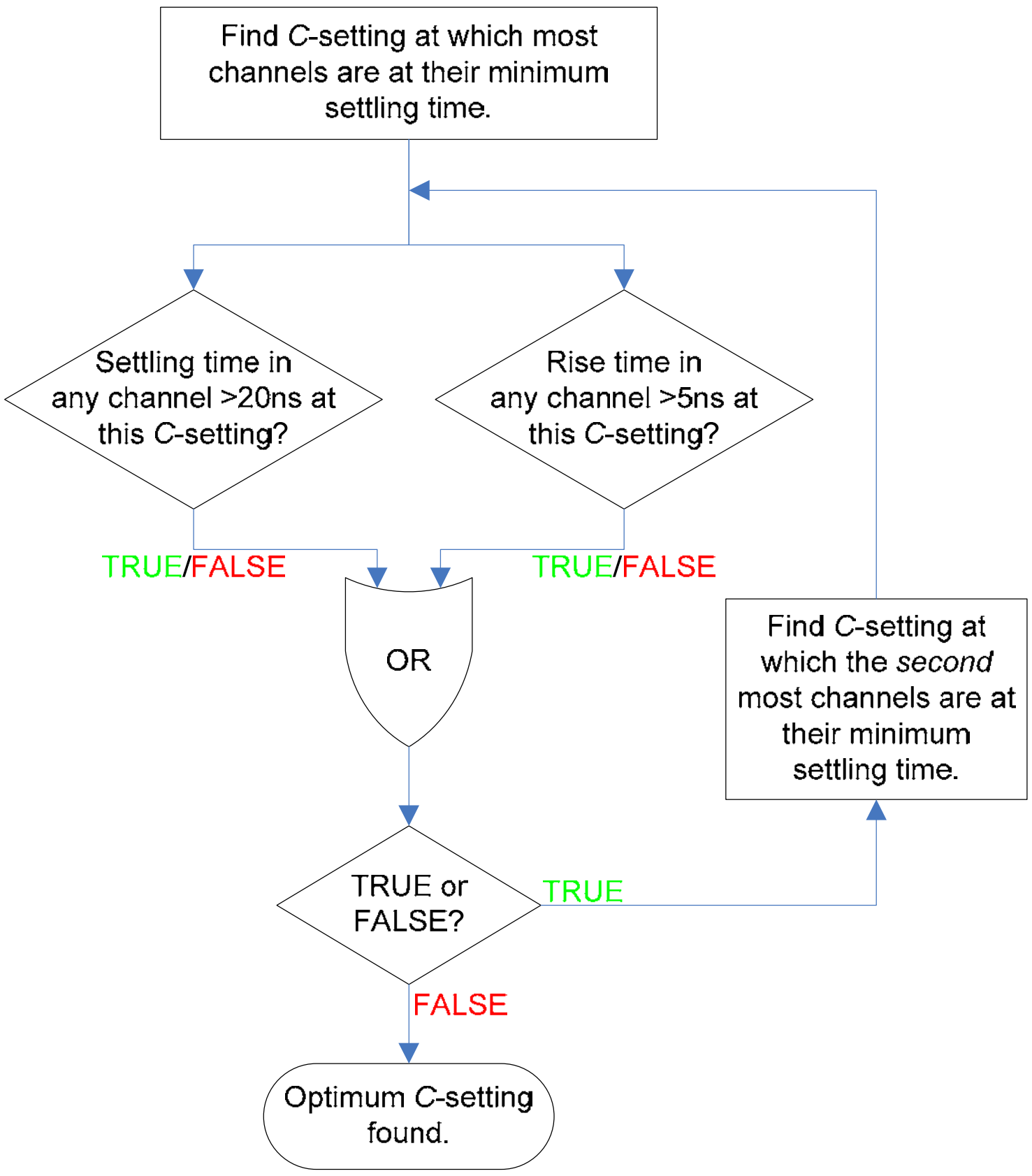

**Figure 2.15:** Flow chart of the algorithm used to determine the optimum capacitance setting of ARx12 modules.

The appropriate matching capacitance can be found from the settling time plots of Figure 2.14, using an algorithm (illustrated in Figure 2.15) that operates on a module by module basis. The algorithm emphasizes the performance in terms of settling time. It maximizes the number of channels that are within specification for each module, and if there are any that exceed it, they are never above 20ns. The rise time requirement (less than 5ns) is more relaxed, and it is there to differentiate between settings that are very close in settling time performance.

Once set to their best capacitance setting, the modules' dynamic performance is optimized. The settling time and rise time histograms of all channels of all modules of the example production batch are shown in Figure 2.16. The mean rise time is 3.43ns, and the settling time meets the ARx12 specification of 15ns for 196 of 240 channels (81.7%), with a mean of 10.63ns. Due to the inherent limited accuracy of the settling time measurement, the performance of the modules is deemed to be compliant with the specifications. The settling times measured during production are those of a complete test link, which includes a laser transmitter. In addition, the settling time specification of the entire link is 20ns, which means the performance of the worst channels is still adequate.

[6] The *C*-setting is the same for all channels in a given ARx12 module (i.e. it is not possible to tune each channel separately in one module).

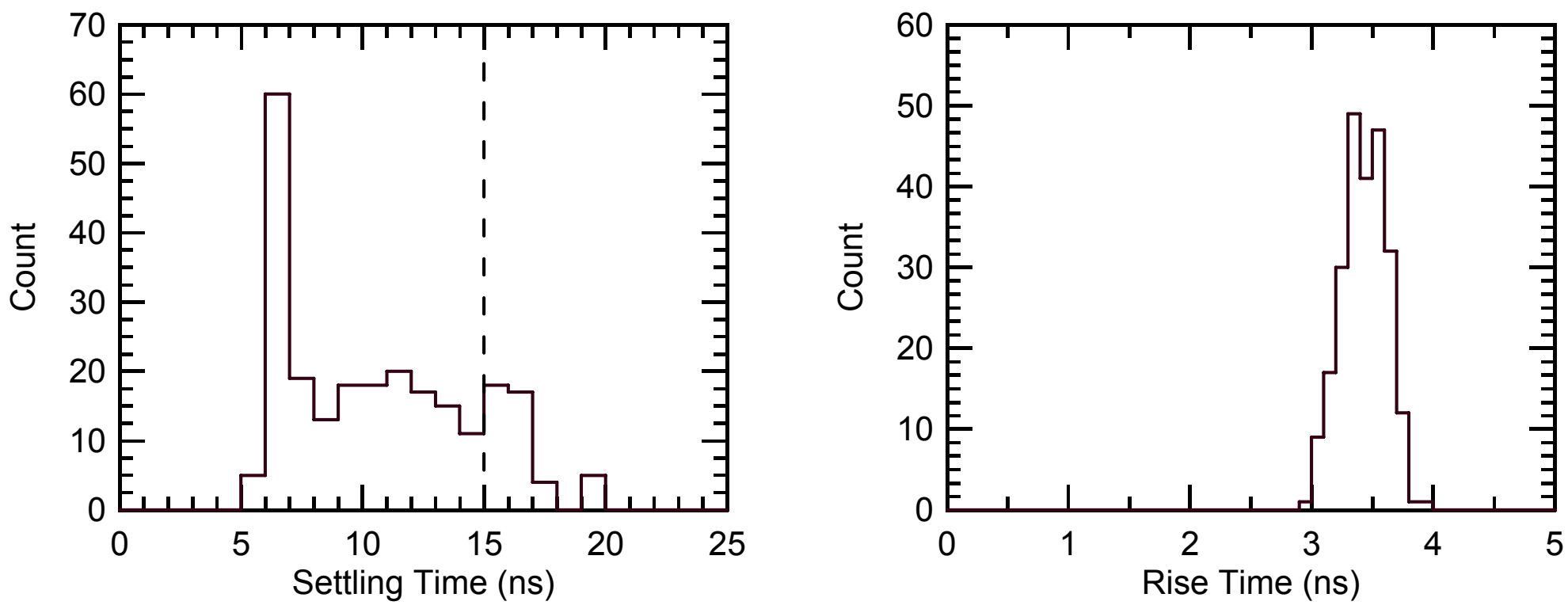

**Figure 2.16:** Histograms of the settling time (left) and rise time (right) for all channels, using each module's optimum capacitance setting.

A correlation between the optimal capacitance setting and the bandwidth of the receivers was found during production testing. This is demonstrated in Figure 2.17, where the optimum capacitance setting is plotted against the average bandwidth of all twelve channels in each module of this batch. It is hoped that this correlation will provide a means of selecting, a priori, the best capacitance setting on a module-by-module basis, based on the manufacturer's bandwidth production test results.

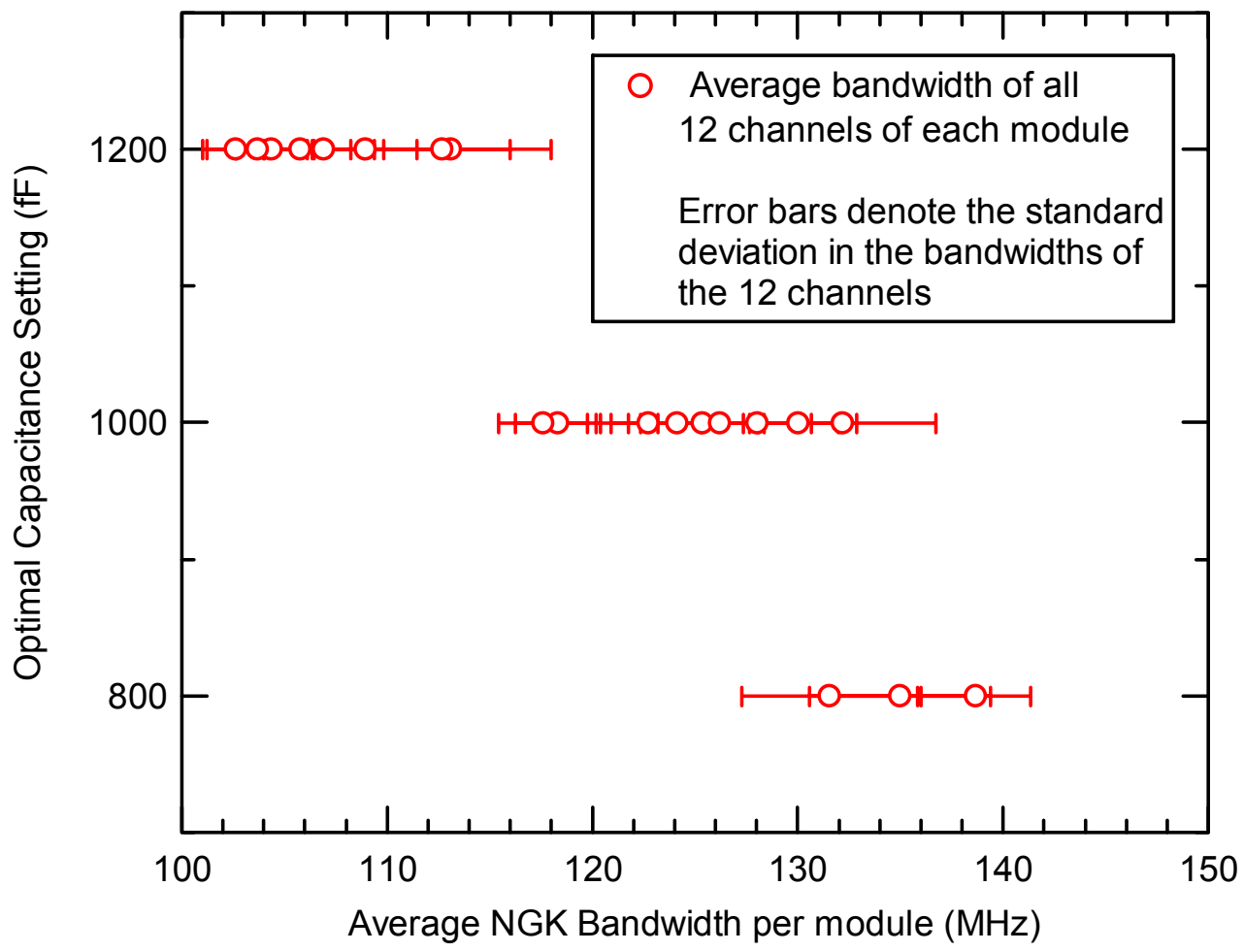

**Figure 2.17:** Correlation between optimal capacitance setting and NGK supplied bandwidth. Switches $C_1$ and $C_2$ were off for the bandwidth measurements.

## 2.4 Performance Verification of the CMS Tracker FED Analog Front-End Electronics

The previous sections gave an overview of the specifications and performance of the individual components comprising the optical link. The optical link has an impact on the performance of the entire readout system, which includes the FED board hosting the optoelectronic receivers. Before signals reach the ADC, they undergo certain conditioning by the FED's analog front-end components. These are essentially a part of the analog portion of the readout link, and hence their dynamic performance is equally important as that of the optical link components..

Testing of the FED has been carried out at CERN in order to assess the dynamic performance of the board's analog front end electronics, as well as of the complete readout system, including the optical link. This section summarizes part of the testing carried out on the FED's analog electronics which is most relevant to this thesis. More details and results can be found in [25]. A pulse generator was used to provide a well-controlled input that allows for accurate measurement of rise and settling times to be made, which is otherwise impossible using pulses (ticks) produced by the APV chip in the final readout system configuration.

A schematic representation of the FED's front end analog electronics is shown in Figure 2.18. Only one signal path (i.e. one readout channel) is shown for illustrative purposes. The analog optoelectronic receiver converts the incoming light into a current that flows through a load resistor, hence producing a single-ended voltage signal. This is then converted to a differential signal by the EL2140C differential driver IC. A voltage divider stage halves the signal levels to match the input range of the ADC (Analog Devices AD9218).

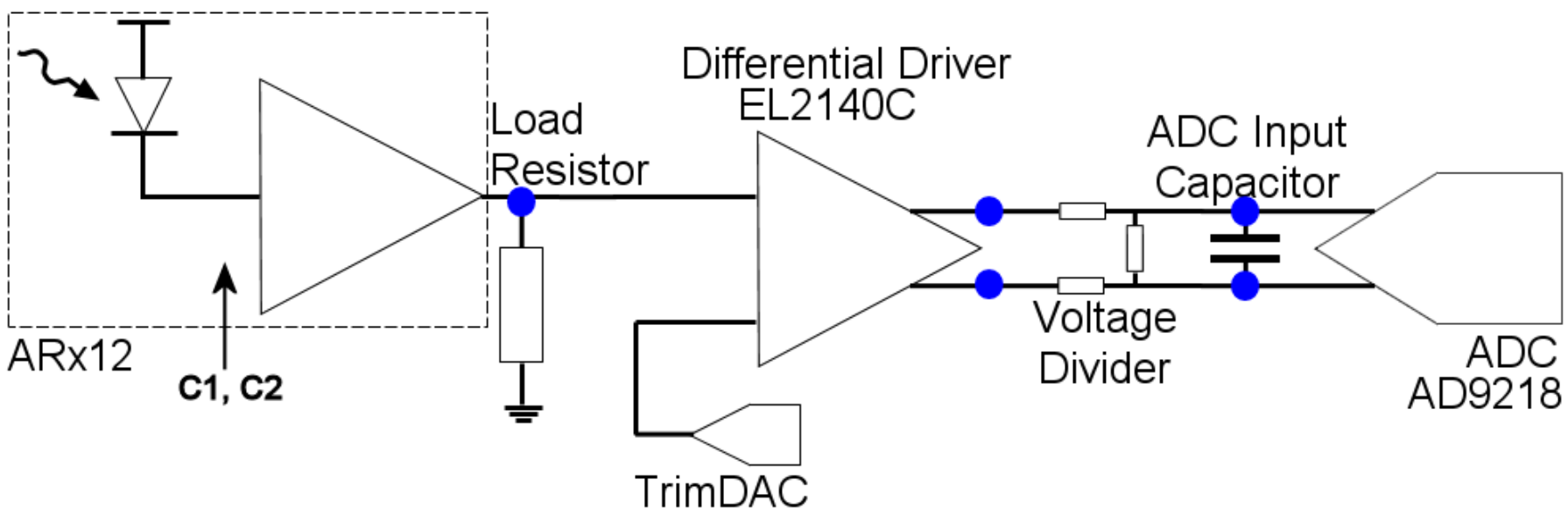

**Figure 2.18:** The CMS Tracker front end analog electronics. The blue circles indicate where the FED was probed.

The dynamic behavior of the ARx12 has been tested rigorously at CERN, both during the qualification and production phases of the device. The vast majority of receiver modules achieve optimum dynamic performance at a compensation capacitance setting ($C_{ARx}$) of 1000fF in the production test fixture. It was therefore necessary to determine whether receivers mounted on FEDs behaved in the same way as in the production test fixture at CERN.

The first set of results on a pre-production FED (version 1) revealed that the differential driver introduces some overshoot to the incoming pulses [25]. It will be shown that modules performing best at 1000fF in the production tests should be set to 800fF when mounted on FEDs.

The production version FED (version 2) was also tested. The FEDv2's front-end analog electronics differ only in the ARx12 load resistor value, which was lowered to 62Ω (from 100Ω on the FEDv1). This was done in order to reduce the overall link gain (since the gain is proportional to this load resistor value) which was necessary to meet the dynamic range specifications of the readout system. The change in load resistor was made on the basis of the work described in Chapter 4. The FEDv2 was tested in order to verify that the new load resistor would not adversely affect the dynamic performance of the readout system.

### 2.4.1 Test Setup

Figure 2.19 shows the test setup used. It should be noted that only one FED channel could be tested at a time. A LeCroy 9210 pulse generator was used to provide a differential input to the AOH. As in measurements made on a reference link in the Tracker Lab, the period was set to 100ns, with 50% duty cycle. 800mV differential amplitude was used with a common mode of 1.25V. The pulse generator was controlled by a portable PC via a USB-GPIB interface. The PC was also equipped with a USB-$I^2C$ for setting the AOH gain and laser bias. The laser was deliberately biased high, since it was impossible to run the usual automatic link tuning procedure. This limitation was due to the fact that, in this setup, it was not possible to interface the AOH to the FEC CCU that is normally used for control purposes with the final system setup (i.e. when the APV is used to send data to the FED). To verify that the optical pulse out of the AOH was not being clipped, it was observed on an oscilloscope using an optical head, and the biasing performed 'by eye'. The AOH gain was set to 5.0mS, the lowest gain setting available. On the FED side, the ARx12's X0-X5 settings were set to the nominal value (switches X0, X2 on). For the electrical probing tests, the pulse generator was set to trigger itself automatically and continuously. Thus, a continuous pulse train was fed to the AOH. Waveforms of the pulses at various test points were recorded on a LeCroy LT354 oscilloscope, using differential and single-ended probes.

For data captured by the FED, the Trigger Sequencer Card (TSC) was used to provide a synchronous trigger to both the FED and the pulse generator. On receipt of a trigger, the pulse generator was set to output a pulse after a delay of 2μs. The FED was in scope mode, capturing data points every 25ns for a time window of 7μs (280 points), thus ensuring that the incoming signal would be visible inside the time frame captured. The FED incorporates a 'fine delay skew', which skews the ADC sampling clock in 32 steps within a 25ns period[7]. The fine delay was used to obtain data points spaced apart by less than 1ns. The FED was programmed to capture 7μs of data (one 'event') for each of the fine delay skew settings. The pulse was then reconstructed by interleaving the data corresponding to the 32 skew settings. The TSC sent 100 triggers per skew setting, and hence 100 events were captured. Each data point was then determined by averaging over all the events.

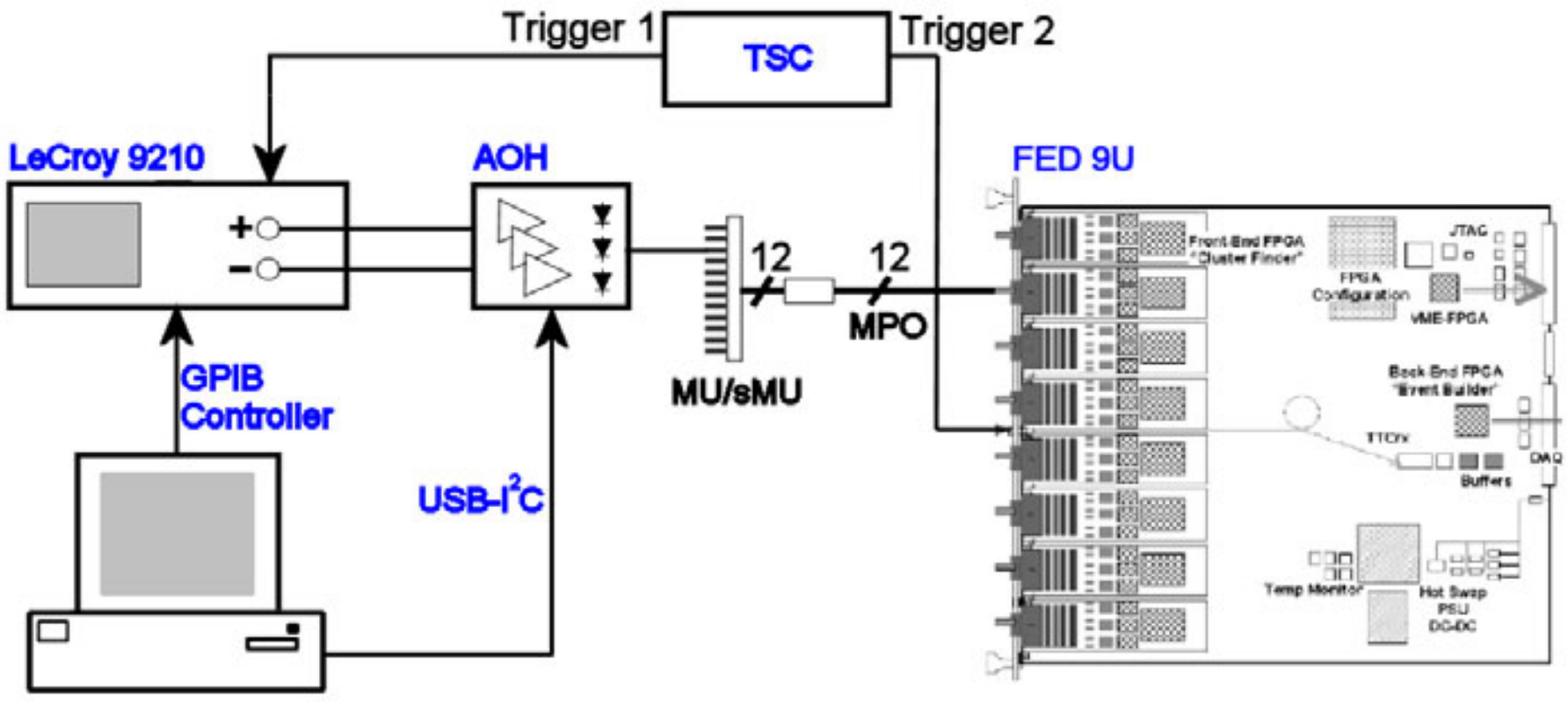

**Figure 2.19:** Test setup used for the dynamic characterization of the FED.

The configuration used to set up the triggers is shown in Figure 2.20. The triggers arrive almost simultaneously to the FED and pulse generator (at ~1μs on the scope trace), while a pulse is output from the AOH 2μs later. This ensured that the pulse was near the middle of the FED's data capture window.

[7] This was the case at the time these tests were performed. The latest FED software maps the 32 fine skew delays to 25 settings, so that each skew setting corresponds to a step of approximately 1ns.

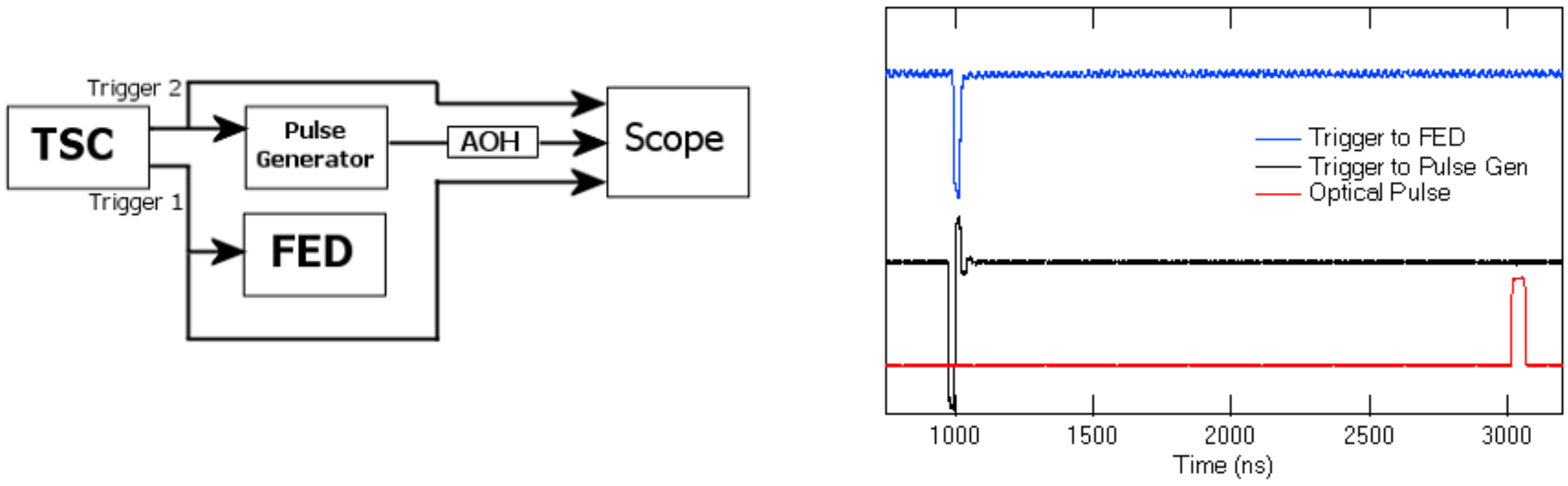

**Figure 2.20:** Setting up the triggers to the pulse generator and FED (left) and oscilloscope traces showing the relative timing of the triggers and of the AOH output pulse (right).

The same procedure was used to test a FEDv2 in order to verify dynamic performance with the new load resistor value of 62Ω. Twelve channels on two different front ends were tested, ensuring that every possible physical orientation of components and routing of tracks was covered.

### 2.4.2 Dynamic Test Measurement Errors

Before discussing the results, it is useful to look at the measurement errors for the dynamic tests, particularly those relating to the settling time calculation. The specification is fairly stringent (the input pulse must settle to within 1% of its final value in 15ns or less), and hence relatively small errors in either the data taking or the normalization of the pulses during analysis may severely affect the settling time calculation.

When reconstructing the input pulses, the FED was set to record only 10 data samples per time delay setting[8]. This was done for measurement speed. As a result of the limited averaging, the noise is on the pulses was measured to be approximately ±3 ADC counts (approximately ±0.5%). However this is virtually insignificant when calculating the settling time. The dominant source of error originates from the shape of the signal generator pulse which was not a perfect square pulse. Consequently, the settling time is heavily dependent on how the pulses are normalized (in particular, what part of the pulse is used to determine its final, settled level). In order to quantify this uncertainty, the normalization window was shifted by ±10ns around the 'ideal' point and the settling time calculated each time. The resulting percentage error was estimated at approximately ±5%.

[8] 10 samples per skew setting were used for the tests made on the FEDv2. The first tests on the FEDv1 were made using 100 points per skew setting.

Of course, this affects the absolute values of the time calculation. Special care was taken throughout all calculations to keep the position of the normalization window a constant time after the rising edge of each pulse. Hence this 'error' should be systematic for all settling times calculated, and the results can therefore be compared safely.

### 2.4.3 Results

FEDv1 RESULTS

The waveforms obtained by probing the FEDv1 are presented in this section, for illustrative purposes. The results of the FEDv2 are presented later, but only settling and rise times are given, without the associated waveforms.

Figure 2.21 (left), shows the electrical input to the AOH, obtained using a differential probe. The resulting optical signal at the output of the AOH was viewed on the scope using an optical head (Figure 2.21, right).

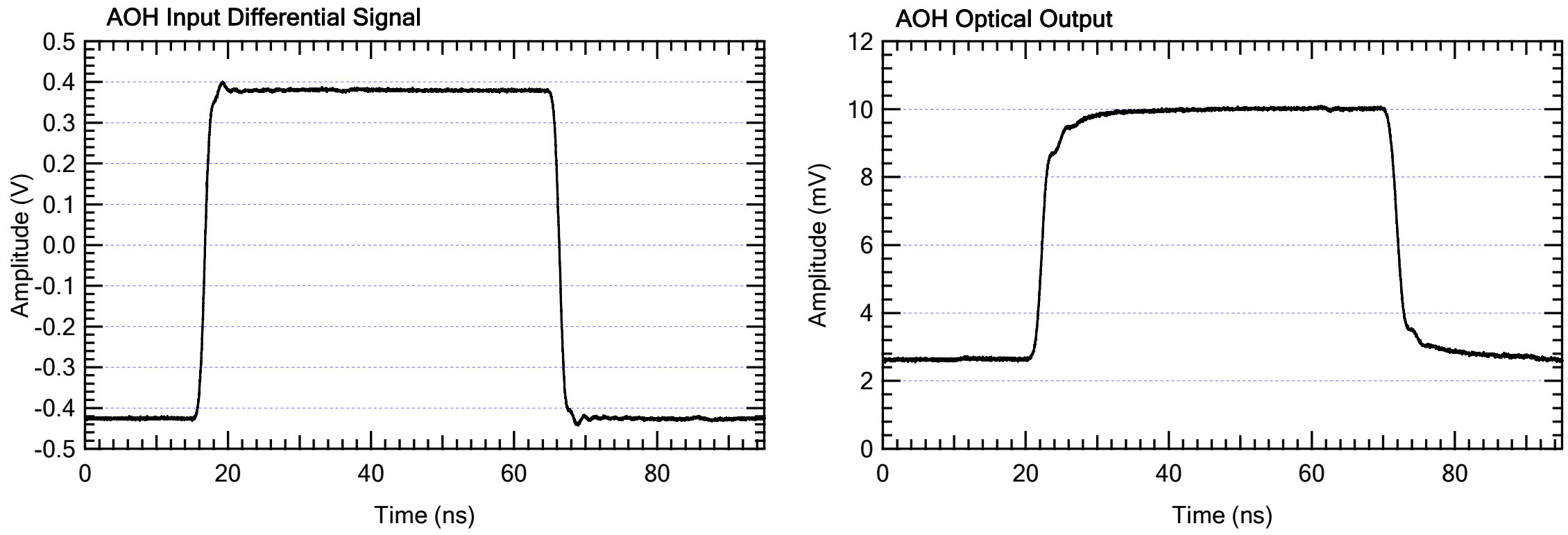

**Figure 2.21:** Differential Input to the AOH obtained using a differential probe (left), and corresponding pulse at the output of the AOH connected to an optical head.

The FED was probed electrically at the following points (Figure 2.18):

- ARx12 load resistor (single-ended).
- Output of the differential driver (differential).
- ADC input capacitor (differential).

Figure 2.22 (left) shows the signals obtained for all four capacitance settings of the ARx12, probed at the load resistor (i.e. the optical receiver's output). In the same figure on the right, the settling and rise times[9] are shown as a function of $C_{ARx}$. The settling time specification of 15ns is illustrated by the dotted line[10]. As expected, the result is very similar to what is seen in the ARx12 production tests performed in the lab at CERN. The effect of increasing the capacitance setting is to speed up the pulse, as is clear by the rise time decrease. The overshoot is also affected, which has a dominant influence on the settling time. Clearly, this particular channel performs best at a capacitance setting of 1000fF, where the settling time is at a minimum and well within the 15ns specification. The vast majority of optical receivers behave in the same way, giving confidence that the tested FED channel is a typical case of what can be expected in the final system. This is illustrated in Figure 2.23, which shows the settling time plots of all channels of all 10 modules coming from the same ARx12 production batch as the receiver on the tested FED [24]. The vast majority of channels perform best at 1000fF.

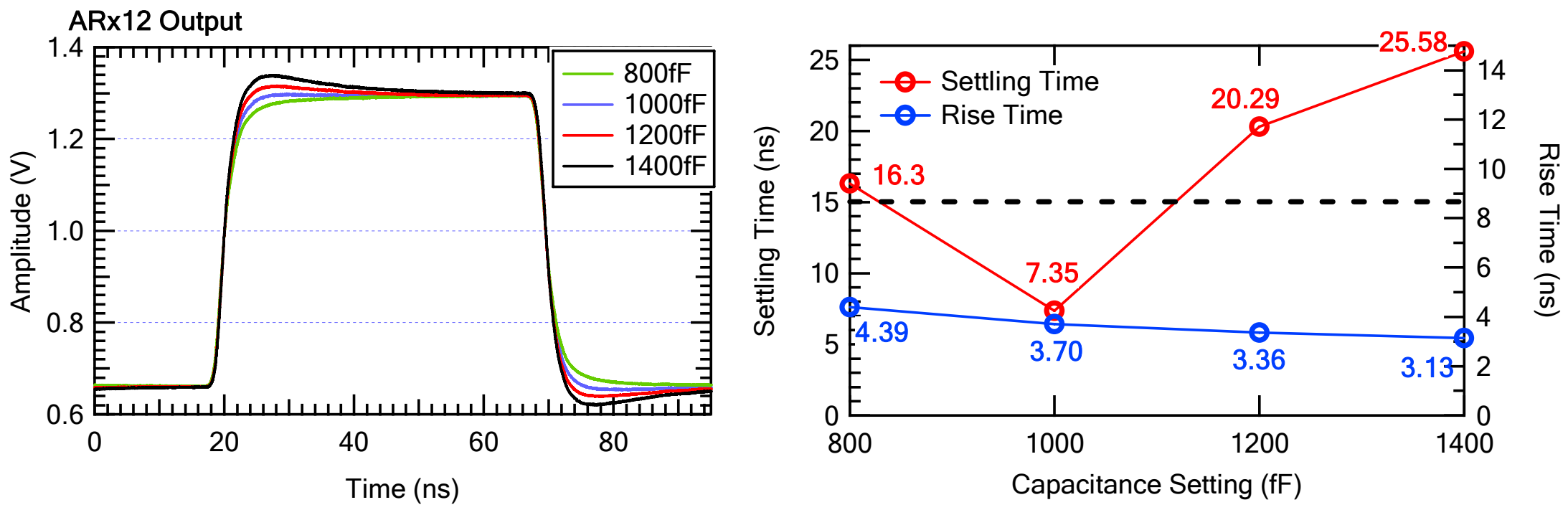

**Figure 2.22:** Single-ended signals at the load resistor of the ARx12 for all four ARx capacitance settings, with the corresponding settling and rise times shown on the right.

[9] The settling time is defined as the time it takes the pulse to settle to within 1% of the final value. The rise time is calculated from 10% to 90% of the pulse's final value.
[10] This is the specification for the ARx12. The optical link maximum specified settling time is 20ns.

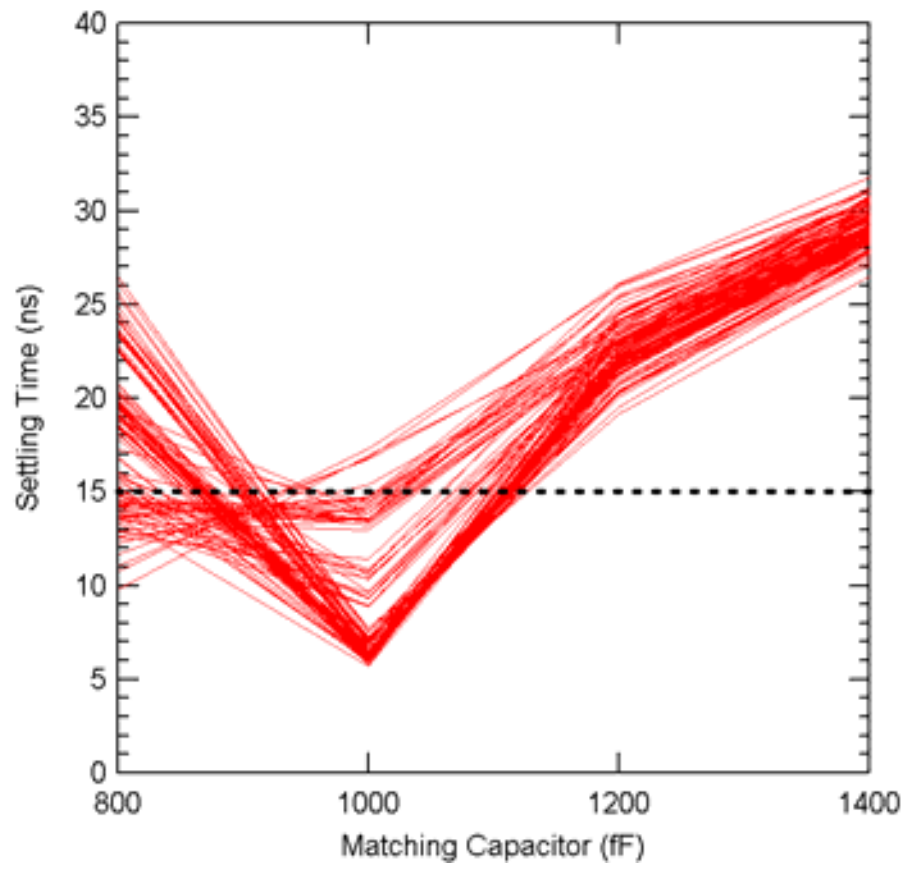

**Figure 2.23:** Settling time plots for all channels of the 10 acceptance test modules belonging to batch R4 [24].

Figure 2.24 shows the waveforms obtained when placing the differential probe on the output of the differential driver IC. The ringing observed after the rising edge of the pulse is probably due to the way the probe was attached to the IC's pins. Extra attached wires (2-3cm long) had to be connected to the probe's ends to enable probing between the tiny pins. An increase in the overshoot of the pulse is clearly visible. This is mirrored in the settling time plot, where the minimum settling time is now achieved at a capacitance setting of 800fF. Using a single-ended probe, the signal on each branch of the differential driver was observed (Figure 2.25). The extra peaking on the differential signal appears to be due to the asymmetry of the individual branch signals.

Probing the input of the ADC proved very difficult, and it was impossible to maintain a consistent contact across the pins of the ADC input capacitor. The results are nevertheless included for completeness (Figure 2.26). Despite the uncertainty of the probing, they do seem to confirm the additional overshoot observed at the output of the differential driver.

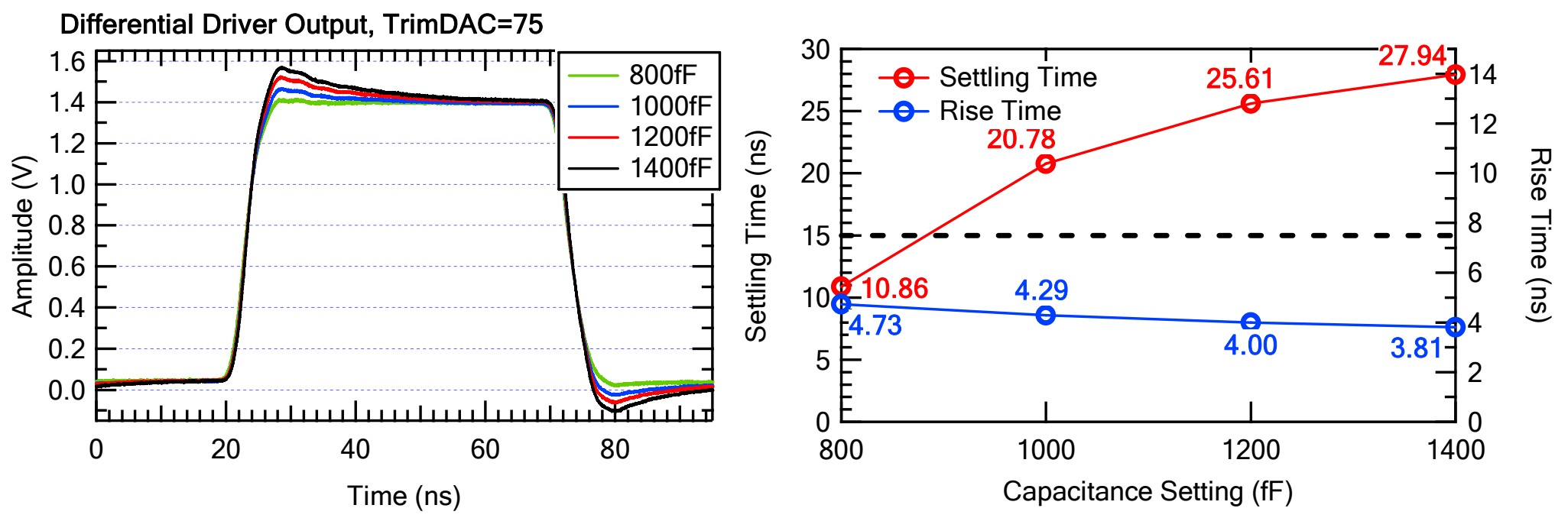

**Figure 2.24:** Signals at output of differential driver IC for all four ARx12 capacitance settings (left), with corresponding settling and rise times (right).

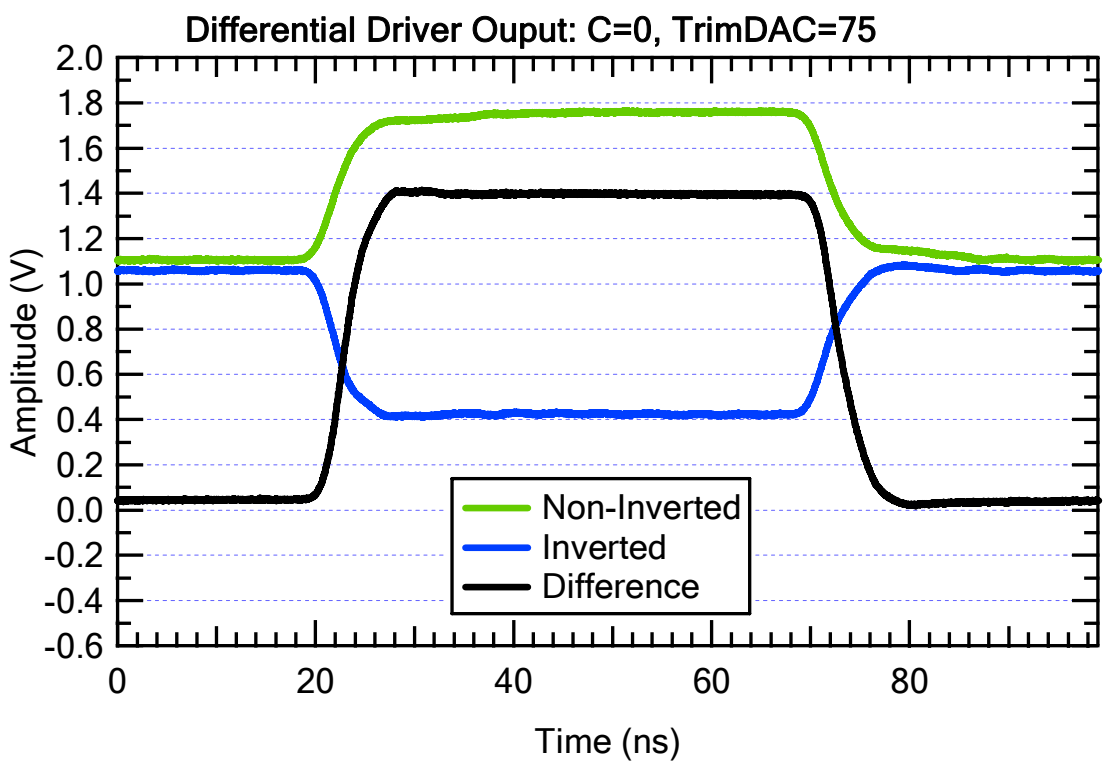

**Figure 2.25:** Signals obtained by probing each branch of the differential driver output using a single-ended probe. Only one ARx12 capacitance setting was used.

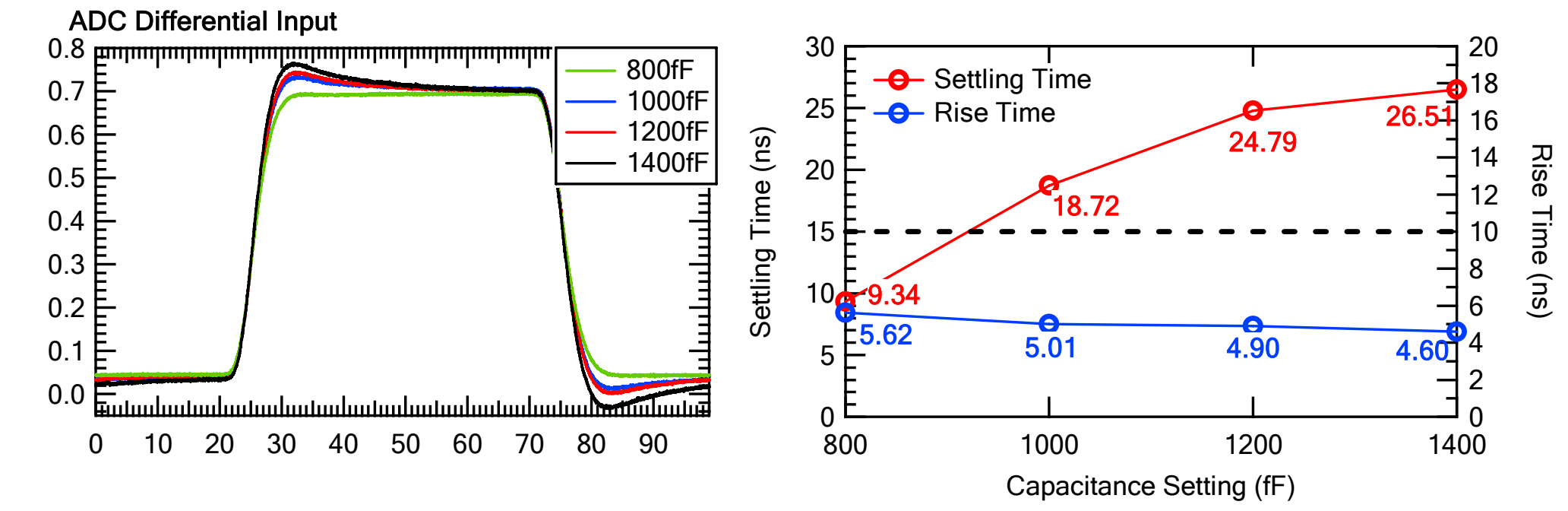

**Figure 2.26:** Signals at the input of the ADC for all four ARx12 capacitance settings (left), with corresponding settling and rise times (right).

After probing electrically, the FED was used to capture data and the incoming pulses were reconstructed as described in section 2.4.1. The results are shown in Figure 2.27, which confirm that the best ARx12 capacitance setting for this channel is 800fF. It is also evident that there is a significant slowdown of the pulses at the ADC. The rise times are now as much as 1.5ns higher than in the preceding stage. This could be attributed to the capacitance seen by the signals due to the 4.7pF capacitor across the differential inputs, as well as the additional internal capacitance of the ADC (typically ~3pF). Moreover, while the analog bandwidth of the ADC is quoted at 300MHz, there is no specification for the slew rate, which could ultimately be a limiting factor for the rise time of the pulses.

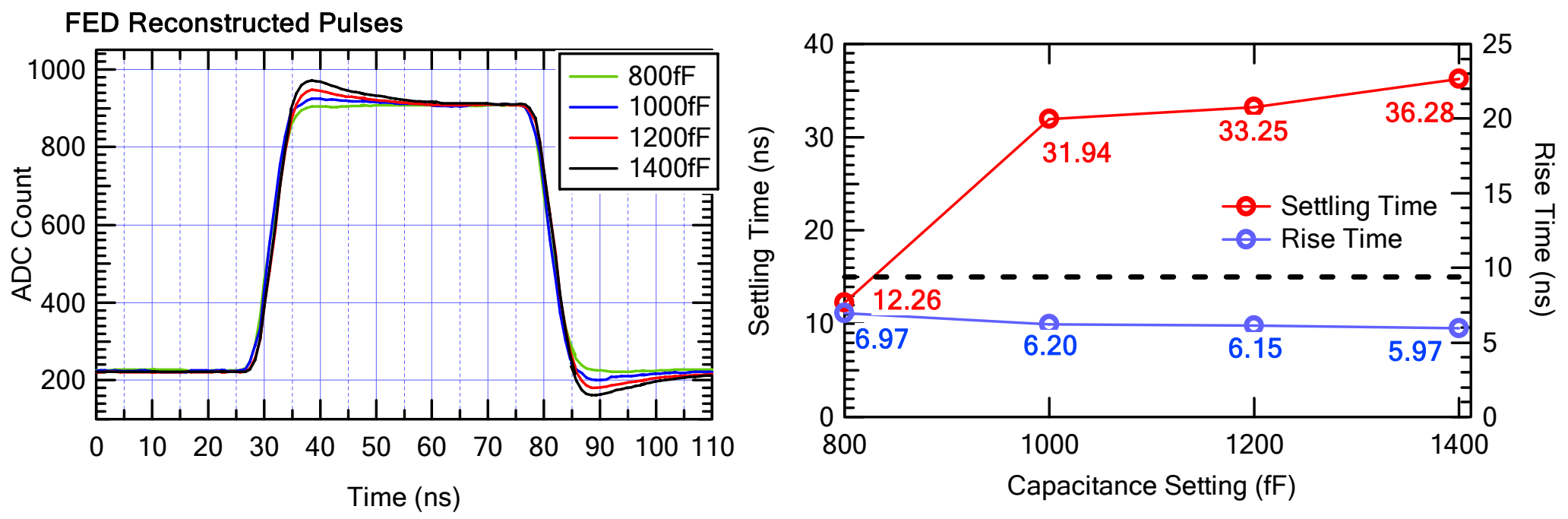

**Figure 2.27:** Pulses obtained by reconstructing captured FED data (left). Settling and rise times are shown on the right.

The results have shown that the overshoot of the input pulses is increased after the differential driver IC stage. Consequently, when selecting the best ARx12 capacitance setting for a given receiver module, the effect of the FED's front end analog electronics should also be taken into account in order to achieve the best possible dynamic performance. The results suggest that modules showing best capacitance settings of 1000fF in the production tests (i.e. when looking at the output of the optical receiver), should be set to 800fF to compensate for the extra peaking introduced by the FED.

### FEDv2 RESULTS

Figure 2.28 is a typical example of the results obtained. Every channel tested performed best in terms of settling time at a capacitance setting of 800fF, as can be seen in Figure 2.29. In addition, all channels meet the 15ns settling time specification. This is consistent with what was observed in the previous results. The average settling[11] and rise times at 800fF are 10.22ns and 5.47ns respectively. The dynamic performance of the FEDv2 is within specification, and no deterioration has been observed as a result of lowering the ARx12 load resistor.

[11] Settling target set to 1% of final value.

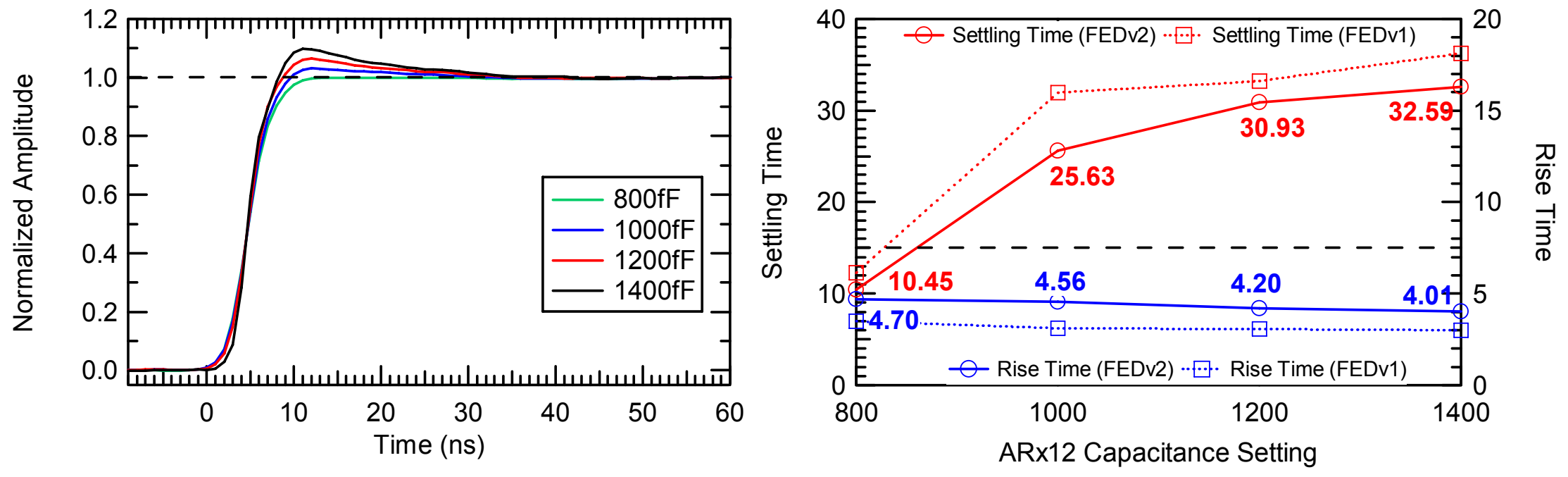

**Figure 2.28:** Typical dynamic behavior in the FEDv2. FED channel 93 (front end 8, ARx12 channel 8) shown. The results from the FEDv1 tests are also shown for comparison (dotted lines).

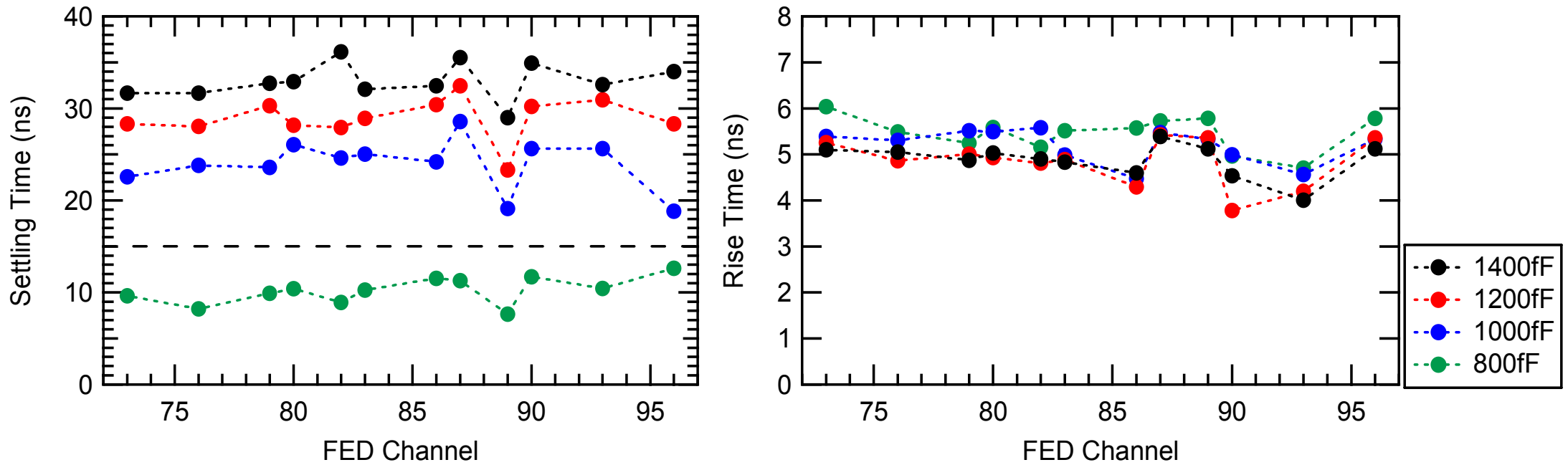

**Figure 2.29:** Settling and rise times for the tested channels as a function of ARx12 capacitance setting.

IMPACT OF CHANGING THE SETTLING TIME TARGET

Settling times for all channels were recalculated for settling targets of 1.5%, 2.0%, 2.5% and 3% of the final pulse value. The results for $C_{ARx}$=1000fF are shown in Figure 2.30.

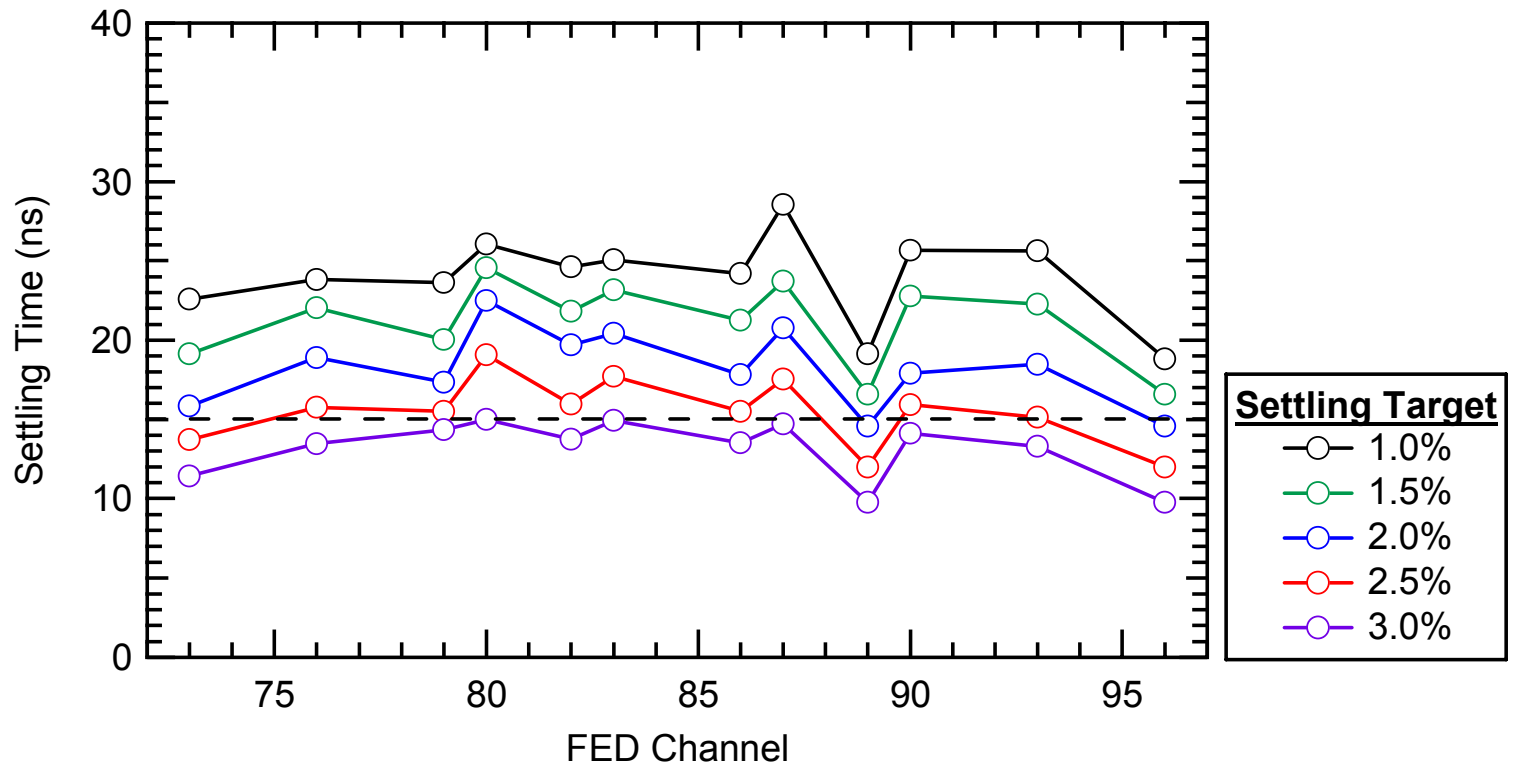

**Figure 2.30:** Settling times for all tested channels, using different settling targets. Only capacitance setting 1000fF is shown.

The significance of these results is that for all channels at $C_{ARx}$=1000fF, the analog pulses are within 3% of the final (desired) value after 15ns. This suggests that the maximum amplitude error in a given readout channel resulting by selecting the wrong (neighboring to the optimum) $C_{ARx}$ setting will likely be less than ~3%.

## 2.5 Conclusions

The components of the optical link have been introduced, and performance metrics that are most relevant to an analog readout system were presented. Results from production tests were used to demonstrate that each component meets its specifications, in the context of the requirements set out for the entire readout link.

The dynamic performance of the analog electronics of the CMS Tracker FED has been tested extensively. A carefully controlled analog input in the form of signals from a pulse generator was used. The test setup allowed fairly accurate dynamic measurements to be carried out, hence allowing comparisons to production test results of individual readout link components. It has been shown that the analog part of the readout system (including the optical link and FED analog electronics) will perform according to specification, in terms of the dynamic performance.

# Chapter 3

## SETTING UP THE OPTICAL LINKS

*The CMS Tracker readout links must be set up correctly for optimum operation. The need for new optical link setup routines became apparent when attempting to study the readout system's performance in test beam conditions, with a large number of deployed links. Due to the limit on the data range of the system imposed by the FED's ADC, it was necessary to develop new algorithms that could calculate any link gain, regardless of whether the transmitted data exceeded the ADC's input range. The new optical link setup routines are presented. The functionality of the readout link and all tuning parameters is also detailed.*

## 3.1 Readout Link Functional Description

The readout link must be set up for correct operation prior to taking physics data. There are various handles that can be used to adjust the static parameters of the link, and they are detailed in this section. In the final system, some of these handles will be hidden from 'regular' users by the software, and will only be accessible by experts. These are referred to as 'fixed' in this document. Parameters that users are 'allowed' to adjust during or between physics data runs will be referred to as 'adjustable' or 'tunable'. Figure 3.1 is a schematic representation of the components that make up the readout link, showing the handles available. Sections 3.1.1 and 3.1.2 describe all fixed and adjustable parameters, while their impact on link operation is illustrated in more detail in section 3.1.3.

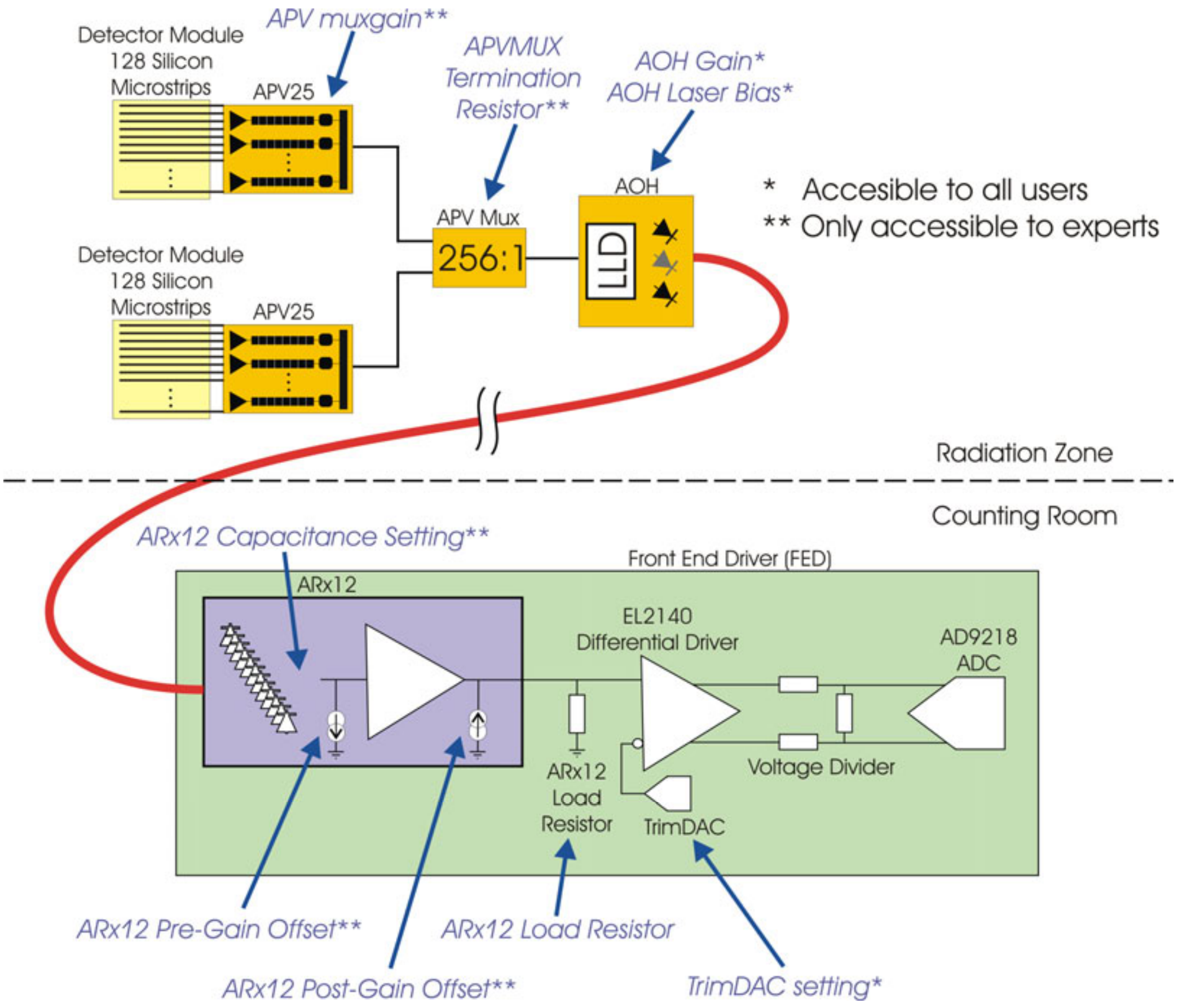

**Figure 3.1:** Readout link components from the detectors to the FED front-end analog electronics, showing one optical channel.

### 3.1.1 Fixed Parameters

APV MUXGAIN

The APV contains a muxgain register that determines the chip's gain [1]. It defines the size of the resistor used at the input of the APV's multiplexer. Five resistors are available, with the middle sized resistor giving the nominal gain of 1mA/MIP (for thin, 320µm detectors). An adjustment up to ±30% (in steps of 15%) is possible using the other resistor values.

The APV's nominal gain has been selected to be consistent with the readout system's dynamic range requirement to accommodate signals up to 3.2MIPs in 8 bits of the FED's ADC (256 FED ADC counts). A 1MIP signal from the detectors results in 1mA being output from the APV, which is then converted to 100mV at the input of the optical link (by the resistors in the APVMUX chip). At the link's nominal gain of 0.8V/V, the corresponding link output is 80mV. This is equivalent to 80 FED ADC counts (since the gain of the ADC is 1count/mV). It follows that, since 1MIP=80 ADC counts nominally, signal sizes up to 3.2MIPs fit in 256 counts (or 8 bits).

APVMUX TERMINATION RESISTOR

The APVMUX chip interfaces between the APV25 chip and the AOH, multiplexing the outputs of two APV25 chips onto a single AOH input. The differential current outputs of the APV25 chips are converted into voltages by internal APVMUX resistors. There are eight 400Ω resistors connected in parallel between each differential input and a reference voltage pad. An 8-bit register controls switches that are in series with each resistor to allow variation of the resultant resistance between 400 and 50Ω.

The resistance was made variable to allow flexibility when interfacing with the other components in the readout chain (namely the APV and LLD chips). A study performed at CERN [2] has shown that the optimum setting is 100Ω per differential input branch. This optimum setting should always be used during regular Tracker operation, since it preserves signal integrity while maintaining the desired signal size as dictated by the readout chain's dynamic range requirement.

### ARx12 Pre-gain and Post-gain Offset

The 12-channel Analog Optoelectronic Receiver (ARx12) contains a two-stage current amplifier after the pin diodes (see Chapter 2). In order to allow tuning of the static transfer function of the device, it was designed with a current sink at the input of the amplifier (pre-gain offset) and a current source at the output (post-gain offset). Both offsets are programmable via $I^2C$ commands.

The pre-gain offset is controlled by four switches and effectively subtracts current from the input to the ARx12 amplifier, which has a specified linear input range of ±150µA. The offset has been designed to be variable such that the input signals can be adjusted to match the amplifier and therefore maximize the linear range of the device. A range of 0 to ~375µA can be selected with a resolution of 25µA. The nominal value that achieves maximum linear range was found to be ~100µA during qualification tests of the device and should remain constant for the duration of the Tracker's operation.

The post-gain offset simply adds current to the output of the ARx12's amplifier and was included in the design to allow adjustment of the signal current flowing into the FED's analog electronics following the optical receiver. A range of 0 to ~9mA is selectable with a resolution of ~2.5mA. The nominal value of the post-gain offset is 0mA.

### ARx12 Capacitance Setting

Tuning of the dynamic performance of the ARx12 is possible through two switches that control the matching capacitance between the pin diodes and input of the receiver amplifier. In order to compensate for mismatches between the two, the matching capacitance was designed to be variable, with four settings available (800, 1000, 1200 and 1400fF).

This is a crucial parameter of the optical link, since capacitance mismatch can cause extremely slow pulse settling times. The data from the Tracker's detectors is essentially analog pulse amplitude modulated (PAM), with the amplitude of each output pulse representing the signal from a single microstrip. Hence the readout system relies on sampling the height of each output pulse at 25ns intervals. It is therefore important to minimize the settling time of the pulses in order to obtain the most accurate value of the signal on each microstrip. It follows that synchronization is crucial for such a sampling system, and fast-settling pulses mean that a larger window in time exists for the optimum sampling instant (i.e. there is a larger tolerance for picking the right sampling point in time).

Extensive studies have been carried out on the dynamic performance as a function of ARx12 capacitance setting. From the work described in Chapter 2, it is most likely that all modules will be set to 800fF, though this is still under consideration.

ARX12 LOAD RESISTOR

The ARx12 load resistor provides another handle for adjusting the gain of the link prior to installation. Its value will be fixed for the duration of the experiment. The gain of the analog optical link is crucial for the readout system, since it affects the dynamic range of the data being captured. While it is obvious that each component in the readout chain directly determines the overall link gain, not all component gains are variable or can be modified easily during production.

To overcome this limitation and provide flexibility during the production phase, the system has been designed with a current-output optical receiver. The current is changed into a voltage by a load resistor, the value of which is directly proportional to the gain of the whole readout link. It is therefore possible to change the overall optical link gain easily (should this be needed), simply by replacing the load resistor. The work described in Chapter 4 determined the optimum value, and is now fixed in the final version FED (FEDv2) at 62Ω.

### 3.1.2 User-adjustable Parameters

There are three parameters that regular users can adjust during regular Tracker operation. These will be set after running automatic setup routines, but can also be changed manually by operators.

AOH (LLD) LASER BIAS

The laser transmitter on the AOH must be properly biased by the LLD chip so that it is never below the lasing threshold during operation. It is also desirable to operate the laser at low current for better noise performance [3]. Moreover, operating at the lowest possible bias current ensures that the receiving amplifier will always be within the linear part of operation (biasing too high could, potentially, drive too much light into the receiver which would in turn cause more current to flow into the ARx12 amplifier). While this could be rectified by increasing the pre-gain offset setting of the receiver, it would require tuning of yet another parameter on a link-by-basis, therefore increasing the complexity of the setup procedure needlessly.

There is no nominal setting for the biasing of the laser transmitters, since they can behave differently from device to device. Hence the bias point is chosen after running the automatic setup procedure. The bias current is programmable via $I^2C$ from 0 to ~22mA, in ~0.45mA steps (0-50 in $I^2C$ steps).

### AOH (LLD) GAIN SETTING

In the final system, only one handle is available to regular users for adjustment of the readout link gain. The LLD was designed with four gain settings (5.0, 7.5, 10.0, 12.5mS, also referred to as gain settings 0 to 3). The goal is to equalize the gain of every link to be as close to 0.8V/V as possible. Equalization of the gain throughout the Tracker readout links ensures uniform performance across all detectors. Of course, due to the finite granularity of the settings available, the final system will exhibit a certain amount of gain spread. More details on this can be found in Chapter 4, where this spread is predicted accurately using production and test beam data.

There is no nominal setting, though the system was designed so that the 7.5mS setting will most often be used when the system is running at the Tracker operating temperature of -10°C. This allows for maximum flexibility, ensuring that both low-gain and high-gain links can be compensated for. In addition, gain reduction induced by radiation damage and component aging can be mitigated using one of the two higher gain settings.

### TRIMDAC

The TrimDAC is connected to the inverting input of the differential driver IC (Figure 3.1). By adjusting its value, the incoming analog signal can be moved to match the FED's ADC range. The TrimDAC setting is determined by calculation in an automatic setup routine.

**Table 3.1:** Summary of readout link parameters.

| Parameter | Accesibility | Nominal Value | Range/resolution |
|---|---|---|---|
| **APV muxgain** | Experts only | ~1mA/MIP | +/-30% from nominal in steps of +/-15% |
| **APVMUX Input Resistor** | Experts only | 100Ω on each branch of differential input ($15_{16}$) | 50Ω to 400Ω All permutations of eight 50Ω parallel resistors. |
| **AOH Gain** | All users | Varies | 5.0, 7.5, 10.0, 12.5mS |
| **AOH Laser Bias** | All users | Varies | 0-50 $I^2C$ steps (~0-22mA) |
| **ARx12 pre-gain offset** | Experts only | ~100μA | 0-375μA in 25μA steps |
| **ARx12 post-gain offset** | Experts only | 0mA | 0-9mA |
| **ARx12 Load Resistor** | Fixed for CMS duration | 62Ω | N/A |
| **TrimDAC** | All users | Varies | 0-255 TrimDAC counts |

### 3.1.3 Detailed link operation

The effect of the various parameters on the relationship between the link input and output is illustrated by Figure 3.2, which shows a typical transfer characteristic. Such a transfer characteristic is obtained by scanning the laser bias current from $I^2C$ steps 0-50 (equivalent to ~0-22mA), with no input from the APV25. At each step, the data at the FED is captured. The FED is set to *virgin raw* mode, which gives 10bits of data in a range of 1.024V. Hence, each FED ADC count is equivalent to 1mV at the output of the link.

MATCHING THE ADC INPUT: TRIMDAC SETTING

As can be seen in Figure 3.2, the output is at a constant level at first (from $I^2C$ steps 0 to 7). At these points there is insufficient bias current for lasing to occur; the laser is below threshold and no light is emitted. This ‘zero-light level’ is determined by the TrimDAC setting. The FED ADC has an input range of approximately ±0.5V (differential) with a 1.1V common mode. On the other hand, the ARx12 provides a single-ended output, with a typical range of ~0.4 to 2V. It is the job of the differential driver IC (Figure 3.1) to convert the signal to differential and adjust the level of the input signal so that it is centered on 1.1V, hence matching the ADC's input requirement. The TrimDAC is connected to the inverting input of the differential driver, while the analog detector data is fed into the non-inverting input. Hence by changing the TrimDAC setting, one can move the zero-light level (and hence the entire analog signal) vertically within the ADC range.

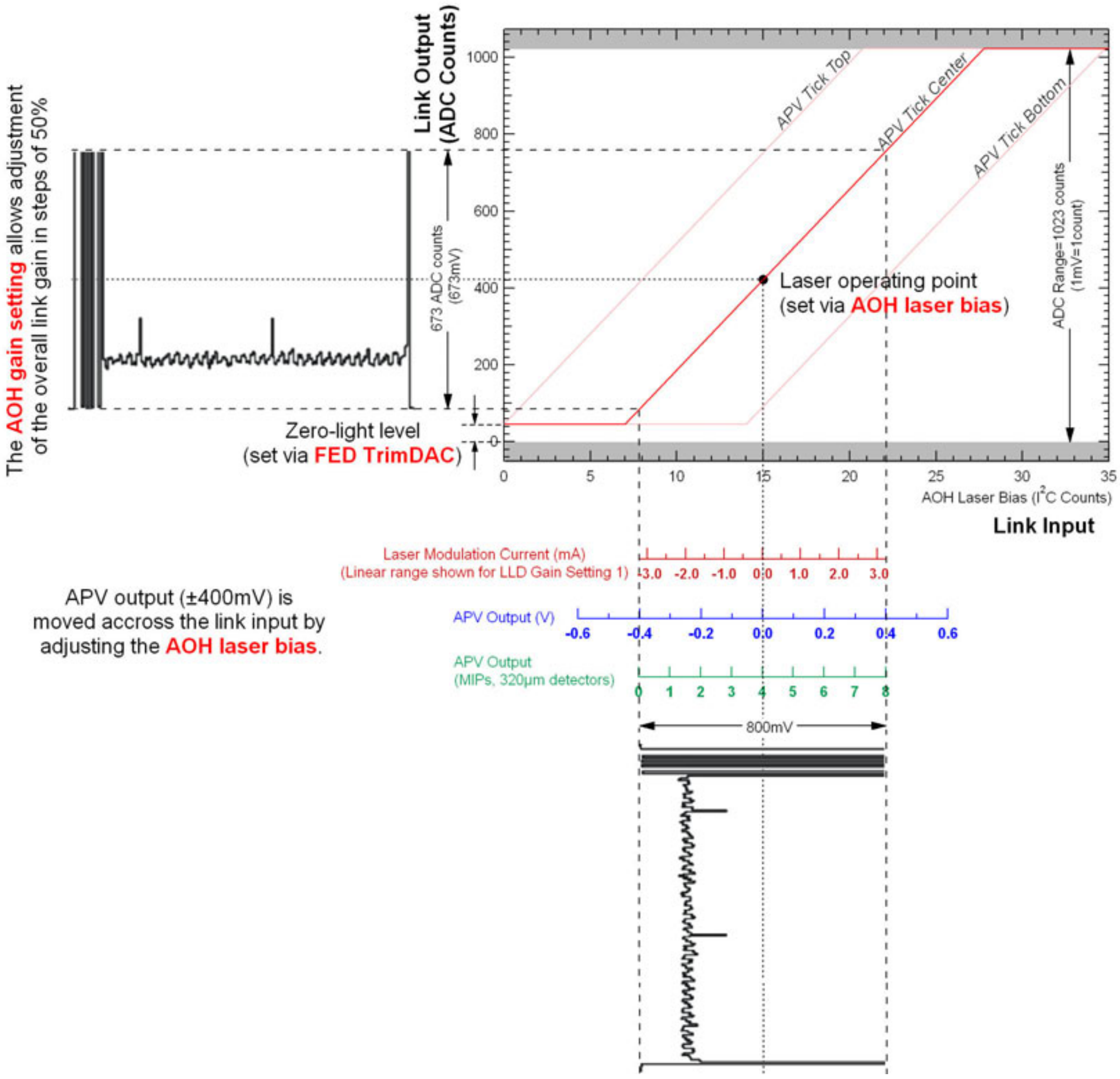

**Figure 3.2:** Typical link transfer characteristic (solid red line). The effect of transmitting a typical APV data frame through the link is also illustrated (light red lines).

It is important that the zero-light level is clearly visible within the ADC range (i.e., it should never be set at 0 ADC counts or below). The reason for this is that for proper optical link configuration it is essential to be able to distinguish the laser threshold. In this case, the laser threshold can be seen at ~7 $I^2C$ steps. If the threshold is not visible, it is impossible to find the optimal bias point of the laser. In the final system, an automatic routine will be used to set the value of the TrimDAC. This must be run with no light coming from the front end (all AOHs off), and should set the zero-light level to ~10-50 ADC counts.

READOUT LINK GAIN AND ARX12 OFFSETS

The gain of the readout link is determined by the slope of the transfer characteristic in Figure 3.2 and the AOH (LLD) gain setting (since the x-axis scale is in $I^2C$ counts). Equation (3.1) shows how the readout link output is related to the link parameters (see Table 3.2 for an explanation of the symbols used).

$$Link_{out} \propto \left( \left[ \left( Laser_{out} \times PIN \right) - I_{phoff} \right] \times ARx_{gain} + I_{outoff} \right) \times R_L \quad (3.1)$$

Where the laser output is given by:

$$Laser_{out} \propto \left( LLD_{gain} \times Link_{in} + Laser_{bias} \right) \times Laser_{eff} \times Fiber_{loss} \quad (3.2)$$

**Table 3.2:** Link parameter symbols used in Equation (3.1) and Equation (3.2).

| Description | Symbol |
|---|---|
| Link Output (mV or ADC counts) | $Link_{out}$ |
| Laser Output Power (mW) | $Laser_{out}$ |
| ARx12 PIN Diode Responsivity (mA/mW) | $PIN$ |
| ARx12 Pre-gain Offset (mA) | $I_{phoff}$ |
| ARx12 Post-gain Offset (mA) | $I_{outoff}$ |
| ARx12 Amplifier Gain (mA/mA) | $ARx_{gain}$ |
| ARx12 Load Resistor (Ω) | $R_{load}$ |
| LLD Gain (mS) | $LLD_{gain}$ |
| Link Input (mV) | $Link_{in}$ |
| Laser Bias Current (mA) | $Laser_{bias}$ |
| Laser Efficiency (mW/mA) | $Laser_{eff}$ |
| Fiber Loss (no units) | $Fiber_{loss}$ |

The slope of the transfer characteristic depends on the efficiency of the laser transmitter, ARx12 pin diode responsivity, ARx12 amplifier gain and the value of the load resistor. While the laser transmitter efficiency and ARx12 pin diode responsivity only affect the slope, the other parameters also affect the zero-light level; any change in their value is multiplied by the ARx12 current offsets, therefore affecting the absolute signal levels.

The slope of the transfer characteristic in Figure 3.2 is not enough to determine the gain of the link, since the input is shown in $I^2C$ counts. The LLD chip modulates the input signals from the APV25 on top of the static transfer characteristic, as can be seen in Figure 3.2. By changing the gain of the LLD, the user effectively changes the relationship between the APV output (the two bottom axes in Figure 3.2) and the link input *in terms of laser bias $I^2C$ steps*. In other words, if the LLD is adjusted to output more current per input mV to the laser transmitter, the bottom axes (APV output) would be expanded in relation to the link input (current, in $I^2C$ steps). The overall gain of the link is therefore the product of the transfer characteristic slope (converted to mV/mA) and the AOH gain setting (mS).

The zero-light level (and hence the absolute signal level) is affected by the pre-gain and post-gain offsets in the ARx12. The pre-gain offset reduces the amount of current coming from the pin diodes in the optical receiver. While adjusting this value will change the level of the signal at the output, this setting should not be used for tuning. Its purpose is to ensure that the incoming signals do not saturate the internal ARx12 receiver on either side of its input range. The setting that results in the largest input linear range was chosen during qualification of the receiver amplifier. Figure 3.3 shows the transfer characteristic of the receiver amplifier obtained with all combinations of pre-gain offset settings (chosen using the four switches on the device). As can be seen, subtracting too much or too little current from the input of the amplifier results in saturation and a smaller input range. This setting will be adjustable only by experts when CMS begins operation.

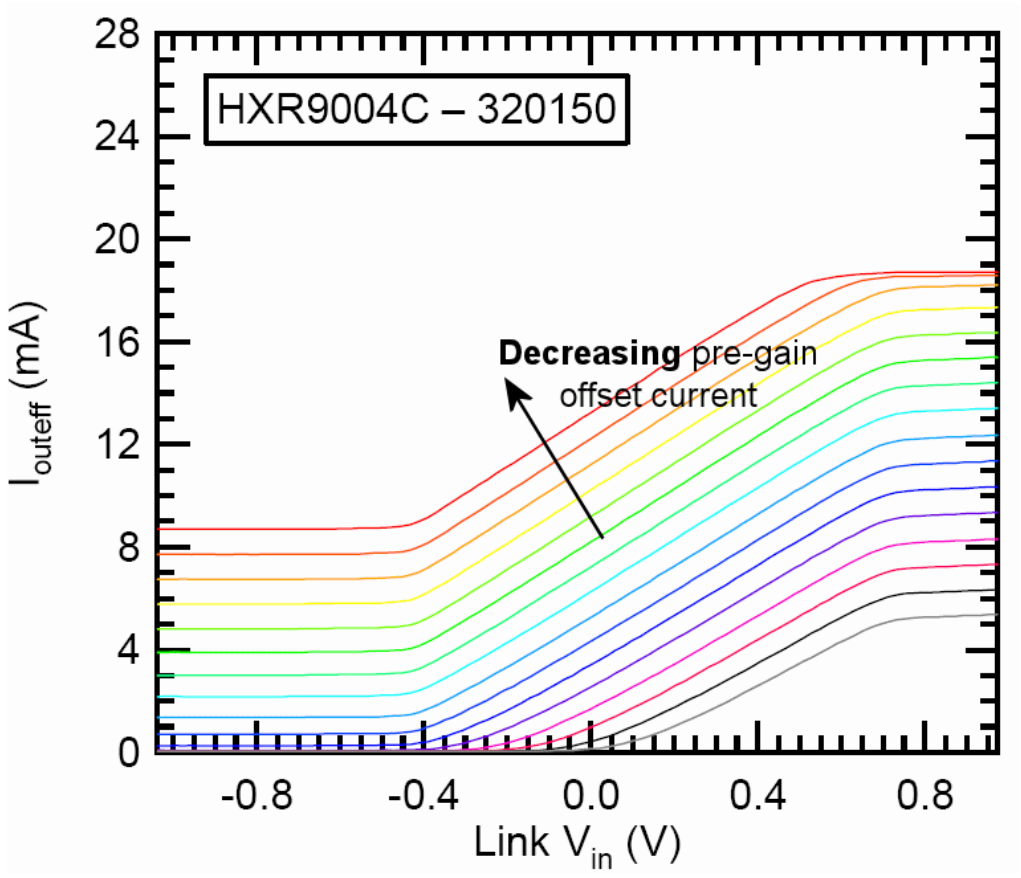

**Figure 3.3:** Receiver Amplifier transfer characteristics as a function of pre-gain offset current [4].

The ARx12 post-gain offset simply adds current to the output, therefore shifting the whole signal. In the final system this setting will be masked to regular users. The default setting is all switches off, meaning no current is added to the output signal.

OPERATING POINT: AOH LASER BIAS

The laser transmitter on the AOH must be biased appropriately to ensure correct operation of the optical links. Pulse amplitude modulation is achieved by the LLD sending a differential signal to the laser transmitter. Modulation occurs by subtracting from and adding to the laser's input current, in response to the differential signal fed to the LLD. The operating point must be chosen so that the bottom of the modulating signal never drives the laser below threshold; otherwise clipping at the bottom of the signal will occur. The operating point is chosen by adjusting the AOH laser bias setting. For each gain setting, there is a different corresponding optimal operating point.

## 3.2 Optical Link Gain and Laser Bias Setup

A new algorithm that has been developed (and will be adapted for the automated routines to set up the optical readout links) is described in this section. Since the gain calculation relies on measuring the height of the APV digital header (or APV tick) which is a constant quantity, the setup routine has been re-designed to deal with the limitation imposed by the ADC input range which clips signals over 1024 counts (~1.024V). Hence link gain can always be calculated, regardless of whether the APV tick fits in the ADC range.

There are two basic considerations that guide the gain and bias selection for the link:

1. The overall gain of each readout link should be set as close as possible to 0.8V/V, in order to optimize the dynamic range spread of the system close to the specification[1].
2. The laser bias current must be set high enough to avoid clipping of the APV signals from the bottom (i.e. the laser transmitter must always be driven above threshold by the modulating APV signals). Moreover, it is desirable to set the bias to the lowest setting that achieves the above requirement, since the noise generally increases with higher laser current. Finally, the operating point of the laser must be such that temperature fluctuations do not force the transmitter to lase below threshold.

### 3.2.1 Algorithm Description

DATA REQUIREMENTS

The APV is first set to output synchronization 'tick marks'. In this mode, 25ns pulses with a height equivalent to 8MIPs (in 320μm detectors) are produced every 3500ns [1]. Since the data from two APV25s are multiplexed by the APVMUX chip, the synchronization ticks from two APVs appear combined to form a 50ns long pulse. The laser bias is scanned from $I^2C$ setting 0 to 50 (~0-22mA), and a frame of data containing one tick, sampled at 40MHz, is collected by the FED. The level (in ADC counts) of the tick's base and top is then determined for each bias point. This must be done for all four gain settings on the AOH, in order to find the gain setting giving an overall link gain closest to 0.8V/V. Once the setup routine has completed, there will be four 'top' and four 'base' curves for each optical link, corresponding to each of the four AOH gain settings (Figure 3.4).

[1] 0.8V/V refers to the target gain required to achieve the specified dynamic range for the Tracker optical links during normal operation. Users will be able to select the target gain (and hence dynamic range) as dictated by their specific needs.

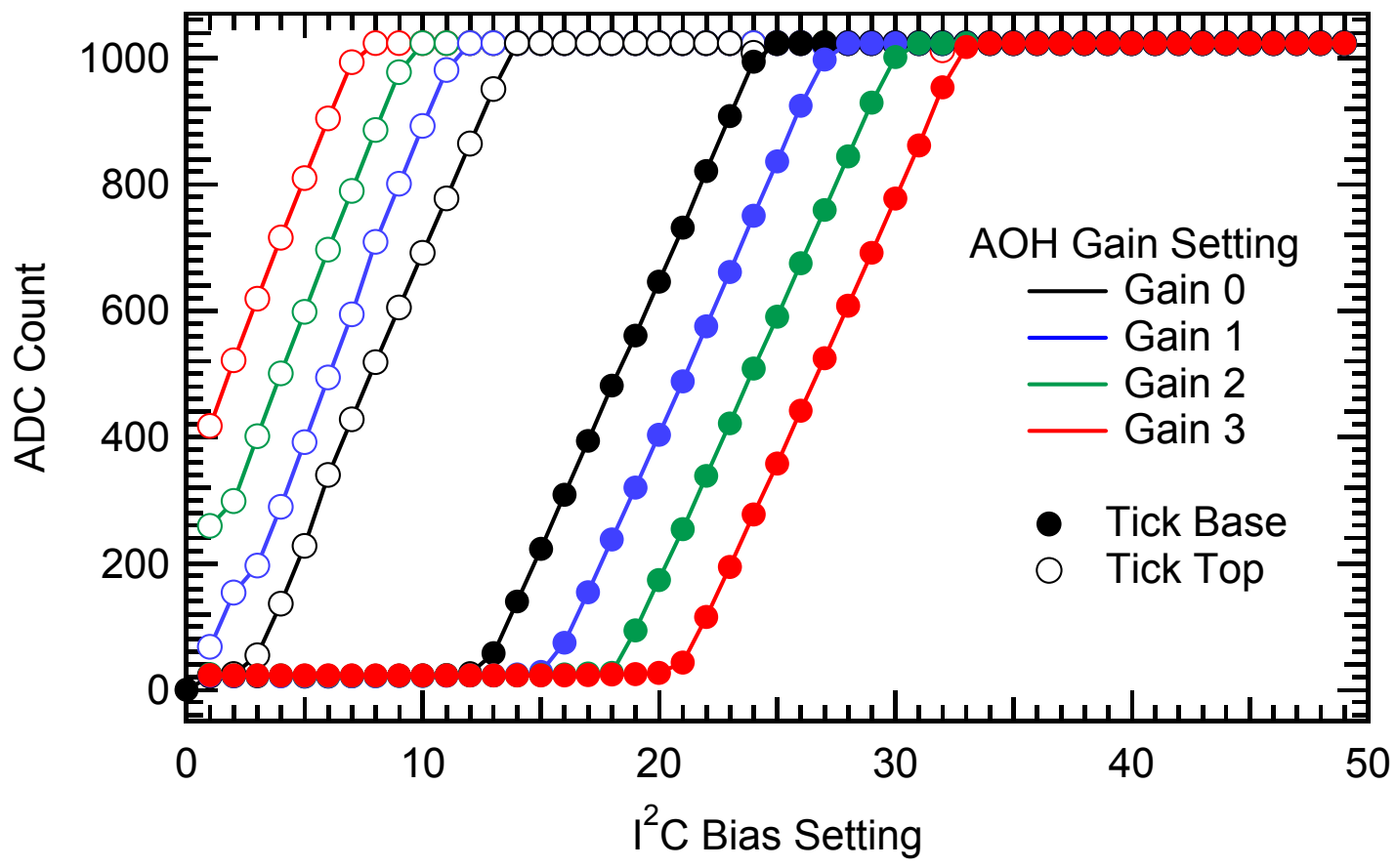

**Figure 3.4:** Typical setup routine data for one optical link, showing top and bottom curves for all four AOH gain settings.

It is important to note that the automatic setup routine for the final system relies on having first adjusted the FED's TrimDAC, which allows level shifting of the incoming signals to match the FED's ADC input range. This adjustment must be done with all the lasers turned off. As explained earlier, the zero light level must be adjusted to be visible within the ADC range (i.e. ~10-50 ADC counts), so that the user can distinguish when the base of the tick 'clears' the laser threshold as the bias setting is increased.

### GAIN CALCULATION

The gain of the readout link is determined by measuring the height of the APV tick at the output of the link. It assumes that the APV digital header is 800mV at the input of the optical link. The algorithm described here works for any gain, even if the APV tick is larger than the ADC range. For the following illustrative example, refer to Figure 3.5 that shows data obtained for one AOH gain setting of one optical link. The level of the tick's base and top is plotted as a function of $I^2C$ (laser) bias setting, for AOH gain setting 2. The data was obtained from a setup run of the TOB Cosmic Rack in the October 2004 test beam at X5, CERN.

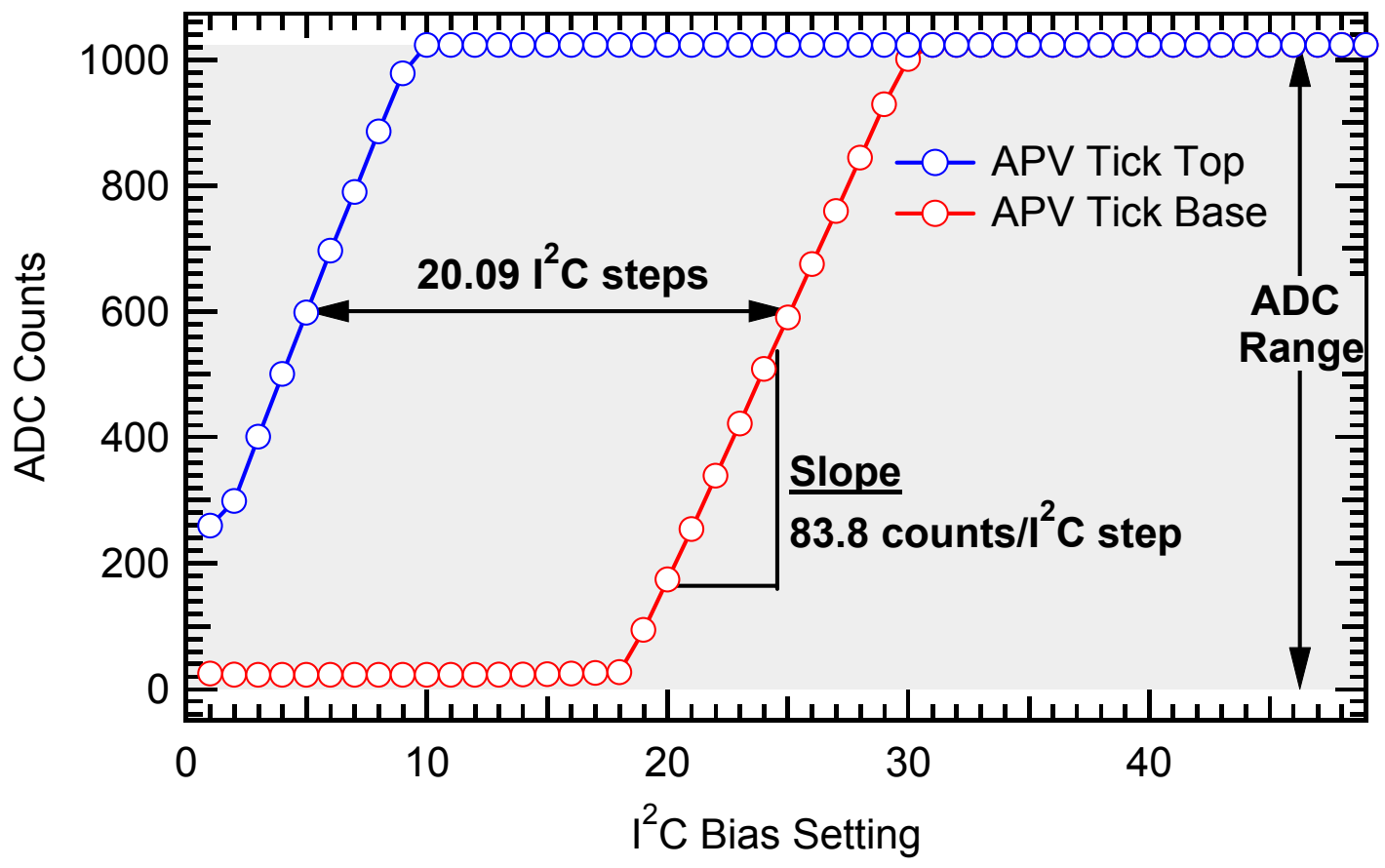

**Figure 3.5:** Example of gain calculation for AOH gain setting 2 of one optical link in the TOB Cosmic Rack.

In the example, the APV tick does not fit in the FED's ADC range. This is obvious from the fact that the top of the tick reaches saturation at $I^2C$ step 10, long before the tick base appears in the ADC range ($I^2C$ step 18). The height of the tick can still be determined by recognizing that a given ADC level is crossed by the tick base a constant number of $I^2C$ steps 'after' the top. Setting this reference level to 600 ADC counts, and by interpolation on the plots of Figure 3.5, the base 'follows' the top by 20.09 $I^2C$ steps. This is essentially the height of an APV tick in terms of input current, in $I^2C$ bias steps. From the slopes of the two plots, this value can be converted to ADC counts: 83.8 ADC counts/$I^2C$ step * 20.09 $I^2C$ steps = 1683.5 ADC counts. Given the height of the APV tick (800mV), the gain of this particular link at AOH gain setting 2 is 1683.5/800=2.10V/V.

### BIAS POINT SELECTION

The selection of the optimum laser operating point should be carried out after the appropriate AOH gain has been selected. With the sole constraint being that the laser is always driven above threshold, the bias selection algorithm should be immune to the effect of saturation at the top of the FED ADC's range due to high link gain, which was the case with the previous setup routine.

The same data used for the gain selection is required. As an example, consider the data shown in Figure 3.5. It is assumed that AOH gain setting 2 was selected by the gain selection algorithm. As explained in section 3.1, the clipping at the bottom of the base curve occurs due to the laser being driven below threshold by the modulating current from the LLD. From the previous discussion, the optimum bias point of operation occurs at the lowest $I^2C$ setting in which the base of the APV tick is clearly visible in the FED ADC's range (this can be referred to as 'base lift-off' to ease subsequent explanations). This ensures that the modulating current is always above laser threshold.

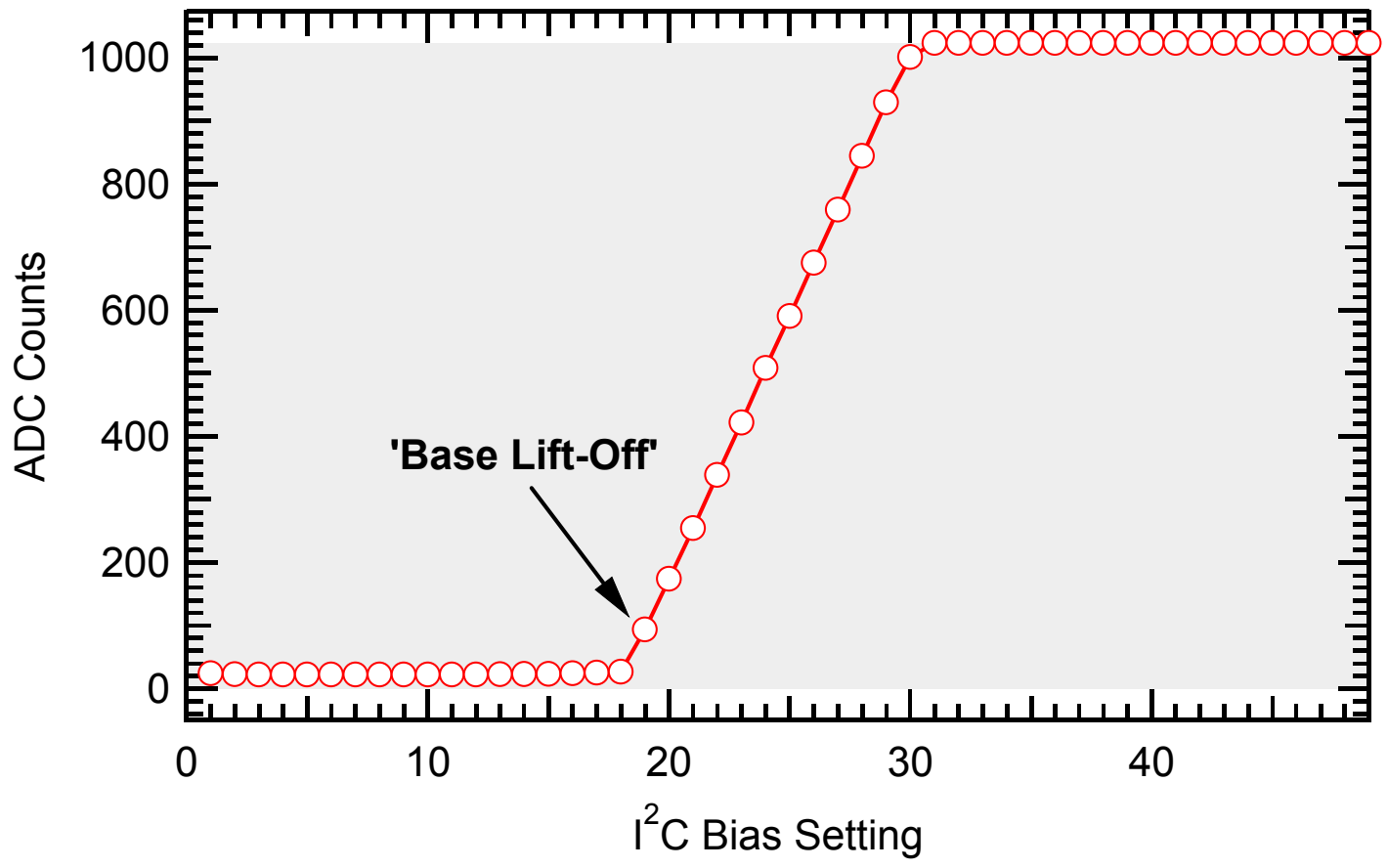

**Figure 3.6:** APV Tick Base Vs Laser Bias Current

It follows that the only data needed to find the optimum bias point is the plot of APV tick base Vs $I^2C$ setting (Figure 3.6). The algorithm needed to locate the optimum bias point from the plot of Figure 3.6 will not be specified exactly, since there can be numerous implementations that achieve similar results. As an example, the algorithm could be based on taking the double derivative of the base versus bias current plot. This yields two distinct 'spikes' indicating changes in gradient of the original plot. The first spike designates the $I^2C$ bias point where 'base lift-off' occurs (Figure 3.7), which can be determined with a simple level-finding algorithm. In this case, the algorithm used for finding the derivatives is such that 'base lift-off' is the first point after the top of the spike. Of course, this is implementation-dependent, and this example is for illustrative purposes only.

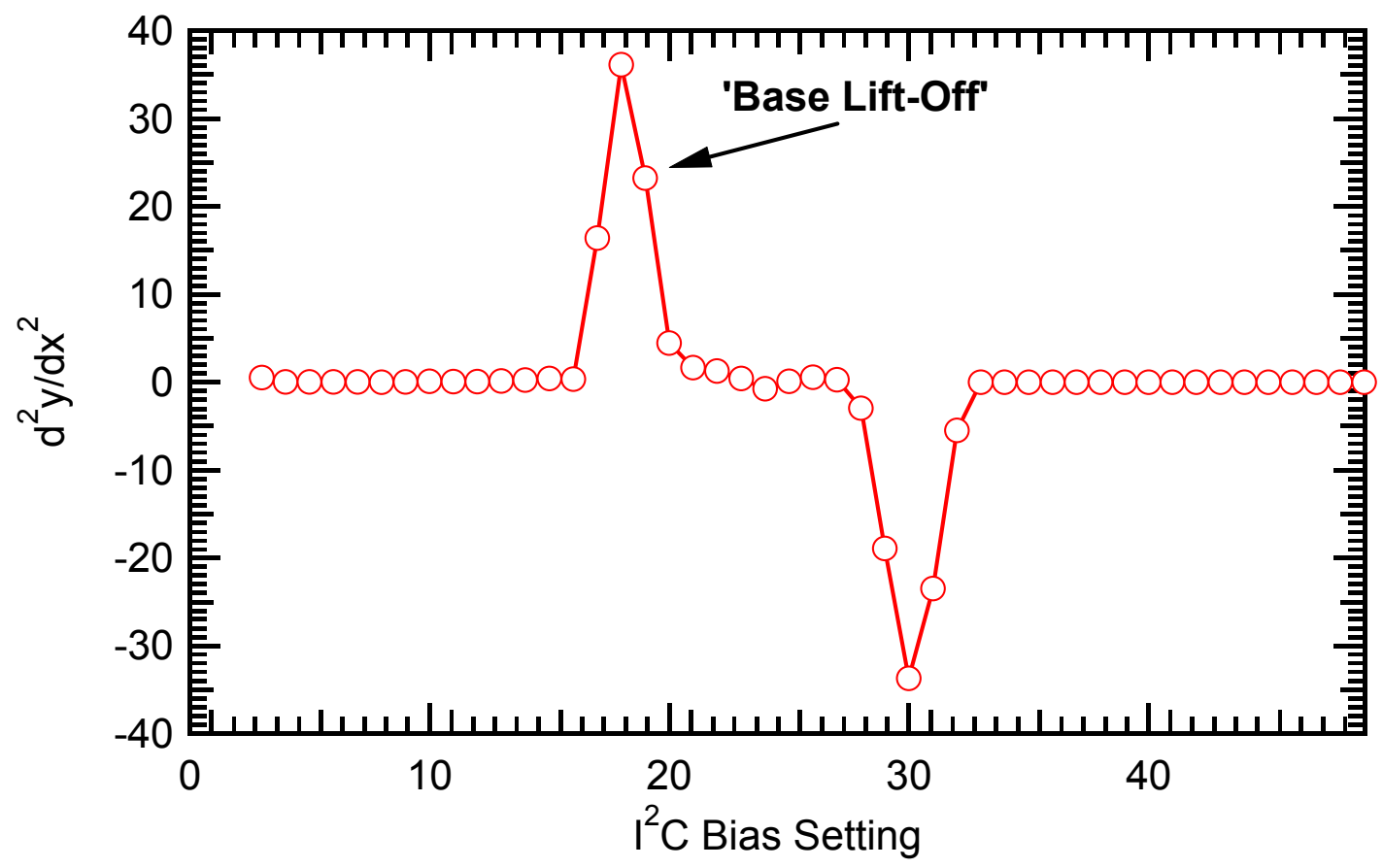

**Figure 3.7:** Double derivative of the APV tick base plot of Figure 3.6

### Bias Point Dependence on Temperature

The optimal operating point of the laser is obviously dependent on the laser's threshold. The threshold itself is affected by temperature. It follows that the temperature must remain fairly stable if the laser threshold (and hence the bias point) is to remain constant during normal operation. Otherwise, since the bias point is chosen just above laser threshold, it is possible that a large enough temperature increase can shift the laser's output below threshold. It is therefore important that the bias point selection algorithm is such that allows sufficient margin above laser threshold to mitigate the effect of temperature fluctuations expected in the final CMS Tracker environment.

Tests performed on the AOH (see Chapter 4 for more details) have shown that the laser threshold increases by ~0.2 $I^2C$ steps/°C for the temperature range between -15°C and -5°C. To calculate the margin required, we will assume the temperature on the laser will remain stable within ±2°C (worst-case scenario) during regular operation[2]. Hence, if the temperature increases by a maximum of 2°C, the bias point should be chosen to be at least 0.4 $I^2C$ steps (2 °C × 0.2 $I^2C$ steps/°C) above the laser's threshold. This can be achieved by first finding the 'base lift-off' point as described previously, and then checking if the margin is met. If not, the next highest $I^2C$ should be selected as the laser's operating point. To illustrate the procedure, consider the example of Figure 3.8. Due to the finite granularity of the

[2] There is no specification in the CMS Tracker TDR regarding temperature stability on the AOH. This could be determined empirically, depending on the cooling system of each Tracker subsystem. The users of each subsystem should then decide on the margin required and incorporate this into their link setup routines, following the example given here.

I$^2$C bias steps, the base lift-off does not necessarily coincide with the laser threshold, as illustrated in the zoomed portion of Figure 3.8. The exact point at which lasing starts (i.e. the threshold) occurs between I$^2$C steps 17 and 18 in the example. This 'real' threshold can be found by fitting a straight line to the base curve and finding its intersection with the zero-light level (~36.5 ADC counts in this case). The equation of the resulting fit is:

$$y = 79.7x - 1377.2 \qquad (3.3)$$

Hence, if one sets $y=I_{th}$=36.5, the laser threshold can be determined in I$^2$C steps:

$$I_{th} = \frac{36.5 + 1377.2}{79.7} = 17.7 \text{ I}^2\text{C steps} \qquad (3.4)$$

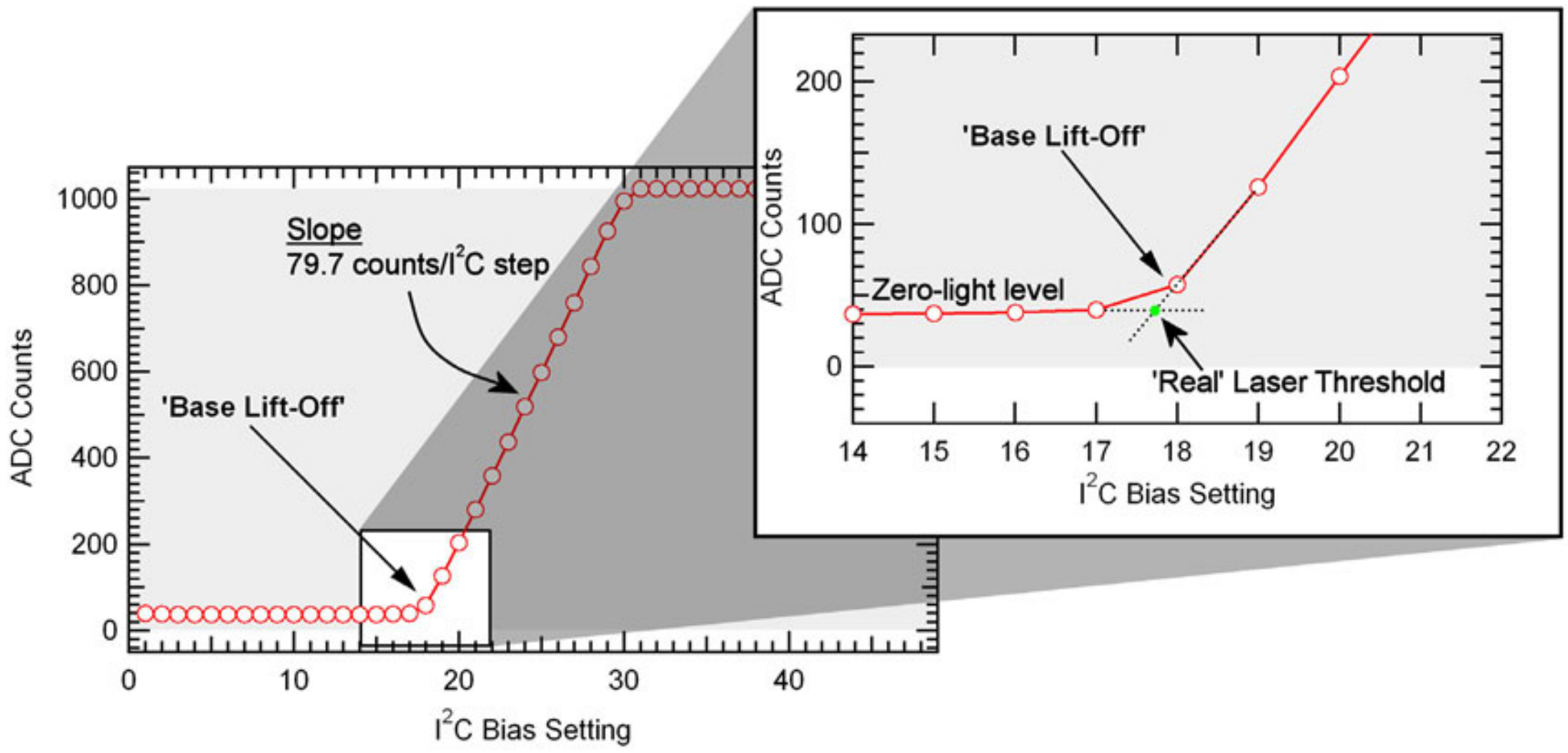

**Figure 3.8:** Showing a typical case where 'base lift-off' occurs very close to the threshold, leaving little margin for mitigation of temperature effects.

The base lift-off point (I$^2$C step 18) is only 0.3 I$^2$C steps above the laser's threshold, in this example. If this were selected as the operating point, it would violate the requirement for a margin of 0.4 I$^2$C steps. If the laser temperature increased by 2°C, it would be driven from below threshold, which is an unwanted situation. Hence, to ensure proper operation the setup routine should select the next I$^2$C step (=19) as the optimum operating point. It should be emphasized that the exact margin of operation should be set according to the thermal stability of the final system, and the one used in this example is for illustrative purposes only.

## 3.3 Summary

A comprehensive description of the readout link's functionality has been given. The link's tuning parameters have been detailed and their impact on the set-up of the optical links was explained.

New algorithms for the set-up of the optical links have been developed. Particular emphasis has been put on being able to calculate the gain regardless of environmental conditions, and a detailed guideline for the selection of optimum laser bias point was given. The new setup routines are currently being incorporated in the online software of the Tracker's readout system.

# Chapter 4

## PREDICTING THE GAIN SPREAD OF THE CMS TRACKER READOUT OPTICAL LINKS

*Approximately 40 000 analog optical links will read out the data from 10 million silicon microstrips in the CMS Tracker. In an analog system, the overall gain affects the dynamic range and resolution of the data being read out. Production is sufficiently advanced to allow the extraction of the real distribution of gain for each component making up the complete optical link. The purpose of this study is to examine the aggregate effect of the individual component gain distributions on the readout system's dynamic range, and its uniformity throughout the thousands of deployed links in the CMS Tracker. To this end, a Monte Carlo simulation based on production test data, and augmented with results from deployed links in real test systems, has been carried out. The results give an estimate of the spread in gain and dynamic range that can be expected in the final system, running at -10ºC.*

## 4.1 Introduction

One of the most important parameters in an analog readout system is the gain, which directly affects the size and resolution of detector signals that can be captured at the output of the link. During normal operation, 8 bits (256 counts) of data will be retained in the readout chain's back-end (i.e. the FED) for transmission to the data acquisition system. The physics goals of the Tracker require that signal sizes up to (the equivalent of) 3.2MIPs from the thin 320μm detectors are visible in the data that is read out. A MIP traversing a detector corresponds to a signal of 100mV at the input of the optical link. Since 1 FED ADC count = 1mV, the gain of the link must be 0.8V/V for a 3.2MIP signal to be visible within 256 counts of FED ADC data (320mV*0.8=256mV). Hence, for uniform performance throughout the Tracker optical links, it is desirable to have an equalized gain spread as close as possible to the specified gain of 0.8V/V.

The overall link gain is determined by the aggregate effect of all constituent components. The work presented in this document will determine the distribution of gains that can be expected in the final CMS Tracker readout system and the effect this has on the dynamic range is explored.

The present study is an extension of the work presented in [1], where the distributions of the individual component gains were assumed to be uniform within their specifications. The previous study was essentially a 'worst-case scenario' as far as the total gain spread is concerned. The main difference in the current study is that quality assurance data from production testing is used to compile histograms of the gain for each component. A Monte Carlo simulation that samples the real data is used to produce the most realistic prediction for the link gain spread to date.

Section 4.2 reviews the work presented at the 10$^{th}$ Workshop on Electronics for LHC and Future Experiments [2], which constituted the first use of component production test data to predict the final system gain spread by simulation. Since production tests take place at room temperature, the first iteration of the simulation could only be used for predicting the gain spread at room temperature. The results obtained via simulation are compared to those from a deployed TEC system in the May 2004 test beam [3].

Section 4.3 details the gain spread observed in optical links deployed in test beam systems, comprising the full readout chain components. The results from the October 2004 test beam using the Tracker Outer Barrel (TOB) Cosmic Rack (CRack) [4] were used to establish the variation of gain with temperature. This was facilitated by the presence of a cooling system in the CRack with the ability to operate at sub-zero temperatures. The accuracy of the gain calculation method is validated using physics data from the test beam.

Having calculated the average variation of link gain with temperature, the simulation of optical link gains has been updated and the results appropriately scaled to give the final prediction of dynamic range spread for the nominal Tracker operating temperature. The results are detailed in section 4.4.

### 4.1.1 Principle of Gain Calculation Method

As described in Chapter 2, the analog detector signals are encapsulated in a frame which includes a digital header (see Figure 2.3). It is important to note that the digital header height is not affected by any of the APV's settings. It is thus a constant for the duration of CMS operation, relative to other parameters in the front-end electronics. The optical link setup routine makes use of this quantity for calculating the link gain (and hence picking the appropriate AOH gain setting). This is essentially an approximation of the real 'particle gain', which is the quantity which is of most importance to the Tracker readout system. The term 'particle gain' refers to the amplification (or attenuation) of the signals arising from particles traversing the detectors.

Furthermore, it is worth noting that the APV header height (800mV) exceeds the specification for optical link linearity of 600mV (differential), referred to the input of the link [5]. Hence, it could be expected that signal compression in the link would compromise the accuracy of any technique based on measurements of APV tick height. The relationship between the particle gain and that calculated from the setup routine is briefly introduced in section 4.3.1. There is no evidence of non-linear degradation in the results obtained.

During an optical link setup run (often referred to as 'gain scan'), the APV outputs synchronization pulses (or 'ticks') [6]. These are identical in height to the digital header. The FED captures the transmitted APV ticks and the height of the tick is measured. Based on the assumption that the tick height is 800mV at the input of the link, and given that each FED ADC count corresponds to 1mV, the overall readout link gain in V/V is determined by simply dividing the output height in ADC counts by 800.

## 4.2 Predicted Optical Link Gain Distribution from Production Data

The ~40 000 optical links built from electronic, optoelectronic and optical components must match the performance requirements of the overall readout system in terms of dynamic range and resolution. The readout links are required to transmit 3.2MIPs with 8bit resolution[1] [1, 5]. Process monitoring of the components' production has yielded a statistically significant set of test data to allow real distributions of the performance parameters to be extracted. The or probability density functions (PDFs) used are shown in Figure 4.1. The overall link gain spread is determined by the aggregate effect of all constituent components' gain distributions.

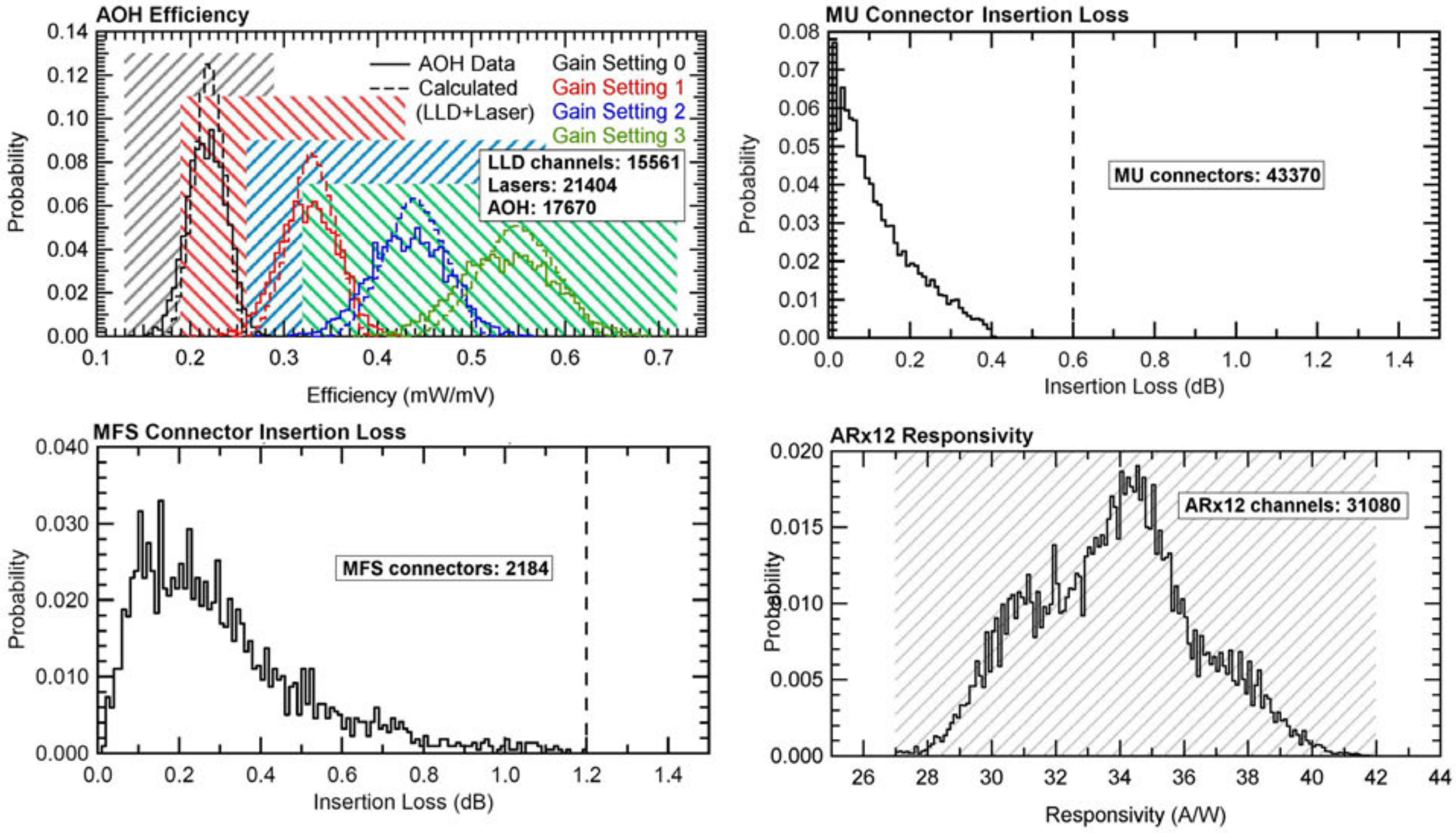

**Figure 4.1:** PDFs of the component gains used in the Monte Carlo simulation. Shaded areas and dotted lines indicate the limits of the specifications. The number of channels used in each case is also indicated.

[1] For 320μm silicon strip detectors.

A Monte Carlo simulation has been produced to compute the complete link gain distribution from the available component production test data. On the transmitting side, the LLD on the AOH was designed with switchable gain settings to compensate for component gain tolerances. Four gain settings are available, with nominal values of 5.0, 7.5, 10.0 and 12.5mS, allowing a certain amount of gain equalization in the final system. The effect of switching on the gain distribution is investigated. The results obtained are detailed in [2], but will be reviewed in this section, along with the method followed.

### 4.2.1 Simulation Method and Room Temperature Results

The simulation model includes the gain spreads of all optical link components, including the insertion loss of the connectors at each patch panel (Figure 4.2). The ARx12 output resistor value is assumed to be 100Ω for this simulation[2]. The inverse transform method [7] is used to sample the gain distributions of each of the components. In each iteration of the Monte Carlo simulation, the samples are multiplied together to obtain four overall link gains, each one corresponding to one of the four AOH (LLD) gain settings. The simulation can be run using equalization (i.e. selecting one of the four gains that is closest to the nominal target gain of 0.8V/V) or by simply selecting the gain corresponding to the same AOH setting for every link. The process is illustrated in the flow chart of Figure 4.3.

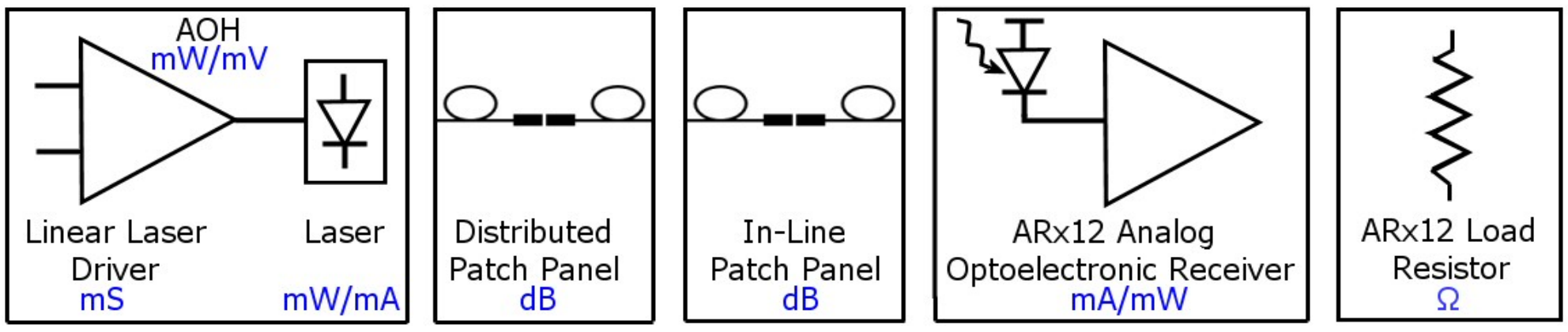

**Figure 4.2:** Optical link components used in the Monte Carlo simulation.

The simulation was first run for each of the four gain settings of the LLD, without any attempt at equalization. The resulting distributions are roughly Gaussian (Figure 4.4). The nominal gain of 0.8V/V is also shown on the plot (dashed line). There is a very large gain accessible range using the available LLD settings. The position of the distributions with respect to the 0.8V/V specification suggests that lower gain links can be better compensated by using higher gain settings. The high-end tail of the gain 0 trace exceeds the nominal gain, and since these links are already at their lowest setting, they cannot be further compensated. Figure 4.4 also shows the relationship between the link gain in V/V and the height of the APV tick in FED ADC counts. The scale assumes that an APV tick is 800mV at the input of the AOH (or 8mA output from the APV), and that the gain of the FED ADC is 1 ADC count/mV [8].

[2] The Monte Carlo simulation was first carried out when the ARx12 load resistor value was 100Ω (i.e. that of the pre-production FEDv1 board).

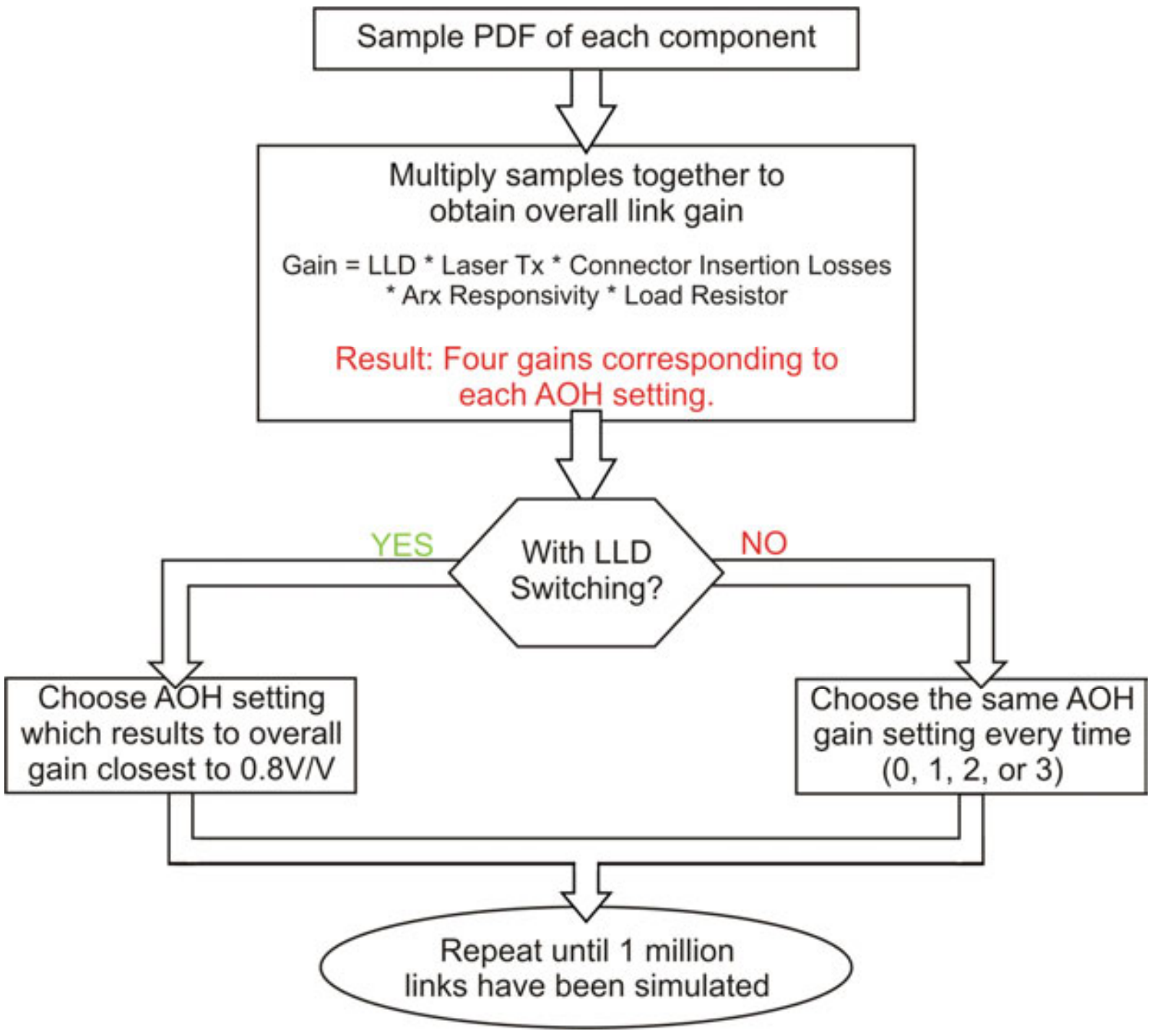

**Figure 4.3:** Flow chart for the Monte Carlo simulation of optical link gain.

The simulation was also run incorporating the ability to switch between LLD settings in order to equalize all gains as close as possible to the nominal value. Figure 4.5 shows the resulting spread. The laser driver switching process can be thought of as cutting into the source distributions (i.e. the single-gain distributions of Figure 4.4) and selecting the slices centered on 0.8V/V. This is best visualized in Figure 4.4, where the switched spread is superimposed on the 'single-gain' distributions (shaded area). Figure 4.5 shows the switched gain in more detail. The distribution has two distinct boundaries equidistant from the target gain value, with the lower limit at ~0.64V/V, and the upper limit at ~0.96V/V.

It should be noted that there are very few links that lie above the upper limit in the switched gain distribution, and are not visible in the plot. This is due to the tail of the Gain 0 distribution that exceeds the limit of 0.96V/V, and represents 0.007% of the links. Since one million link gains were simulated, these are statistically insignificant as far as the real system is concerned, where there are comparatively much fewer links (40 000). These outliers are already at their lowest gain setting, and cannot be further compensated. Clearly, it is not an ideal situation to have any part of the Gain 0 distribution above the upper limit.

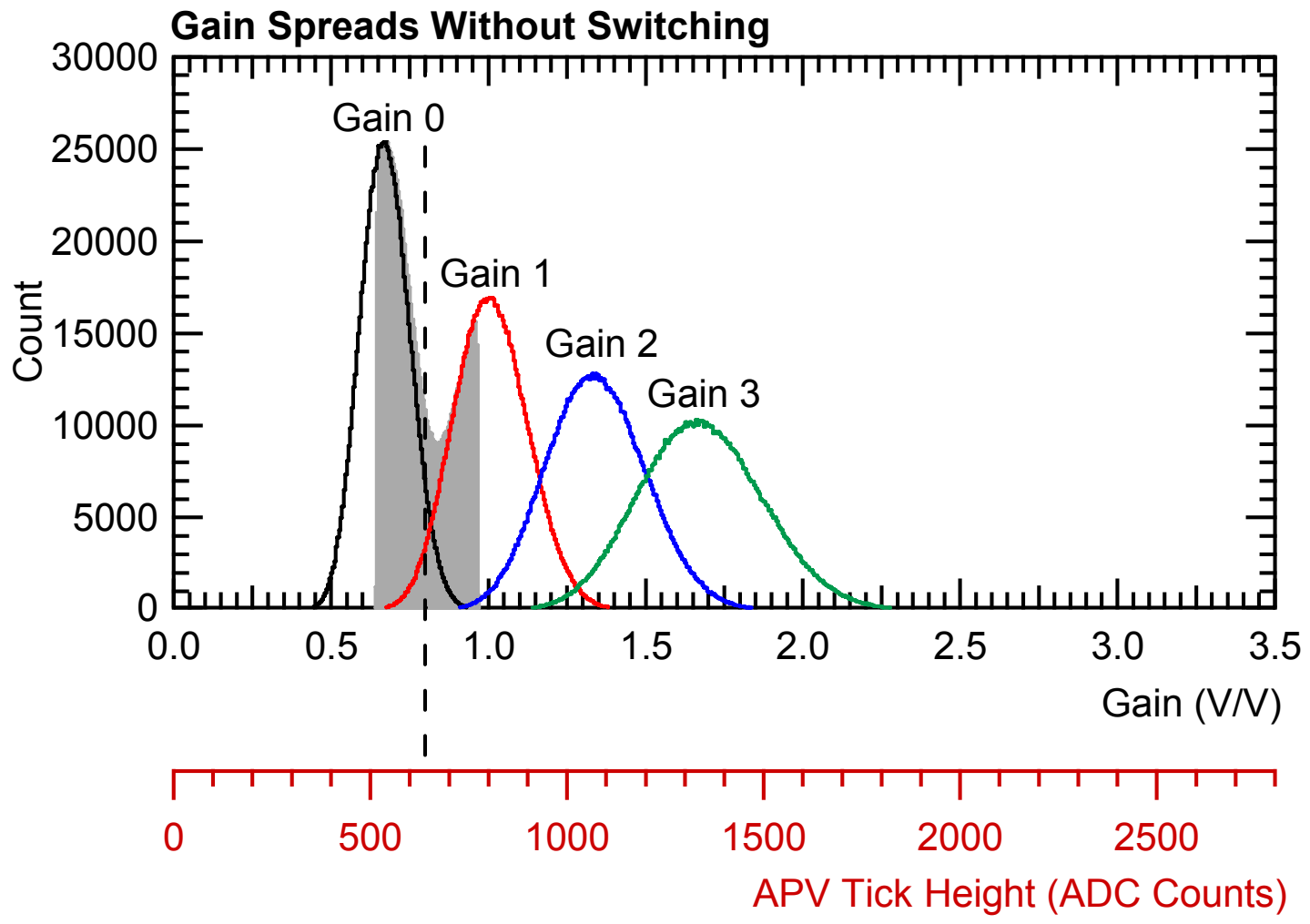

**Figure 4.4:** Showing the 'single gain' spread distributions predicted by simulation using real production data without switching of the LLD. The shaded area shows the distribution resulting after equalization using the 4 available gain settings.

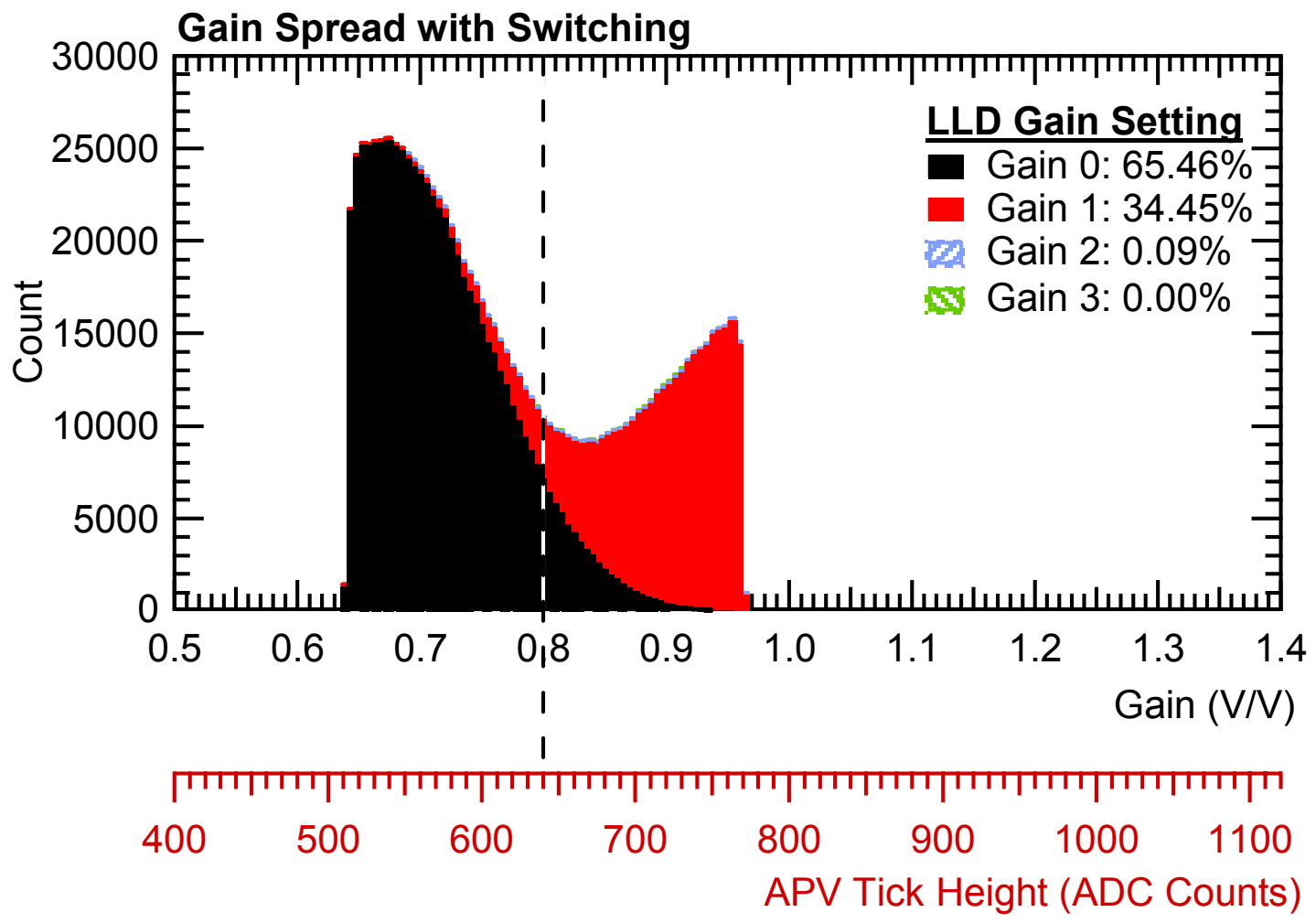

**Figure 4.5:** Equalized link gain distribution obtained by switching of the LLD, showing the contributions from each gain setting.

The maximum spread of the equalized gain distribution can be determined analytically given the simple switching algorithm that selects the gain closest to the target value. For the following analysis the upper extent of the equalized distribution is denoted by $u$, while the lower limit is $l$ (in V/V). We begin by imposing the restriction that the high-end tail of the gain 0 distribution (see Figure 4.4) must be lower than $u$ and the low-end tail of the gain 3 distribution must be above $l$. In addition, we will assume that the dominant gain settings used when equalizing are settings 0 and 1, since this gives the maximum spread (this will occur when equalization around the target gain is achieved by using mostly these two settings). Since the equalization algorithm selects the gain closest to the target value of 0.8V/V, it follows that the extents of the equalized distribution will be equidistant from the target:

$$\begin{aligned} &0.8 - l = u - 0.8 \\ &\Rightarrow u + l = 1.6 \end{aligned} \qquad (4.1)$$

The resolution of the gain settings determines the worst-case difference between the lowest and highest equalized link gains. From production test data of the LLD chip [2] it is known that the four available gain settings are (on average) 5.38, 8.06, 10.74 and 13.41mS (see Figure 4.6). Hence, the maximum ratio of successive gain settings occurs between gain setting 0 and gain setting 1, and is 1.5 (8.04/5.38~1.5). Any link that has a gain less than $l$ clearly will have to be set to the next highest gain setting to achieve a gain closer to 0.8V/V. Similarly, any gain over $u$ must be set to the next lowest gain setting. This implies that, at worst, the relationship between $u$ and $l$ is given by:

$$u = 1.5l \qquad (4.2)$$

Equations (4.1) and (4.2) can be solved simultaneously to obtain l=0.64V/V and u=0.96V/V. It has been shown that, as long as the high-end tail of the gain 0 distribution is below 0.96V/V and the low-end gain of the gain 3 distribution is above 0.64V/V, the equalized gain distribution will lie between 0.64 and 0.96V/V.

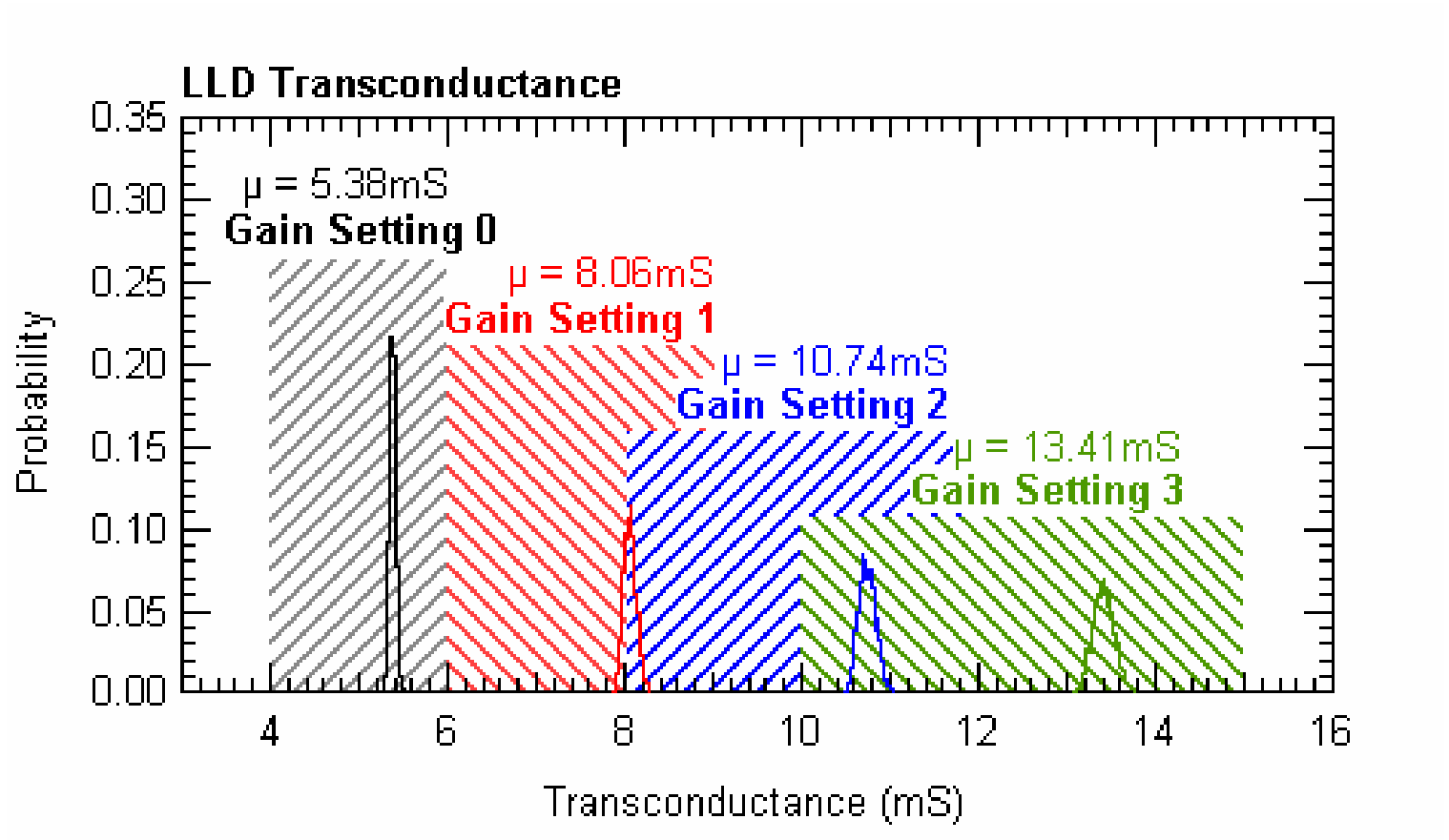

**Figure 4.6:** PDFs of Linear Laser Driver transconductance for each of the four gain settings selectable on the AOH. Shaded areas indicate the corresponding specification boundaries

Having established the gain distribution, the spread in dynamic range can be determined. The impact of the switched link gain spread on the dynamic range of the readout system is illustrated in Figure 4.7. After the analog signals are transmitted through the optical links, they undergo digitization by a 10-bit, 1.024V input-range ADC [8] on the FED in the counting room. Digitization and processing (including pedestal subtraction) is performed, and the two MSBs of the data are discarded[3]. It is therefore useful to look at the signal size that can be transmitted by the system using 8 bits, before clipping occurs. The signal size is in terms of electrons and is easily related to MIPs for both thin and thick detectors. Hence the dynamic range figure of merit can be expressed in electrons/8bits or MIPs/8bits[4].

[3] This is the case in the current FED firmware; in the future it may be possible for the users to select which data bits they wish to retain in the captured data.
[4] Note that these two quantities depend on the detector thickness. This is reflected in the additional axes of Figure 4.7, where (an approximate) relation between signal size in electrons and MIPs for thin and thick detectors is given.

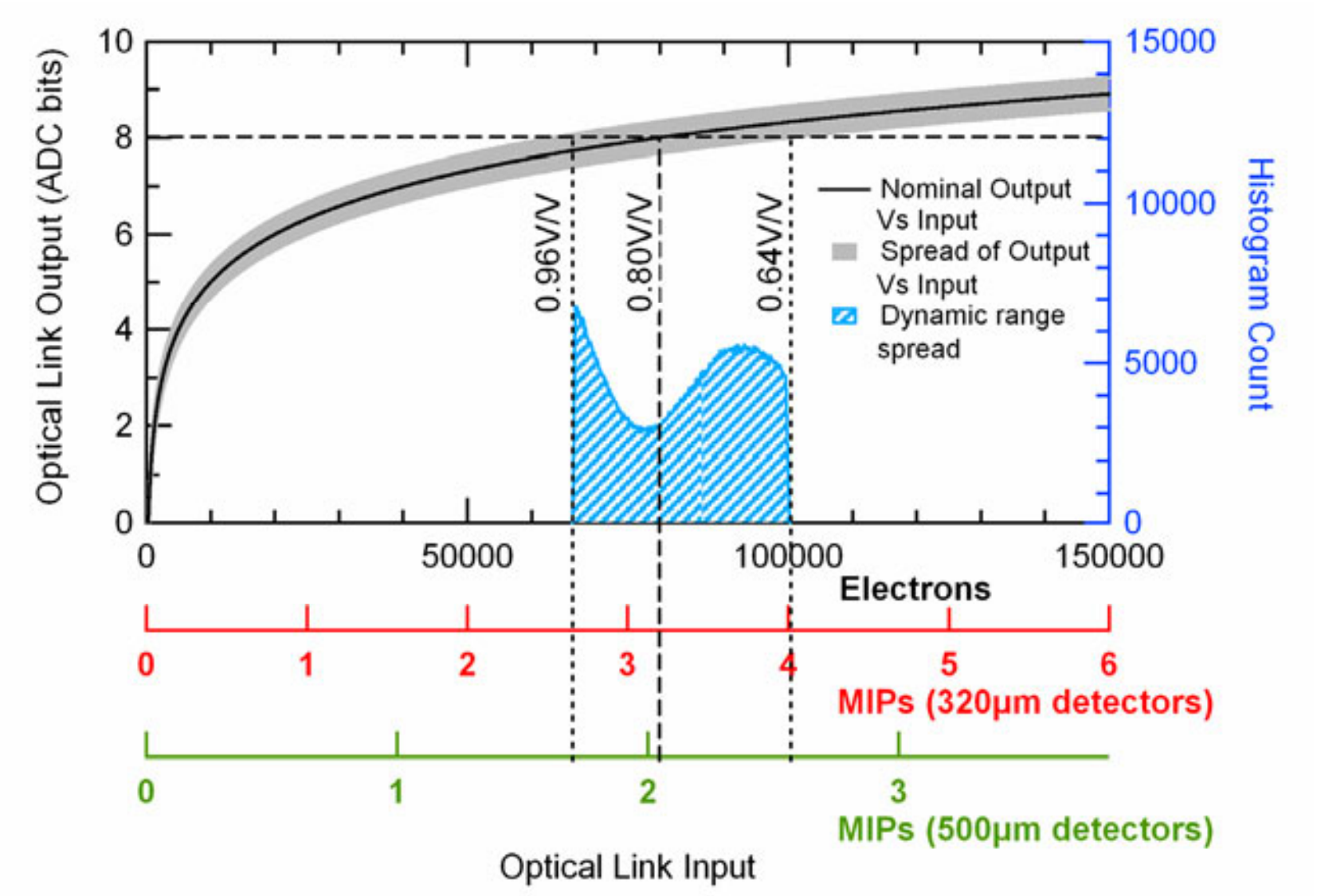

**Figure 4.7:** Showing Optical Link Output vs Input in ADC bits (left axis). The histogram (right axis) shows the spread in dynamic range.

The dynamic range spread can be determined from Figure 4.7, where the signal size at the link's output (in ADC bits) is plotted against the signal size at the link's input (in electrons and MIPs). By interpolation, it is possible to determine the signal sizes that can be accommodated in 8bits of the FED's ADC (horizontal dotted line). Hence, at the specified gain, signal sizes up to 80 000 electrons (3.2 MIPs) can be transmitted in their entirety. The shaded area around this line corresponds to the full range of gains predicted by the simulation, using equalization. Again, by interpolation, the maximum signals that fit in 8bits will range from ~66 000 to 100 500 electrons.

On the same figure, the spread in dynamic range (in electrons/8bits and MIPs/8bits) is also shown in the form of a histogram. There are a few statistically insignificant links with a dynamic range between 63 200 and 66 500 electrons/8bits. These are the same simulated links having gains over 0.96V/V in the switched gain spread (Figure 4.5). Ignoring these, the dynamic range of the links will lie between 66 500 and 100 500 electrons/8bits.

The signal sizes can also be interpreted in MIPs, assuming that 1 MIP produces 25 000 electrons in thin detectors, and 39 000 in thick detectors (see bottom axes of Figure 4.7). For thin detectors, the maximum signal sizes will be between 2.65 and 4 MIPs/8bits. The corresponding range for thick detectors is 1.7 to 2.6 MIPs/8bits.

### 4.2.2 Comparison to the Gain Spread from Deployed Optical Links at Room Temperature

The gains of real optical links deployed in the May 2004 test beam by the Tracker End Cap (TEC) subsystem [3] were compared to the results obtained by simulation. The TEC system link gains were calculated using the data obtained by the automated setup runs.

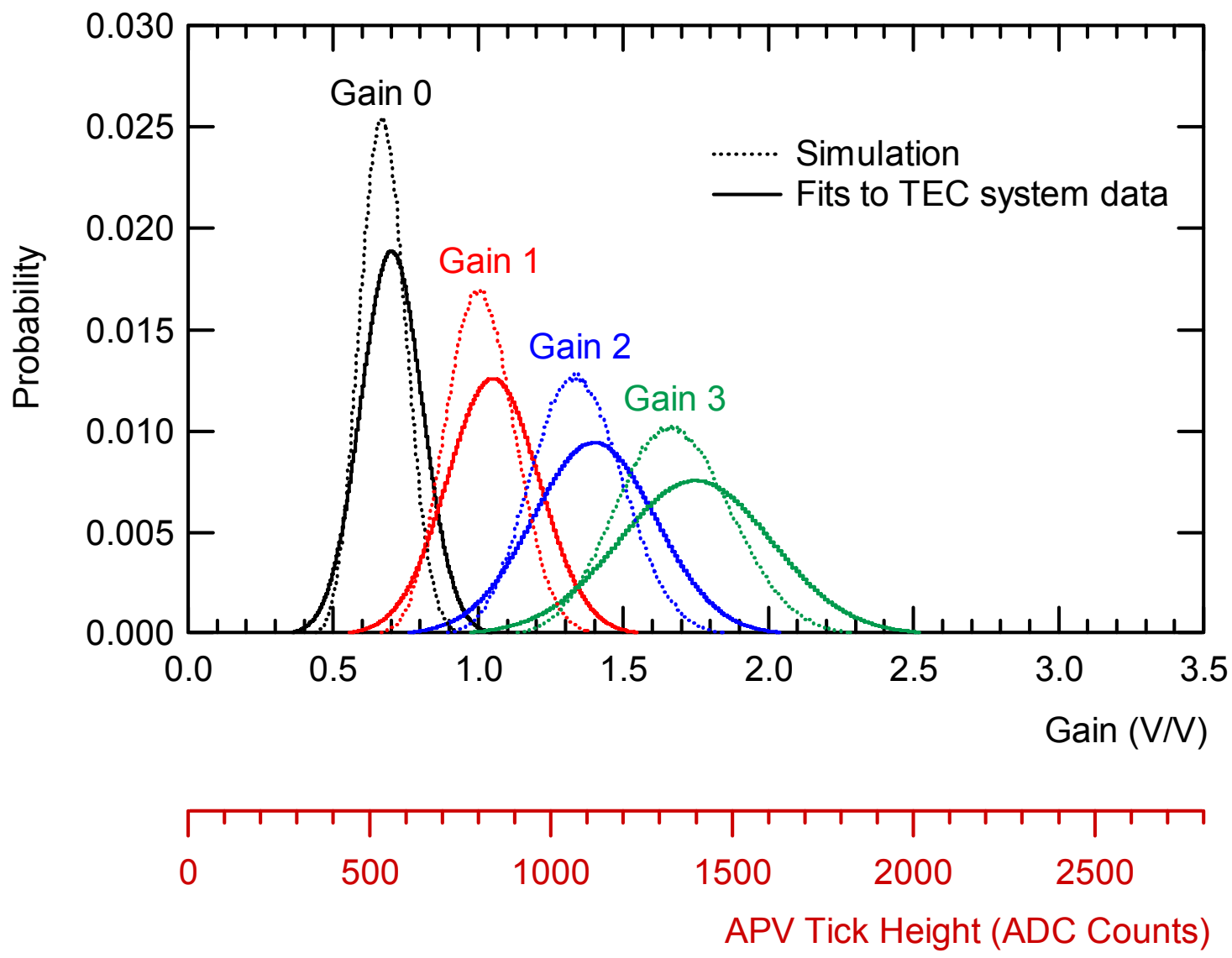

**Figure 4.8:** Comparison of the readout link gain distributions obtained by simulation and from the TEC subsystem in the June 2004 test beam.

The gains of 121 TEC optical links with no cooling were determined and histograms obtained for each AOH gain setting. Figure 4.8 shows the single-gain distributions obtained by simulation (dotted lines) and the Gaussian fits to the histogrammed data from the TEC system (solid lines). It is immediately obvious that the real system gains are slightly higher: The mean of the simulated gain 0 distribution is 0.67, compared to 0.69 from the TEC data. In addition, the spread is higher in the real links (standard deviation=0.15 versus 0.10). This is not surprising, given that there are components of the readout chain that have not been simulated (e.g. APV and APVMUX chips, as well as passive and active components on the Front End Driver board (FED) analog front end). The results show that there is good correlation between the simulation and the data obtained from real systems.

## 4.3 Measured Gain Distribution from Deployed Optical Links

While production testing of the link's constituent components takes place at 'room temperature' at several locations, the CMS Tracker will be operated at -10ºC (hybrid temperature) [9]. Hence, it is instructive to understand how the readout link gains –and hence the dynamic range– will be affected by the cold environment. The simulation based on production test data alone is not enough to quantify the optical link gain variation with temperature, since the production tests take place at room temperature.

The test beam in October 2004 at X5, CERN, presented an opportunity to study a significant number of real deployed links. The results presented in this section are from 99 operational links in the Tracker Outer Barrel CRack [4]. The cooling system of the CRack allowed the temperature dependence of gain to be studied. The gain calculation relies on data obtained by the online setup routines. It should be noted that the optical receiver's (ARx12) load resistor value was 100Ω (for the FEDs used in the test beam).

### 4.3.1 Accuracy of the Online Gain Calculation Routine

If the setup routine is to be used to draw useful conclusions about the gain spread, its accuracy must first be evaluated. The gain calculated by the online setup routine is derived from the APV header height at the output of the link. Therefore, its relationship with 'real' signals due to particles traversing the silicon detectors in the Tracker depends on sensible assumptions made about the output (and gain) of the APV. Clearly, the objective of the setup routine is to estimate the real particle gain as accurately as possible, using the most stable metric available.

The real particle gain of the readout links was measured by taking physics data on the CRack in the October 2004 test beam at CERN, X5. Histograms of the cluster signal in ADC counts at the output of each readout link in the CRack were obtained for three different runs. By performing a Landau fit for each link, the Most Probable Value (MPV) of the signal induced by a particle traversing the detectors was found. Assuming all detectors in the test system are the same, the value of the MPV for each link will depend on the readout system's gain. In order to have a standardized metric that is independent of the type of detector, all gains are calculated with reference to the signal size produced by a minimum ionizing particle (MIP) traversing the silicon detectors perpendicularly. This is referred to as 'MIP gain' in this document.

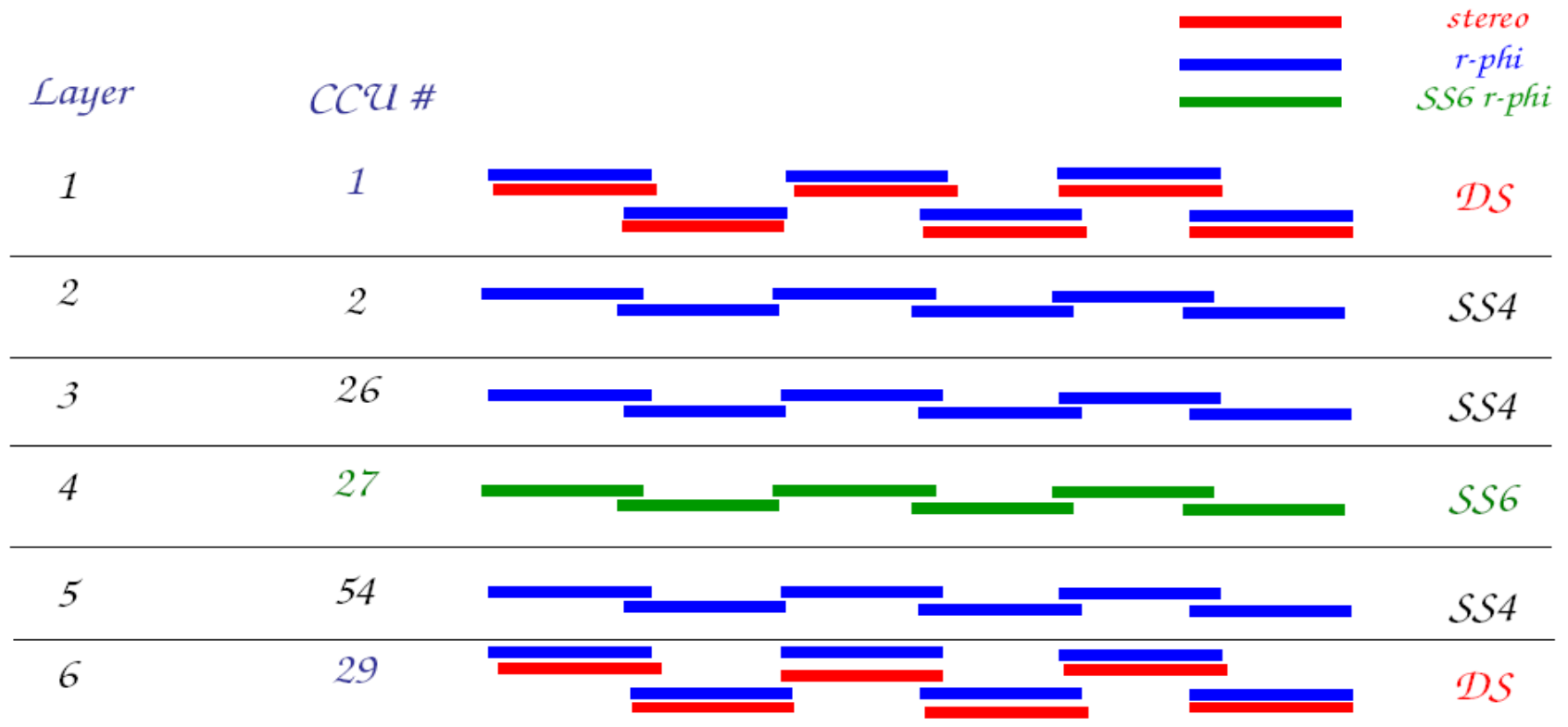

**Figure 4.9:** CRACK layers in the October 2005 test beam [4].

METHOD

The CRack consisted of 48 detector modules with 102 optical channels (of which 99 were operational). The modules were arranged in 2 double-sided and 4 single-sided layers (Figure 4.9). For each physics run, cluster signal histograms were compiled, with each histogram corresponding to one APV. Since the data from two APVs are multiplexed onto each optical link, there are two signal size histograms for a given optical link. The only difference between the two is the gain of the particular APV chip (since all other components are, of course, the same). In the following gain calculations, the two histograms per optical link were essentially combined by taking the average of the two gains calculated.

Data from one APV are shown here to illustrate the method used. The cluster signal histogram is shown in Figure 4.10. The MPV was determined by fitting a Landau curve, with the fit error also being reported.

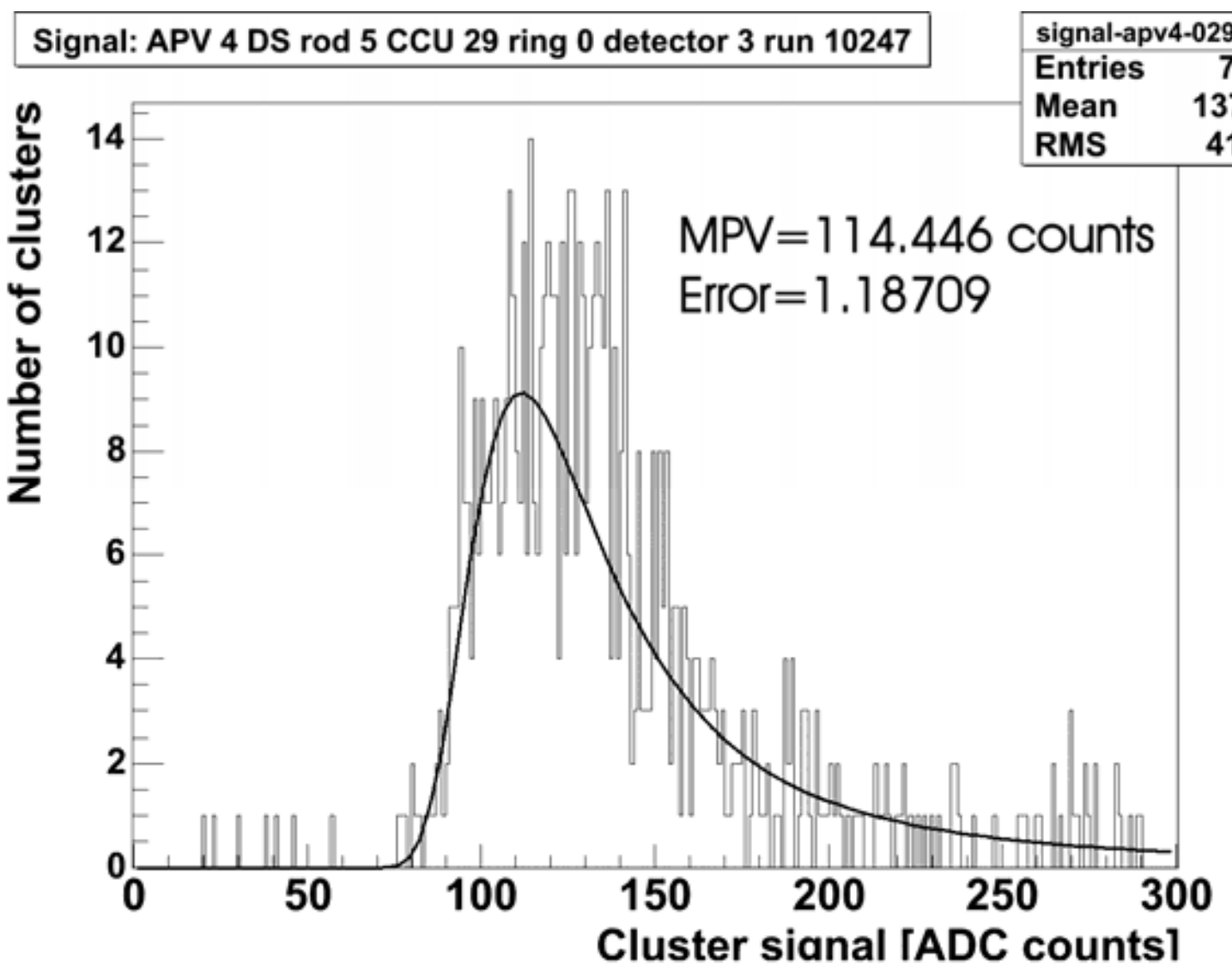

**Figure 4.10:** Cluster signal size histogram for one link.

In the example above, the signal cluster size has an MPV=114.5 ADC counts. However, this signal does not exactly correspond to the signal from a MIP, for the following reasons reasons:

- In all CMS Tracker test beams up to October 2005 the APV gain setting was not set to the nominal value corresponding to ~1.56mA/MIP for thick (500 µm) detectors and ~1mA/MIP for thin (320µm) detectors. Hence, the signal sizes seen in the test beams are approximately ~15% lower than what is expected in the final system, with the nominal APV setting. This factor is taken into account in the calculation.
- The particles in the X5 beam were not minimum ionizing. In the case of the pion beam, the momentum was 120GeV/c with a spread of ~1%. The muons originate from the decay of pions in the last 100m of the beam. Their momenta range between ~60 and 120GeV/c. Figure 4.11 shows the expected energy loss (dE/dx) of particles traversing various thicknesses of silicon, as a function of βγ. For both muons and pions at the momenta used in the test beam, the energy loss lies on the flat part of the dE/dx curve. Hence, relative to a MIP passing through 320µm of silicon, the particles at X5 deposited ~8% more energy.

After accounting for these factors, it is possible to calculate the signal size in ADC counts that would be produced by a MIP crossing a silicon microstrip detector in the CRack. In the example of Figure 4.10, this would yield 121.4 ADC counts. To enable a comparison with the gain calculated by the automatic routine based on APV header height, we can convert this number to V/V. The height of the APV header is 8mA, and this corresponds to 800mV at the input of the AOH. At the nominal APV setting of 1.56mA/MIP (thick detectors), the header height corresponds to a signal of 8/1.56=5.1MIPs for the detectors in the CRack. Hence, 1MIP=156mV at the input of the link. Since, at the link's output, 1 ADC count = 1mV, the gain of the example link is 121.4/156=0.78.

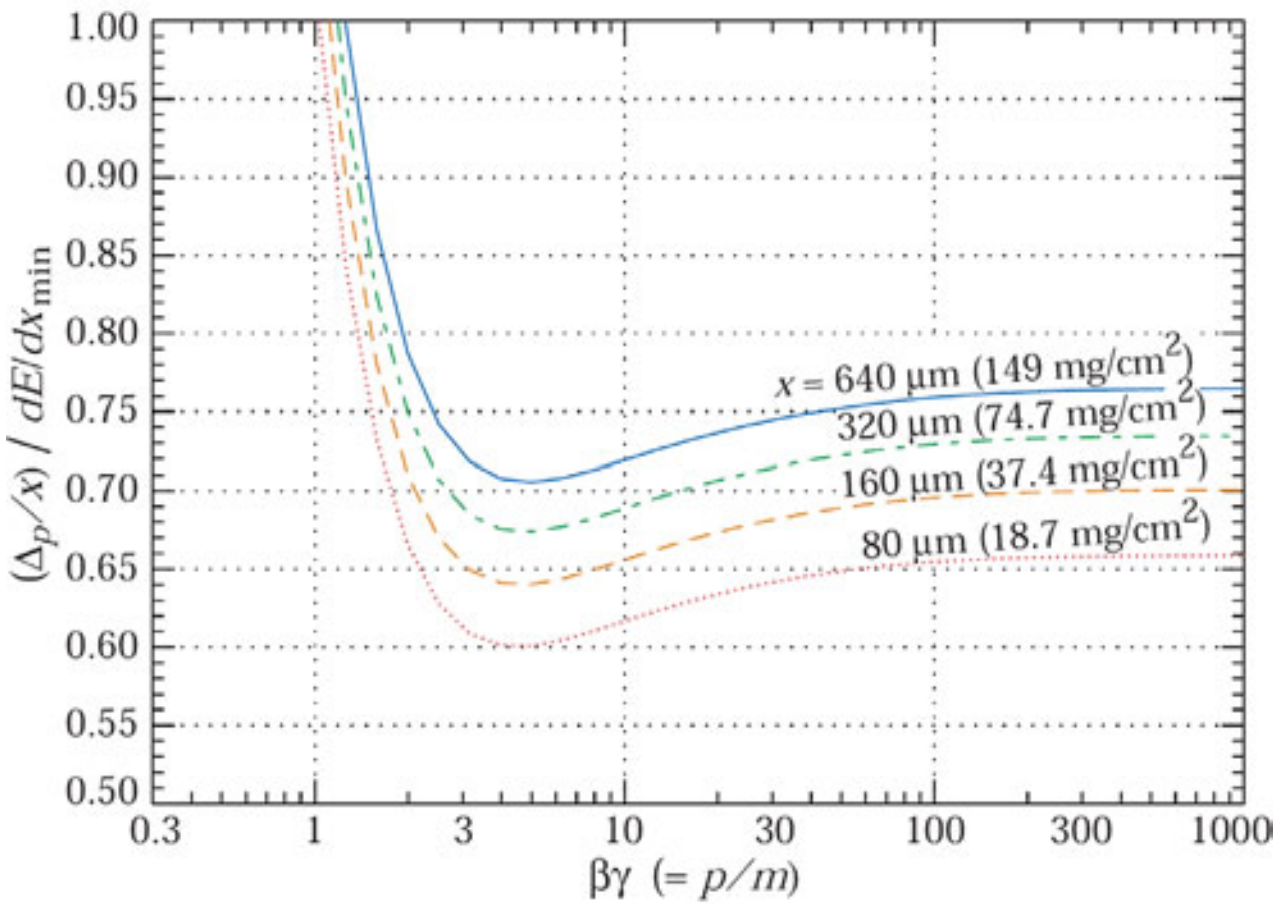

**Figure 4.11:** Most probable energy loss in silicon, scaled to the mean loss of a minimum ionizing particle, 388 eV/µm [10].

Not all detector modules were hit by beam. Hence it was not possible to calculate the gains of all optical links in the CRack system. Only modules with a sufficient number of hits were selected for analysis (~44 APVs depending on the particular run). While this may seem like a small number of links, it is clearly adequate for evaluating the accuracy of the setup routine. The statistics are not, however, enough to be used for analyses such as determining the variation of gain with temperature.

RESULTS

Table 4.1 shows the three physics runs and the corresponding conditions that were used to compile signal size data. In each case, the optical links were set up using the automated routines for laser biasing and gain calculation, thus enabling a direct comparison with the MIP gain under identical conditions. Figure 4.12, Figure 4.13, and Figure 4.14 are comparison plots of the gain calculated using the MIP signal size against the gain calculated from the setup routine using the APV headers, for each run. The error bars denote estimated calculation error for each method. The MIP gain error was derived by the error on the Landau fit, while the setup routine error was set to a constant ±5%[5].

**Table 4.1:** Physics runs used for compiling signal size histograms.

| Run number | Particles | Air Temperature (ºC) | APV mode |
|---|---|---|---|
| 10247 | Muons | -12 | Peak |
| 10329 | Pions | -13 | Deconvolution |
| 10628 | Muons | -25 | Deconvolution |

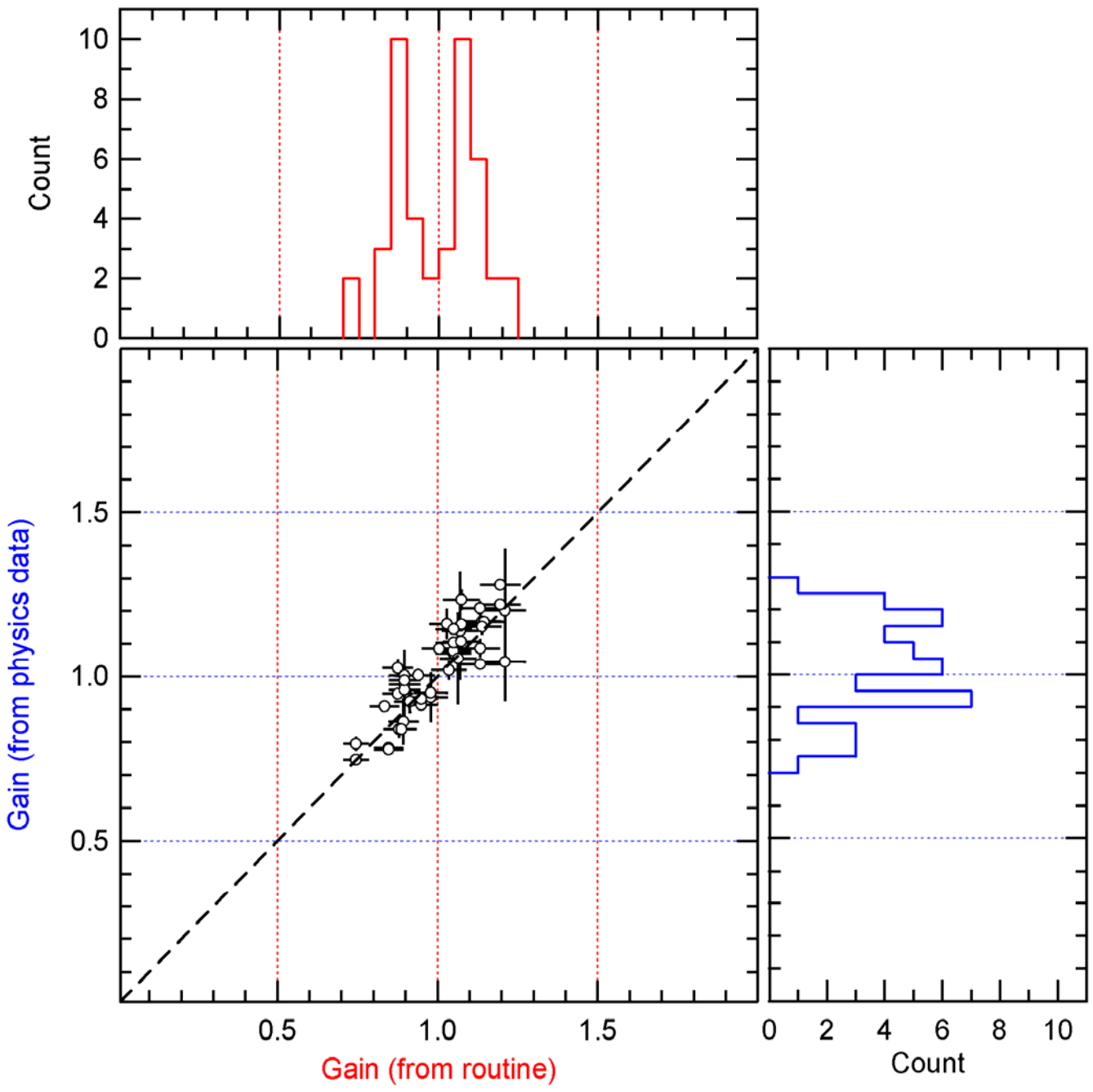

**Figure 4.12:** Comparison of gains calculated from physics data and the automated setup routine for run #10247 (muons, $T_{air}$=-12 ºC, peak mode).

[5] Due to the number of factors affecting the setup routine's gain calculation, it is not straightforward to come up with an exact value for the error. This is an (over-)estimation of the error, based on extensive experience with the setup of the link from test beams.

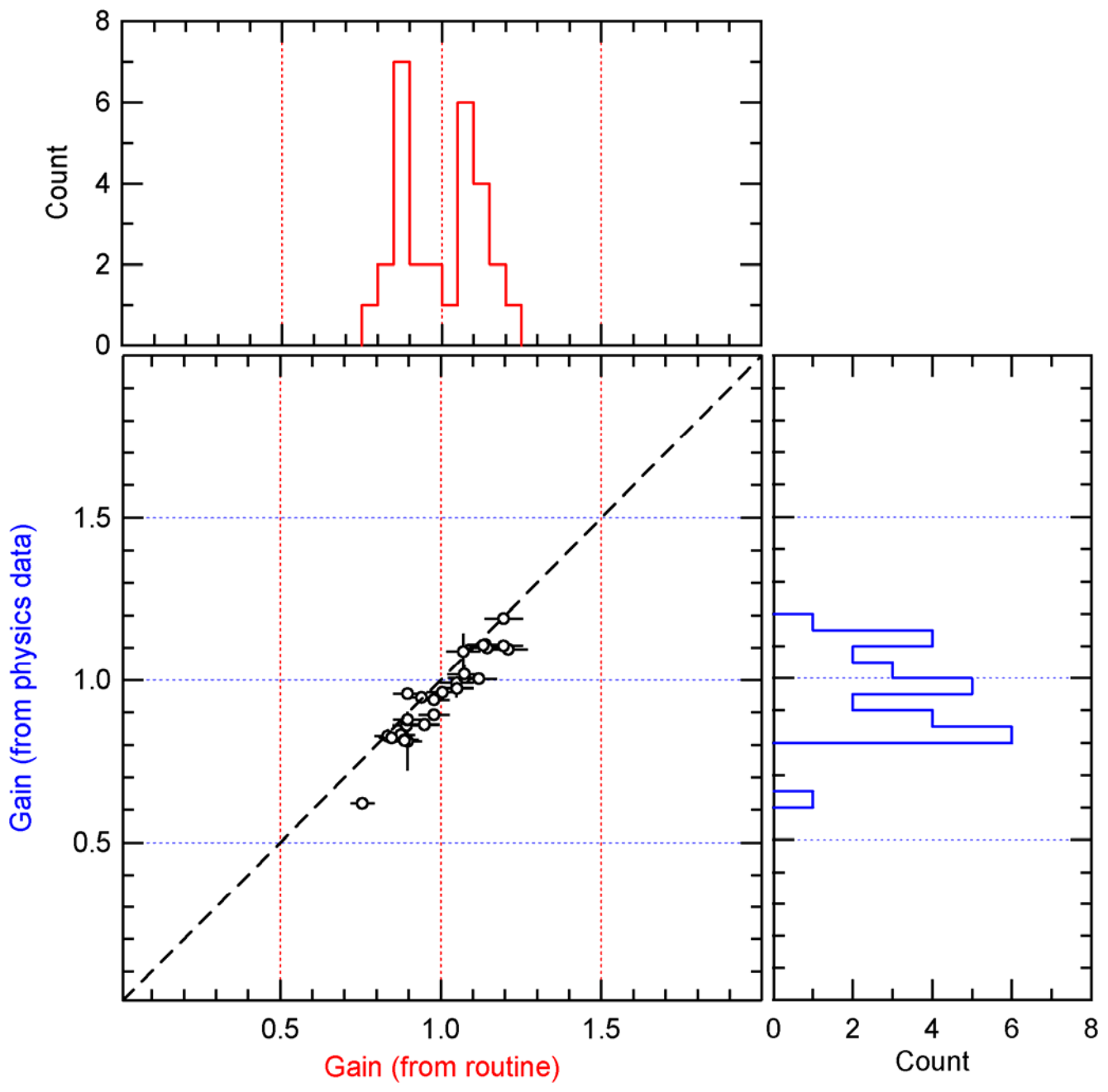

**Figure 4.13:** Comparison of gains calculated from physics data and the automated setup routine for run #10329 (pions, $T_{air}$=-13 ºC, deconvolution mode).

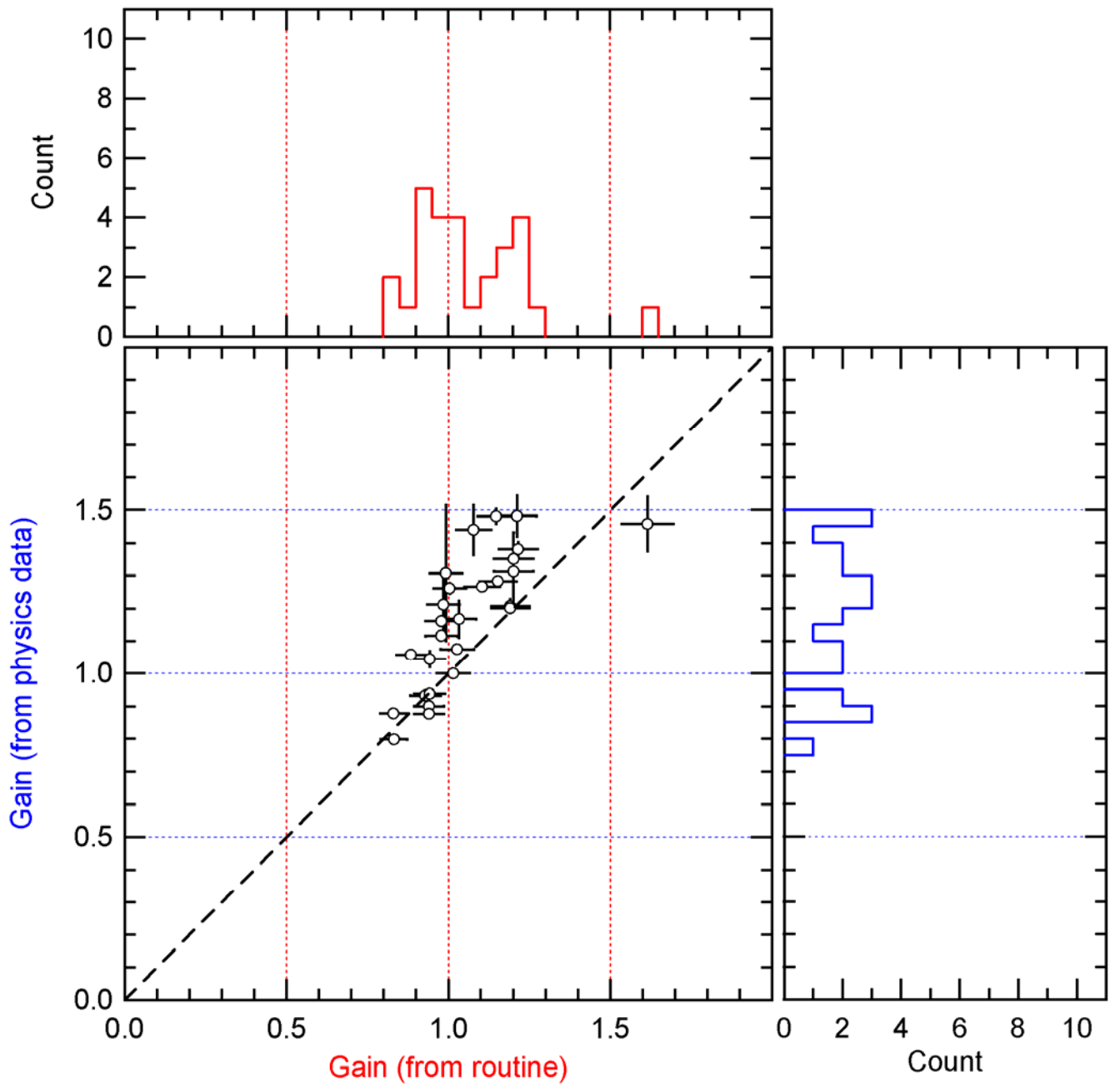

**Figure 4.14:** Comparison of gains calculated from physics data and the automated setup routine for run #10628 (muons, $T_{air}$=-25ºC, deconvolution mode).

Runs 10247 (Figure 4.12) and 10329 (Figure 4.13) show very good correlation between the MIP gain and setup routine gain. The correlation at -25ºC (Run 10628, Figure 4.14) is not as good, with the MIP gain being, on average, 10% higher. It is not unexpected that, with changing temperature, the two gain calculations are different; it is known that the APV header amplitude changes differently with temperature than the chip pulse height (and hence the signal size due to the energy deposition of a particle) [11]. However, for a given decrease in temperature, and using the APV parameters suggested in [12], one would expect the digital header height to increase by ~8-9% over a ΔT of 35ºC, compared to ~5-6% for the particle gain. While this is contradictory to what is observed in these results, the exact APV settings for each of the runs in this analysis are not known, while the document on recommended APV operation in the cold was not available until after the October 2004 test beam. Hence this could explain the larger discrepancy in the correlation plot of Figure 4.14.

From the results obtained, the accuracy of the setup routine gain calculation can be deemed satisfactory. A certain amount of offset is expected when compared to the ‘real’ MIP gain, but this should be less than 10%, in the worst cases. This is relatively small compared to the granularity of the AOH gain settings (50% steps). Hence, when equalizing the gain of the links throughout the Tracker, the gain error induced by the setup calculation should have a negligible effect.

Finally, the setup routine gain calculation was purposely designed to depend on a link input quantity (the APV digital header height) that is a constant regardless of other front-end detector hybrid settings (namely APV parameters that affect the chip’s gain). It follows that the accuracy of the gain values calculated by the setup routine then depends on the particular conditions and parameters chosen by the users. The setup routine is therefore provides a good estimate of the gain in each analog readout link. After having set-up the optical links with the automated routines, users can obtain signal size histograms from real physics data to fine tune the readout link gains as required.

Finally, the setup routine gain calculation was purposely designed to depend on a link input quantity (the APV digital header height) that is a constant regardless of other front-end detector hybrid settings (namely APV parameters that affect the chip’s gain). It follows that the correlation of the gain values calculated by the physics data and the setup routine then depends on the particular conditions and parameters chosen by the users. The setup routine therefore provides a good estimate of the gain; after having setup the optical links with the automated routines, users can obtain signal size histograms from real physics data to fine-tune the gains as required.

### 4.3.2 Optical Link Gain in the CRack as a Function of Temperature

The air temperature in the CRack was used for this study, since it is approximately a constant for all modules, and should track proportionally with the variation of hybrid temperatures. The gain distributions obtained from four setup runs of the 99 operational CRack optical links are shown in Figure 4.15. In each plot the dotted line represents the distribution resulting from setting all links to AOH gain setting 0. The solid line is the distribution obtained when attempting to equalize all links to the same target value of 0.8V/V[6] (the target is indicated by the dashed line on the x-axis).

The threshold of a laser changes very predictably with temperature. This property is exploited in order to verify that the CRack's AOH temperatures change in proportion to the system's air temperature. The laser bias point refers to the optimum operating point of the AOH (as determined by the setup runs described in Chapter 3) and this changes in direct proportion to the laser's threshold.

The threshold's dependence on temperature, $T$, is given by [13]:

$$I_{th} = I_{th}(0) \cdot \exp(\Delta T / T_0) \qquad (4.3)$$

$T_0$ and $I_{th}(0)$ are known as the characteristic temperature and current. Taking the natural logarithm:

$$\ln(I_{th}) = \ln(I_{th}(0)) + \Delta T / T_0 \qquad (4.4)$$

Hence the natural logarithm of the threshold current (and therefore the bias point) varies proportionally to laser temperature.

Figure 4.16 shows the average bias point (for each of the four CRack setup runs of Figure 4.15) versus the reported air temperature (top), as well as the natural logarithm of the average bias point versus air temperature (bottom). Clearly the two plots follow the relationships of Equations (4.3) and (4.4). This implies that the temperature of the lasers on the AOHs in the CRack was indeed proportional to the reported air temperature of the system, giving confidence in the results obtained in this section. While the CRack air temperature is not the same as the AOH temperature, the *relative* temperature change is clearly the same. This relative change can therefore be used to estimate the gain increase that would be expected in the optical links.

[6] The target gain of 0.8 is the specification for the optical readout links of the CMS Tracker.

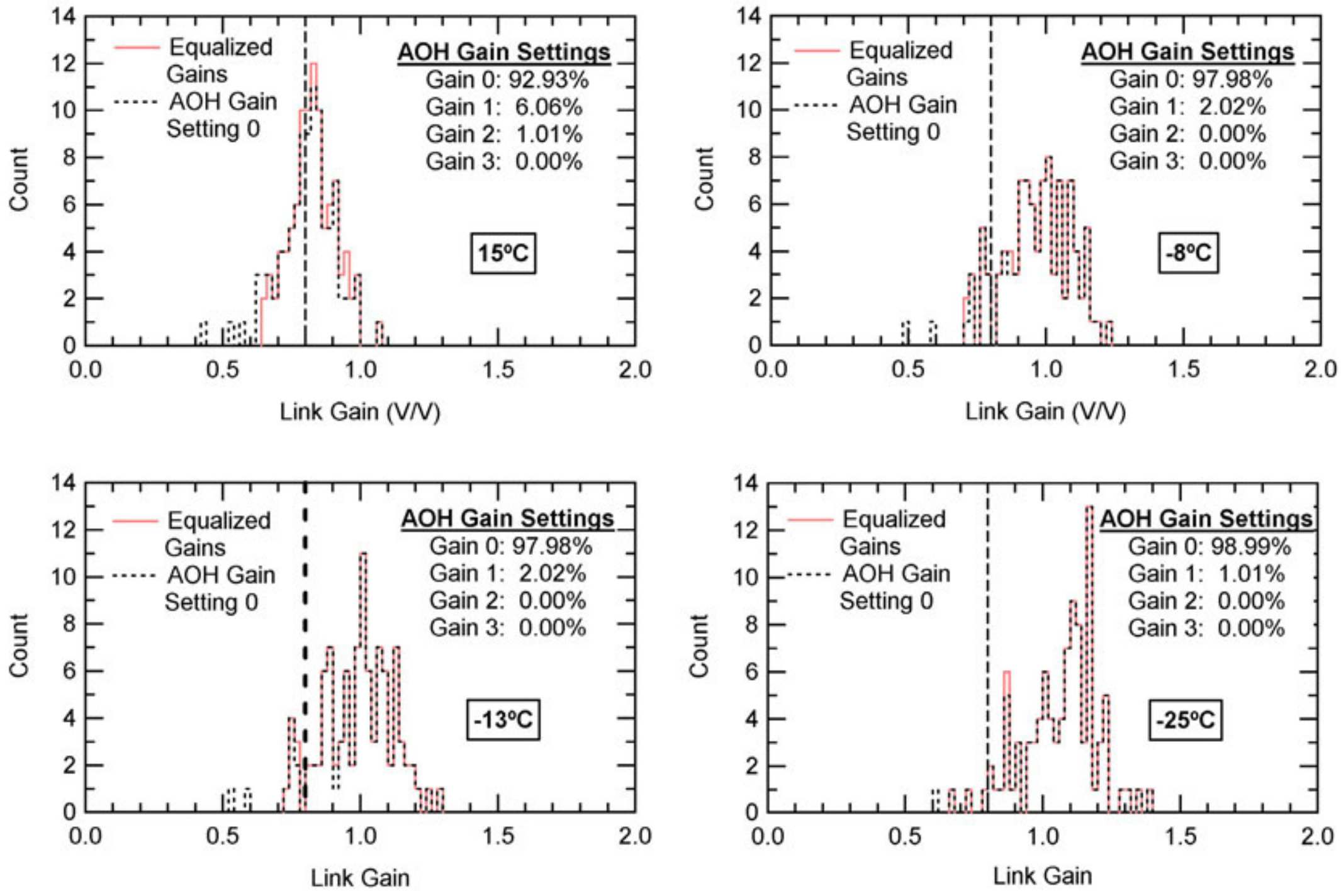

**Figure 4.15:** Readout link gain spreads for the 99 CRack links in the October 2004 test beam, for four different temperatures

Figure 4.17 shows the gain-temperature correlation plot obtained from the four setup runs analyzed. Each point on the plot corresponds to the mean gain of all 99 links for that particular run (only AOH gain setting 0 is considered for calculating the mean).

The results reveal an obvious trend of increasing gain with lower temperature. The spread of the gain distribution also seems to become larger. Even at 15ºC and with all links set to AOH gain setting 0, the gains of the links are on the high end (Figure 4.15). As a consequence, when attempting to equalize to 0.8V/V, the vast majority of the links (93%) have to be set to the lowest gain setting in order to be closest to the target gain. Indeed, the mean gain is 0.81V/V at this temperature, allowing very little flexibility for compensating high gain links. As the temperature is lowered the problem of high gain gets even worse, as can be seen in Figure 4.15. For temperatures below 15ºC the mean gain of the CRack links is over 0.8. With a load resistor of 100Ω, high gain links cannot be compensated for, resulting in large gain spread in the system.

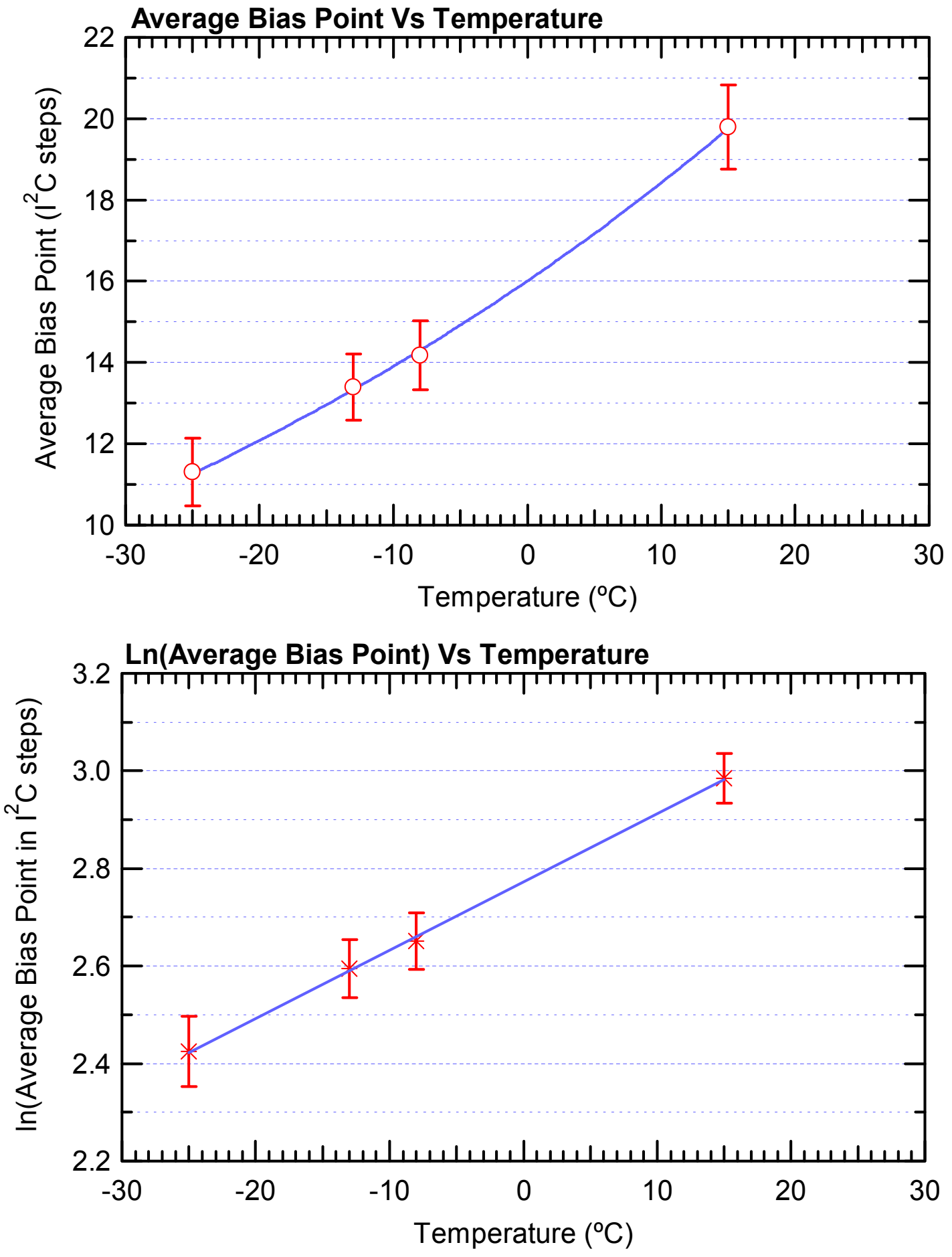

**Figure 4.16:** Average laser bias point versus temperature (top) and ln(average laser bias point) versus temperature (bottom). The error bars denote the standard deviation of the data sets.

By fitting a straight line through the points in Figure 4.17 the average gain change with temperature is found to be -0.0064 $°C^{-1}$. For a temperature change of +25 to -10ºC ($\Delta T$=35ºC) the gain increases by ~30%. The corresponding change in APV tick height is from ~595 to 774 ADC counts, at gain setting 0. This result can be compared to previous measurements made on the individual front-end components of the readout links that will also be cooled. While the temperature dependence of laser efficiency is not necessarily linear, it has been shown that the efficiency generally increases with temperature. In [14] the Mitsubishi laser transmitters showed an average efficiency increase of 0.22%/ºC. For a $\Delta T$=35 ºC, this translates to a gain increase of ~8%. The APV digital header changes by about ~8-9% when cooled by the same amount [12]. Finally, there are no data regarding the gain of the APVMUX [15] and LLD chips. However, one would also expect an increase in gain with lower temperature. Hence, the 30% change obtained from the CRack test beam results seems to be reasonable for the full readout chain gain. The result obtained here is obviously only valid for the CRack, since it depends on the particular implementation of the cooling system. Nonetheless similar environmental conditions are expected in CMS.

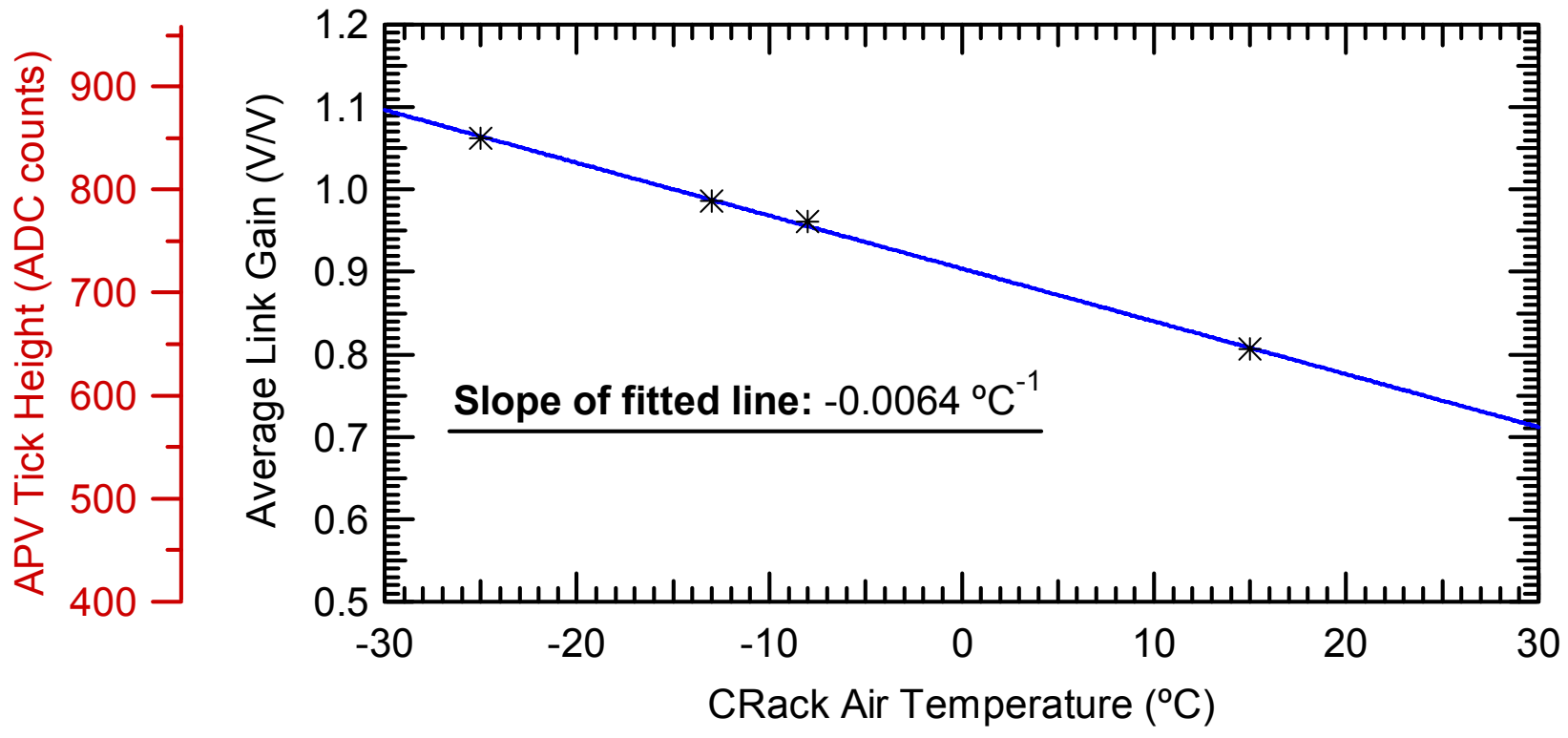

**Figure 4.17:** Average link gain vs CRack air temperature for all four setup runs analyzed.

## 4.4 Predicting the Final Gain Spread in the CMS Tracker Optical Links at the Nominal Operating Temperature

From Figure 4.17, the temperature change from +25°C to -10°C corresponds to an average gain increase of ~30%. This increase was incorporated into the Monte Carlo simulation in order to obtain the best estimate of gain spread that can be expected in the final system during operating conditions. It should be noted that, while laser transmitter efficiency generally shows an increase in gain with lower temperature, this varies from device to device and is due to the unpredictable variation in laser-fiber coupling efficiency [14]. Hence operation at different temperatures implies a varying spread of laser transmitter slope efficiency, and therefore of the aggregate link gain. Since there are no significant statistics on the laser efficiency change with temperature, the effect of temperature on the spread of the efficiency has been ignored in this study.

Figure 4.18 shows the non-equalized, single-gain distributions obtained via simulation for low temperature. The mean of the gain 0 distribution is above the target value of 0.8V/V. Virtually all links would have to be set to AOH gain setting 0 when attempting to use equalization. This is illustrated in Figure 4.19 which also indicates the percentage of links set to each gain setting. It is also noteworthy that the total spread of the equalized distribution is higher, with the lowest gain at 0.64 and the highest at 1.32V/V. The low end is the same as in the room temperature case, while the high end corresponds to the tail of the gain 0 distribution (i.e. high gain links for which there is no lower AOH gain setting).

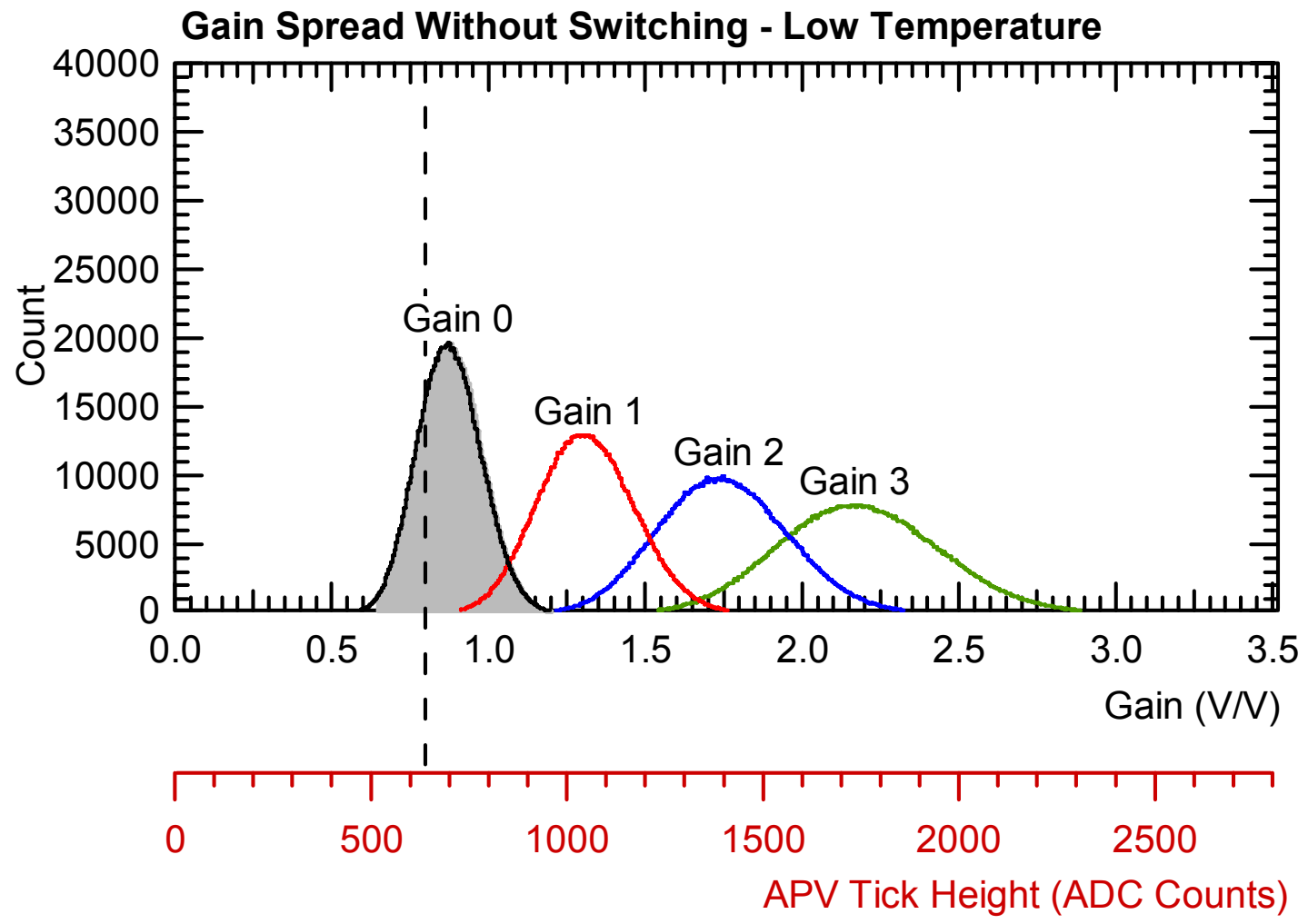

**Figure 4.18:** Showing the 'single gain' spread distributions predicted by simulation without switching of the LLD, at -10ºC. The shaded area shows the distribution resulting after equalization using the 4 available gain settings.

The high gains observed can be attributed to the fact that the system was designed assuming uniform spreads in the connector losses, within their specifications (0-0.6dB for the MU and 0-1.2dB for the MFS connectors). In reality, the insertion losses of the connectors are far better (Figure 4.1). In addition, the LLD transconductance is on the high end of its specification (roughly 7% higher than the nominal).

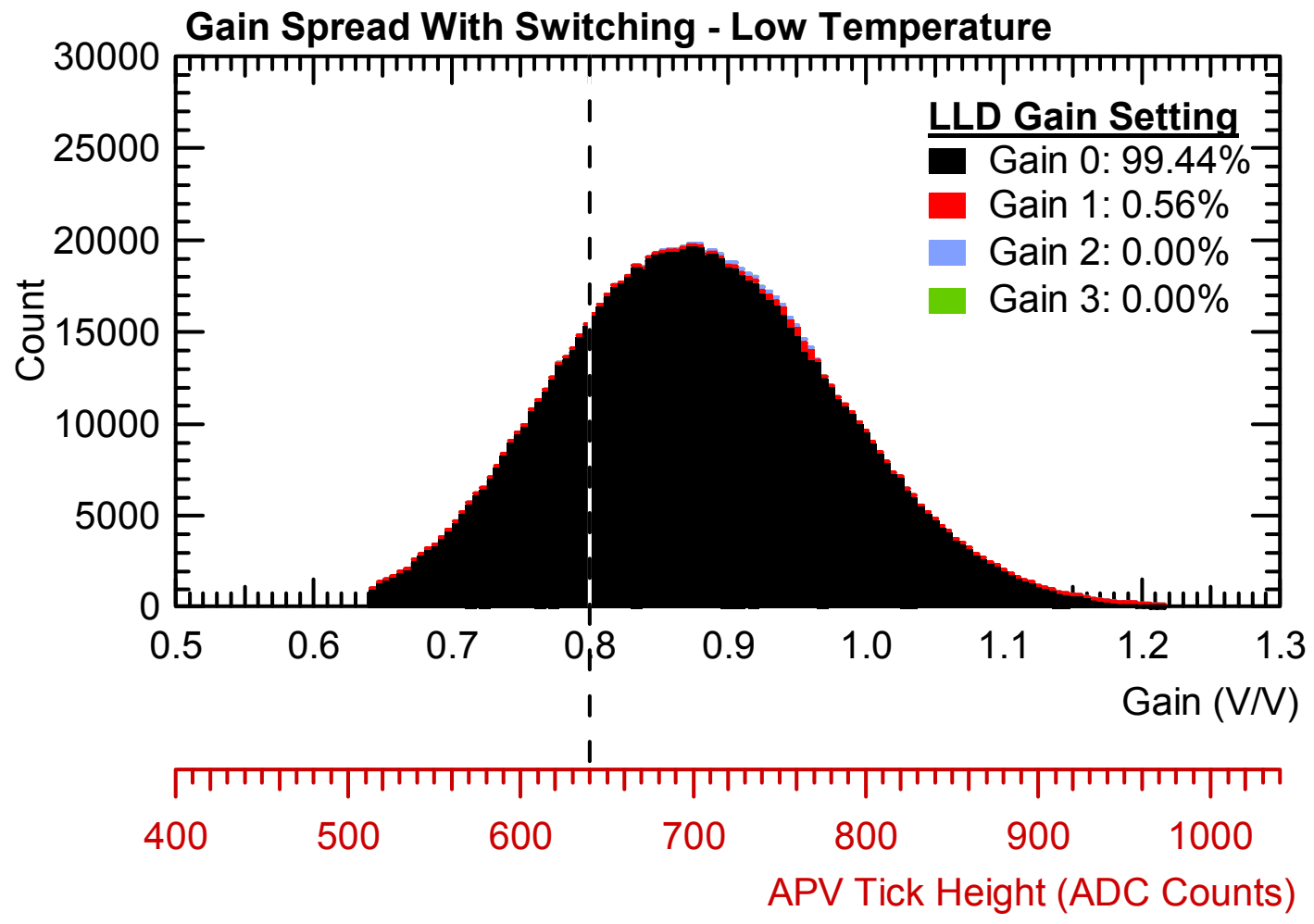

**Figure 4.19:** Equalized link gain distribution at -10ºC obtained by switching of the LLD, showing the contributions from each gain setting.

The results clearly show that the gain of the optical links is too high, for low temperature and a load resistor value of 100Ω. This is not an ideal situation, and the ability to equalize the optical link gains is completely lost. The effect on the dynamic range is illustrated in Figure 4.20. Due to the inability to equalize the gains, the dynamic range spread is significantly increased when compared to room temperature, with the low-end of the distribution tail at ~48 500 electrons/8bits (~1.9 MIPs/8bits for thin detectors) and the high-end at ~100 000 electrons/8bits (4 MIPs/8bits for thin detectors). For thick detectors, the corresponding distribution ranges from 1.2 to 2.6 MIPs/8bits.

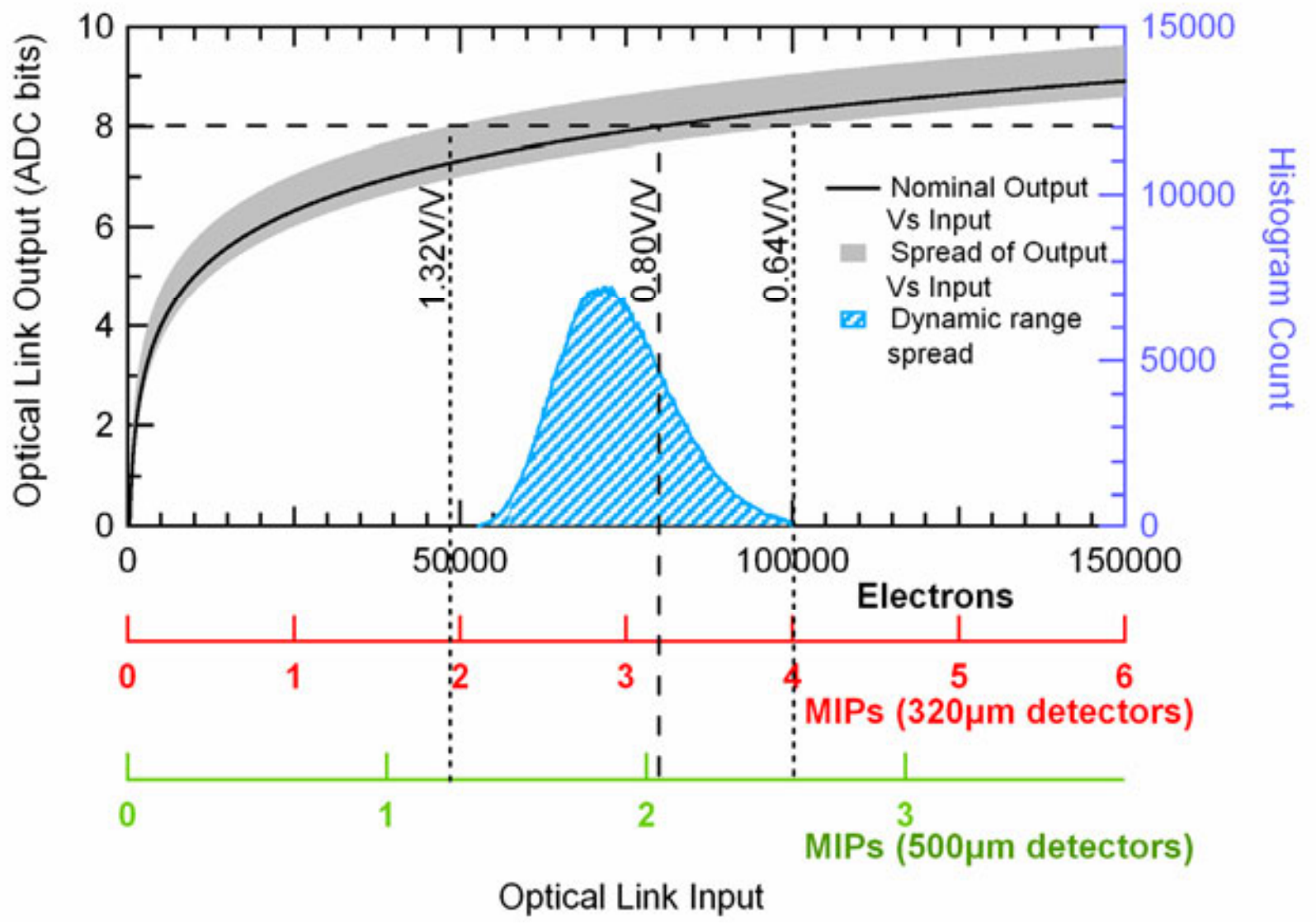

**Figure 4.20:** Showing Optical Link Output vs Input in ADC bits (left axis) for -10 ºC. The histogram (right axis) shows the spread in dynamic range.

### 4.4.1 Gain Compensation: Changing the ARx12 Load Resistor

While, the optoelectronic receiver's (ARx12) load resistor value will be fixed for the duration of the experiment, it provides a handle for adjustment of the overall link gain before the value is frozen. Using the Monte Carlo simulation, the effect of changing the load resistor can be observed. The new value must be such that it allows recovery of the dynamic range lost due to temperature effects (~30%). In addition, even at room temperature the vast majority of links have to be operated at one of the extreme AOH gain settings (setting 0). This does not allow much equalization flexibility, and hence it is desirable to further adjust the gain so that AOH setting 1 is the most common setting. The resistor value that achieves both of the above objectives was found via simulation to be 62Ω. The single-gain and equalized distributions are shown in Figure 4.21 and Figure 4.22, while Figure 4.23 shows the dynamic range spread that can be expected with a load resistor of 62Ω. The results show that the ability to equalize the gains at the nominal Tracker operating temperature can be recovered with this load resistor value. The equalized gain spread lies between 0.64 and 0.96V/V, as expected. Furthermore, the most frequent AOH gain setting will be setting 1, allowing sufficient flexibility for compensation of both low and high gain links. This change in the load resistor has been implemented on the production version of the FED (version 2).

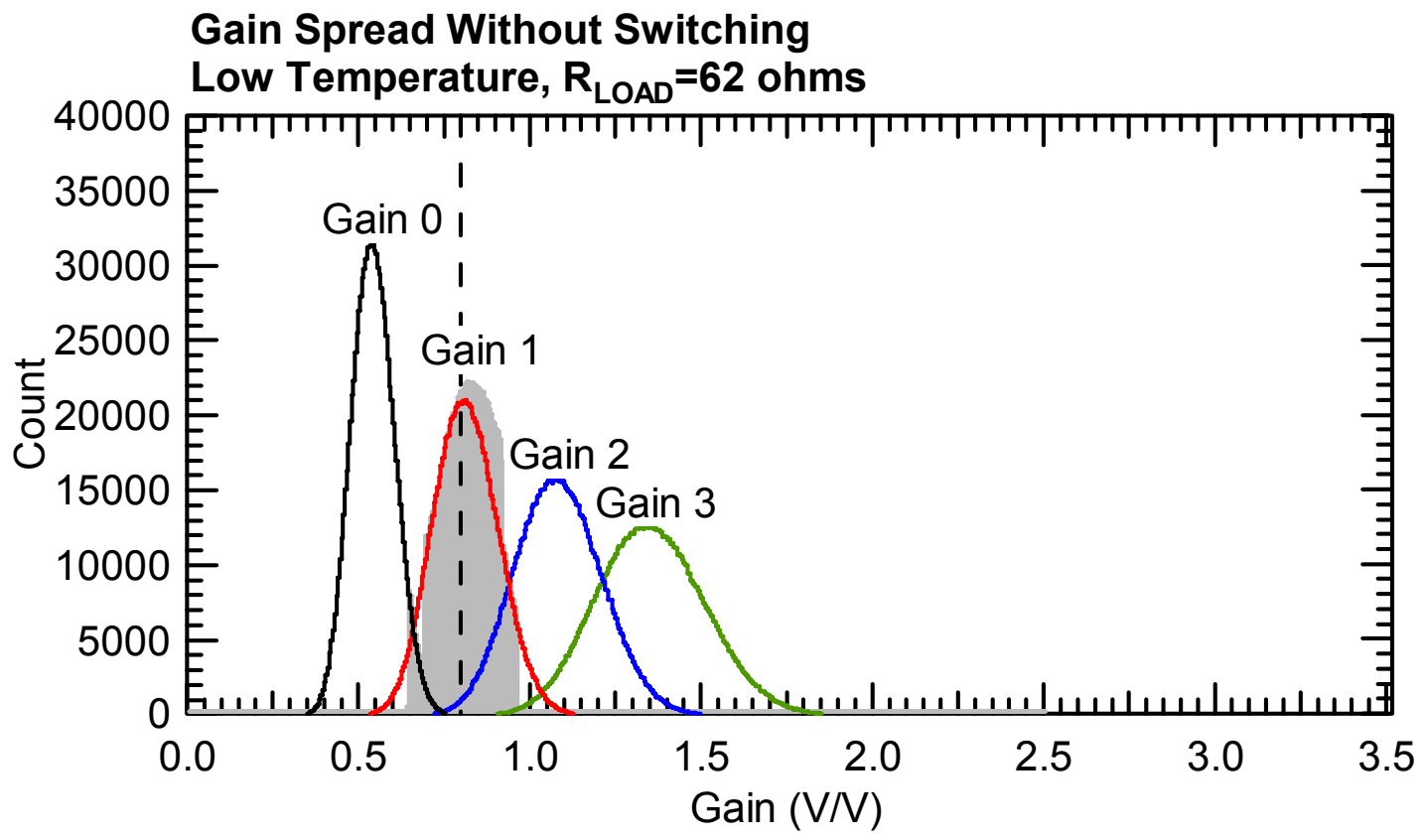

**Figure 4.21:** Showing the 'single gain' spread distributions predicted by simulation without switching of the LLD, at -10ºC and an ARx12 load resistor of 62Ω. The shaded area shows the distribution resulting after equalization using the four available gain settings.

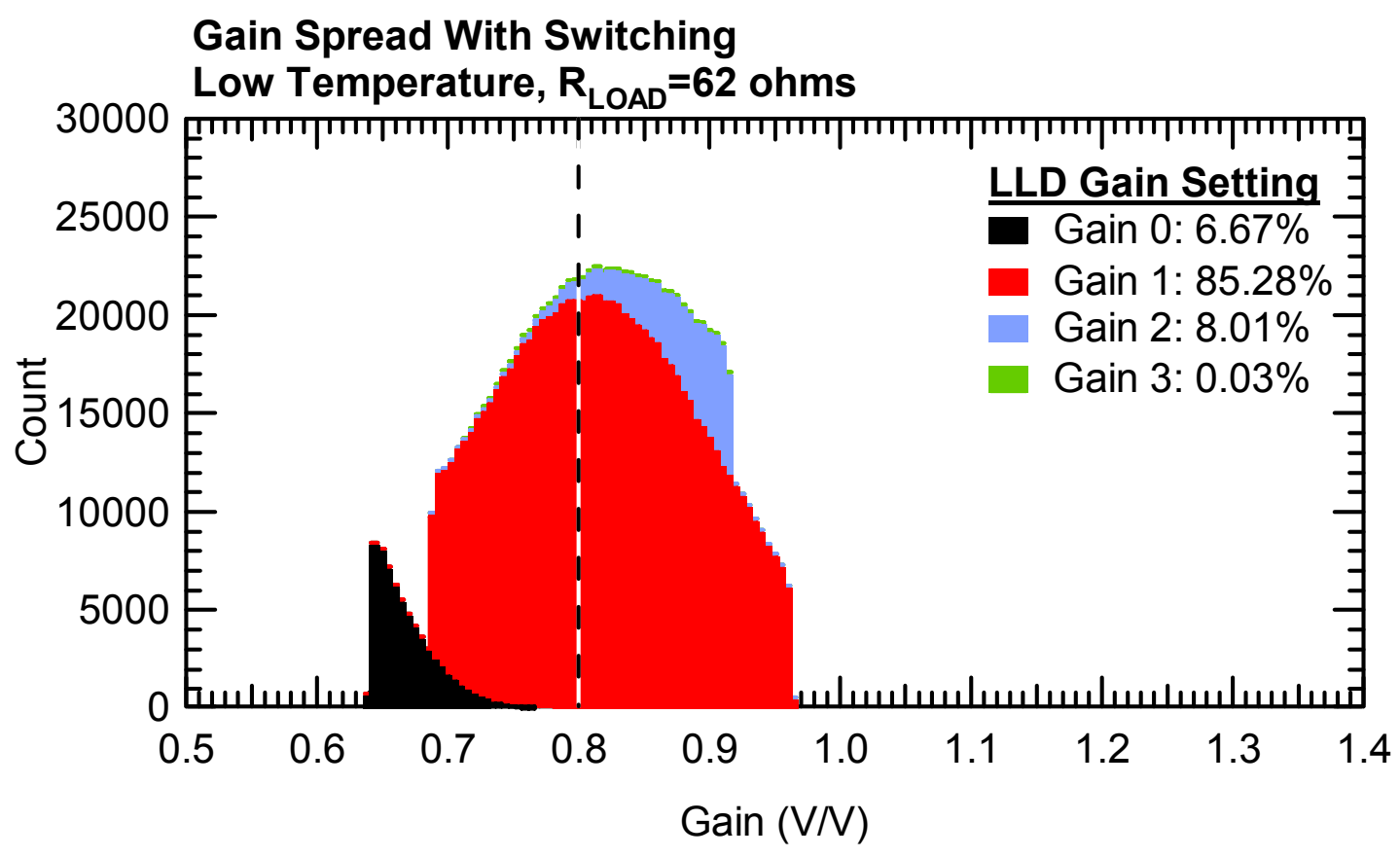

**Figure 4.22:** Equalized link gain distribution at -10ºC with switching of the LLD and an ARx12 load resistor of 62Ω, showing the contributions from each gain setting.

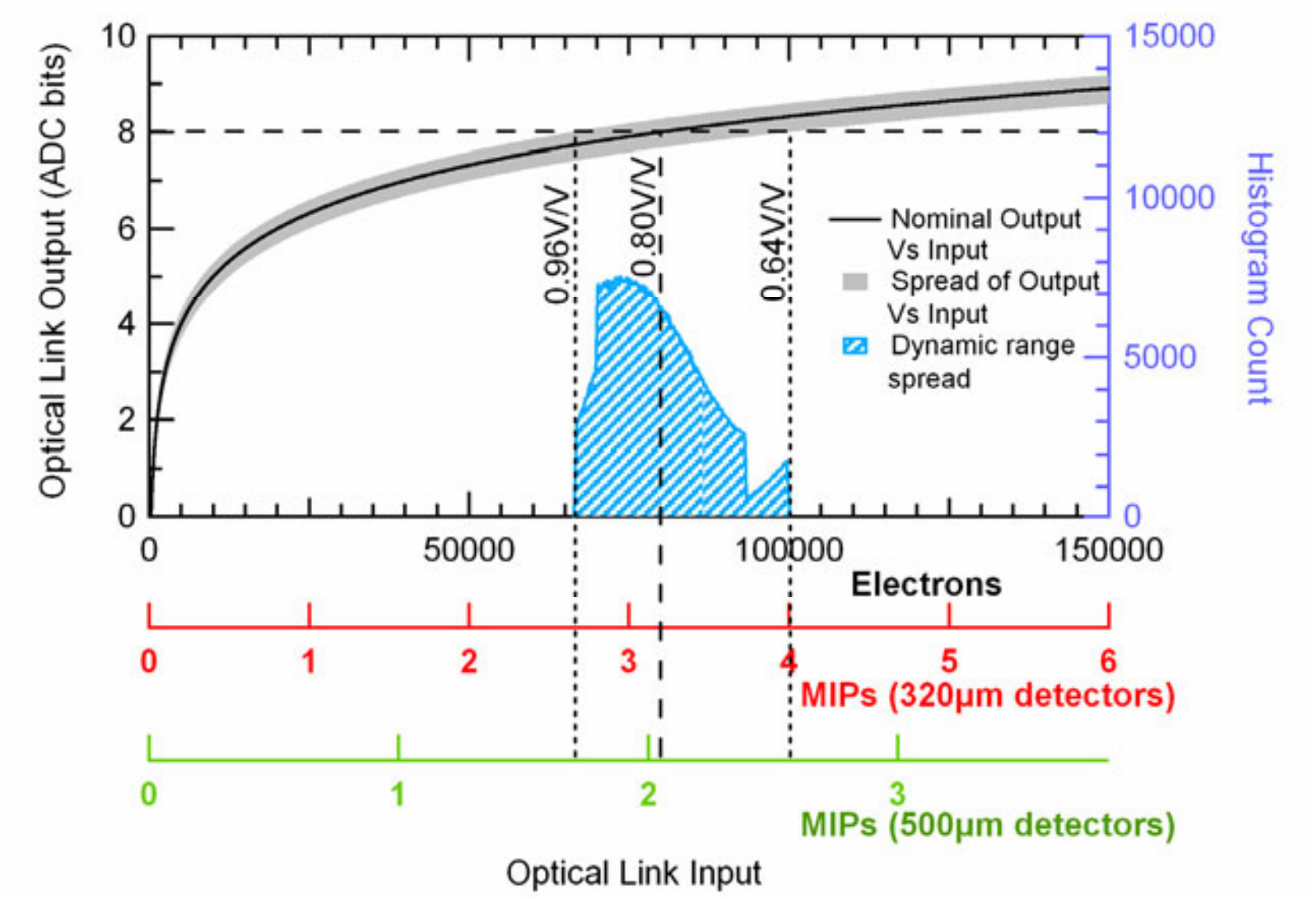

**Figure 4.23:** Showing Optical Link Output vs Input in ADC bits (left axis) for -10ºC an ARx12 load resistor of 62Ω. The histogram (right axis) shows the spread in dynamic range.

Table 4.2 gives the simulated limits on the APV tick heights that will be observed after equalization at various temperatures, with a load resistor of 62Ω. Results are shown for the full range of data, as well as for 98% of the links in order to eliminate the effect of statistically insignificant distribution tails.

The expected mean, minimum and maximum APV tick heights after equalizing with the AOH gain settings are given for various air temperatures from +25 to -20°C. The corresponding usage of AOH gain settings in the equalization process is also shown in Table 4.2. The final column indicates the percentage gain increase that is expected for the non-equalized case (i.e. the average increase of link gain that would be expected if only a single AOH gain setting is used). It should be noted that the air temperature/gain relationship is only exactly valid for the CRack system, though it may be very similar to other systems. In any case, the range of temperature (and corresponding gain) explored shows that equalization is possible in every case, and that it will result in the gain range calculated above.

**Table 4.2:** Simulation results for various temperatures.

| Air Temperature (°C) | Tick Height after Equalization (ADC Counts) | | | | AOH Gain Settings Used (%) | | | | Average Gain Change (%) |
|---|---|---|---|---|---|---|---|---|---|
| | | Min | Mean | Max | Gain 0 | Gain 1 | Gain 2 | Gain 3 | |
| 25 | Full Range | 472 | 640 | 752 | 0 | 20 | 70 | 10 | 0.0 |
| | **98%** | **551** | **640** | **728** | | | | | |
| 20 | Full Range | 492 | 639 | 768 | 0 | 33 | 62 | 5 | 4.3 |
| | **98%** | **550** | **639** | **729** | | | | | |
| 15 | Full Range | 508 | 637 | 772 | 0 | 46 | 52 | 2 | 8.6 |
| | **98%** | **550** | **637** | **730** | | | | | |
| 10 | Full Range | 508 | 635 | 772 | 0 | 59 | 40 | 1 | 12.9 |
| | **98%** | **550** | **635** | **731** | | | | | |
| 5 | Full Range | 508 | 636 | 772 | 0 | 70 | 29 | 1 | 17.2 |
| | **98%** | **549** | **636** | **739** | | | | | |
| 0 | Full Range | 508 | 640 | 772 | 1 | 79 | 20 | 0 | 21.5 |
| | **98%** | **535** | **640** | **752** | | | | | |
| -5 | Full Range | 508 | 644 | 772 | 3 | 84 | 13 | 0 | 25.9 |
| | **98%** | **520** | **644** | **759** | | | | | |
| -10 | Full Range | 508 | 649 | 772 | 7 | 85 | 8 | 0 | 30.2 |
| | **98%** | **517** | **649** | **762** | | | | | |
| -15 | Full Range | 508 | 651 | 772 | 11 | 84 | 5 | 0 | 34.5 |
| | **98%** | **515** | **651** | **764** | | | | | |
| -20 | Full Range | 508 | 650 | 772 | 18 | 79 | 3 | 0 | 38.8 |
| | **98%** | **514** | **650** | **765** | | | | | |

### 4.4.2 Errors and Limitations

The relationship between gain and temperature was explored using data from real systems with a significant number of readout links deployed. The automated setup routines were used for extracting the gains of the links and their accuracy was verified using real physics data (section 4.3.1). While the results show good correlation between the 'real' particle gain in the physics runs and the gain calculated by the routine, there is some remaining uncertainty. As far as the setup routine gain calculation is concerned, disagreement between runs was observed even in the case where the (reported) air temperatures were the same.

Certainly the most obvious source of error regarding the conclusions that can be drawn when comparing different systems is the uncertainty of the front-end temperature. The CRack air temperature was consistently used in all measurements, but this does not necessarily coincide exactly with the hybrid or sensor temperatures. The results rely on the fact that the *relative change* in the temperature of the electronics is similar to that of the air temperature. We have shown that this is the case in the results of section 4.3.2. Clearly, the conclusions reached in this document are exactly valid only for the CRack, but are expected to be similar for all sub-systems in the final CMS Tracker.

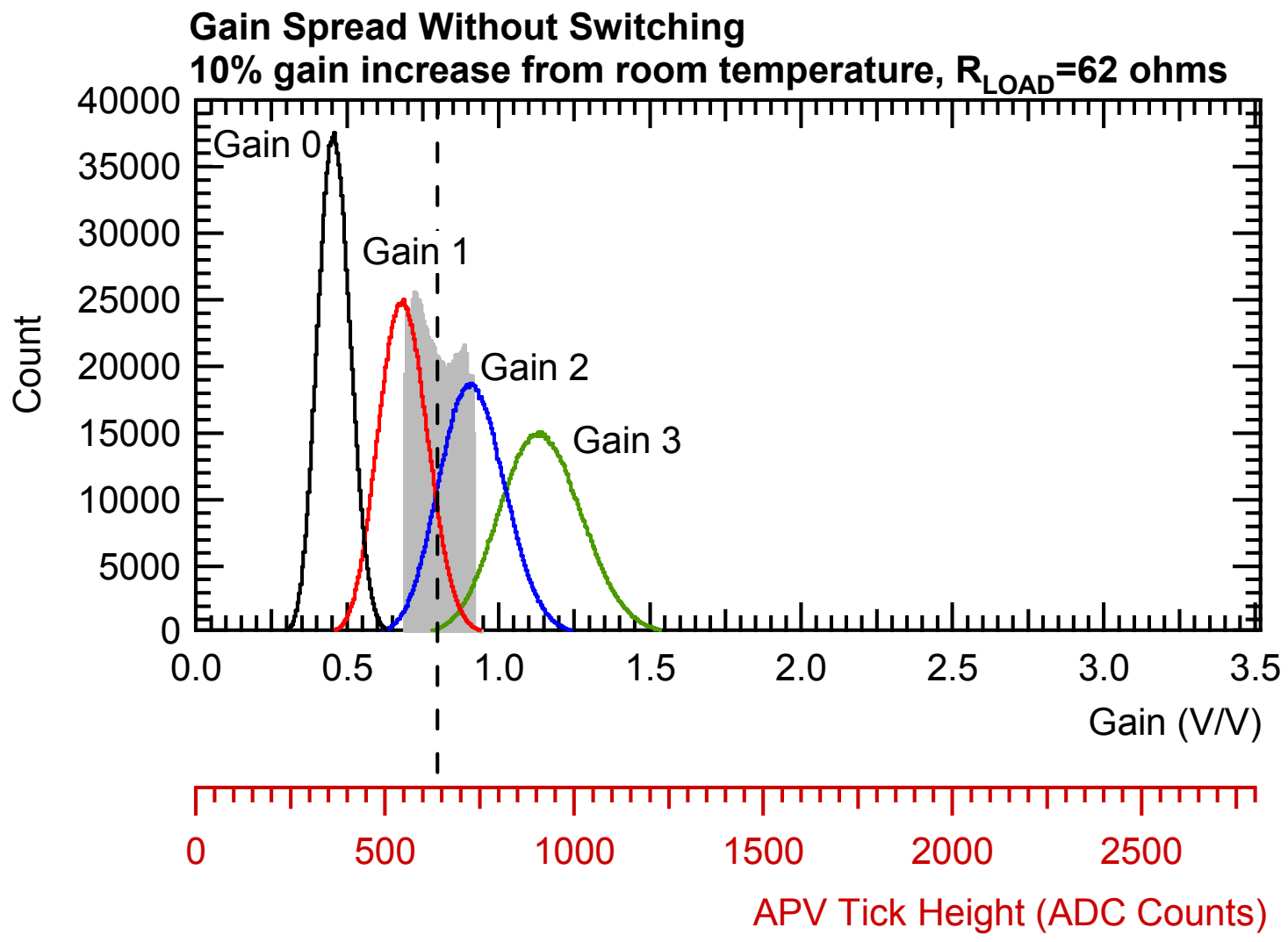

**Figure 4.24:** Illustrating the effect on the single-gain and equalized (shaded histogram) distributions of a 10% gain increase (relative to the room temperature simulation and with a load resistor of 62Ω).

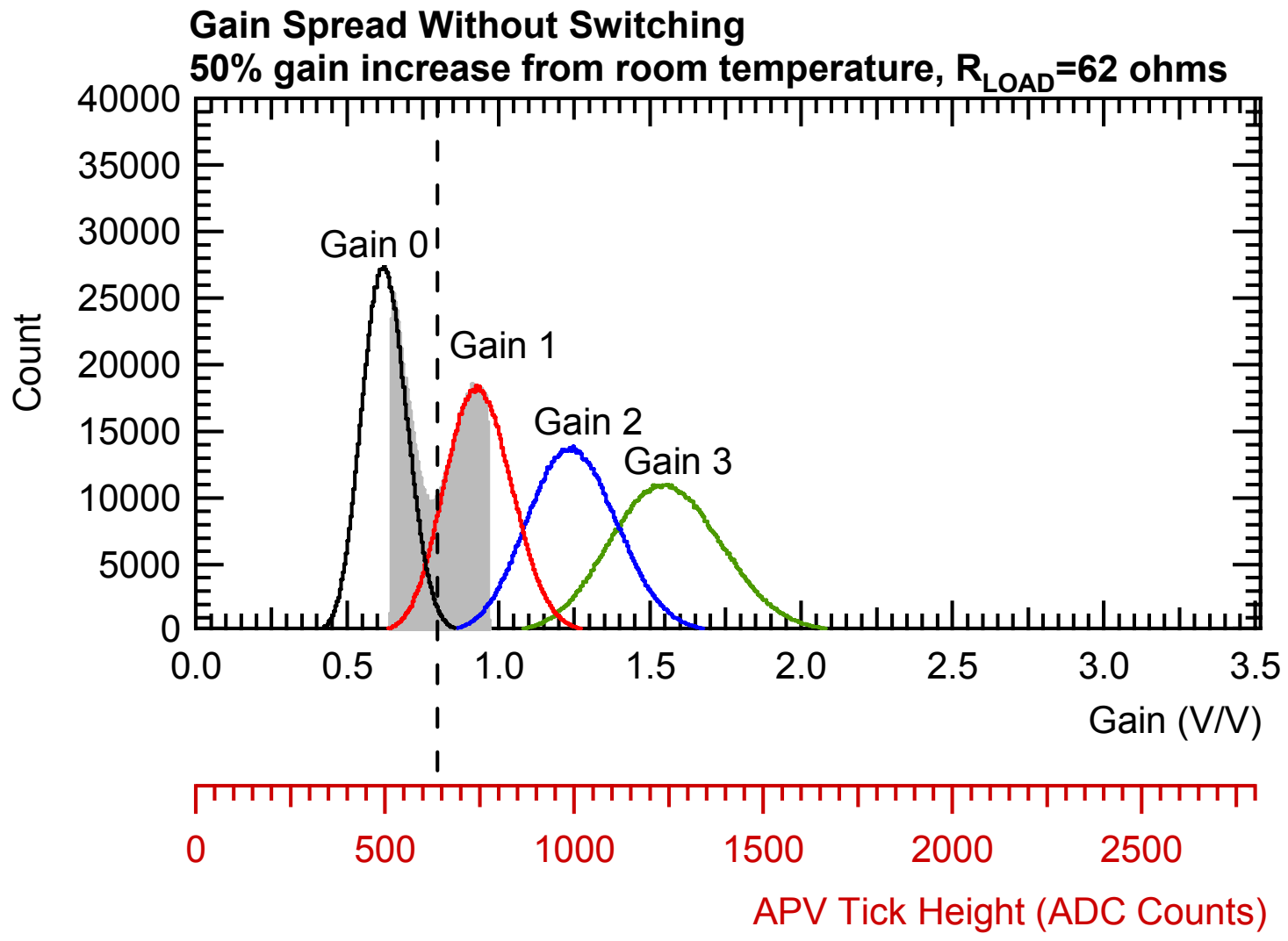

**Figure 4.25:** Illustrating the effect on the single-gain and equalized (shaded histogram) distributions of a 50% gain increase (relative to the room temperature simulation and with a load resistor of 62Ω).

The effect that the above errors have on the conclusions drawn by this study can be easily seen using the Monte Carlo simulation of the link gain. Recognizing that the choice of load resistor value required to recover lost dynamic range hinges on the temperature-gain dependency result, it is possible to simulate what would happen to the link gains (with 62Ω load resistors) if the *assumed* gain change with temperature is off by a large amount. Figure 4.24 shows the single-gain distributions obtained via simulation, assuming the average gain increase at nominal Tracker operating temperature is only 10% compared to room temperature. Figure 4.25 shows the results assuming a gain increase of 50%. In both cases, the single-gain distributions are such that, with a target gain of 0.8V/V, the link gains can be equalized within the 0.64 to 0.96V/V window. The results give confidence in the choice of load resistor value, since they show that even if the gain-temperature relationship has been calculated wrongly, gain equalization (and hence the dynamic range spread) will not be affected.

The physics runs used in evaluation of the setup routine accuracy were limited in the statistics (i.e. number of particles hitting the detectors). In order to draw a useful conclusion regarding the correlation between the gain calculated by the setup routine and that from physics data, one should have sufficient readout links to compare. Since most detectors in the CRack were not in the beam, there were fewer readout links that could be used in the first place. In addition, a decision had to be made about the minimum number of entries a cluster size histogram should have to obtain a decent Landau fit. This had to be chosen low enough so that enough links could be included in the calculations (20-40 links), but high enough to obtain a relatively low error on the fit. Therefore the gains calculated from the physics runs should be used cautiously, in order to make a rough comparison with the gains obtained by the setup routine.

Finally, the actual spread of the single-gain distributions obtained via simulation is expected to be larger in a real system. This is due to the components not simulated (notably the APV and APVMUX chips, and the analog front end of the FED). However, with the change in load resistor implemented, the spread on each single-gain distribution would have to roughly double (compared with that predicted by the simulation) before the equalized spread can no longer be contained between the 0.64-0.96V/V window. Therefore the conclusions drawn by the simulation are still most likely to be exact, despite the slightly optimistic spread in the single-gain distributions.

## 4.5 Conclusions

The gain spread that can be expected in the CMS Tracker analog optical readout links has been studied by means of a Monte Carlo simulation relying on production data. The dependence on temperature of optical link gain was determined by measurements made on the TOB CRack in the October 2004 test beam at X5, CERN. A linear relationship between gain and temperature was assumed and incorporated into the Monte Carlo simulation, in order to predict the gain spread at the nominal CMS Tracker operating temperature of -10°C.

The results at low temperature showed that the gains of the readout links calculated by simulation using real production data were too high, and the ability to use the AOH gain switch for equalization near the target gain value was completely lost. Consequently, the spread in dynamic range became unacceptably high. The simulation was used to determine the value of optoelectronic receiver load resistor that was needed in order to lower the gains and hence recover the lost dynamic range of the system. The new value (62Ω) was implemented in the production version of the FED board.

An analysis of potential errors concerning the temperature dependence of optical link gain was introduced. It has been shown that even with a large error in the expected gain change at low temperature, the new load resistor value will ensure that all gains in the Tracker's readout optical links will lie within the 0.64-0.96V/V window after equalization. Hence, once equalized, the spread in dynamic range of the final readout system will be from 2.7 (1.7) to 4 (3.6) MIPs/8bits for thin (thick) detectors (assuming that the 8 LSBs of the data captured at the FED's ADC are retained). The corresponding spread in APV tick heights is 512 to 768 ADC counts.

# Chapter 5

# Upgrading the Tracker Optical Links: Digital Modulation

*The possibility of upgrading the CMS Tracker optical links using bandwidth efficient digital modulation is investigated analytically in this Chapter. The aim is to estimate the data rate achievable using the current optical link components, augmented with additional components to perform digital modulation/demodulation. An introduction to some fundamental concepts of digital communications is made, followed by the data rate calculation. The analysis is dependent on assumptions made about the noise performance of the optical link, based on original specifications. The results obtained are purely illustrative, to enable a gentle introduction to the method. A more accurate computation is made in Chapter 6, where laboratory tests were made using real digital signals.*

## 5.1 Introduction

The current CMS Tracker optical links employ analog Pulse Amplitude Modulation (PAM) at 40MS/s. The Signal to Noise Ratio (SNR) of the system is specified so that the link has an equivalent digital resolution of at least 8 bits. Hence, the analog modulation scheme is akin to digital baseband PAM using 256 distinct levels (8 bits) at 40MHz. The equivalent data rate is 320Mbits/s (=8×40MHz).

The next iteration of the CMS Tracker will be operated in the Super LHC (SLHC) environment, and will have to cope with significantly increased data rates due to the tenfold increase in luminosity that is foreseen [1, 2]. In contrast to the telecoms industry where the optical fiber and its installation drive the cost of a transmission system, it is the cost of the optoelectronic components that represents a large fraction of the CMS Tracker electronics budget. The high cost of development of new components able to match the physical and environmental constraints of a high energy physics experiment provides the motivation behind re-using the current link components. Hence, it is proposed to convert these links to a digital system in order to achieve higher data rates using the current analog link components. Additional components on either side of the existing links would be required to perform the necessary digitization, digital transmission and reception. The main constraints are the bandwidth of the link and available signal power, limited by the transmitter hybrid (AOH) and the optoelectronic receiver amplifier (ARx12). Therefore, a bandwidth efficient digital modulation scheme is required to achieve transmission at Gbit/s rates. The feasibility of such a conversion must therefore be explored in terms of performance that can be achieved and implementation complexity. It is the former which is considered in this thesis.

The purpose of this work is to introduce the potential application of advanced digital communication techniques in a future upgrade of the CMS Tracker readout optical links. The concept involves using one or more digitally-modulated radio frequency (RF) sinusoidal carriers in order to make efficient use of the available bandwidth. In this chapter some basic concepts of digital communications are introduced, and an analytical approach to calculating the maximum data rate possible over the current optical links using digital modulation is presented. The calculation is based on a simplified model of the optical link, where the noise is assumed to be AWGN, with the noise power spectral density (PSD) being derived by the specifications. The results shown here are only included to aid understanding and illustrate the procedure for assessing the performance of a generalized digital communication system. The data rate calculation method presented in this chapter is also applied in Chapter 6, where more realistic results are obtained from laboratory tests.

## 5.2 Digital Modulation Basics

### 5.2.1 Overview

A basic introduction to digital communication concepts is required for understanding Chapters 5 and 6. It is not possible to cover all relevant topics of such a diverse field of study, and hence, only the basic concepts directly applicable to the subsequent analysis will be outlined. The reader is encouraged to consult the multitude of excellent digital communication texts in existence, such as [3].

The main components of a generic digital communication system are shown in Figure 5.1. In the case of the CMS Tracker, high energy particles traverse a silicon microstrip detector, thus producing analog signals. This is represented by the *analog information* block in Figure 5.1. Currently, the analog signals from the detector are collected by the APV chip which stores the sampled values in an analog pipeline memory. In addition to the sampling operation, quantization of the analog signals would also be required for a future upgrade (*sampler/quantizer*). The digitized samples are then sent through the *source encoder*, which map the values to codewords. For example, the samples of a 1024-level ADC could be converted into 10-bit words. In the case of a readout system, this could also include Forward Error Correction (FEC).

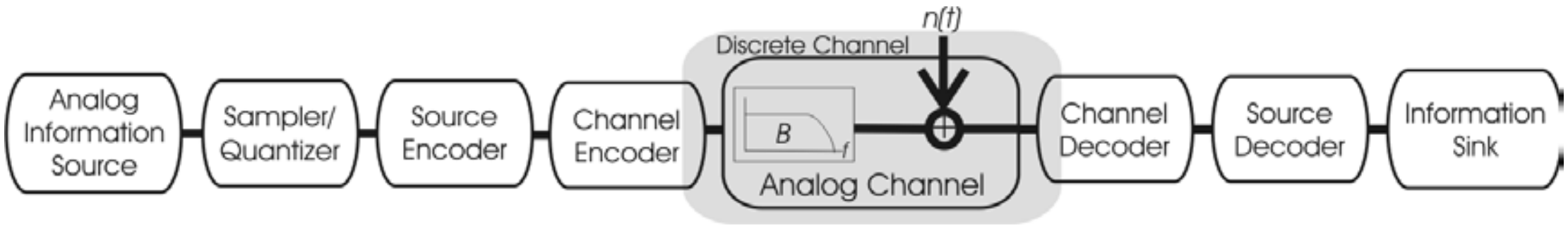

**Figure 5.1:** The basic building blocks of a digital communication system. In this thesis only the channel encoder/decoder blocks are considered.

The job of the *channel encoder* is to convert the codewords into suitable analog signal waveforms for transmission over the communication channel. The channel is the medium over which modulated signals can propagate. For example, in the case of a wireless network such as Wi-Fi, the channel is the free space between a transmitter and receiver through which electromagnetic waves travel. In the CMS Tracker's case, the components of the optical link constitute the analog channel.

On the receiving end of the channel, the reverse process is followed. The *channel decoder* interprets received signals into codewords, and the original data stream is restored at the *source decoder*. The data can then be stored and processed.

### 5.2.2 The Shannon Capacity

In his famous 1948 paper [4], Claude E. Shannon presented his channel coding theorem. This postulates that for a channel with information capacity, $C$, all rates below capacity are possible. More precisely, for all rates $R<C$, there exists a sufficiently complicated code allowing transmission with an arbitrarily small probability of error. Conversely, error-free transmission cannot be achieved if $R>C$. While Shannon's work proves the existence of codes/modulation that achieve capacity, it does not provide the means to construct them.

The capacity of a bandlimited, AWGN communication channel is the fundamental limit to the achievable transmission rate. The capacity, $C$, of a channel (in bits/s) is defined as a function of the bandwidth (in Hz) and Signal to Noise Ratio (SNR):

$$C = W \log_2(1+\mathrm{SNR}) \qquad (5.1)$$

The SNR can be expressed as:

$$\mathrm{SNR} = \frac{P_{avg}}{W \cdot N_0} \qquad (5.2)$$

*Where $P_{avg}$ is the average transmitted signal power, and $N_0$ is the double-sided noise power spectral density (PSD).*

By defining the normalized channel capacity as $C/W$ (in bits/s/Hz), Equation (5.1) becomes:

$$\frac{C}{W} = \log_2\left(1+\frac{P_{avg}}{W \cdot N_0}\right) \qquad (5.3)$$

It is useful to relate the normalized channel capacity as a function of $E_b/N_0$, i.e. the ratio of bit energy to noise PSD[1]. The energy per bit is related to the average transmitted power by the data rate by:

$$P_{avg} = CE_b \qquad (5.4)$$

Hence, Equation (5.3) can be written as:

$$\frac{C}{W} = \log_2(1 + \frac{C \cdot E_b}{W \cdot N_0}) \qquad (5.5)$$

Rearranging:

$$\frac{E_b}{N_0} = \frac{2^{C/W} - 1}{C/W} \qquad (5.6)$$

Equation (5.6) is graphically depicted in Figure 5.2. An interesting consequence of Shannon's Capacity Theorem is that there is a lower bound on $E_b/N_0$, below which reliable data transmission is impossible. It is straightforward to show that the so-called *Shannon Limit* occurs at -1.6dB.

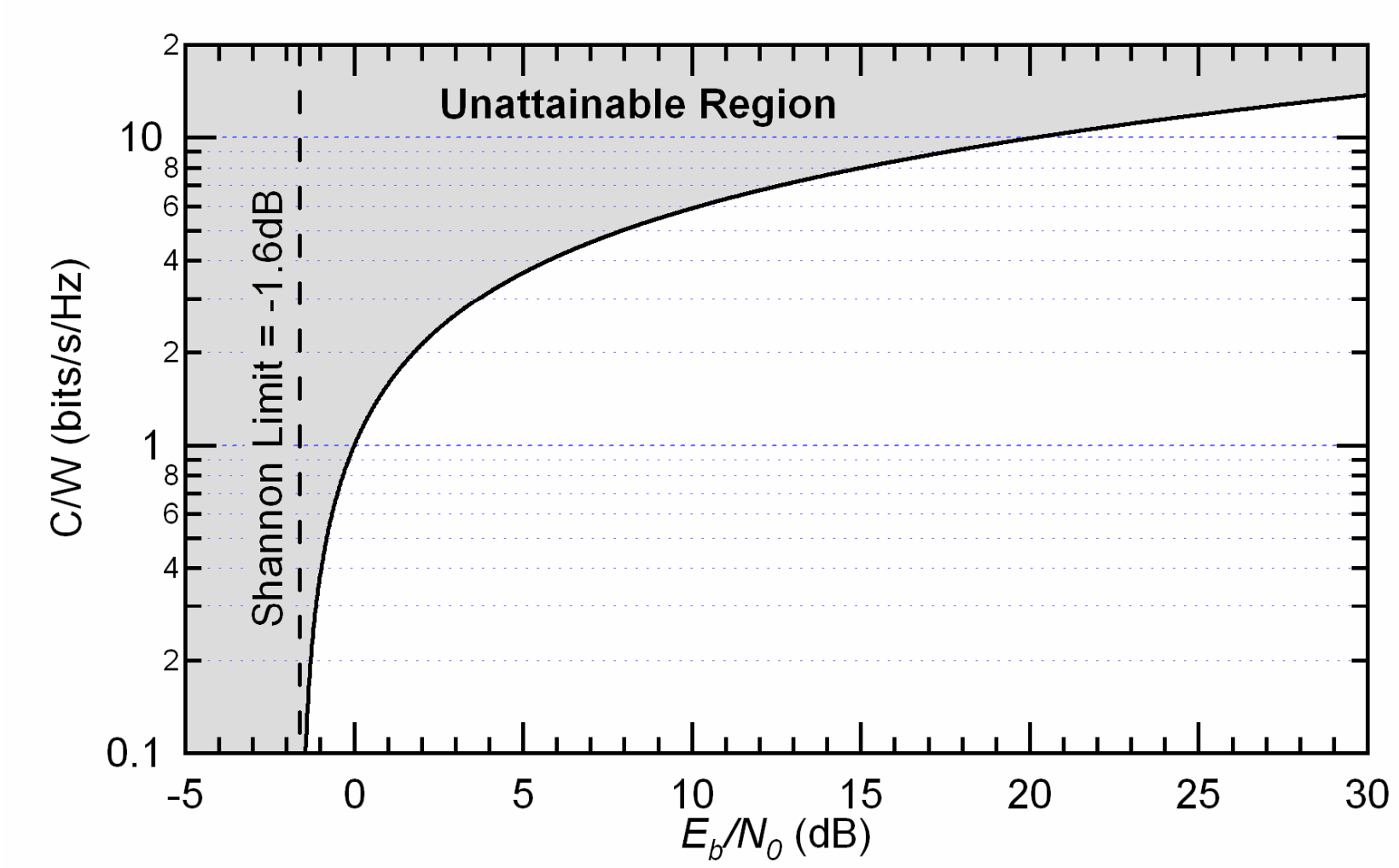

**Figure 5.2:** Normalized channel capacity as a function of SNR per bit.

### 5.2.3 Digital Modulation for Spectral Efficiency: *M*-QAM

OVERVIEW

In multi-level (or *M*-ary) modulation schemes, each transmitted symbol conveys $\log_2(M)$ bits of information. Binary transmission can be considered a subset of *M*- ary, with *M*=2. *M*-ary schemes are used in bandwidth limited channels, where it is beneficial to trade off SNR (and therefore error rate) for increased spectral efficiency.

[1] This is essentially the same as the SNR per bit, and the two terms will be used interchangeably. The SNR per bit is simply the SNR divided by the number of bits/symbol in a given modulation scheme. i.e., if the scheme employs $M$ symbols, the number of bits is $\log_2(M)$, and hence the SNR per bit is $\text{SNR}/\log_2(M)$.

Modulated signals can be either baseband or passband. Examples of baseband transmission include binary Non Return to Zero (NRZ), Manchester, as well as *M*-ary Pulse Amplitude Modulation (PAM)[2]. These schemes produce signals whose spectral content is near DC.

In passband modulation, the baseband information is modulated on top of a higher-frequency sinusoidal carrier. In passband PAM, the amplitude of the carrier is varied in response to the baseband information signal. The phase of the carrier can also be varied, as is the case with Phase Shift Keying (PSK). Quadrature Amplitude Modulation (QAM) combines PAM and PSK, to transmit more bits/symbol. This thesis focuses on QAM, which is described in more detail in the rest of this section.

## QUADRATURE AMPLITUDE MODULATION

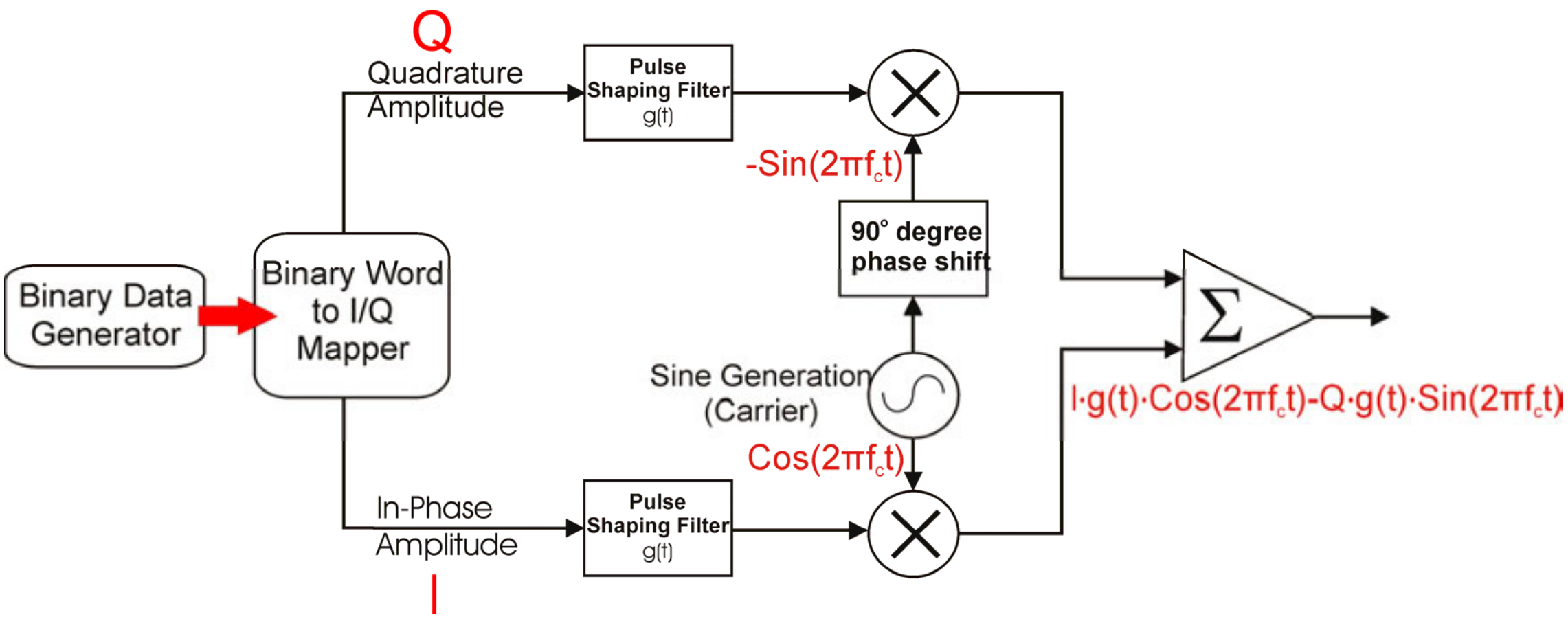

**Figure 5.3:** Illustrating QAM signal generation.

Combined amplitude and phase modulation is achieved by simultaneously impressing two separate *k*-bit symbols on two quadrature sinusoidal carriers (Figure 5.3) [3]. Each branch in Figure 5.3 is essentially the amplitude modulation (PAM) of the corresponding carrier. A symbol consisting of 2*k* binary bits is clocked in to the I/Q mapper at the required symbol rate. Half of the bits are translated to a quadrature amplitude, *Q*, and half to an in-phase amplitude, *I*. After

2 In the passband domain, PAM is also known as Amplitude Shift Keying (ASK).

pulse shaping, these are modulated on the sin and cos carriers respectively, and summed at the output. The resulting signal waveforms can be expressed as:

$$s(t) = \mathrm{Re}\left[(I + jQ)g(t)e^{j2\pi f_c t}\right], \quad for\ 0 \le t \le T$$
$$= I \cdot g(t)\cos(2\pi f_c t) - Q \cdot g(t)\sin(2\pi f_c t) \qquad (5.7)$$

*Where g(t) is the signal pulse and I and Q are the information-bearing signal amplitudes of the quadrature carriers.*

Equivalently, *s(t)* can also be expressed as [3]:

$$s(t) = \mathrm{Re}\left[Ve^{j\theta} g(t)e^{j2\pi f_c t}\right]$$
$$= Ve^{j\theta} g(t)\cos(2\pi f_c t + \theta) \qquad (5.8)$$

*Where* $V = \sqrt{I^2 + Q^2}$ *and* $\theta = \tan^{-1}(Q/I)$.

Equation (5.8) is a more intuitive way of showing that QAM signals are generated by combined amplitude and phase modulation. In *M*-QAM, each of the *M* symbols is denoted by a particular combination of values of *I* and *Q*, and any combination of in-phase and quadrature amplitudes can be selected.

### THE CONSTELLATION DIAGRAM

It is convenient to view QAM symbols on a polar (or IQ) plot. This is essentially the complex envelope of the modulated carrier, sampled at the symbol period. A symbol can be plotted as a vector in the complex plane, with each axis representing the in-phase and quadrature amplitudes. Figure 5.4 shows a QAM symbol corresponding to *I*=3 and *Q*=5 plotted in the IQ plane. This symbol is represented by a sinusoid with magnitude 5.83 (arbitrary units) and phase shift of 59° (with respect to the carrier).

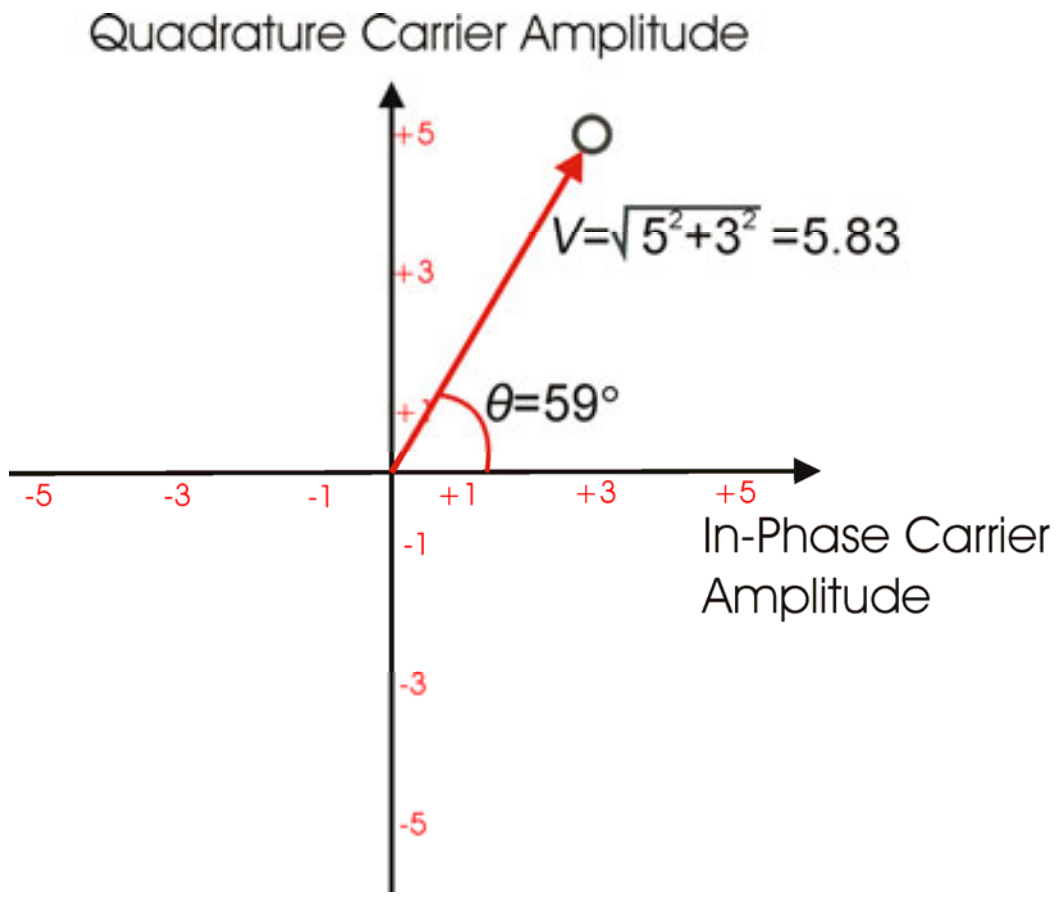

**Figure 5.4:** Illustrating a single symbol from an *M*-QAM system plotted in the IQ plane.

The entire symbol alphabet of a QAM system can be visualized in the IQ plane, in what is termed a constellation diagram. A receiver uses the constellation points as the reference points for transmitted symbols corrupted by noise. It estimates what was actually transmitted by selecting the point on the constellation diagram which is closest (in Euclidean distance) to that of the received signal. The constellation diagram allows a straightforward visualization of this process, and can give insight into various aspects of the system's design and performance. Hence, in a noisy channel, the distance between the constellation points is what determines the error rate.

The arrangement of the symbols in the constellation diagram (i.e. the $I$ and $Q$ amplitudes assigned to each symbol) is a design parameter and should be tailored to the particular application's requirements. In this thesis only the common rectangular (and cross) constellations will be considered (Figure 5.5).

It is easy to see that the error rate in *M*-QAM will depend on the distance between adjacent symbols. The minimum distance in all constellations shown is 2 (in terms of the arbitrary scale shown). It follows that for a given noise power, all of these QAM schemes would have the same SNR. Hence 8 bits/symbol can be transmitted using 256-QAM (i.e. 6 times more throughput than with 4-QAM) and one would expect the same SNR in the system. However this comes at the expense of higher average transmission power. 256-QAM requires ~19dB higher average signal energy compared to 4-QAM to preserve the minimum distance. Table 5.1 shows the average transmission energy required for a minimum distance of 2 between symbols in *M*-QAM [5]. As a rule of thumb, approximately 3dB more energy is needed for each additional bit.

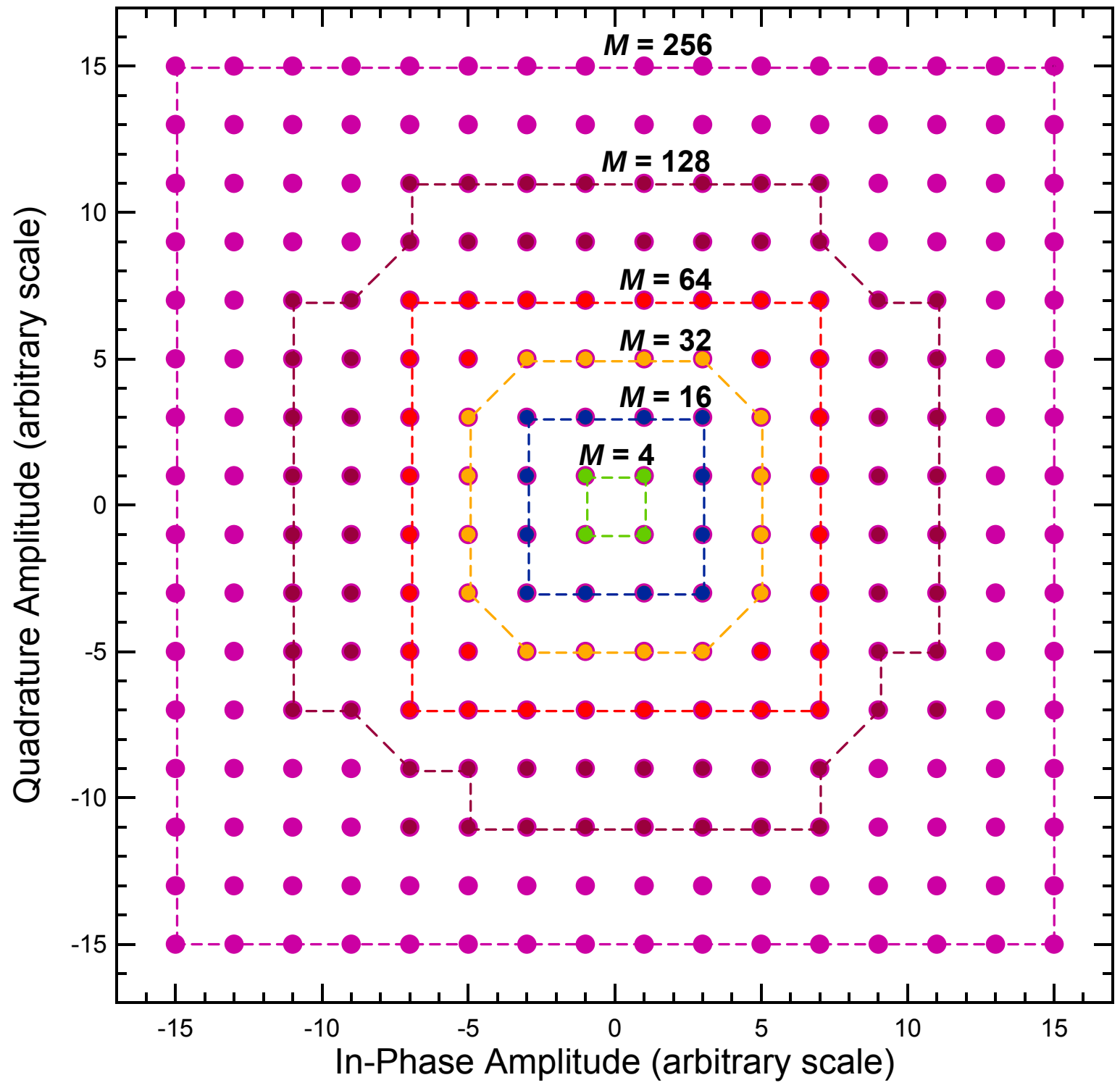

**Figure 5.5:** Constellation diagrams for rectangular and cross QAM. $M = 4$, 16, 32, 64, 128 and 256 shown. $M$=8 has been omitted for clarity.

**Table 5.1:** Average signal energy required for the transmission of the QAM schemes shown in Figure 5.5.

| $M$ | Bits/Symbol ($\log_2 M$) | Energy $E$ | $10\log_{10}E$ |
|---|---|---|---|
| 4 | 2 | 2 | 3.0 |
| 8 | 3 | 4.73 | 6.8 |
| 16 | 4 | 10 | 10.0 |
| 32 | 5 | 20 | 13.0 |
| 64 | 6 | 42 | 16.2 |
| 128 | 7 | 82 | 19.1 |
| 256 | 8 | 170 | 22.3 |

### QAM Error Rate

The upper bound in the symbol error rate (SER) of rectangular QAM is given by [3]:

$$P_e \approx 2erfc\left(\sqrt{\frac{3 \cdot SNR}{2 \cdot (2^b - 1)}}\right) \qquad (5.9)$$

*Where b is the number of bits/symbol in the scheme (i.e. $b=\log_2 M$) and $SNR=b\cdot E_b/N_0$.*

The complementary error function, *erfc*, is given by:

$$erfc(x) = \frac{2}{\sqrt{\pi}} \int_x^\infty e^{-t^2} dt \qquad (5.10)$$

The relationship between bit error rate (BER) and SER ultimately depends on the particular bit (to codeword) mapping used. However, the BER can be approximated by making some simplifications. For SNR>>1, it can be assumed that only adjacent symbol errors occur. If Gray encoding is employed, it follows that only a single bit error occurs for every symbol error. Hence, the following relation holds:

$$BER \approx \frac{SER}{\log_2 M} \qquad (5.11)$$

### EVALUATING MODULATION SCHEMES

Figure 5.6 shows the SER and BER of QAM for four different values of *M* as a function of SNR per bit, calculated from Equations (5.10) and (5.11). This plot is referred to as the *error probability plane* for QAM, and describes the possible operating points available with this type of modulation. The trade-offs available are illustrated by the arrows in Figure 5.6.

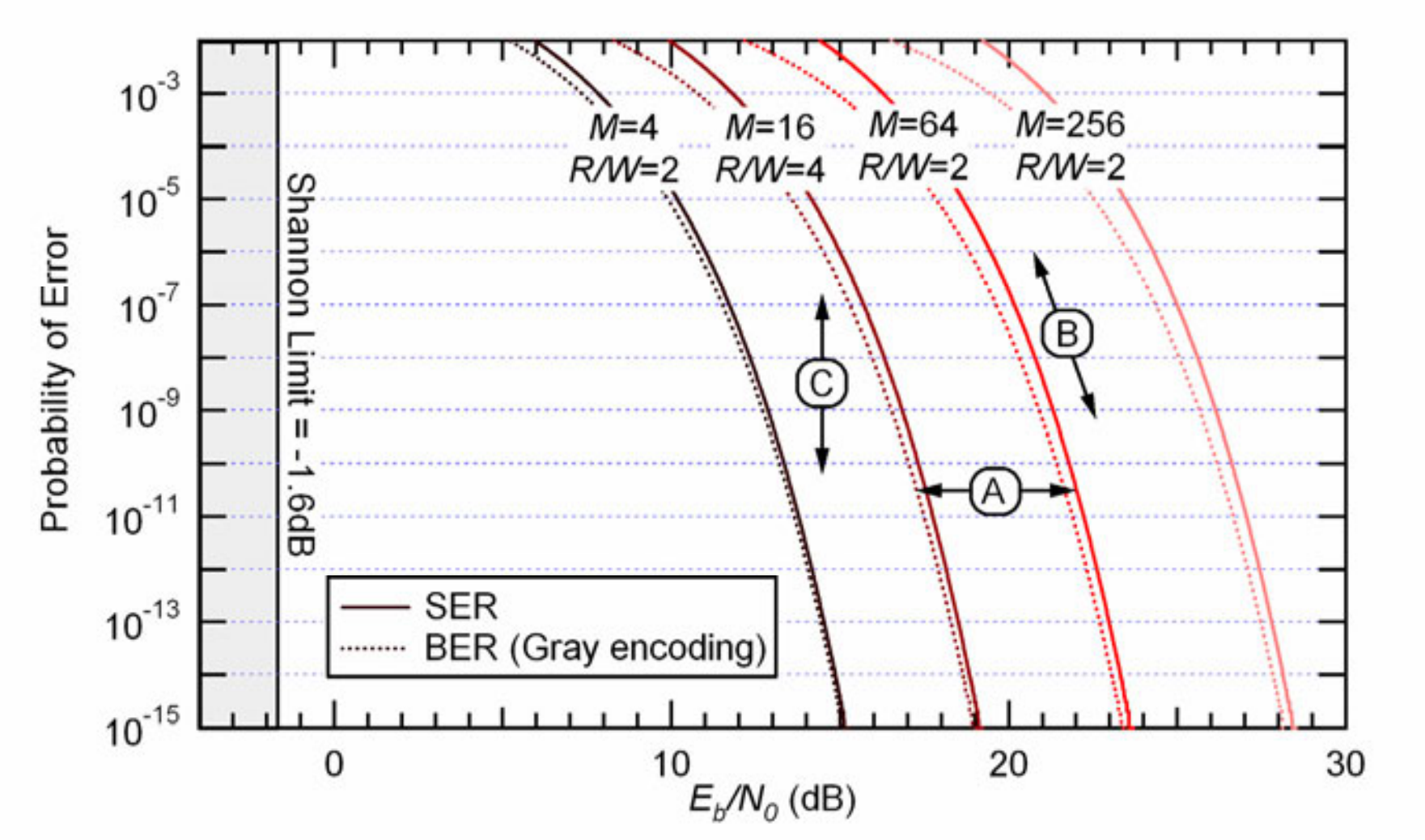

**Figure 5.6:** SER and BER (with Gray encoding) for QAM as a function of the SNR per bit.

The line denoted by A in Figure 5.6 shows the trade-off between the SNR per bit, $E_b/N_0$, and bandwidth efficiency, $R/W$, for a fixed BER. In qualitative terms, increasing the bits/symbol improves bandwidth efficiency, but requires more SNR to preserve the probability of error. For a fixed bandwidth, the BER can be reduced at the expense of $E_b/N_0$ (line B). This makes sense, since increasing the SNR per bit improves the BER, for a given modulation scheme. Finally, for fixed $E_b/N_0$, bandwidth efficiency may be increased by selecting higher order modulation scheme (i.e. increase $M$). This, of course, comes at the expense of inferior error rate performance. It is worth noting that movement along B only requires changing the available SNR per bit, which normally involves simply varying the transmission power. On the other hand, the trade-offs involving movement along A and C require a change in modulation scheme.

The information contained in Figure 5.6 can also be extracted from the QAM *bandwidth efficiency plane*, where the normalized rate ($R/W$) is plotted against the SNR per bit ($E_b/N_0$). This is illustrated in Figure 5.7, with the same tradeoff lines shown previously in the error probability plane. The plane includes the familiar Shannon capacity, which sets the theoretical limit to what is achievable in an AWGN channel. On the error probability plane (Figure 5.6) contours of equal bandwidth were essentially plotted. Here, curves of equal error probability are shown, though obviously only the points corresponding to a realizable modulation scheme are truly significant. The plane is a useful tool for assessing tradeoffs involving fixed BER, which is a common design constraint.

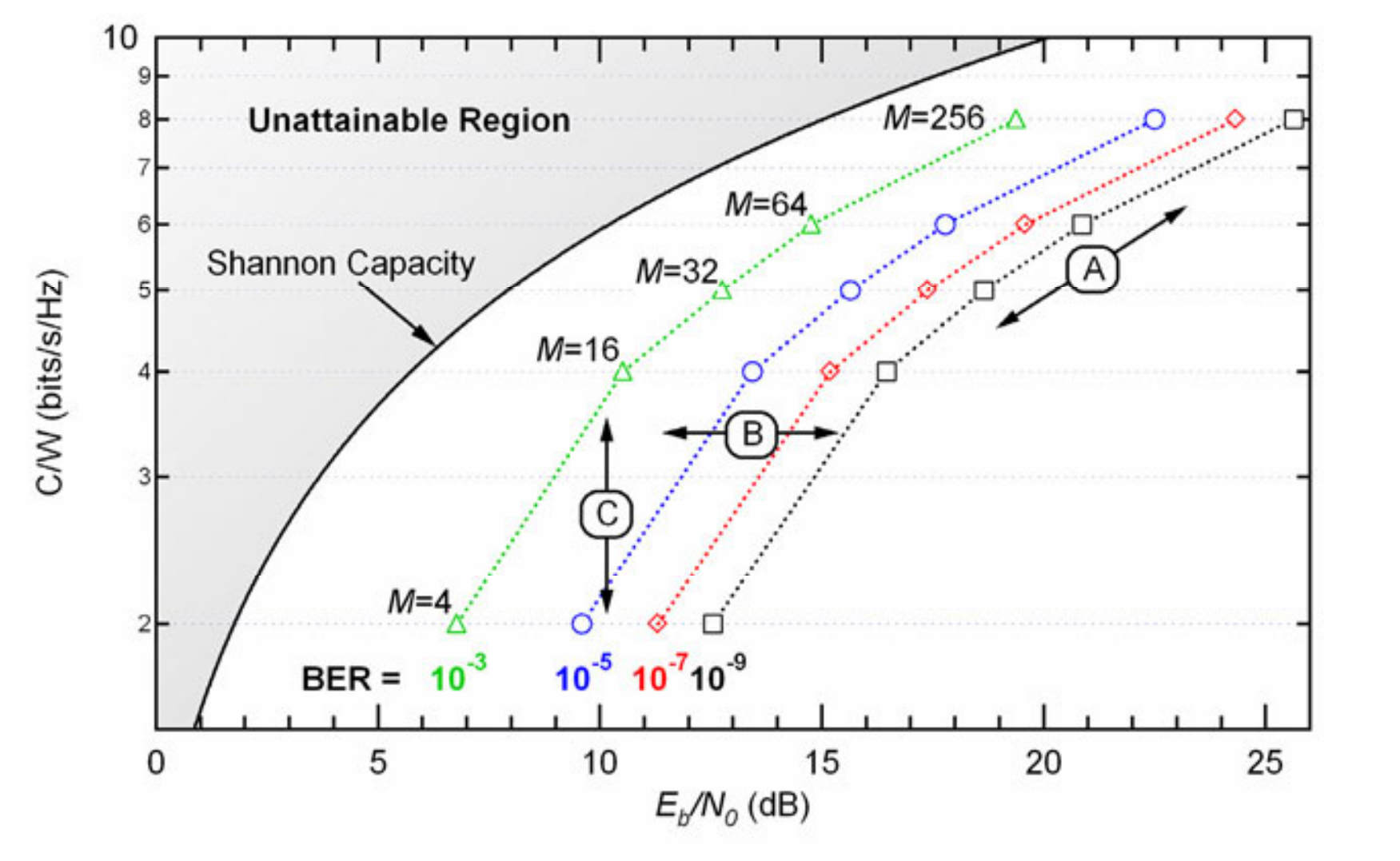

**Figure 5.7:** Bandwidth efficiency plane for QAM signals.

The bandwidth efficiency plane provides a powerful means for the comparison of different digital modulation schemes. Several common schemes will be compared, and their bandwidth requirements will be considered first.

For PSK signals, the required bandwidth is that of the equivalent baseband signal pulse (i.e. *g(t)* in Equation (5.7)). Hence the occupied bandwidth depends on the detailed characteristics of *g(t)*. To simplify the comparison, assume that *g(t)* is a pulse of duration $T$ and that its bandwidth $W$ is roughly equal to $1/T$. For a scheme with $b$ bits/symbol and a data rate $R$, $T=b/R=(\log_2 M)/R$. Hence:

$$W = \frac{R}{\log_2 M} \qquad (5.12)$$

It follows that, for a fixed rate $R$, increasing $M$, decreases the required channel bandwidth. Bandwidth efficiency (or normalized data rate), measured in bits/s/Hz, is simply the bit rate to bandwidth ratio:

$$\frac{R}{W} = \log_2 M \qquad (5.13)$$

In single-sideband (SSB) PAM, a channel bandwidth of $1/2T$ is required for signal transmission. Since $T=k/R=(\log_2 M)/R$:

$$\frac{R}{W} = 2\log_2 M \qquad (5.14)$$

Hence PAM-SSB has twice the bandwidth efficiency of PSK.

In QAM, there are essentially two PAM orthogonal carriers. Hence the rate is twice as high as in PAM-SSB. The signal, however, must be transmitted with double sideband (DSB), making the bandwidth efficiency of QAM and PAM-SSB the same.

Figure 5.8 shows $R/W$ against $E_b/N_0$ for QAM, PAM-SSB, PSK, Differential PSK and FSK, for a BER of $10^{-9}$ (assuming Gray encoding is used and Equation (5.11) holds). Also plotted is the normalized capacity of the AWGN channel, obtained from Equation (5.6). This gives an idea of how well these modulation schemes perform relative to the theoretical limit of reliable data transmission.

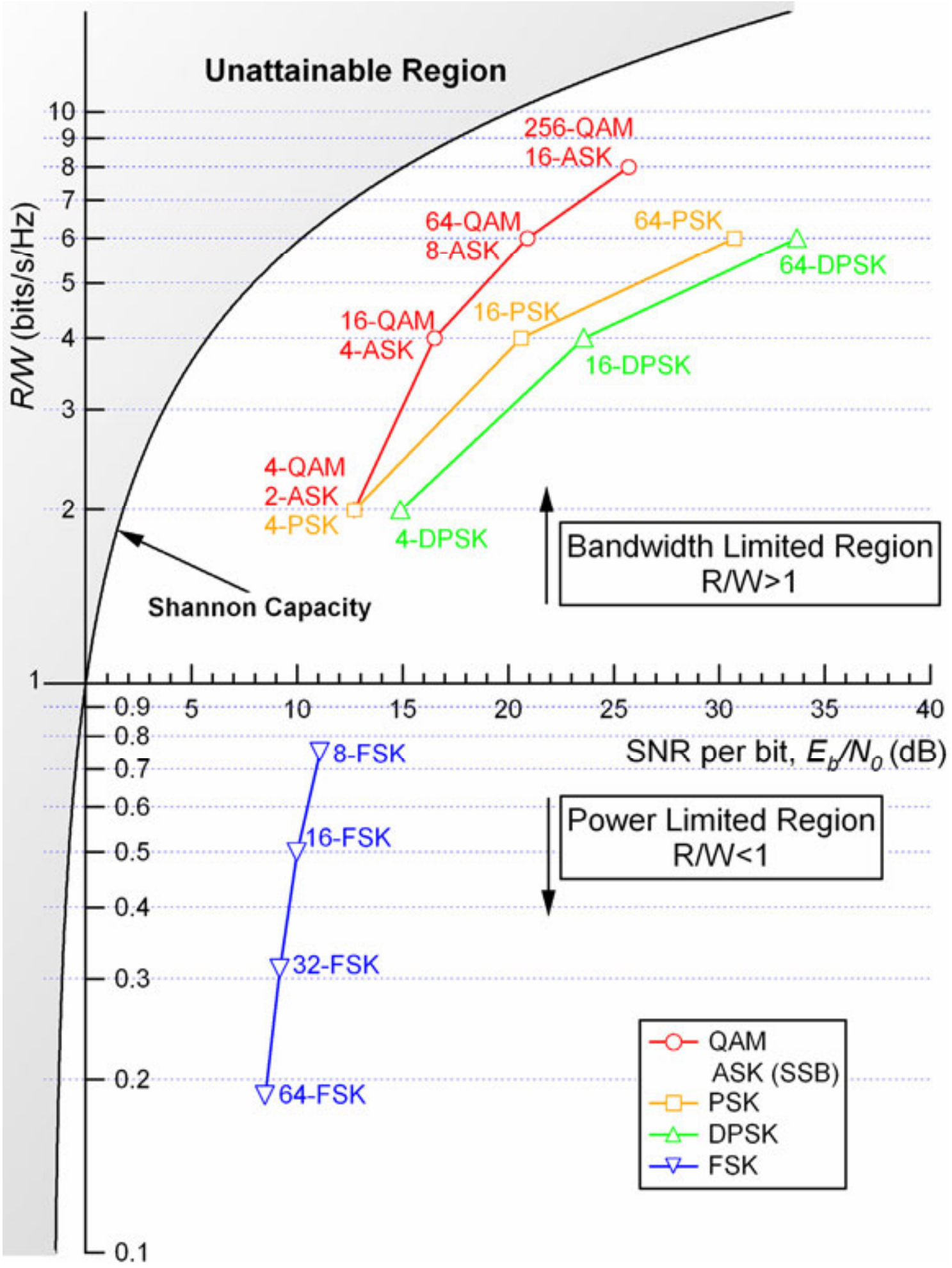

**Figure 5.8:** Comparison of several modulation schemes for a BER of $10^{-9}$.

It is immediately obvious that increasing *M* results in better bandwidth efficiency for PAM, QAM and PSK. The higher data rate, however, requires higher SNR per bit. This suggests that these schemes are more suited to bandlimited channels where there is sufficient SNR to support higher order schemes, therefore maximizing $R/W$.

On the other hand, FSK signals yield $R/W \leq 1$. As $M$ increases, $R/W$ decreases due to the need for more channel bandwidth to accommodate the $M$-ary orthogonal signals. Conversely, the SNR per bit required decreases with increasing $M$. It follows that FSK is more appropriate for power-limited channels where bandwidth is plentiful.

### 5.2.4 Multi-Carrier Modulation

The concept of using multiple carriers to transmit information has its roots in the 1950s and 1960s, and is based on Frequency Division Multiplexing (FDM). The idea involves dividing the channel into several smaller sub-channels and placing a modulated carrier (normally QAM) in each. Initially, filtering was used to completely separate adjacent sub-channels and avoid interference. The concept of allowing spectral overlap between carriers can be traced as far back as the1960s [6], though it was not until the 1990s that the first commercial systems became viable. This was mainly due to the advent of powerful digital signal processors that could actually synthesize the sum of modulated waveforms using the Fast Fourier Transform (FFT) [7].

Multi-carrier modulation can nowadays be found in a variety of applications, such as Asymmetric Digital Subscriber Line (ADSL), Wi-Fi (IEEE 802.11) and the European Telecommunication Standards Institute (ETSI) standards for Digital Audio Broadcast (DAB) and Digital Video Broadcasting-Terrestrial (DVB-T). All of these are based on the principle of orthogonally spaced carriers. The term Discrete MultiTone (DMT) is used in the case of wireline systems such as ADSL, while Orthogonal Frequency Division Multiplexing (OFDM) is more often encountered in the wireless world. It is frequently the case that OFDM implies equal constellation sizes across all carriers. A variant scheme where this is not the case is termed *Adaptive* OFDM. When referring to the generic multi-carrier scheme (which may or may not include uniform constellations) the terms OFDM, multi-carrier and multi-tone will be used interchangeably throughout this text.

There are several benefits associated with OFDM. Firstly, if the sub-channel spacing is made small enough (i.e. by using a large number of carriers in a given channel bandwidth), each carrier will be subject to a flat frequency response, hence avoiding InterSymbol Interference (ISI). This eliminates the need for equalization at the receiver, which is compulsory for a single-carrier system in a frequency selective channel. In addition, the data stream is divided among the carriers into several, low bit rate streams. This means that the symbol has considerable duration in time, thus exhibiting greater immunity to impulse-like noise. Finally, by dividing the channel into many sub-channels, it is possible to optimally allocate the transmission power in the channel, as described by Shannon in [8]. This so-called ‘water-filling’ technique is described later.

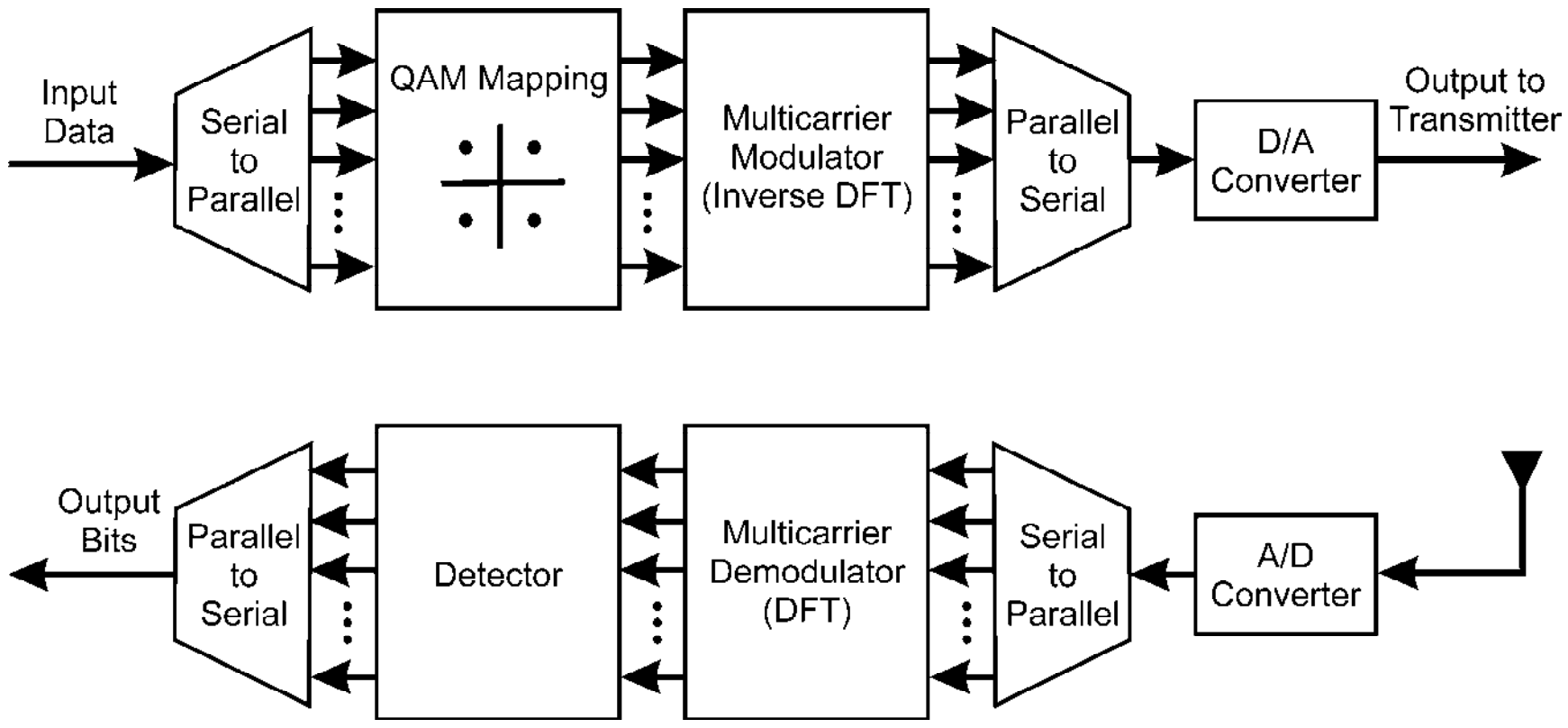

**Figure 5.9:** Multicarrier communication system.

Figure 5.9 shows the basic multicarrier communication link [3]. The FFT algorithm is used to efficiently compute the DFT and synthesize the signal at the transmitter and to demodulate the received signal at the receiver. The input data stream is segmented into frames of $N_f$ bits by the serial-to-parallel buffer. The $N_f$ bits are divided into $\tilde{N}$ groups, where the $i^{th}$ group is assigned $\tilde{n}_i$ bits. It follows that:

$$\sum_{i=1}^{\tilde{N}} \tilde{n}_i = N_f \qquad (5.15)$$

Each of the $\tilde{N}$ groups can be encoded separately, and hence the number of bits output from the encoder for the $i^{th}$ group is $n_i \geq \tilde{n}_i$. In this thesis we will consider the OFDM system as being comprised of $\tilde{N}$ independent $M$-QAM carriers, with $M = 2^{n_i}$ symbols. The complex-valued information symbols in each sub-channel are denoted by $X_k$, $k$=0, 1, … , $\tilde{N}-1$. The inverse FFT (IFFT) is used to modulate the $\tilde{N}$ carriers by the information symbols $\{X_k\}$. In order to ensure that the IFFT yields a real sequence equivalent to $\tilde{N}$ QAM carriers at the output, we create $N=2\tilde{N}$ information symbols by defining:

$$X_{N-k} = X_k^*, \quad k=1, \ldots, \tilde{N}-1 \qquad (5.16)$$

The symbol $X_0$ is split into two real parts, where $X_0' = \mathrm{Re}[X_0], X_{\tilde{N}} = \mathrm{Im}[X_0]$. The $N$-point IFFT then gives:

$$x_n = \frac{1}{\sqrt{N}} \sum_{k=0}^{N-1} X_k e^{j2\pi nk/N}, \qquad n=0, 1, \ldots, N-1 \qquad (5.17)$$

Where $1/\sqrt{N}$ is simply a scale factor making the output independent amplitude of the number of carriers. The sequence $\{x_n,\ 0 \leq n \leq N-1\}$ corresponds to the samples of the sum $x(t)$ of $\tilde{N}$ sub-carrier signals, expressed as:

$$x(t) = \frac{1}{\sqrt{N}} \sum_{k=0}^{N-1} X_k e^{j2\pi kt/T}, \qquad 0 \leq t \leq T \qquad (5.18)$$

*Where T is the symbol duration.*

Hence the carriers are centered at the frequencies $f_k = k/T$, $k$=0, 1,…, $\tilde{N}$. It is this frequency spacing that accounts for the orthogonality of the sub-channels, and allows ISI-free demodulation.

At the receiver, the demodulated sequence can be expressed as:

$$\hat{X}_k = C_k X_k + \eta_k, \qquad k=1, \ldots, N-1 \qquad (5.19)$$

*Where* $\{\hat{X}_k\}$ *is the output of the N-point DFT demodulator,* $C_k$ *is the frequency response of the channel at* $f_k$ *and* $\eta_k$ *is the additive noise corrupting the signal.*

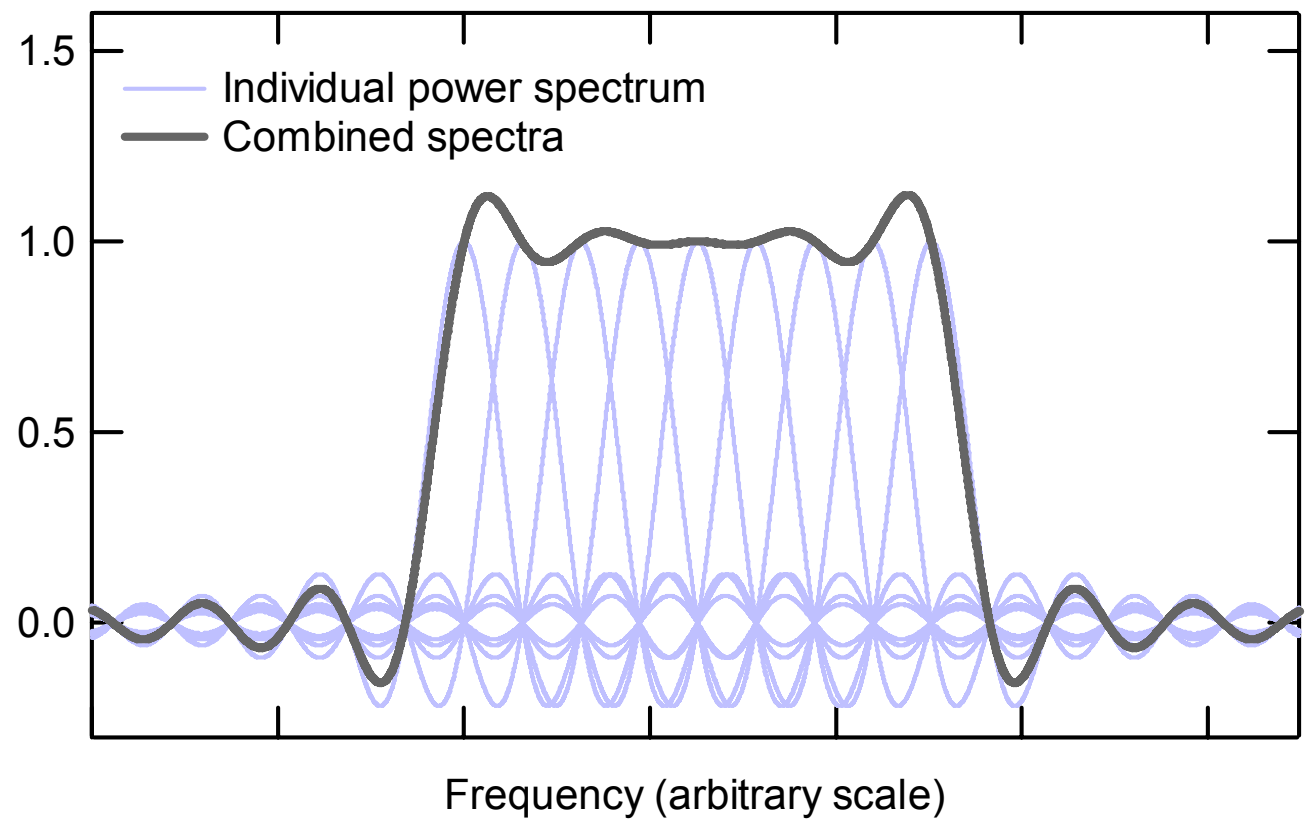

**Figure 5.10:** Illustrating the individual and combined power spectra of an 8-carrier OFDM system.

The principle of orthogonality in OFDM can be illustrated by considering a simple example of a system with $\tilde{N}$=8. Each carrier is assigned 1 bit, by its presence or absence in the output spectrum. The power spectra of the individual carriers have the form sinc($x$)=sin($x$)/$x$ [9]. This is demonstrated in Figure 5.10 where the individual and summed power spectra of 8 carriers spaced at $1/T$ is shown. In the frequency domain, the sinc functions' side lobes produce overlapping spectra. Each carrier's frequency spectrum peaks at its center frequency and goes to zero at all integer multiples of $1/T$. The OFDM receiver can effectively demodulate each carrier independently because, at the peaks of each of these sinc functions, the contributions from all other carrier sinc functions are zero.

### 5.2.5 Capacity of the Frequency Selective Gaussian Noise Channel

In 1949, Shannon determined the power allocation to maximize the capacity in the frequency selective channel corrupted by Gaussian noise [8]. Shannon was not referring specifically to a multi-carrier system, but rather to the distribution of transmission power as a function of channel frequency. Nevertheless, his result indirectly demonstrates an advantage of using multiple carriers: If the channel bandwidth is divided into a theoretically infinite number of sub-channels, it is possible to achieve Shannon's capacity-maximizing distribution by appropriate allocation of transmission power to each carrier. Shannon's analysis of what he termed the 'arbitrary noise' channel capacity follows.

The frequency selective channel which adds Gaussian noise can be thought of as a filter with transfer function $Y(f)$ in series with an AWGN source at the output (see section 5.3.2). If the AWGN source has a constant noise PSD $K$, then the noise power spectrum at the output will be $N(f)=K|Y(f)|^2$. By dividing the channel into a large number of sub-channels in the frequency range from, say, 0 to $W$, the frequency response will appear flat in each part. It is then possible to use the AWGN channel capacity equation (section 5.2.2) for each sub-channel.

If the total transmitter power, $P$, is distributed among all sub-channels according to $P(f)$, then the following is true:

$$P = \int_0^W P(f)df \qquad (5.20)$$

The total capacity of all AWGN sub-channels is then given by:

$$C_1 = \int_0^W \log_2\left(1+\frac{P(f)}{N(f)}\right)df \qquad (5.21)$$

The maximum rate of transmission is found by maximizing $C_1$, subject to the total power constraint given in Equation (5.20). This is achieved by maximizing:

$$C_1 = \int_0^W \left[\log_2\left(1+\frac{P(f)}{N(f)}\right)+\lambda P(f)\right]df \qquad (5.22)$$

The condition for this is given by:

$$\frac{1}{N(f)+P(f)}+\lambda=0 \qquad (5.23)$$

It follows that *N(f)+P(f)* is a constant, adjusted so that the total power is equal to *P*. This is referred to as the 'waterline', and is illustrated in Figure 5.11, which is a plot of noise power versus frequency for a hypothetical channel. The consequence of this result is that more power is assigned where the noise is low, and less power where the noise is high, which is intuitive. For an available channel bandwidth *W*, the two waterlines shown correspond to two choices of *P*. The optimal power distribution is simply the difference between the (constant) lines and the noise power.

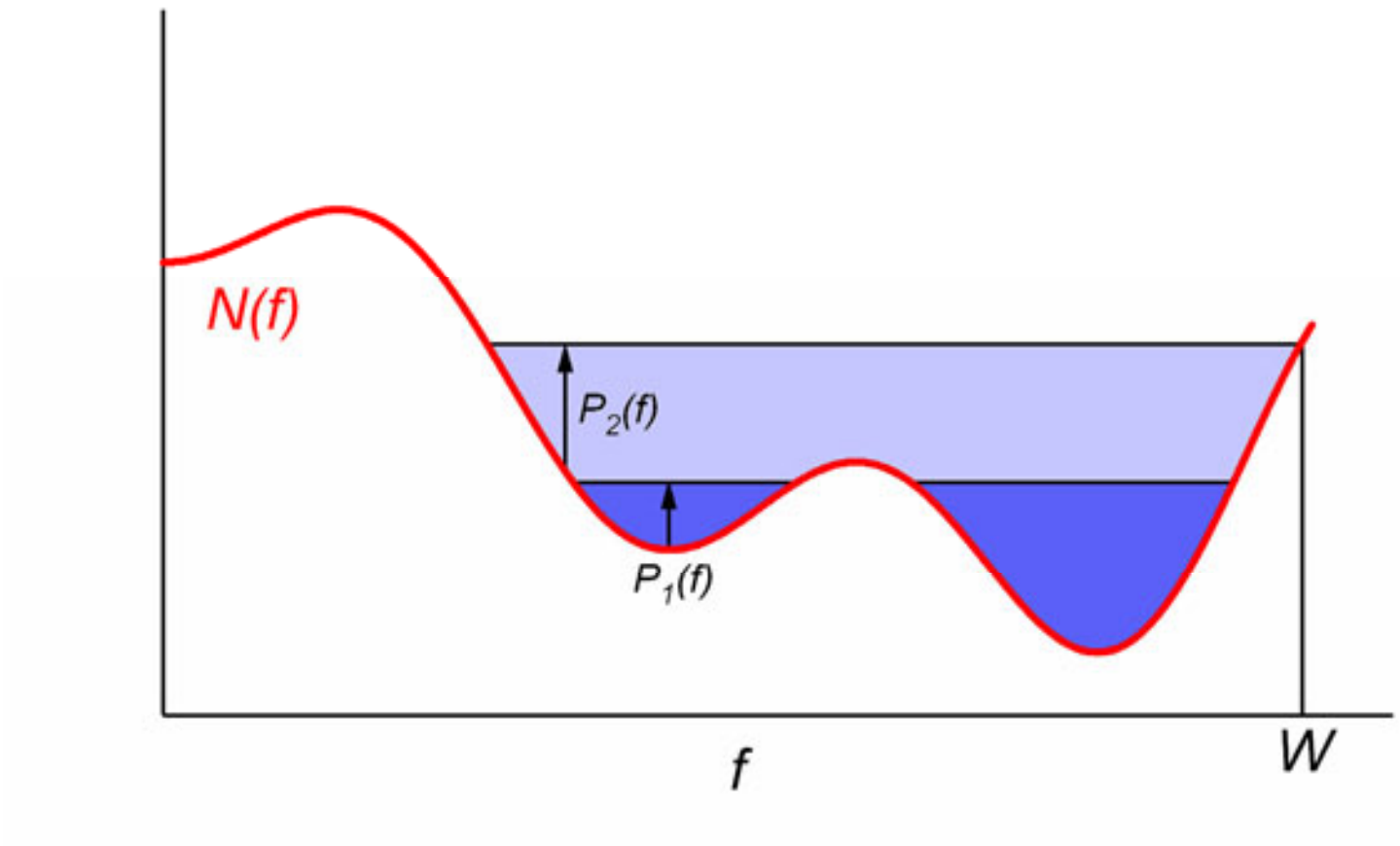

**Figure 5.11:** Illustrating 'water-filling' to obtain the best distribution of transmitter power in a channel with Gaussian noise that varies with frequency as *N(f)*.

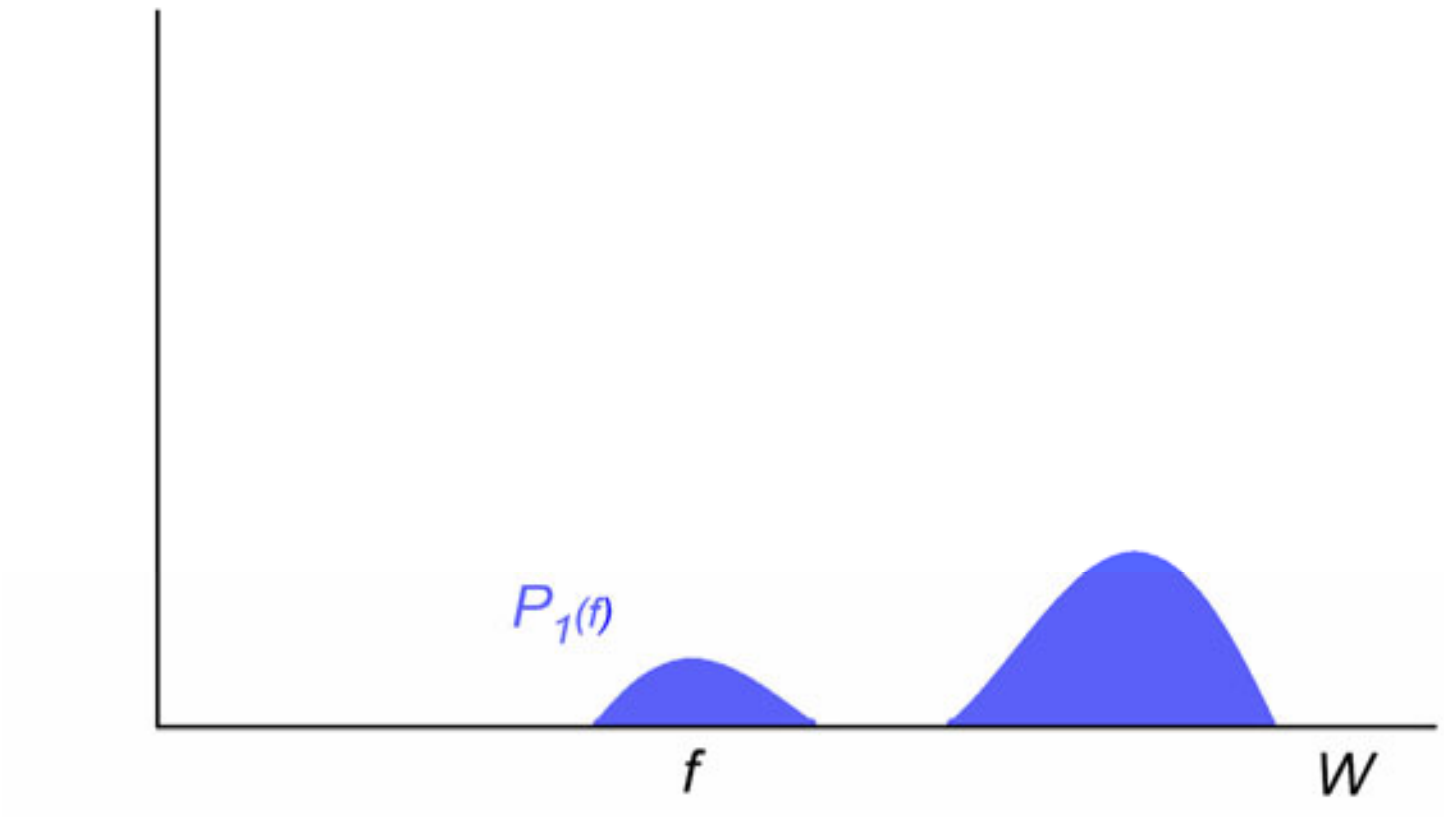

**Figure 5.12:** Optimal transmission spectrum for maximizing channel capacity in the example of Figure 5.11, for the waterline corresponding to $P_1(f)$.

Clearly, it is not always possible to have constant *N(f)+P(f)*, since this would yield negative power at some frequencies (i.e. if the *P* is small enough). It turns out that the best distribution in this case is one where *N(f)+P(f)* is constant wherever possible, and *P(f)* is zero at all other frequencies. Figure 5.12 illustrates the best distribution of transmission power for the channel noise plotted in Figure 5.11, for the lower of the two waterlines ($P_1(f)$).

### 5.2.6 Power and Bit Loading in Multi-Carrier Modulation

Loading algorithms [10] compute the number of bits/symbol ($b_n$) and energy ($E_n$) for each and every sub-channel in a multi-carrier system. In general, loading algorithms can be categorized into types: Those that attempt to maximize the data rate for a given power constraint (*rate-adaptive)*, and those that minimize the error rate, at a fixed data rate (*margin-adaptive)*. In this thesis only the former case is considered.

Loading algorithms may produce $b_n$ that have fractional parts or be very small. Such small or fractional $b_n$ can complicate encoder and decoder implementation. Often, $b_n$ is constrained to integer values, leading to sub-optimal solutions which are easier to implement. In addition, the minimum number of bits/symbol assigned to each carrier is normally ≥ 1.

OPTIMAL LOADING

The capacity of the frequency selective Gaussian noise channel presented in the previous sections has obvious implications for systems employing multiple-carriers. In a scheme with *N* carriers, as $N\rightarrow\infty$, it is possible to achieve the transmission power distribution that maximizes data rate. This is accomplished by distributing the available power to each carrier in the manner shown previously. This is termed *waterfilling*. Several optimal and near-optimal waterfilling algorithms exist (examples can be found in [10]).

DEFAULT LOADING

The default loading scheme is one where the available energy is allocated equally among all carriers. The two constraints are the total energy, and the system's target error rate. Hence, assuming that the information carried by all carriers is equally important, then both the power and the error rate are constants for all carriers in the system. A typical default loading scheme is summarized in the following steps [11]:

1. Load power $P_k=P_{tot}/N$.

   $P_{tot}$ is the total system power and $N$ is the initial number of carriers.

2. Determine the SNRs of all carriers. Hence calculate the bit/symbol assignment, $b_n$, for each carrier[3].

3. Determine the set of carriers, which, for the given target error rate, are assigned $b_n$ < 1bit/symbol.

4. 'Turn off' the carrier with the smallest $b_n$. Redistribute its power to the rest of the carriers. Note that $N_{new}=N_{old}-1$.

5. Repeat until all carriers have $b_n \geq 1$.

As illustrated by Figure 5.7, the total bit rate of a QAM multi-carrier system, $R_b = \sum_n b_n$ , is logarithmically dependent on the transmission power. On the other hand, bit rate is directly proportional to bit loading (i.e. the number of assigned bits/symbol). Hence the aggregate data rate varies far more quickly by a change in the bits/symbol assigned to each sub-channel, than by a change of sub-channel power. From Figure 5.7, the SNR has to be increased by ~2.5-3dB for an increase of 1 bit per QAM symbol at BERs of ~$10^{-5}$-$10^{-9}$, corresponding to a doubling in power in each sub-channel. The consequence of this is that one would expect the default loading scheme to achieve performance which is relatively close to the optimal one achieved by waterfilling. This was demonstrated by Daly in [11], where Willink's test channel [12] was used to compare various different loading schemes. The default algorithm resulted in over 99% of the bit rate achieved with optimal power allocation (for a target symbol error rate of $10^{-5}$).

# 5.3 Digital Link Upgrade Path

## 5.3.1 Overview

An overview of the concepts that are required for the conversion of the analog optical links to a digital system is given here. It should be noted that the scope of this thesis is limited to investigating the data rate possible using a digital modulation scheme that maximizes the bandwidth efficiency of the link.

[3] The number of bits/symbol that can be assigned to a QAM carrier for a given error rate and SNR can be found by re-arranging Equation (5.9). The exact bit-loading process used in the subsequent analysis is described later, in section 5.4.3.

The development should be treated as that of a digital communication system, where the transmission channel is the existing analog optical link (Figure 5.1). The areas of interest are summarized in the following topics.

### 5.3.2 The Channel: The Current CMS Tracker Analog Optical Link

In a digital communication system, the analog channel is the medium over which digitally modulated signals are transmitted. The frequency selective channel which adds white Gaussian noise is a commonly used model for analysis of digital communication systems (Figure 5.13). In later sections this simplified channel model will be used to represent the Tracker optical link.

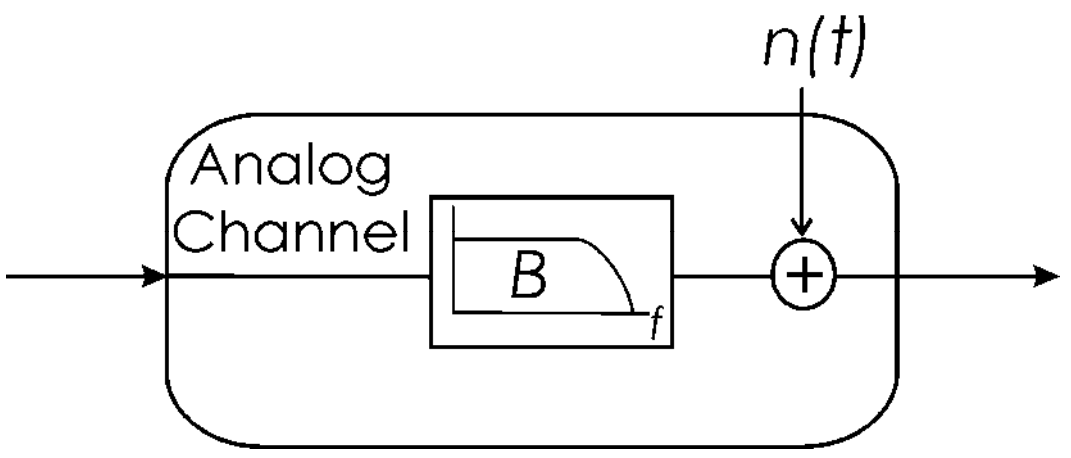

**Figure 5.13:** The frequency selective AWGN channel.

The frequency response of a complete optical link was measured using a Spectrum Analyzer (Figure 5.14). A circuit based on a differential driver IC was used at the spectrum analyzer output in order to adapt it to the AOH input. The response of this circuit was subtracted from the response of the whole link.

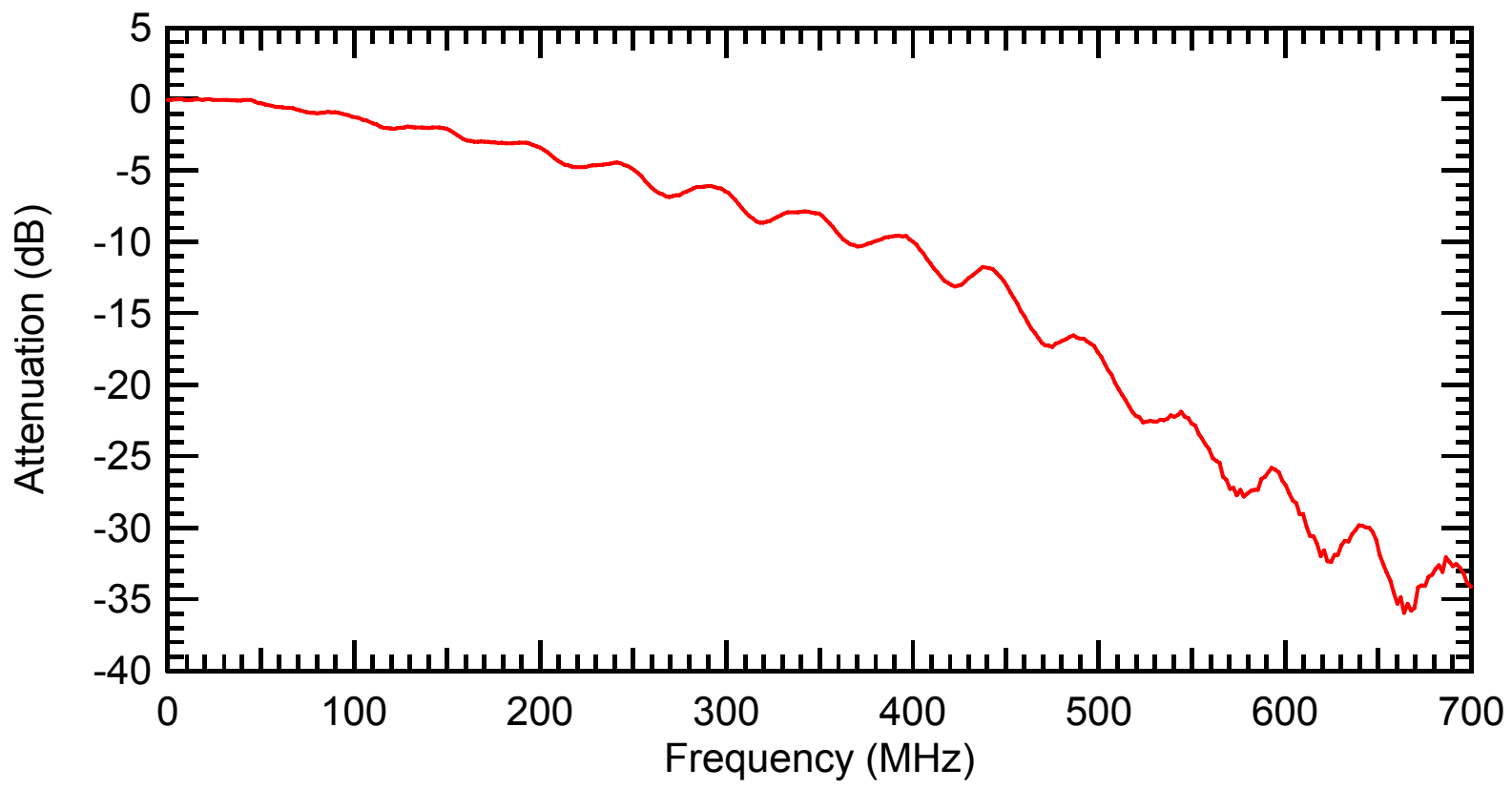

**Figure 5.14:** Frequency response of the analog link. The y-axis is the normalized signal power attenuation.

The noise of the link has been specified over 100MHz for a Signal to Noise Ratio (SNR) of 48dB [13]. In later sections, a constant noise PSD is assumed for all frequencies. Given the SNR specification, the value of the noise PSD, $N_0$, is $\sim 1.5 \cdot 10^{-13}$W/Hz (normalized to a maximum input power of 1W).

### 5.3.3 Channel Encoding - Digital Modulation

A system based on QAM is envisaged, as this provides good bandwidth efficiency. The choice of a single or multi-carrier QAM system depends on practical limitations. Multi-carrier schemes are used in wireless (e.g. Wi-Fi) and ADSL applications, where the channel either exhibits frequency fading, or does not necessarily have a fixed frequency response. A single QAM carrier system could be used in a future digital system upgrade, since the frequency response of each link is a constant. This approach has two major drawbacks: Firstly, the modulator with a huge symbol rate (~400-500MS/s) would be required to fully exploit the available bandwidth. Secondly, the channel's frequency response is not flat, and an equalizer would be required at the receiver to combat the resulting Inter Symbol Interference (ISI). The equalizer would very likely employ an adaptive algorithm to adjust its coefficients in order to match the channel response (since not all links will exhibit exactly the same frequency response). At the Gbit/s data rates envisaged, this would require powerful and complex hardware.

On the other hand, a multi-carrier system based on OFDM would not require equalization at the output, as long as the symbol rate of each carrier was low enough (i.e. sufficient number of carriers are used). Instead, such a system would require an adaptive algorithm for bit and power allocation, depending on the channel characteristics. A drawback of an OFDM-type solution is that more complex hardware would be required on the transmitting end, since an FFT has to be performed. This is not ideal in the context of a HEP experiment where power consumption inside the detector volume is an issue.

In terms of the achievable data rate, OFDM can –in principle– be considered equivalent to a single QAM carrier system. In practice, due to higher PAPR, the transmission power in a multi-carrier system must be 'backed off' to avoid signal clipping that can cause SNR degradation. While coding techniques exist for mitigating these effects, the two systems are not necessarily identical as far as the data rate is concerned.

### 5.3.4 Channel Encoding - Forward Error Correction

Forward Error Correction (FEC) can be used to detect and correct errors in the transmitted data stream. A Bose-Chaudhuri-Hochquenghem (BCH) class code (such as Reed-Solomon) can be employed. The overhead introduced by the extra bits inserted into the data stream is offset by the error-correcting capability of the code. The code used must therefore strike a balance between the two, and will be a function of the data rate and BER achieved with the chosen modulation method. FEC and modulation scheme can also be combined, as in Trellis Coded Modulation (TCM).

### 5.3.5 Source Encoding

A novel approach –for a High Energy Physics (HEP) application– to increasing the information rate of the detector readout link would be to use compression prior to transmission. Research into compression codes and associated algorithms has gone a long way since Shannon first gave birth to the Source Coding Theorem in 1948. Realizable codes that achieve compression very close to the Shannon limit have emerged. This will allow sending more information bits per transmitted symbol, and hence more data can be transferred for a given channel capacity. Compression essentially removes redundancy from the data stream; received bit errors therefore become more severe. A balance needs to be reached between the compression factor and effective BER of the final system. FEC and compression encoding can co-exist, and may be considered together at this stage.

The code used may be one that does not require knowledge of the source statistics (e.g. Lempel-Ziv), or an optimal, variable-length code -such as Huffman- that can achieve compression very close to the source entropy. For the current CMS Tracker, the source statistics are well-known (or can be predicted to a high degree of accuracy) from production testing on detector hybrids. Compression is already employed for physics data collected from experiments, and the corresponding research is a good starting point to investigate the benefit of data compression before transmission. The aim is to demonstrate whether or not gains in transmission efficiency can offset the increased system complexity and power consumption in a future digital system.

### 5.3.6 System-level Considerations

Power consumption at the front end (detector side) electronics is the main constraint in the readout link design. As a guideline, the current CMS Tracker will dissipate 2.4mW per detector channel, with a total 10 million detector channels. Each optical channel carries data from 256 detector channels, and the power consumption is ~134mW per optical channel. New IC technology and scaling will mean lower power consumption, but more processing at the front end (digitization, etc) and more channels will scale similarly. Hence, the overall power constraint is not expected to change in the future.

The CMS trigger rate determines the frequency at which (blocks of) data are sent through the optical links. For future links, this is expected to remain at ~100kHz. Encoding and processing at the front end must be performed fast enough to deal with the shortest time at which triggers can arrive. At present, this is 75ns (three times the clock period), but no safe assumptions about the future can be made at this point. A pipeline memory buffer may also be used to extend the maximum required processing time.

### 5.3.7 Error Rate Requirements

The BER requirements of a future readout link for the upgraded Tracker are currently unknown, since the new version of the sub-detector has not been designed yet. Nevertheless, this is a fundamental specification for a communication link and the data rates calculated in subsequent sections are only significant when given with their corresponding error rates.

The target error rate should be chosen based on the effect that errors have on the quality of the data being read out. For example, in the case of a Tracker-like detector, the performance metric could be the efficiency of track reconstruction. The exact relationship of error rate and efficiency also depends on the format of the readout data. If, for example, the data is encapsulated in frames, a single bit error could possibly render a whole data frame corrupt and unusable. Then, it is the frame error rate that will –to some extent– affect the track finding algorithm and, ultimately, the tracking efficiency.

Clearly, in order to accurately estimate the required BER, one needs to take into account all the relevant detector and readout system parameters and perform simulations to relate error rate to efficiency. Since these are currently unknown, a range of target error rates (based on specifications of current readout links) are used in the calculations made in this thesis. Moreover, the effect of BER on the detector and readout system performance has never been carried out for past systems. Instead, typical values used in industry are normally adopted by the HEP community for digital links.

The CMS ECAL sub-detector is read out by gigabit optical links (GOLs), capable of 800Mbit/s and 1.6Gbit/s data rates, with a BER specification of $10^{-12}$ [14]. This is also the case for the 80Mbit/s CMS Tracker digital control link, while the 40Mbit/s ATLAS SemiConductor Tracker (SCT) readout link's BER specification is $10^{-9}$ [15]. It is likely that a future link will require similar error rates. In addition, forward error correction (FEC) will certainly be included to increase performance.

Commercial optical fiber systems are commonly specified for BERs between $10^{-9}$ and $10^{-12}$. This is often achieved through the use of FEC codes. For example, in ITU-T G.709 is a recommendation based on a Reed-Solomon RS(255,239) code. This is capable of reducing the BER to $10^{-15}$ from an input BER of $10^{-4}$. While this thesis does not take into account possible FEC codes when calculating the potential performance of a future upgrade, the improvement that can be accomplished is well-known. The intention is to provide a more generally applicable result that is not affected (or biased) by the particular implementation, especially the choice of encoding. Therefore, a range of target error rates (down to $10^{-9}$) are used as inputs to the data rate calculations, and the use of FEC is implied to improve performance to below $10^{-12}$.

## 5.4 Maximizing the Bandwidth Efficiency of the CMS Tracker Analog Optical Links: An Analytical Approach

### 5.4.1 Overview

In this section, the achievable data rate using a QAM-based digital system with the current optical link components is determined. A multi-carrier system will be considered, though the results are directly applicable to a single-carrier scheme. In fact, the analysis makes use of a single-carrier equivalent to the multi-carrier system to illustrate the equivalence.

The analysis is based mainly on the work of Cioffi [7, 10]. Peak power issues in OFDM are discussed first. The concept of the SNR gap [7, 11] that determines the number of bits/symbol (and therefore the data rate) for a single QAM carrier is then introduced. Finally the multi-channel SNR [10] is presented. This is a measure that defines an equivalent single carrier performance metric for a multi-carrier system, and this simplifies the calculation of the achievable data rate.

### 5.4.2 OFDM Peak Power Issues

A major drawback of OFDM is the high peak-to-average power ratio (PAPR) of the transmitted signal [16-19]. The peak signal occurs when there is coherent addition of the highest amplitude symbols of each sub-carrier. For a QPSK-OFDM system, the peak power is upper bounded by the number of carriers, *N*, times the average power. Clearly, PAPR is a problem for the transmitting and receiving amplifiers, since they must accommodate these large signals in order to avoid signal clipping which leads to interference, and therefore bit-errors. This is achieved by lowering (or 'backing off') the average power of the transmitted signal, so that the signal peaks fit in the amplifiers' input ranges. Power consumption of a power amplifier depends largely on the peak power, and hence accommodating occasional large peaks leads to low power efficiency. This so-called power 'backoff' leads to lower signal power, and hence lower available SNR for data transmission.

As an example, consider the case of a multi-carrier system with a PAPR=128. The backoff required for completely avoiding signal distortion would be $10 \cdot \log_{10}(128)=21$dB. However, the resulting SNR reduction would lead to a much lower data rate for a given BER. Practical OFDM systems do not employ such a large backoff, at the expense of a tolerable degradation in SER. The chosen backoff of the transmitted signal is thus a compromise between the contradictory desires to achieve a large average transmission power on the one hand and a low distortion due to clipping of the signal on the other [16].

Moreover, high signal peaks in OFDM occur relatively rarely. Hence it is the distribution of the OFDM signal's PAPR that is of interest when trying to quantify the effect of signal compression. The distribution of PAPR is the subject of a lot of research, and various (approximate) analytical solutions have been suggested. For illustrative purposes, one of these is presented here. In [17], Ochiai and Imai have come up with a simplified approximation of the cumulative distribution function (cdf) for the crest factor (i.e. the peak to average amplitude) of an OFDM signal, valid for large *N*:

$$F_C(r) \approx \exp\left[-\sqrt{\frac{\pi}{3}} N r e^{-r^2}\right] \qquad (5.24)$$

*Where r = crest factor.*

From Equation (5.24), the complementary cdf as a function of PAPR can be plotted for various *N* (Figure 5.15). It can be seen that for *N* =100, the probability of the PAPR exceeding 15dB is less than $10^{-11}$.

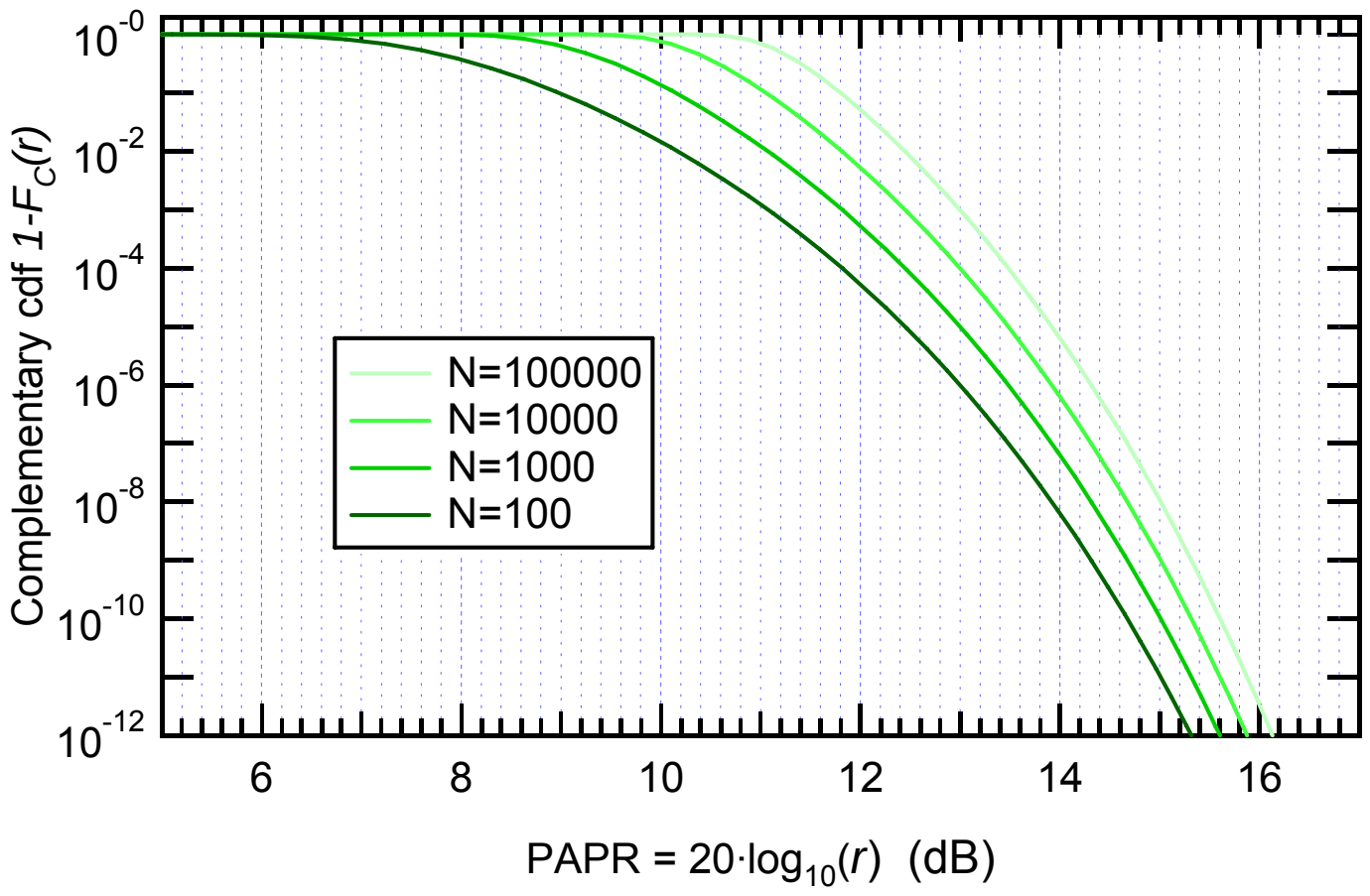

**Figure 5.15:** Complementary cdf of the PAPR for various values of *N*.

Given the signal statistics, as well as the link transfer characteristic, one could potentially calculate the BER penalty incurred due to signal clipping and select the optimum power backoff for a particular implemented system. Amplifier compression is a non-linear effect, making this analysis rather complex. It is also implementation-specific and hence will not be considered in this thesis.

Moreover, the aim of this work is to determine the *maximum* data rate, regardless of the number of carriers used or the modulation schemes selected. A future system could, for example, employ only a few carriers which would greatly reduce the need to for a large power backoff, or employ PAPR reducing techniques. The data rate will therefore be calculated for a range of backoff values, ignoring any resultant BER degradation. The results will then act as a reference against which a future system design can be compared to.

### 5.4.3 Bit Allocation Using the SNR Gap

'SNR gap analysis' [7, 11] will be described here for a single sub-channel (i.e. one QAM carrier) as an introduction to the concepts relevant to multi-carrier system design. The equations needed for calculating the maximum bit-rate in a single-carrier QAM system are given.

The number of bits that can be loaded onto a carrier for a given error probability can be determined from:

$$b = \frac{1}{2}\log_2\left(1+\frac{SNR}{\Gamma}\right) \quad \text{[bits per dimension]} \qquad (5.25)$$

*Where Γ is the 'SNR gap'.*

For a complex carrier with two dimensions (i.e. QAM):

$$b = \log_2\left(1+\frac{SNR}{\Gamma}\right) \quad \text{[bits]} \qquad (5.26)$$

When Γ=1 (0dB), Equation (5.25) is equal to the Shannon Capacity (Equation (5.1)). Any reliable and implementable system must transmit at a rate below capacity, and hence Γ is a measure of loss with respect to theoretically optimum performance.

Let the probability of symbol error in QAM, $P_e$, be given by the upper bound given by Equation (5.9). Rearranging by solving for the number of bits/symbol:

$$b \approx \log_2\left(1+\frac{SNR}{\frac{2}{3}\left[erfc^{-1}(P_e/2)\right]^2}\right) \qquad (5.27)$$

This is the same as Equation (5.25), with Γ given by:

$$\Gamma \approx \frac{2}{3}\left[erfc^{-1}(P_e/2)\right]^2 \qquad (5.28)$$

Hence, Γ can be calculated from Equation (5.28) and the required target BER. Substituting Γ into Equation (5.26) then gives the number of bits/symbol, *b*, that can be loaded onto a QAM carrier. To obtain the data rate for this carrier one simply multiplies *b* by the symbol rate, *SR*.

It should be noted that the performance of an uncoded system is considered in this thesis, with no performance margin. The SNR gap can be calculated to include the effects of coding and the design margin. In dB, $\Gamma_{\text{coded}} = \Gamma_{\text{uncoded}} + \gamma_c - \gamma_m$. For example, in a system where a code with a coding gain of 4dB is used, Γ is reduced by 4dB. In practical terms, this would mean that the code reduces the SNR gap, hence allowing transmission of more bits/symbols with the same error rate (see Equations (5.25) and (5.26)). Conversely, $\gamma_m$ increases the SNR gap, and is a design parameter chosen so that performance is guaranteed within a certain margin of the maximum.

The SNR gap is graphically illustrated in Figure 5.16, where the number of bits/symbol for QAM is plotted versus SNR. The Shannon capacity is also plotted, according to Equation (5.1). The various plots corresponding to different error rates are parallel to the theoretical limit, with the constant separation being equal to the SNR gap. As expected, Γ increases with lower probability of error.

The SNR gap is a measure that can be used for any modulation scheme; it indicates the departure of the scheme from the Shannon Capacity. Hence it offers a convenient way of performing bit-loading from Equation (5.26), as long as the error probability equation of the particular modulation scheme is known.

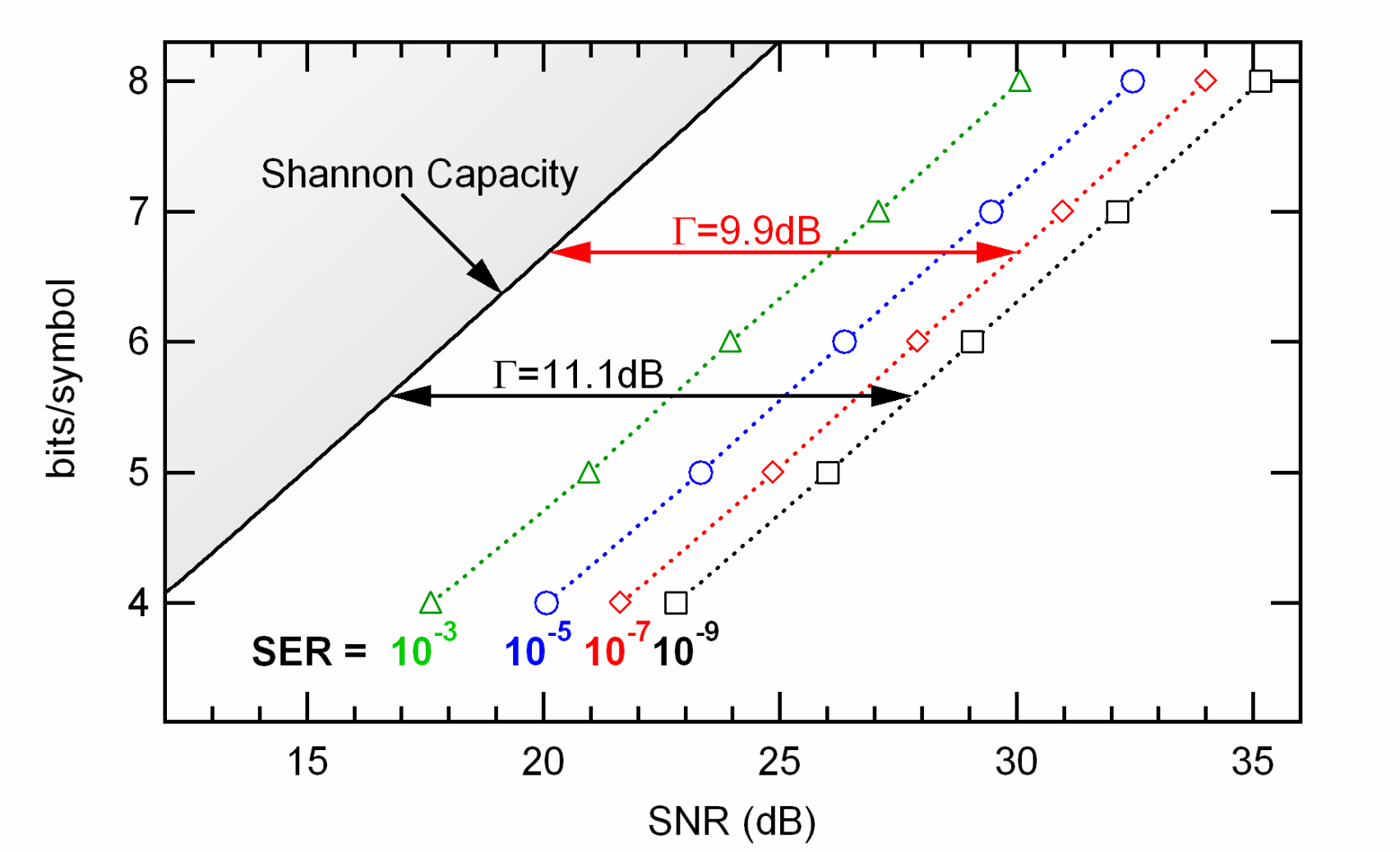

**Figure 5.16:** Illustrating the concept of the SNR gap. The number of bits/symbol for QAM are plotted as a function of the SNR, for various SERs.

### 5.4.4 Multi-carrier QAM Data Rate

The analysis of the previous section can be extended to an OFDM system and assuming that the multiple carriers occupying sub-channels are independent to each other. Hence the aggregate performance of the individual QAM carriers in their sub-channels gives the performance of the multiple carrier system in the complete channel.

The total BER in an OFDM system is simply the weighted average of the individual sub-carrier BERs. It is assumed that all carriers carry equally important information, hence requiring the same BER in each. It follows that the SNR gap, Γ, is the same for all carriers. Given the SNR in each sub-channel, the individual carrier bit rates can be determined as described in the previous section. This process is termed bit-loading. The total bit rate is found by simple addition. In equation form:

$$b_{tot} = SR \cdot \sum_{k=1}^{N} \log_2\left(1 + \frac{SNR_k}{\Gamma}\right) \qquad (5.29)$$

*SR is the symbol rate per carrier, or, equivalently for OFDM, the channel spacing.* $SNR_k$ *denotes the SNR for the* $k_{th}$ *carrier.*

## 5.4.5 Power Allocation

The SNR in each OFDM sub-channel depends on how the total power budget is allocated. Since, in this analysis, the noise is assumed to be white (constant PSD, $N_0$ W/Hz across the whole bandwidth of the channel), the SNR in any given sub-channel will then be directly proportional to the power allocated to the corresponding carrier.

The easiest approach is to allocate equal power to all carriers (i.e. default loading). Let $N$ be the number of carriers, and $P_{total}$ be the total power constraint. The power allocated to the $k^{th}$ carrier is then $P_k=P_{total}/N$. Assuming the generalized AWGN channel described previouly, the SNR in the $k^{th}$ sub-channel can then be obtained from:

$$SNR_k = \frac{P_k \cdot |H_k|^2}{SR \cdot N_0} \qquad (5.30)$$

*Where* $|H_k|$ *is the sub-channel gain, SR is the carrier's symbol rate (it is assumed that this is equal to the sub-channel bandwidth, and the same for all carriers) and* $N_0$ *is the double-sided noise power spectral density.*

While this is a non-optimal solution (see section 5.2.4), it greatly simplifies the analysis. Moreover, as explained earlier, the loss observed compared to an optimally loaded channel is minimal. A negligible effect on the results and on the conclusions reached is therefore expected.

### 5.4.6 Equivalent Single-Carrier Metric: The Multi-Channel SNR

A constant SER (and hence SNR gap) across all sub-channels is assumed. This allows for the use of a single performance measure to characterize the multi-channel transmission system. This measure is a geometric SNR ($SNR_{m,u}$) [7, 10] and can be compared directly to the detection SNR of a single-carrier system employing (perfect) equalization (if the channel is frequency selective).

For a set of $N$ parallel channels of symbol rate $SR_k$ each, the bit rate is:

$$\begin{aligned} b &= \sum_{k=1}^{N} SR_k \cdot \log_2\left(1+\frac{SNR_k}{\Gamma}\right) \\ &= SR_k \log_2\left(\prod_{k=1}^{N}\left[1+\frac{SNR_k}{\Gamma}\right]\right) \\ &= N \cdot SR_k \cdot \log_2\left(1+\frac{SNR_{m,u}}{\Gamma}\right) \end{aligned} \quad (5.31)$$

$SNR_{m,u}$, the multi-channel SNR for a set of parallel sub-channels is defined by:

$$SNR_{m,u} = \left[\left(\prod_{k=1}^{N}\left[1+\frac{SNR_k}{\Gamma}\right]\right)^{1/N} - 1\right]\cdot\Gamma \quad (5.32)$$

The multi-channel SNR characterizes all sub-channels by an equivalent single AWGN channel that achieves the same data rate. In the calculation, Equations (5.31) and (5.32) will be used to estimate the achievable data rate of an OFDM system given the noise PSD and measured frequency response of the Tracker optical links (see section 5.3.2).

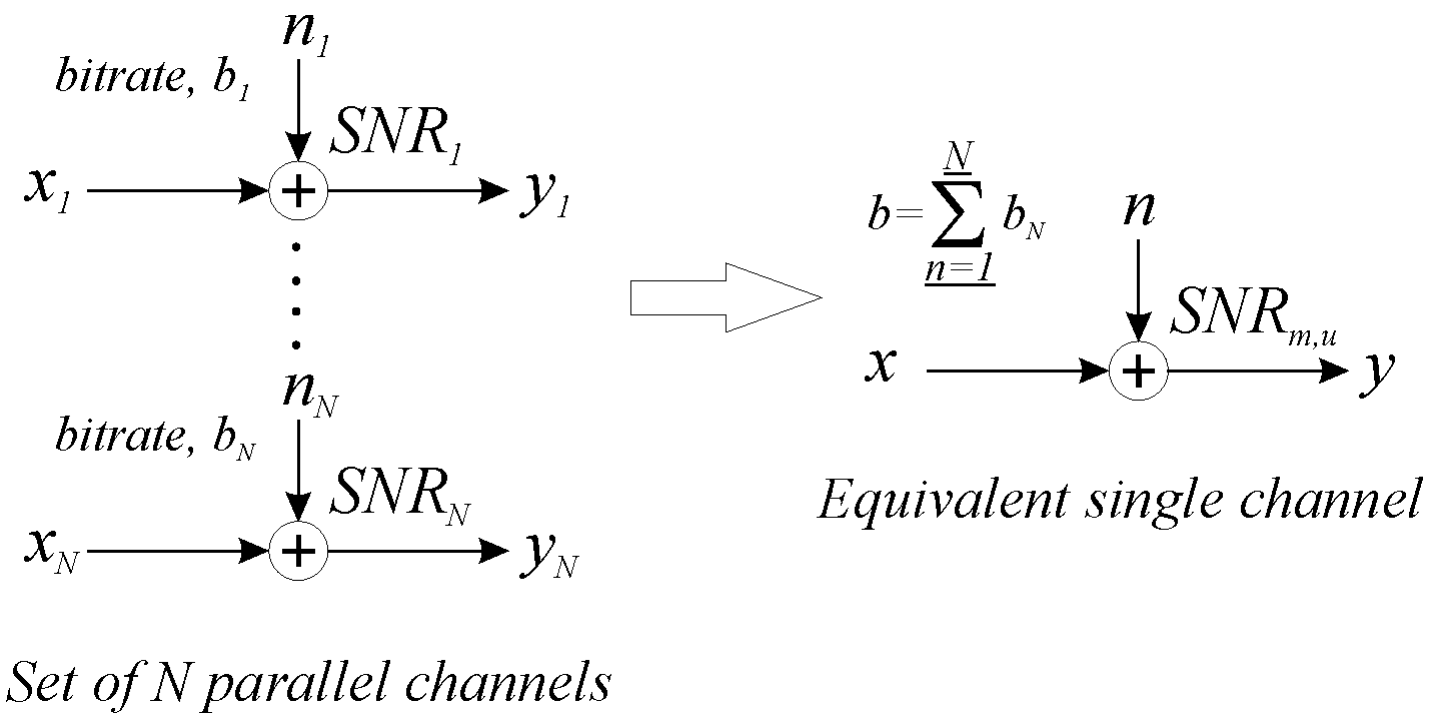

**Figure 5.17:** Illustrating the concept of a single-carrier equivalent to a multi-channel system.

### 5.4.7 Available Bandwidth

To calculate the total data rate of the system, the total frequency range in the channel that will be occupied by the QAM carriers needs to be determined. In OFDM, the sub-channel bandwidth (i.e., carrier spacing) is determined by the chosen symbol rate. Let the channel be divided into a large number of sub-channels, say $N$=256, with equal power allocated to each sub-channel. Since $N$=256, the total occupied bandwidth will be $N \cdot SR$ (where $SR$ is the symbol rate, and is the same for every carrier). To aid the analysis, all 256 carriers in the system will be required to transmit data. The goal, therefore, is to determine the maximum symbol rate in a 256-carrier OFDM system, which results in all carriers having at least 1 allocated bit.

The low-pass response of the channel facilitates the calculation, since only the highest frequency at which the 256$^{th}$ carrier can transmit 1bit/symbol needs to be found. The SNR required for transmission of 1bit/symbol can be calculated by rearranging Equation (5.26):

$$SNR = \Gamma \cdot (2^b - 1) \qquad (5.33)$$

Setting $b$=1 yields:

$$SNR = \Gamma \qquad (5.34)$$

$\Gamma$ is given by Equation (5.28)[4]. Considering the example where the target SER is $10^{-9}$, the required SNR for transmitting 1bit/symbol is ~13dB. Hence, the symbol rate which results in $SNR_{256}$=13dB needs to be determined. Looking at Equation (5.30), it can be seen that $SNR_k$ depends on the channel frequency response (Figure 5.13), as well as $SR$ itself, which is the unknown quantity to be calculated. This can be done iteratively, by increasing $SR$ at each step and calculating $SNR_{256}$. The process is illustrated in Figure 5.18.

[4] While this is valid for QAM, it does not apply to the case of a carrier with $b$=1 (i.e. BPSK), since the probability of error is different. Nevertheless, this is a detail that has a negligible effect on the final result, since there will be only a few carriers with $b$=1. The error is of the order of a few Mbit/s, and is acceptable given that the approximation greatly simplifies the calculation.

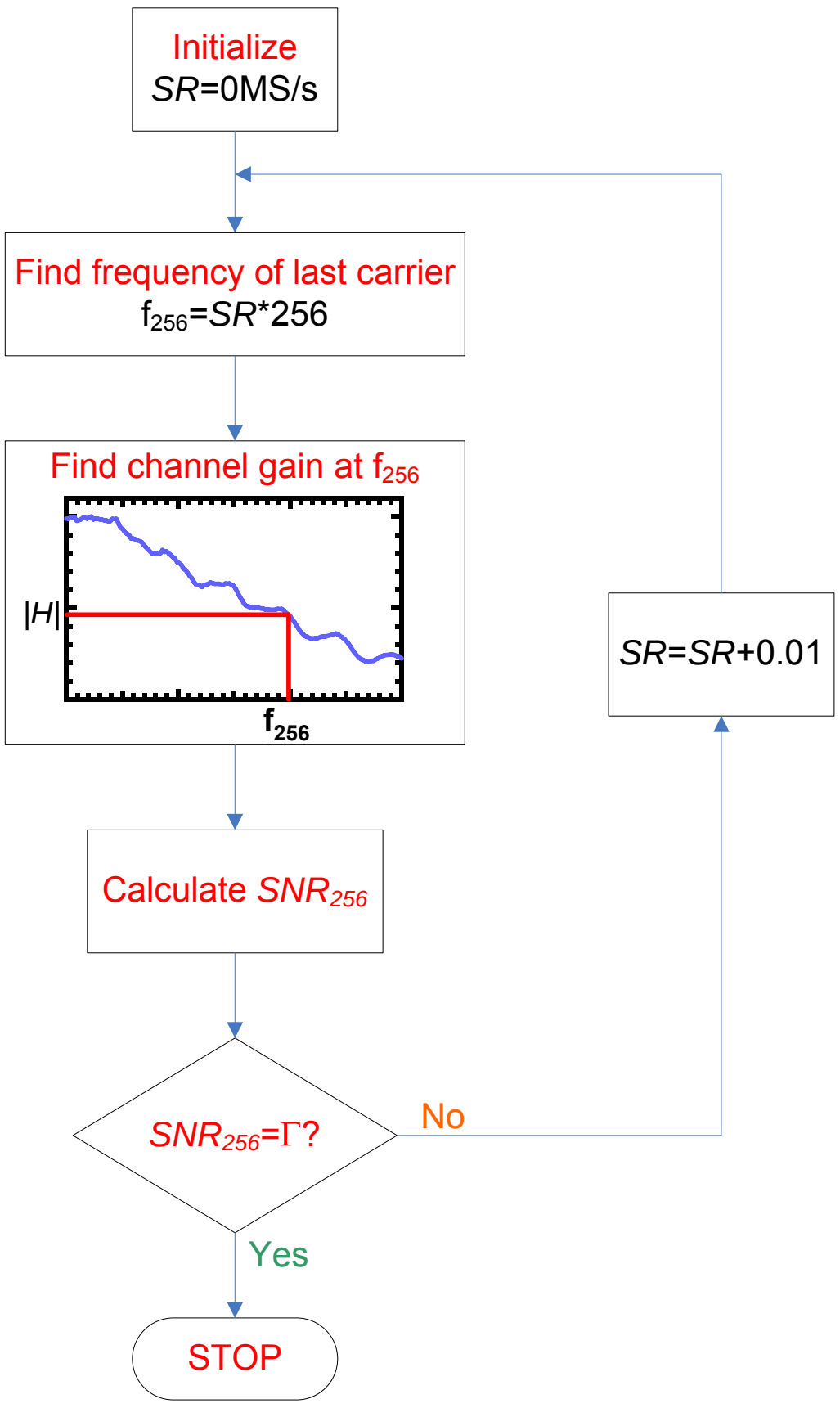

**Figure 5.18:** Algorithm used to determine the available bandwidth.

### 5.4.8 Results

All the information necessary for calculating the data rate in the multi-carrier system is now known. The 256 QAM carriers will be centered at the following frequencies, with the symbol rate determined using the method of section 5.4.7:

$$f_k = k \cdot SR \qquad for\ k = 1\ to\ 256 \qquad (5.35)$$

Using Equation (5.30), and assuming uniform power allocation across all carriers, as in section 5.4.5, $SNR_k$ can be determined for $k$=1 to 256. As an example, consider the SNRs obtained for a target error rate of $10^{-9}$, and no power backoff (Figure 5.19). Plugging the values of $SNR_k$ into Equation (5.32), the multi-channel SNR, $SNR_{m,u}$, is obtained. In the SER=$10^{-9}$ example, $SNR_{m,u}$=20.65dB. The system's data rate can then be determined from Equation (5.31), resulting in 3.16Gbit/s.

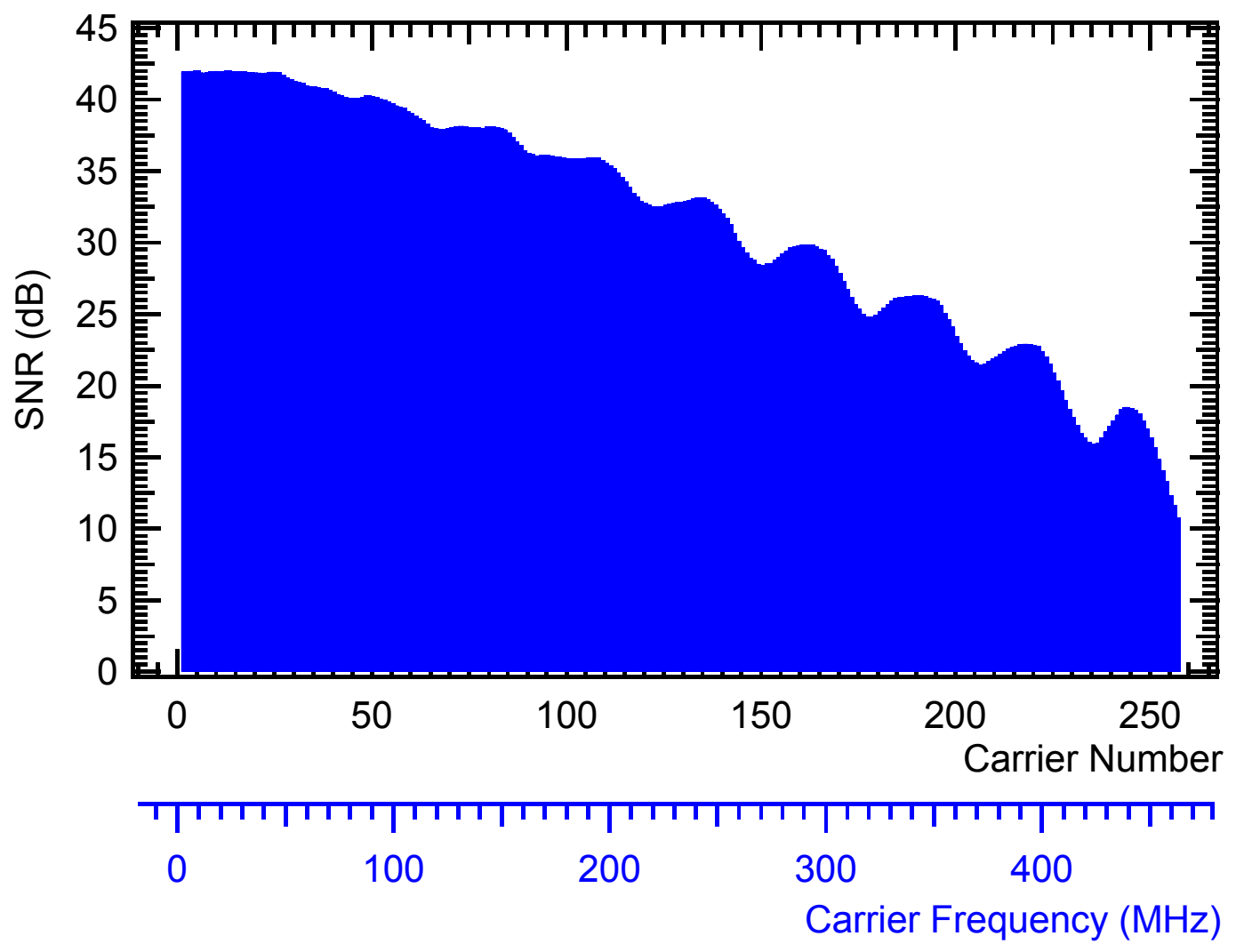

**Figure 5.19:** Showing the SNRs of 256 QAM carriers, spaced at 1.8MHz. Uniform power distribution has been employed.

The same method was used to estimate the achievable data rate using QAM over the CMS Tracker Analog optical links, for various target error rates. Of course, the accuracy of the results depends on the accuracy of the assumptions made. In order to produce an unbiased first-order estimate and evaluate the effect of errors, the main design parameters have been varied to obtain a range of data rates.

The total power available to the system was estimated given the linear input range to the AOH (transmitter). The peak signal that can be sent through the link is assumed to be 600mV. Non-linearities within and outside this input range are not considered in the analysis. The peak power has been normalized to 1W, and this corresponds to using the full AOH input range. A number of power backoff values, from 0 to 12dB, have been used.

Finally, the noise PSD is an estimate based on specifications and is assumed to be constant across all frequencies. The simplified AWGN channel model is used. This probably over-estimates the noise, and hence it is the cautious approach. A range of constant noise PSD values was used in order to determine the effect that a wrong estimate can have on the achievable data rate.

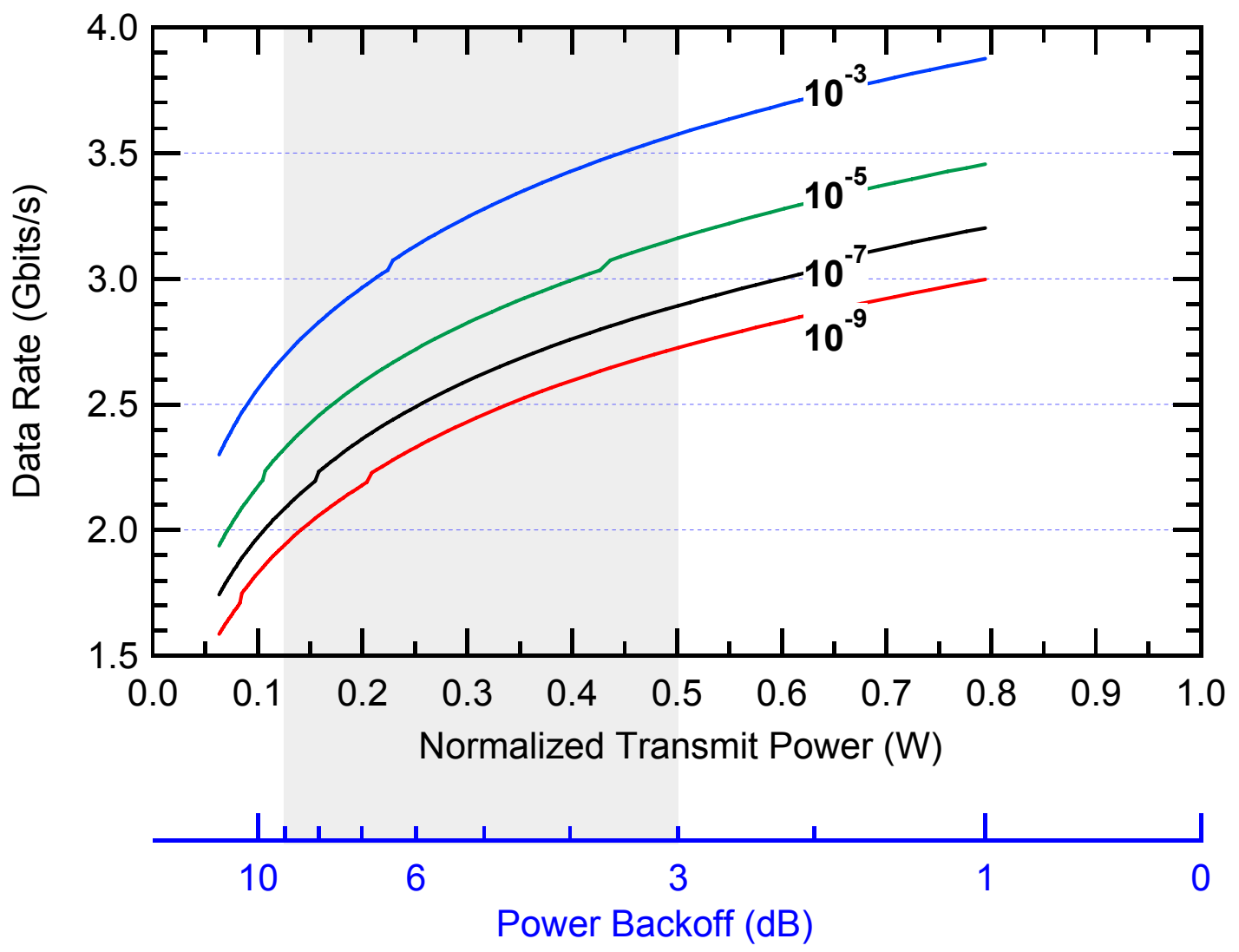

**Figure 5.20:** Data rate as a function of transmitted signal power (and power backoff in dB) for various target SERs.

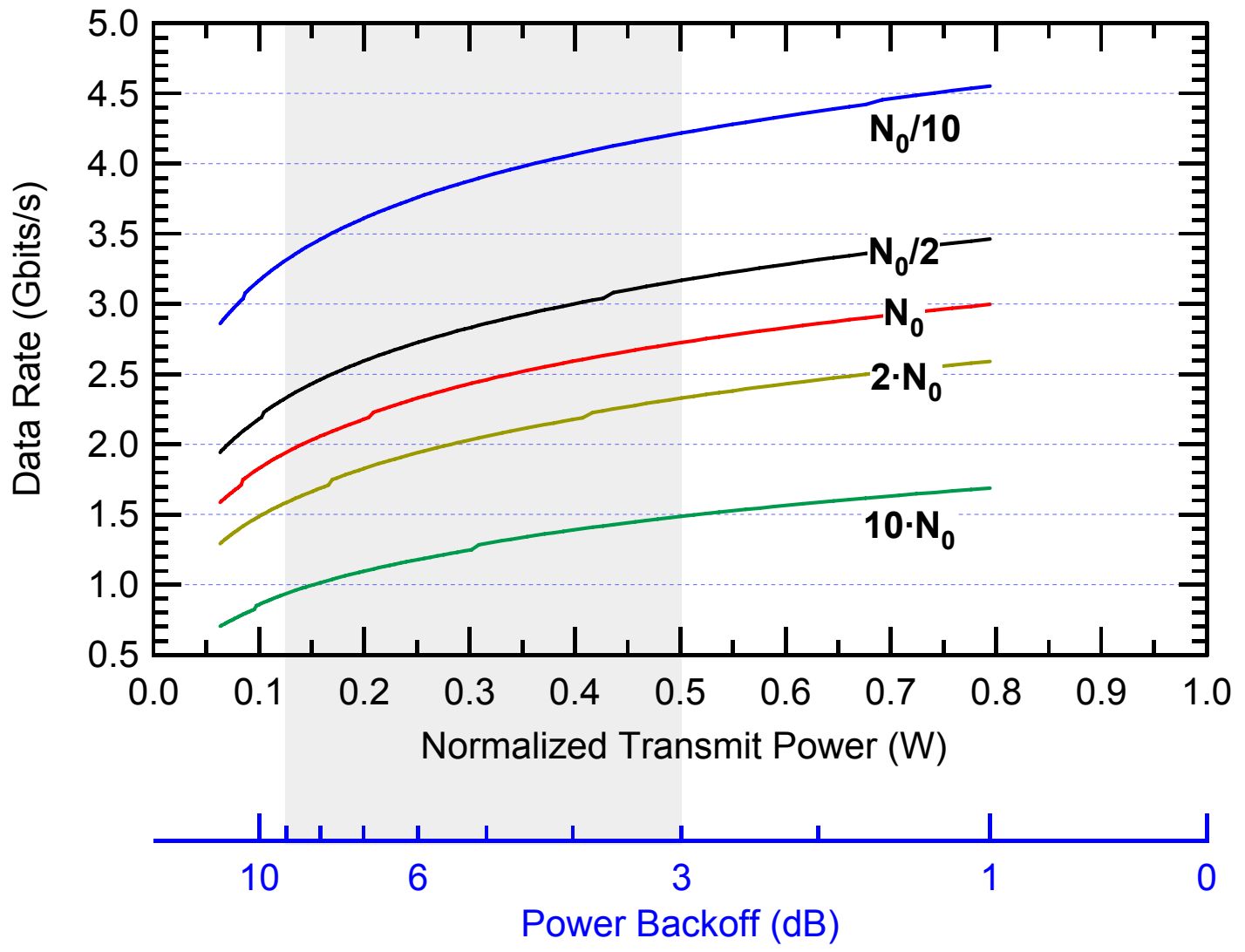

**Figure 5.21:** Data rate as a function of transmitted signal power (and power backoff in dB) for various noise PSDs.

Figure 5.20 shows the data rate against normalized transmitted signal power (and power backoff in dB). The results are shown for four different target SERs. The shaded area corresponds to the backoff range of 3-9dB, which is the range the system is likely to operate in. At a SER of $10^{-9}$, the data rate in this region of operation ranges from ~1.9 to 2.7Gbits/s. The corresponding bandwidth efficiency is ~4.5-6.5bits/s/Hz. This can be compared to the bandwidth efficiencies of Wi-Fi (~3bits/s/Hz), and ADSL (typically up to ~6bits/s/Hz).

In order to assess the impact of the possible error in the average noise PSD used in the calculations, the data rate for a fixed error rate (BER=$10^{-9}$) was computed as a function of transmitted signal power for various noise PSDs. The results are shown in Figure 5.21. The center curve corresponds to the noise PSD, $N_0$, calculated from the specification of the optical link. Results were obtained for higher noise ($10 \cdot N_0$ and $2 \cdot N_0$) as well as lower noise ($N_0/10$ and $N_0/2$).

The above calculations do not include all effects that may be encountered in a practical system. Phase noise, non-linearities in the link transfer function and frequency instabilities can adversely affect the system's performance. It follows that performance of any advanced digital communication scheme is heavily dependent on the actual hardware implementation.

Another source of uncertainty is the frequency response measurement made on the optical link, due to the limited accuracy of the instrument used, as well as the presence of components that would not be present in a future implementation; a differential driver circuit was used before the AOH, while the ARx12 sat on a board that included output buffer amplifiers.

## 5.5 Conclusions

The purpose of this chapter was to introduce the basic concepts of digital communications relevant to a possible future upgrade of the CMS Tracker optical links. The analytical method for calculating the achievable data rate using a bandwidth efficient modulation scheme has been described. Preliminary results based on test data and specifications of current link components have been included, and where necessary, extrapolations have been made to account for unknown characteristics of the link. The assumptions made are typical of the simplified models used in digital communications textbooks, and the corresponding results should therefore be treated with care. A more accurate result for the data rate is obtained in Chapter 6, where laboratory tests using real QAM signals transmitted through a Tracker optical link are described.

# Chapter 6

## Upgrading the CMS Tracker Optical Links: Laboratory Tests

*The possibility of upgrading the CMS Tracker optical links using bandwidth efficient RF digital modulation has been investigated analytically in Chapter 5. In this Chapter the first experiments verifying the viability of using a QAM-OFDM digital modulation scheme in the analog links are presented. The analytical calculation is augmented with real data, and the best possible estimate for the maximum data rate is presented.*

# 6.1 Introduction

## 6.1.1 Motivation

In Chapter 5, an analytical approach to determining the capacity of the CMS Tracker optical link was presented. Experimental tests are needed, not only to test the validity of the assumptions made previously (and hence the conclusions reached), but also to demonstrate that the concept is viable and worth pursuing. Given the time constraints and complexity issues, the implementation of a prototype QAM-based digital communication system that can achieve data rates of several Gbit/s over the current CMS Tracker readout optical link is not a realistic option at this early stage of development. Commercial QAM (and OFDM) modulators with such a high throughput do not exist. In order to experimentally assess whether such a system is feasible and determine the maximum theoretical data rate over the current analog optical link, a different approach is necessary.

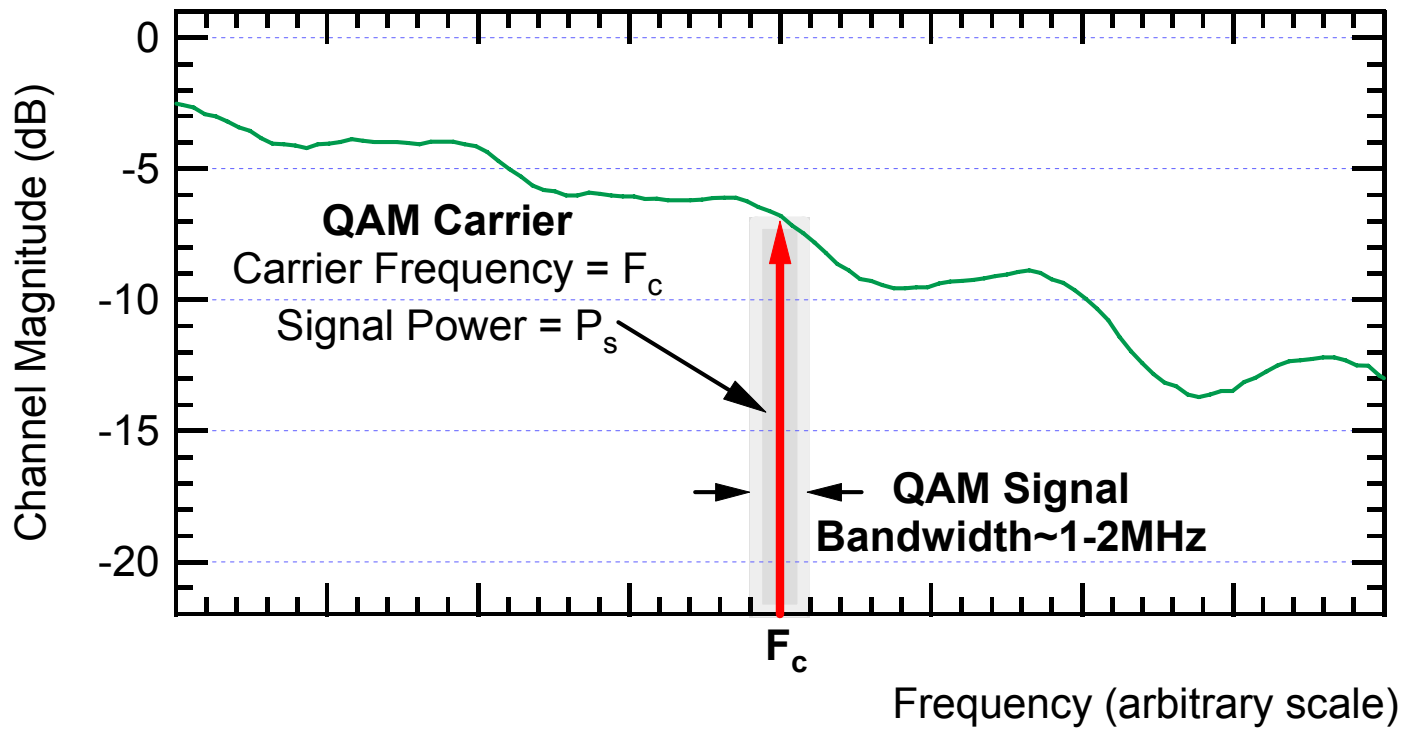

**Figure 6.1:** Illustrating a 1MS/s QAM carrier transmitted through a channel with an arbitrary frequency response. $F_c$ can be varied to across the entire frequency of the channel.

As far as a future upgrade based on QAM modulation is concerned, the system could be implemented using a single or multiple carriers (as in OFDM). While there are practical limitations that will govern this decision, for analytical purposes the two systems can be considered to be equivalent in terms of data rate[1] [1]. Indeed, in the previous chapter the multi-carrier system's data rate calculation was simplified using an equivalent single-carrier system. One could consider a multi-carrier (OFDM) system with hundreds of low-symbol rate (1MS/s) QAM carriers covering the entire frequency range of the channel (~500-700MHz). The idea is to implement a test system which can send (and receive) a single, 1MS/s QAM-modulated RF carrier through the optical link. If the carrier frequency is changed for each test and swept through the entire frequency range, each carrier of the OFDM system is effectively tested independently (Figure 6.1). Of course, the available transmission power needs to be appropriately allocated to each carrier with the total power constraint of the channel, in order to reach useful conclusions. Hence a potential multi-carrier implementation is assessed by the aggregate performance of all single-carrier tests.

[1] This ignores the limitations due to Peak to Average Power Ratio (PAPR) which is normally an order of magnitude higher in a typical OFDM implementation, compared to a single carrier QAM system.

### 6.1.2 Estimating Error Rate and Signal to Noise Ratio of QAM Signals

A basic explanation of the underlying concepts behind the experimental tests is necessary to aid understanding of the subsequent sections. The aim is to determine the number of bits/symbol that can be allocated to a QAM carrier, for a given target error rate and transmission power. The most obvious way of achieving this experimentally would be to perform bit error rate (BER) tests for each carrier. For a given input power, the number of transmitted bits/symbol for the carrier under test would be increased until the specified BER was reached.

While BER is the best way of verifying a system's performance, it requires fairly complex hardware and can be very slow when the target BER is low. As a guideline, for a target BER of $10^{-9}$, $3 \cdot 10^{9}$ bits need to be transmitted through the link with no errors for a 95% confidence level in the result. The tests to be carried out will involve individual, low symbol rate carriers. For example, a 1MS/s carrier with 5bits/symbol transmits 5Mbits/s, requiring 10 minutes to complete the test for an error rate of $10^{-9}$. This is unacceptably slow, given that the tests will involve sweeping the carrier frequency, input power and bit assignment, hence requiring hundreds of individual BER tests.

#### ERROR VECTOR MAGNITUDE

Error vector magnitude (EVM) is a modulation quality metric widely used in digital RF communications [2]. It is sensitive to any impairments that affect the amplitude or phase of a demodulated signal. The error vector is defined as the vector difference between the reference signal (i.e. the ideal constellation points) and the measured (demodulated) signal in the IQ plane (Figure 6.2). EVM is the root mean square (rms) error vector over time, and is usually normalized, either with respect to the *outermost* symbol magnitude (Equation (6.1)) or to the *average* symbol magnitude (Equation (6.2)) of the respective QAM constellation. By convention, it is expressed as a percentage.

$$EVM_{norm_max} = \frac{\text{Error Vector (rms)}}{\text{Maximum QAM Signal Amplitude}} \cdot 100\% \qquad (6.1)$$

$$EVM_{norm_avg} = \frac{\text{Error Vector (rms)}}{\text{Average QAM Signal Amplitude}} \cdot 100\% \qquad (6.2)$$

Regardless of the normalization method used, EVM can be related to the digital SNR. When normalized to the average symbol magnitude, EVM is related to the SNR by:

$$EVM_{norm_avg} = \frac{1}{\sqrt{SNR}} \qquad (6.3)$$

When normalized to the maximum symbol amplitude, the peak to average power ratio (PAPR) of the constellation used must be taken into account. In this case, $EVM_{norm_max} \sim PAPR \cdot EVM_{norm_avg}$, and equation (6.3) can be re-written as:

$$EVM_{norm_max} = \frac{1}{PAPR \cdot \sqrt{SNR}} \qquad (6.4)$$

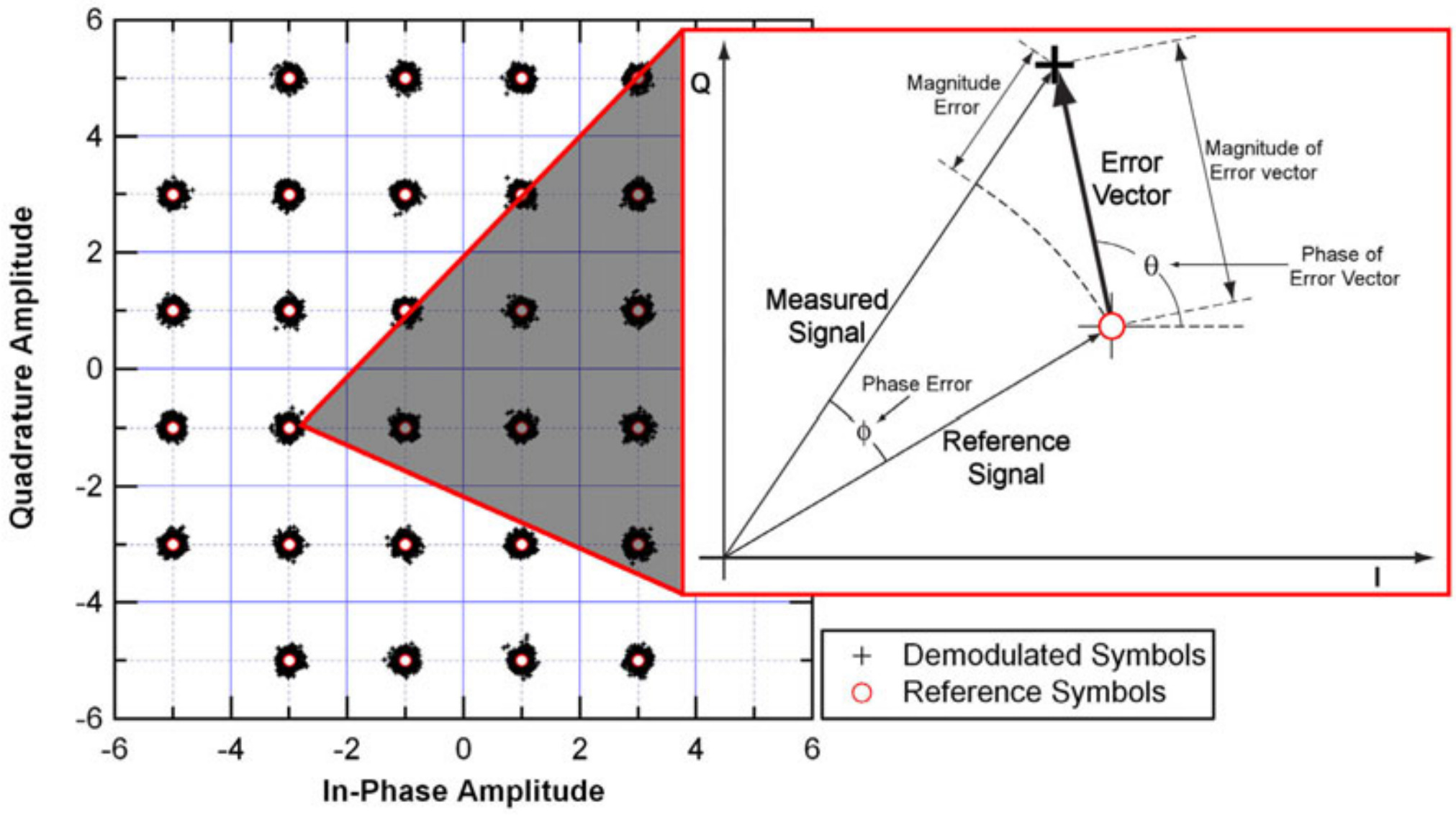

**Figure 6.2:** Illustrating the error vector and related quantities. The IQ plot shows 40 000 demodulated symbols belonging to a 32-QAM constellation in an AWGN channel.

### ERROR RATE FROM EVM

It has been shown how the SNR in a QAM digital communication system can be derived from an EVM measurement. The significance of this is that SNR can be used to relate EVM to the error rate. For a system corrupted by Additive White Gaussian Noise (AWGN), the symbol error rate[2] in QAM is given by [3]:

$$P_e \approx 2erfc\left(\sqrt{\frac{3\cdot SNR}{2\cdot(2^b-1)}}\right) \qquad (6.5)$$

*Where erfc is the complementary error function and b the number of bits per symbol (e.g. 4 for 16-QAM).*

Equation (6.5) is graphically depicted in Figure 6.3 for various modulation schemes, where the probability of error is plotted against SNR. Figure 6.4 shows the corresponding relationship with $EVM_{norm_avg}$, calculated from equation (6.3).

The advantages of using EVM measurements to estimate the error rate, rather than a direct BER test, are that of speed and lower complexity. There is no need to decode the demodulated symbols. EVM can be accurately estimated using only a few thousand symbols, in contrast to a BER test that requires sending and receiving a large number of bits to allow an accurate calculation.

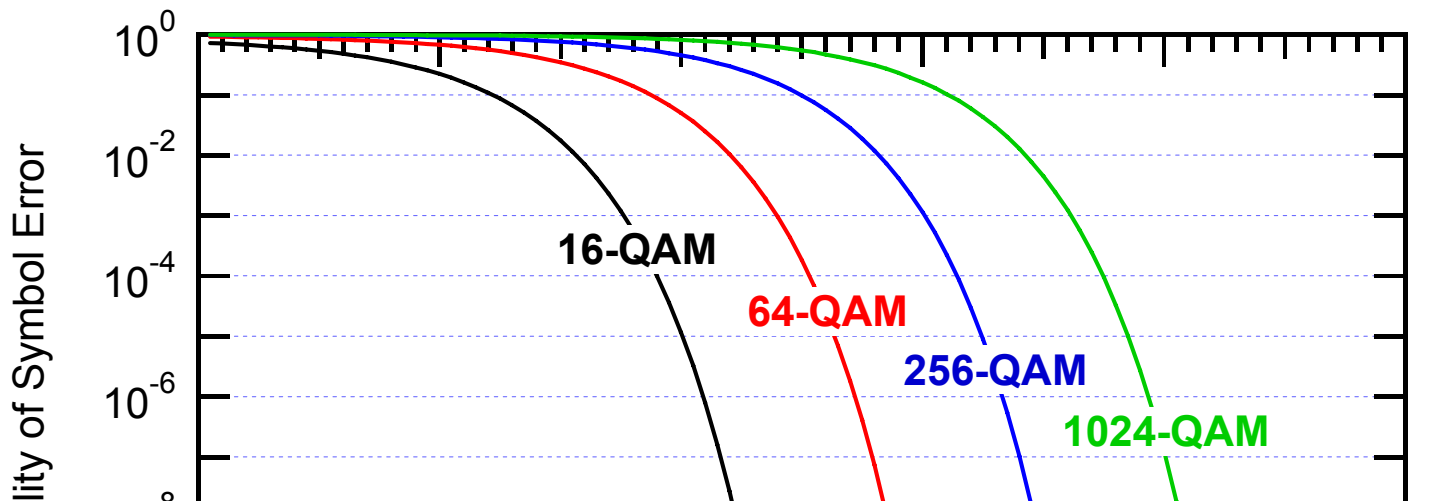

**Figure 6.3:** Probability of symbol error in QAM as a function of SNR.

[2] The relationship between Bit Error Rate (BER) and symbol error rate depends on the bit-mapping and the QAM constellation, as explained in Chapter 5. In any case, the BER is upper bounded by the symbol error rate multiplied by the number of bits/symbol. This discussion is beyond the scope of this chapter, and we will limit ourselves to using the symbol error rate.

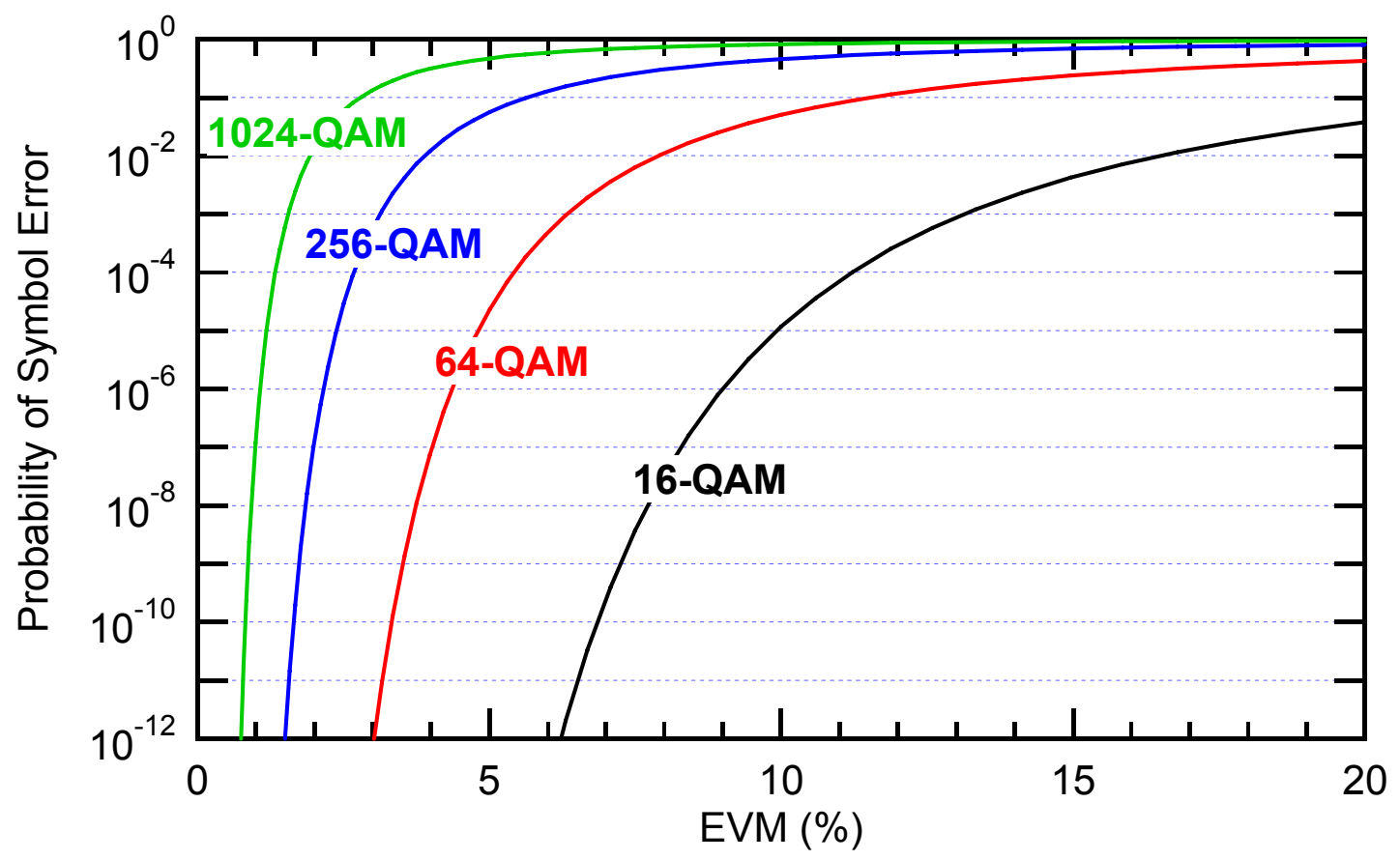

**Figure 6.4:** Probability of symbol error in QAM as a function of $EVM_{norm_avg}$.

### 6.1.3 Objectives

The main objective of the tests described in this chapter is to determine, experimentally, the data rate achievable using bandwidth efficient digital modulation in a link based on the current optical link components. To achieve this, the following questions need to be addressed:

1. What is the total available transmission power?
2. How many bits/symbol can be allocated to each of the QAM carriers across the entire frequency range of the link, given the total available power and target error rate?

For a given target error rate, the SNR of a QAM carrier determines the number of bits/symbol that can be allocated to it. Since the SNR can be calculated from the EVM, it follows that the EVM as a function of carrier frequency and transmission power is all the information needed to answer the above questions.

## 6.2 Test Setup

### 6.2.1 Hardware

The test system must be capable of the following:

1. Symbol rate of at least 1MS/s (a few MS/s desirable).
2. Carrier frequency up to at least 1GHz.
3. QAM constellations up to 256 symbols (i.e. 8 bits/symbol).

An Agilent E4438C ESG vector signal generator [4] capable of producing QAM modulated carriers was used as the transmitter, with a PSA4440A spectrum analyzer acting as the demodulator [5, 6].

### THE MODULATOR: AGILENT E4438C ESG VECTOR SIGNAL GENERATOR

The signal generator used in the tests is capable of producing modulated carriers up to 6GHz, well above the useable bandwidth of the Tracker optical link. Output RF power up to over ~10dBm is possible, which is a lot higher than the maximum input for the optical link. The instrument includes an optional baseband IQ generator capable of QAM modulation up to 8bits/symbol (256-QAM) with a maximum symbol rate of 50MS/s.

### THE DEMODULATOR: AGILENT PSA E4440A SPECTRUM ANALYZER

The spectrum analyzer selected as the receiver is capable of analyzing signals up to a center frequency of 26.5GHz, with an analysis bandwidth of 8MHz. The optional digital demodulation hardware was included, and together with optional firmware, allowed demodulation of QAM signals up to 256-QAM. The analysis bandwidth limitation meant that symbol rates of only a few MS/s could be used in the tests. Still, this is acceptable performance for the tests.

The PSA instrument is able to perform numerous measurements on received QAM signals. Given the center frequency of the incoming signal, as well as the type of modulation and filtering used, the PSA is able to determine various important performance metrics from the received IQ constellation, which is displayed in real time. The peak and rms EVM values are calculated automatically, and given the modulation type, the instrument can also output the corresponding SNR of the received QAM carrier.

### THE TEST SETUP

An optical link consisting of production versions of the AOH and ARx12 was used. One AOH channel's fiber pigtail was mated to one of the connectors on a 12-channel MU/MPO12 patch cord, with the MPO connector mated to the ARx12. The AOH requires a differential input, while the signal generator has a 50Ω single-ended RF output. An interface based on the high-bandwidth AD8351 differential driver was therefore employed. On the receiving end, the spectrum analyzer was operated in ac-coupled input mode, receiving the single-ended output from the ARx12 directly. 50Ω coaxial cables with type-N and BNC connectors were used for the electrical connections between the instruments and components. A Mac computer running Labview and equipped with GPIB provided control for the AOH and ARx12, with an Agilent 34970A switch used to select the appropriate settings on the optical receiver. Finally, a GPIB-equipped PC was used for communication with the signal generator and spectrum analyzer. The test setup is shown schematically in Figure 6.5, and photographed in Figure 6.6 and Figure 6.7.

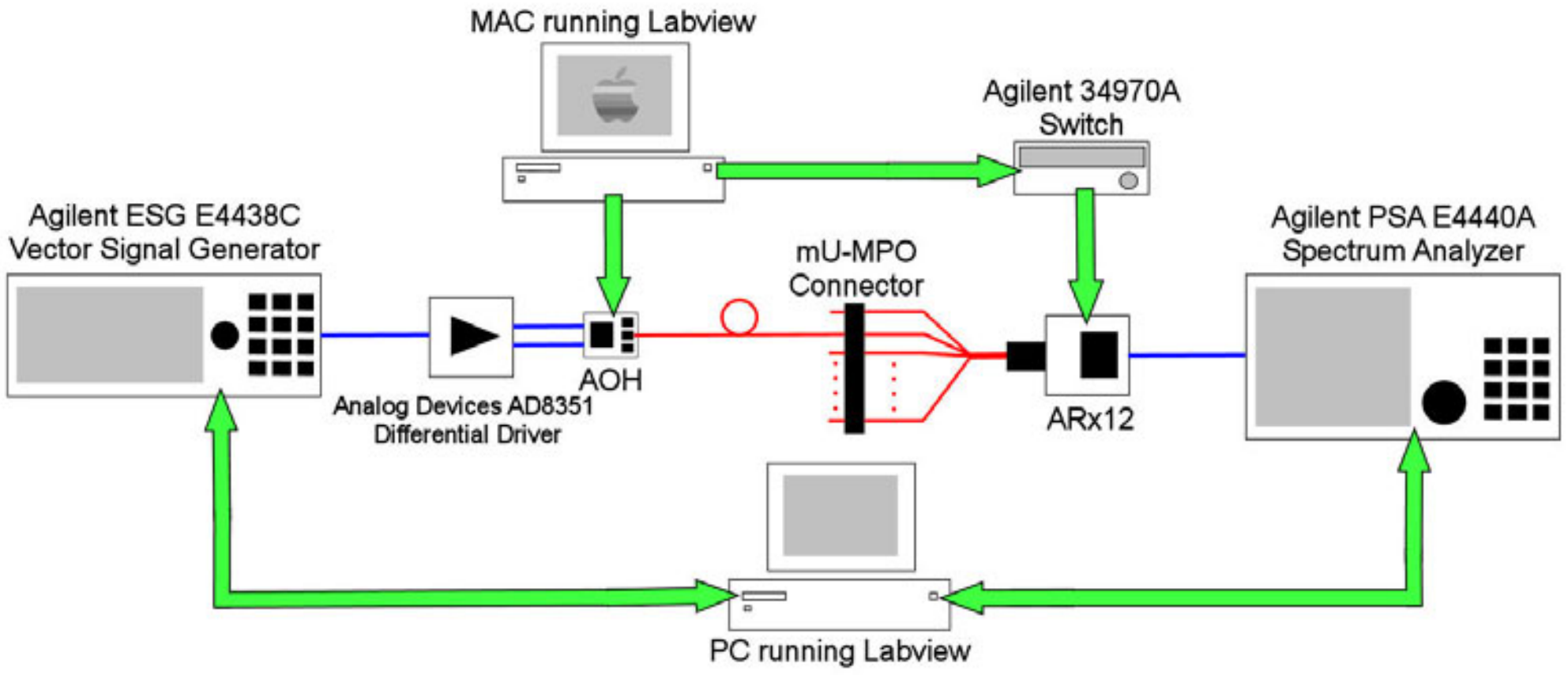

**Figure 6.5:** Showing the test setup used. Control and data acquisition (indicated by the arrows) was performed using GPIB-equipped Mac and PC computers.

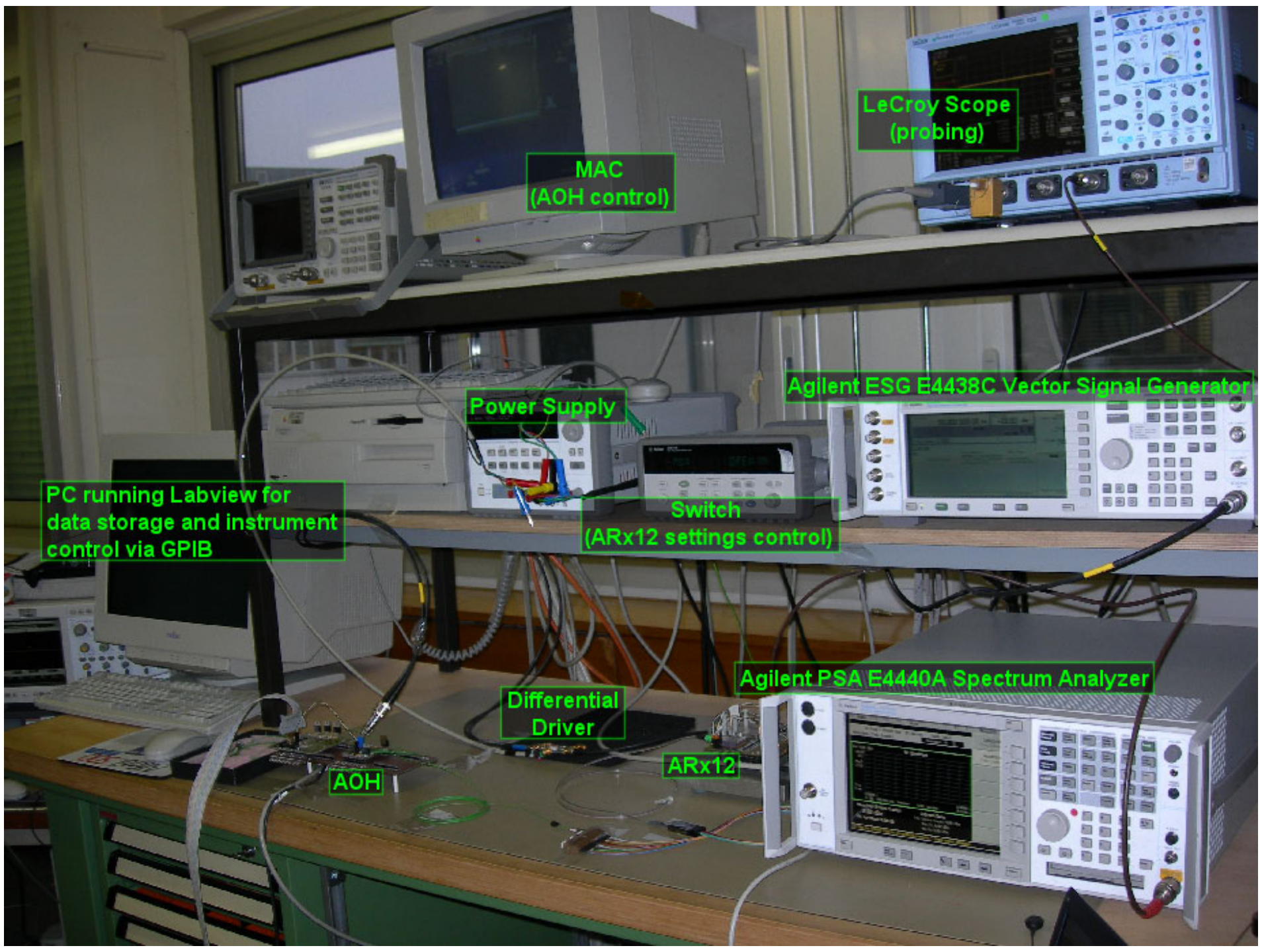

**Figure 6.6:** Photo of the test bench.

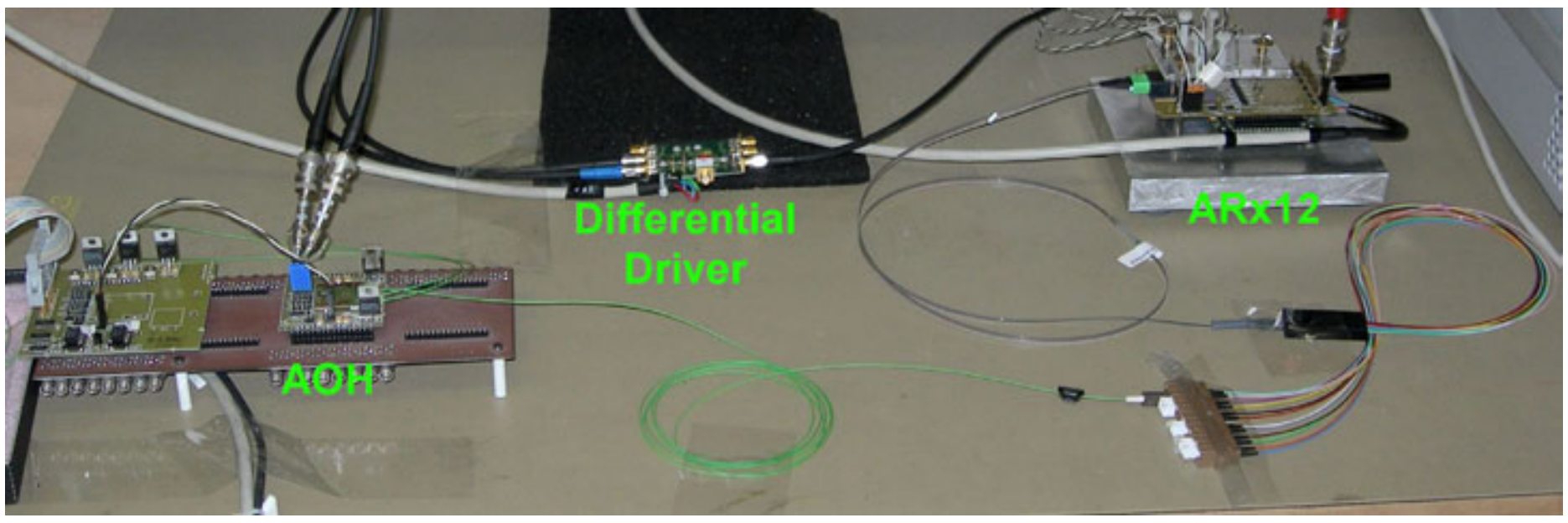

**Figure 6.7:** Close-up of the optical link.

### INPUT POWER AND AMPLITUDE

In the subsequent sections, the transmission power quoted always refers to the output from the ESG E4438C vector signal generator. This is not the same as the input power to the optical link, since the test setup includes additional electronics in the form of the differential driver interface board. In addition, the power quoted refers to the *average* signal power, not the maximum[3].

For completeness, and to aid understanding, the input to the AOH was probed using a differential probe, while a 4-QAM signal was transmitted through the link. The carrier frequency was set to 10MHz, with a 1MS/s symbol rate. Root Nyquist pulse shaping (alpha=1) was used in the transmitter. For reasons of convenience, the same two symbols were transmitted through the link continuously. This made it possible to trigger on twice the symbol period and use averaging to remove the noise due to the probing.

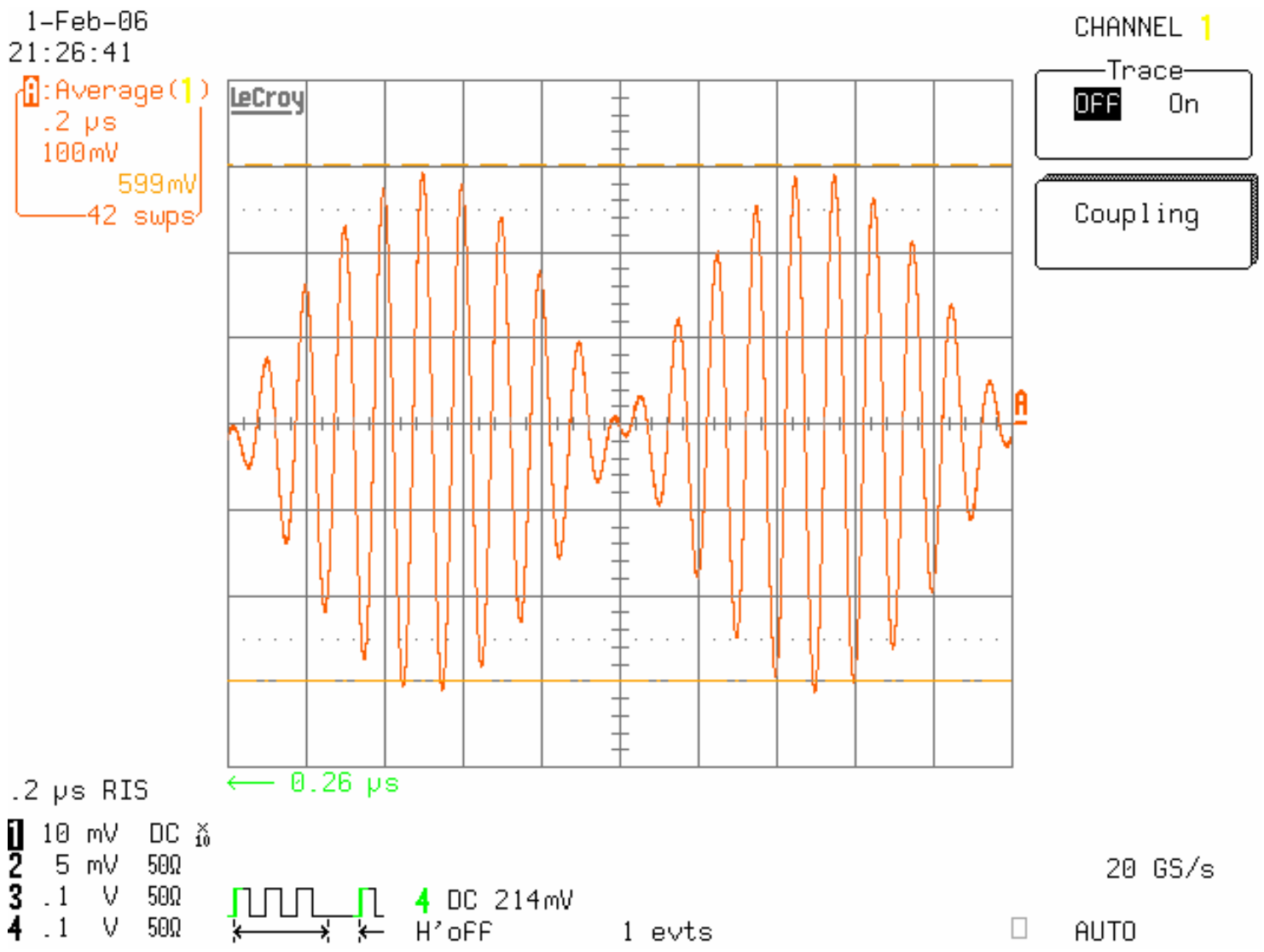

3 The E4438C was used in a mode where the output was regulated so that the power value reported was that of the average signal power.

**Figure 6.8:** Showing an oscilloscope screenshot of a 4-QAM signal, obtained by probing the input of the optical link with a differential probe. The input power reported by the signal generator is -3.5dBm.

Figure 6.8 shows the 4-QAM signal corresponding to an output of -3.5dBm from the signal generator. The markers on the oscilloscope's display show that this power level corresponds to a maximum signal amplitude of ~600mV differential at the link's input, which is also the specified linear range of the AOH. Figure 6.9 is the same signal, this time transmitted with a power of -5.5dBm. The maximum amplitude is 480mV. In terms of power, this is 2dB lower than that of the signal in Figure 6.8, as expected.

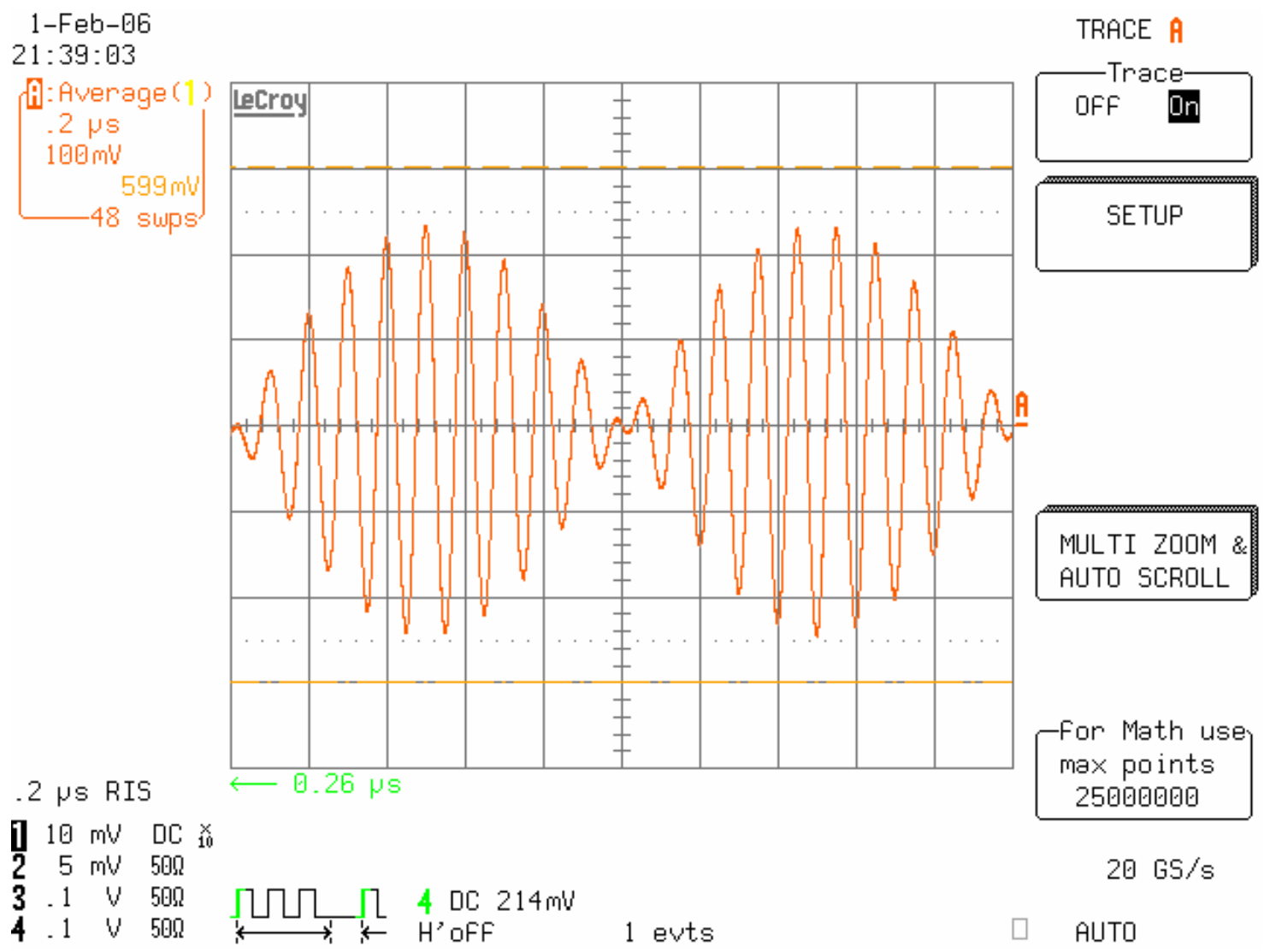

**Figure 6.9:** Showing an oscilloscope screenshot of a 4-QAM signal, obtained by probing the input of the optical link with a differential probe. The input power reported by the signal generator is -5.5dBm.

## 6.2.2 Noise

If the SNR and error rate of a given QAM carrier are to be determined from the EVM, it is important to determine whether or not the noise in the test system is indeed Gaussian (see section 6.1.2 for the relevant theory). Several IQ plots were taken at various frequencies and input powers. The results obtained were similar for every case where there was no amplifier saturation. A typical case is presented in this section.

A 32-QAM signal with a symbol rate of 1MS/s and a carrier frequency of 40MHz was passed through the link. 40 000 symbols were retrieved by the spectrum analyzer. All symbols are plotted on the IQ plot of Figure 6.10, left. In order to obtain maximum statistics on the noise, the received IQ points were amplitude normalized relative to their ideal symbol points and merged into one. Moreover, in the subsequent tests, the EVM calculated is an average over all received symbols, not one particular constellation point. Hence, studying the noise in this way is consistent with the test method.

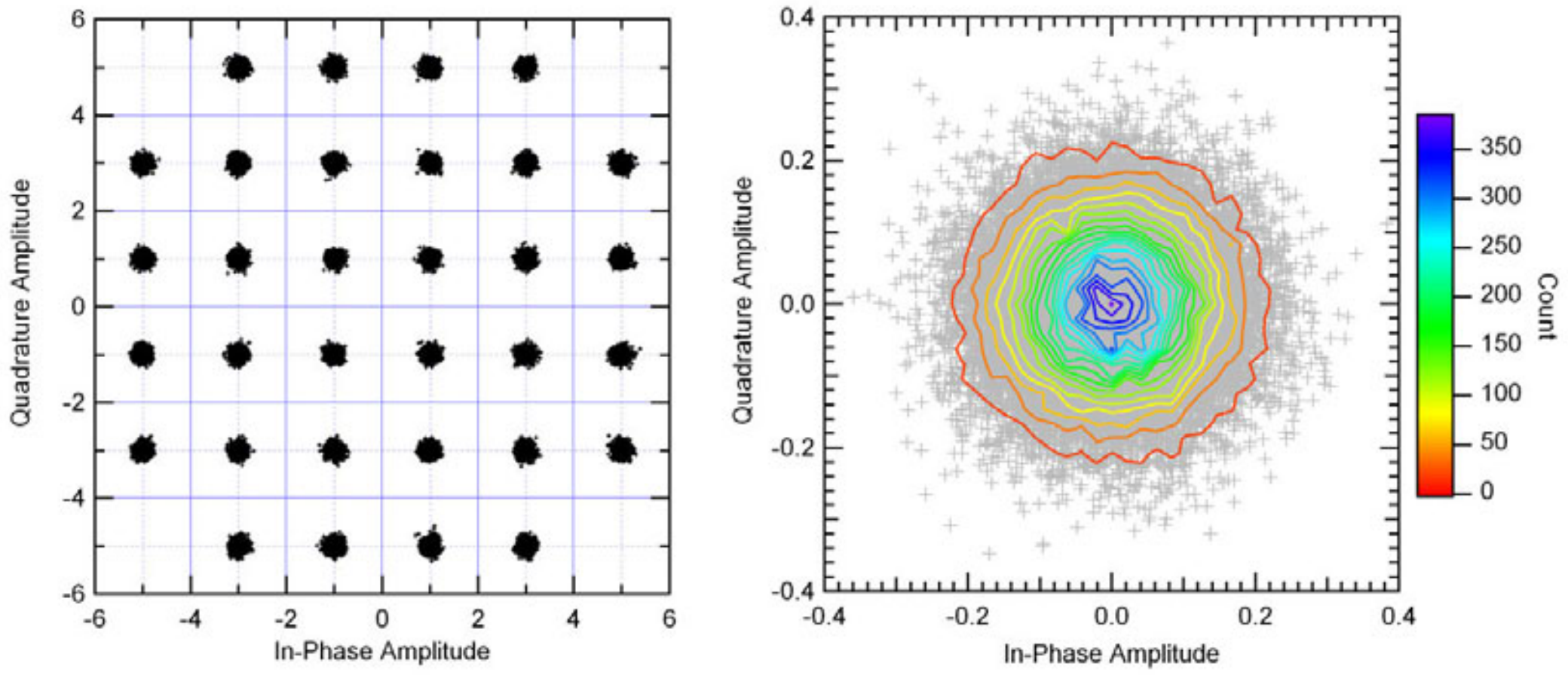

**Figure 6.10:** 32-QAM signal corrupted by noise in the link and test system. 40 000 symbols are plotted in the IQ plot (left). After normalization, all 40 000 symbols were merged into one (right).

Figure 6.10, right, shows all 40 000 symbols plotted together, after amplitude normalization. The received points were histogrammed in two dimensions, and the resulting contour plot is superimposed in Figure 6.10, right. The histogram is also shown in the three-dimensional plot of Figure 6.11. The quadrature and in-phase components of the received symbols have also been histogrammed separately, and these are presented in Figure 6.12 and Figure 6.13. The equations of the Gaussian fits to each are also given. It is clear from the results that the nature of the noise is indeed Gaussian, and therefore the error rate of a QAM carrier can be estimated from the EVM.

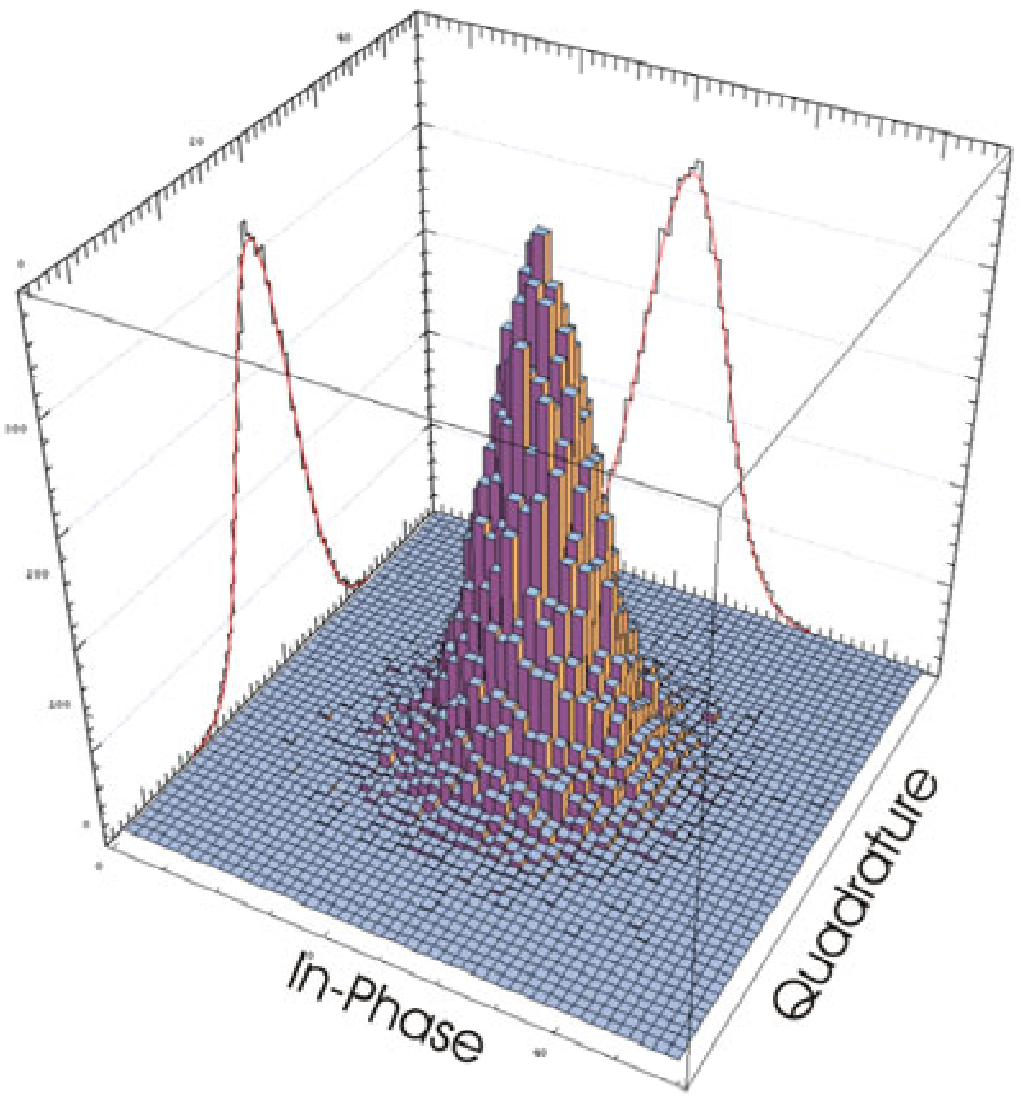

**Figure 6.11:** Three dimensional representation of the two-dimensional histogram of the 40 000 received IQ points in the example of Figure 6.10.

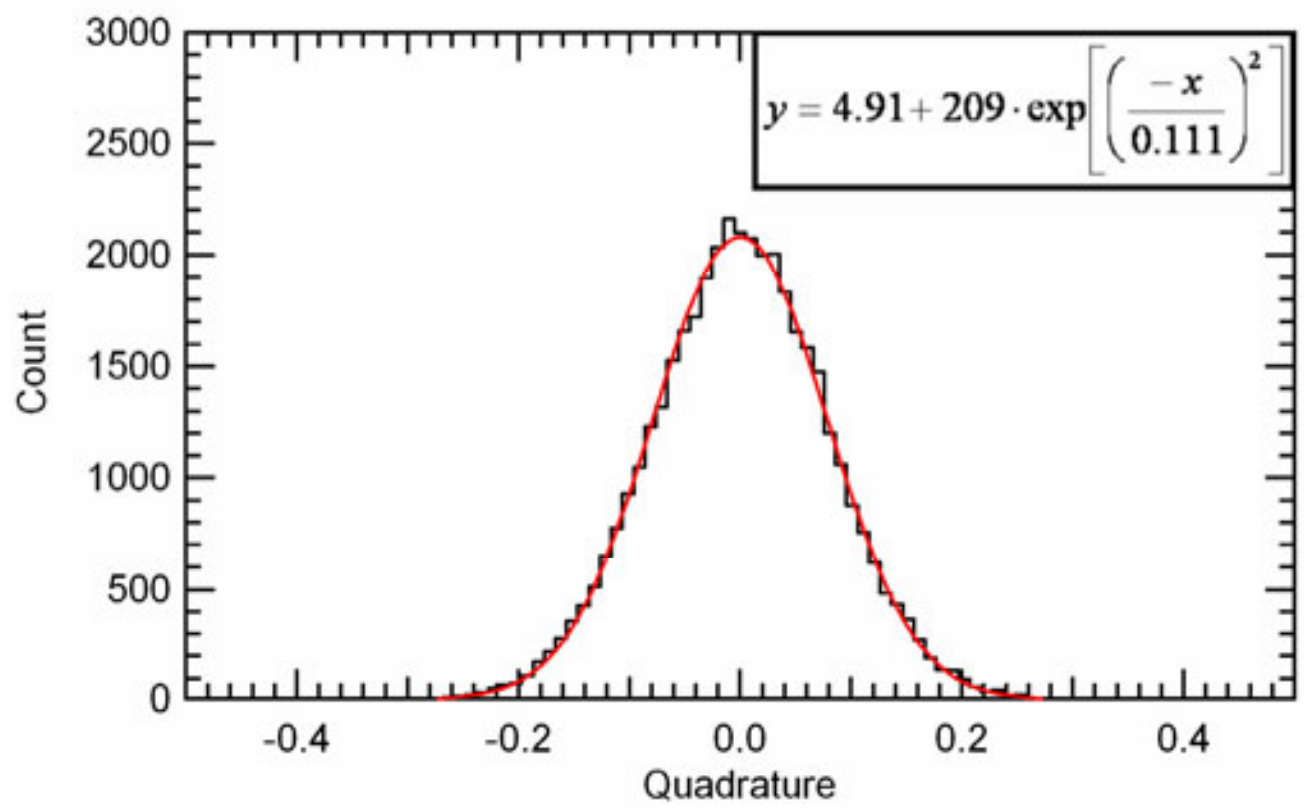

**Figure 6.12:** Histogram of the quadrature component of the 40 000 received IQ points.

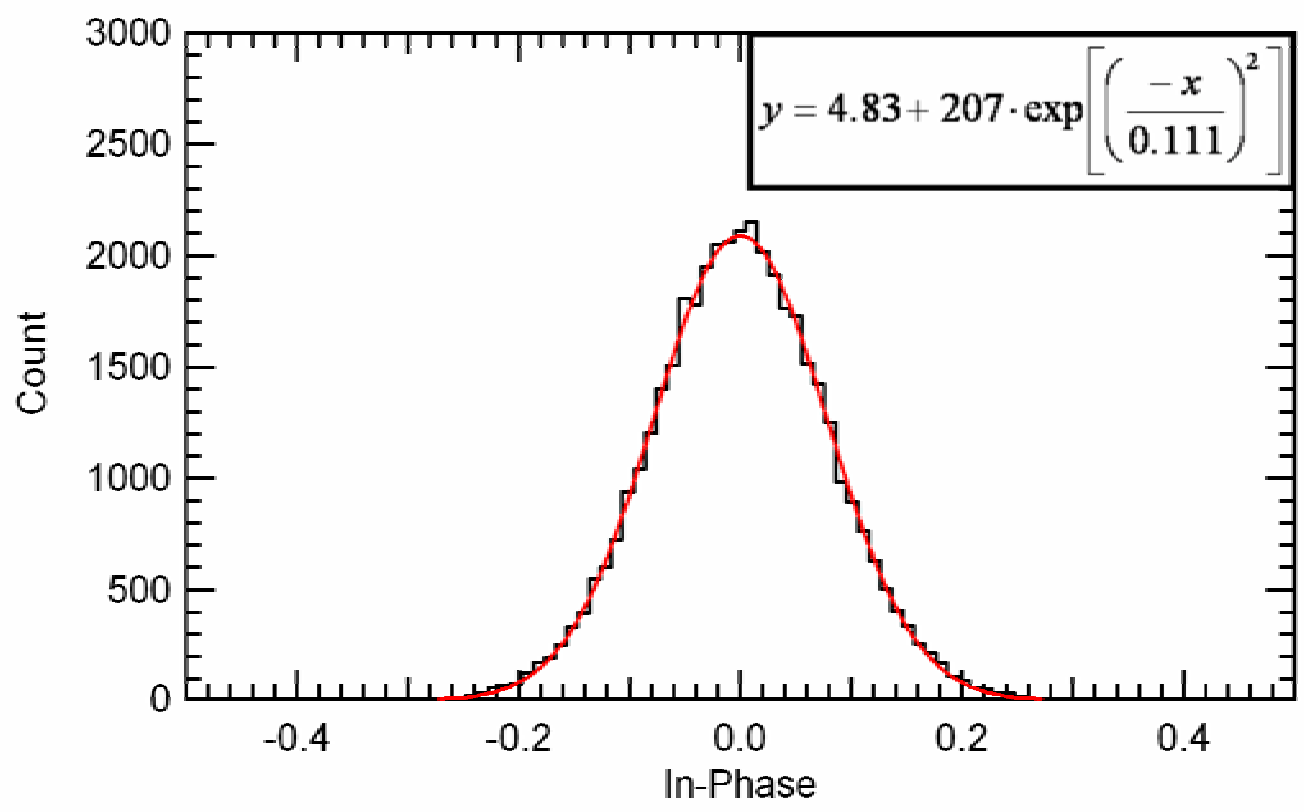

**Figure 6.13:** Histogram of the in-phase component of the 40 000 received IQ points.

### 6.2.3 Channel Frequency Response

The frequency response of the optical link was measured by sweeping the frequency of a sinusoidal signal and measuring the power at the output with the spectrum analyzer. Figure 6.15 shows the normalized result obtained for two different input powers.

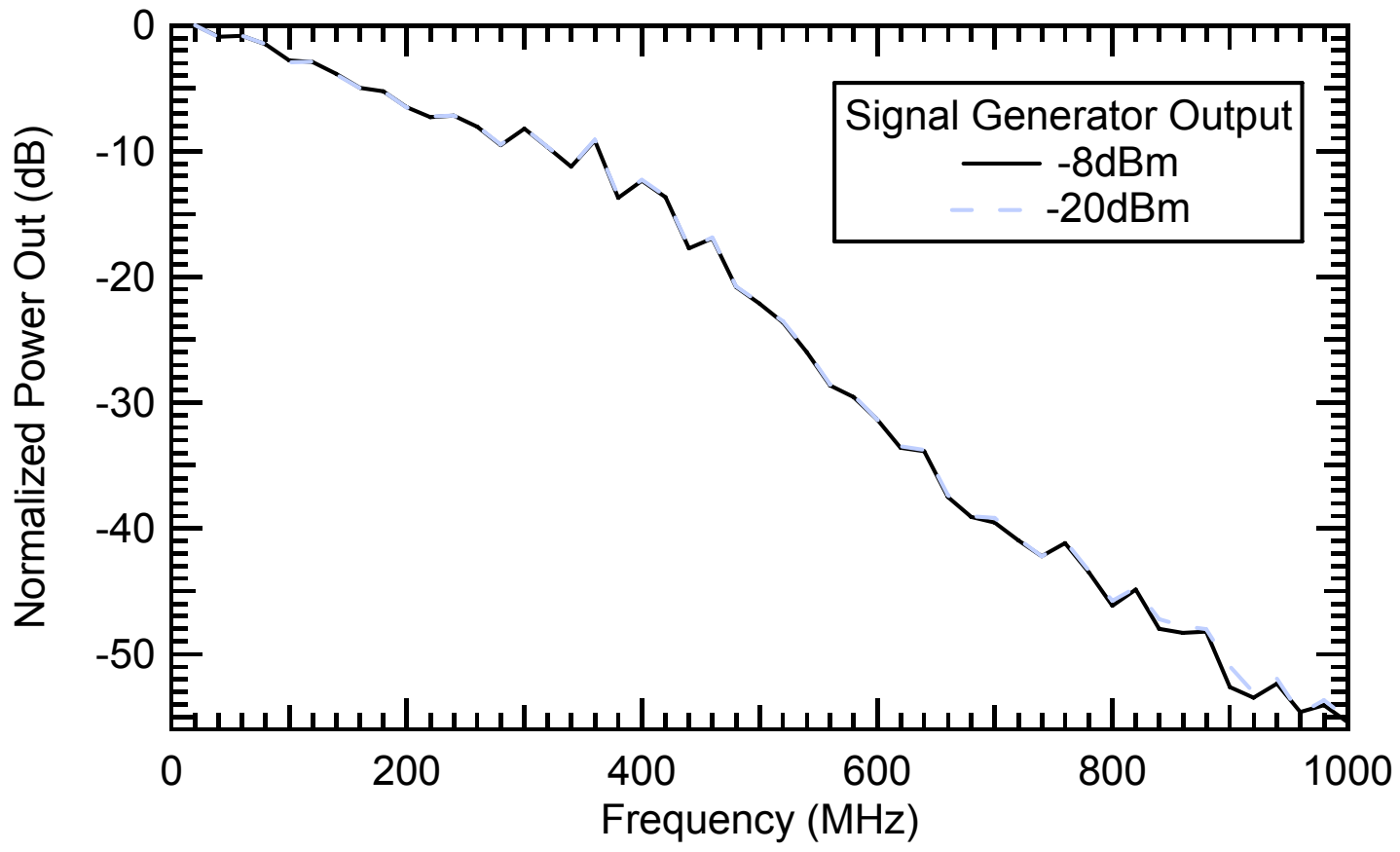

**Figure 6.14:** Attenuation exhibited by the test system and the optical link, measured using two different input powers.

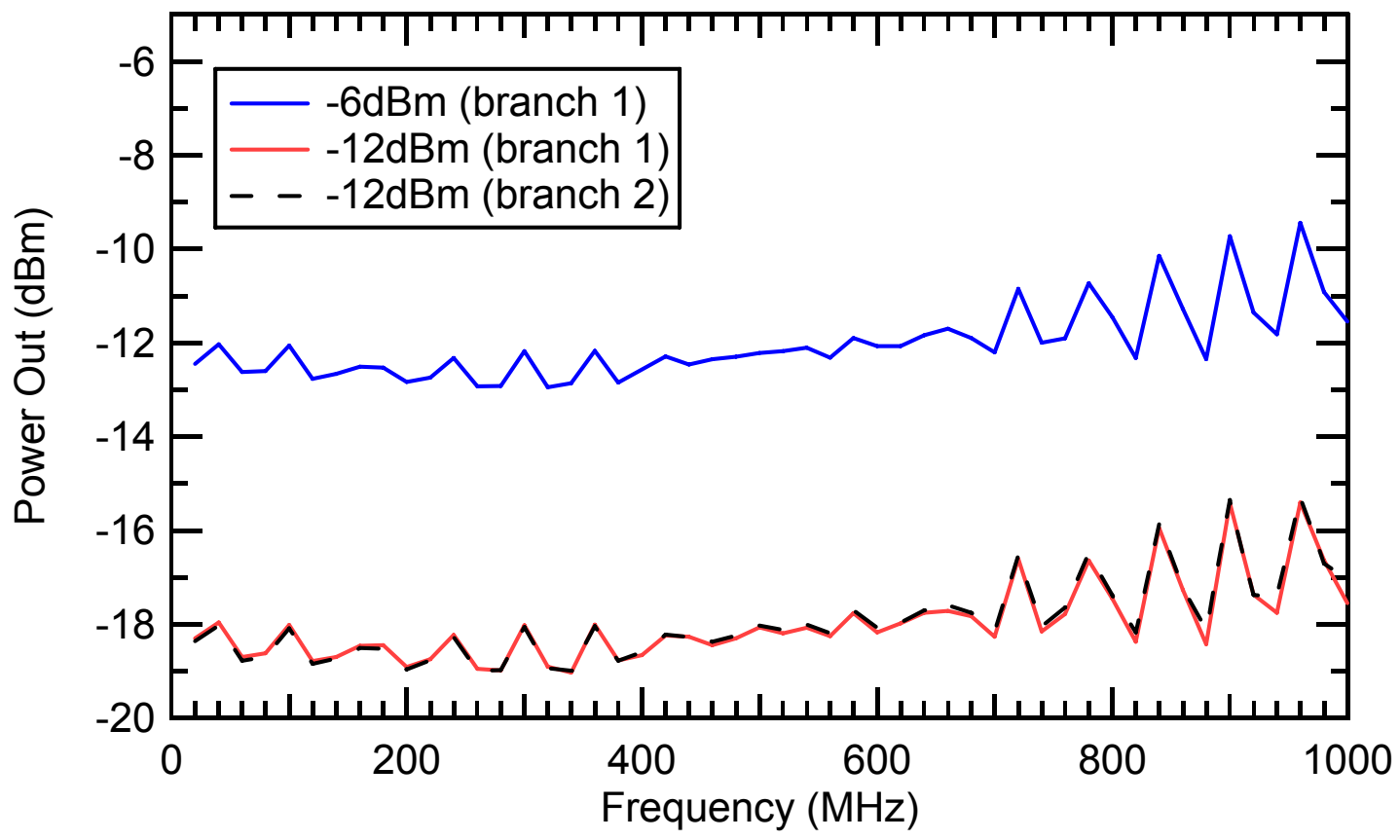

**Figure 6.15:** Attenuation exhibited by the test system only, measured using two different input powers. Only one branch of the differential driver output could be tested at a time.

The above result includes the response of the test system as well (differential driver, cables, etc.). In order to verify that the test system is not adversely affecting the response, the optical link components were removed, and the differential driver outputs were connected directly to the spectrum analyzer, one at a time. Figure 6.15 Shows the frequency responses obtained for two different input powers. Both output branches of the driver were tested. The response is relatively flat, except for the ringing observed after ~700MHz, which may be due to small impedance mismatches. It is clear that the test system does not limit the bandwidth of the entire setup, hence allowing exploration of the full bandwidth of the optical link.

### 6.2.4 Test System Limits

Before proceeding with the main tests, it is important to understand the limitations imposed by the test system itself. The objective of this test is to determine the SNR (or, equivalently, the EVM) of the test system, without the optical link. This represents the maximum performance that can be observed in any link under test. To this end, the optical link was removed from the setup, and one of the differential driver outputs was connected straight to the spectrum analyzer. A 1MS/s 4-QAM signal was passed through the link, modulated with pseudo-random data provided from the vector signal generator's internal baseband generator. Root nyquist pulse-shaping with filter alpha=1 was used. The carrier frequency was swept from 20 to 980MHz, in 20MHz steps, for eight different values of signal power. At each frequency point, the EVM and corresponding SNR was determined by the spectrum analyzer, and averaged over 2 000 received symbols.

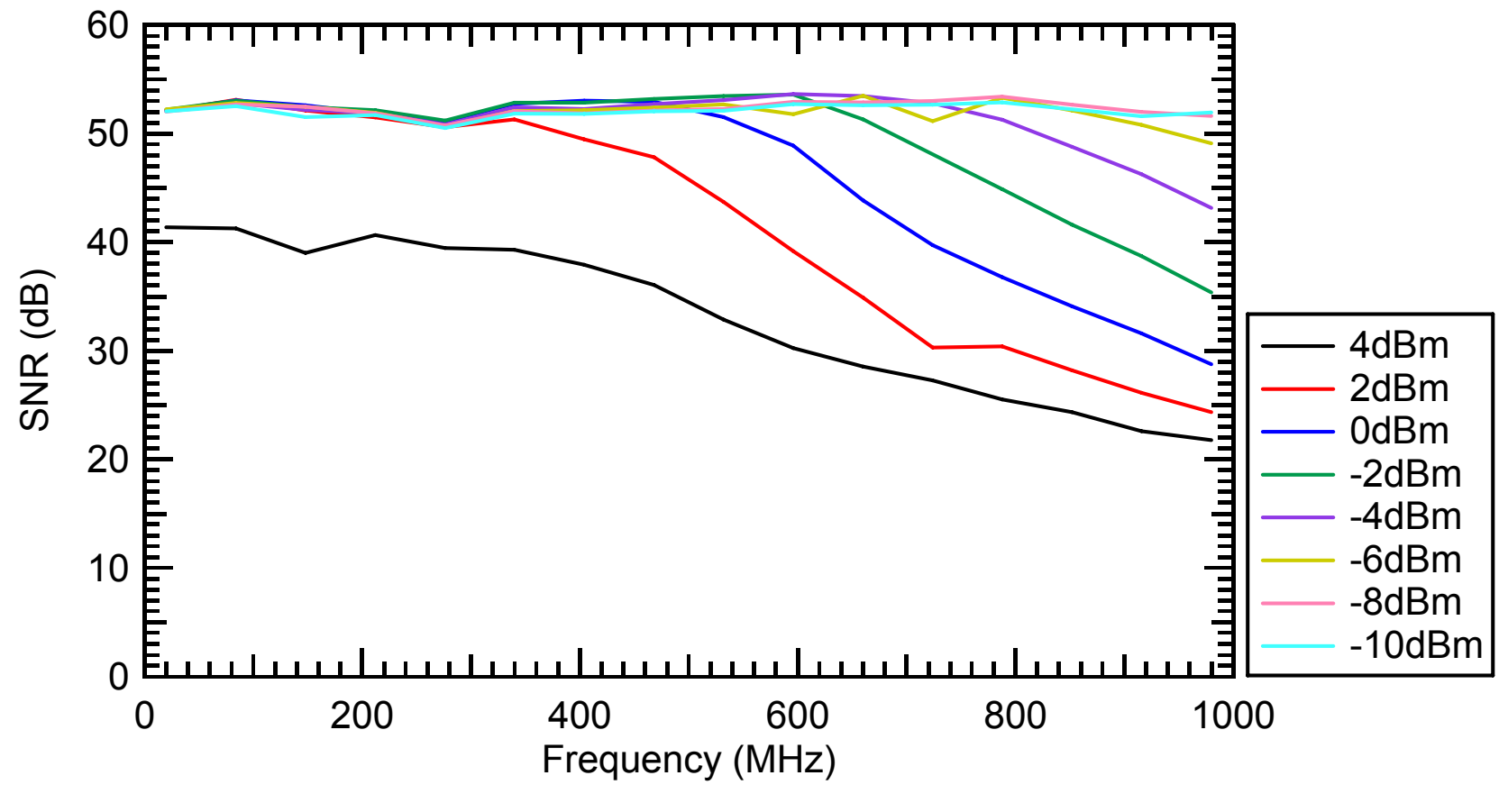

**Figure 6.16:** SNR as a function of frequency for the test system without the optical link, for several input powers. The values of power refer to the output of the signal generator.

Figure 6.16 shows the results obtained in terms of SNR. There is an obvious degradation in performance which is frequency-dependent with power values over ~ -6dBm. The degradation can be attributed to amplitude clipping occurring at the differential driver. This is illustrated in Figure 6.17 and Figure 6.18 that show the differential driver's output, captured by an oscilloscope. The figures show two 4-QAM symbols transmitted with a power of 0dBm at symbol rates of 1MS/s, but at different carrier frequencies[4]. At a carrier frequency of 10MHz (Figure 6.17), there is no obvious signal degradation. At 850MHz, there are signs of amplitude saturation both on top and on the bottom of the signal. Finally, Figure 6.19 shows that for an input power of 4dBm, saturation occurs even at low carrier frequency (10MHz), as would be expected from the SNR degradation observed in the SNR plot of Figure 6.16.

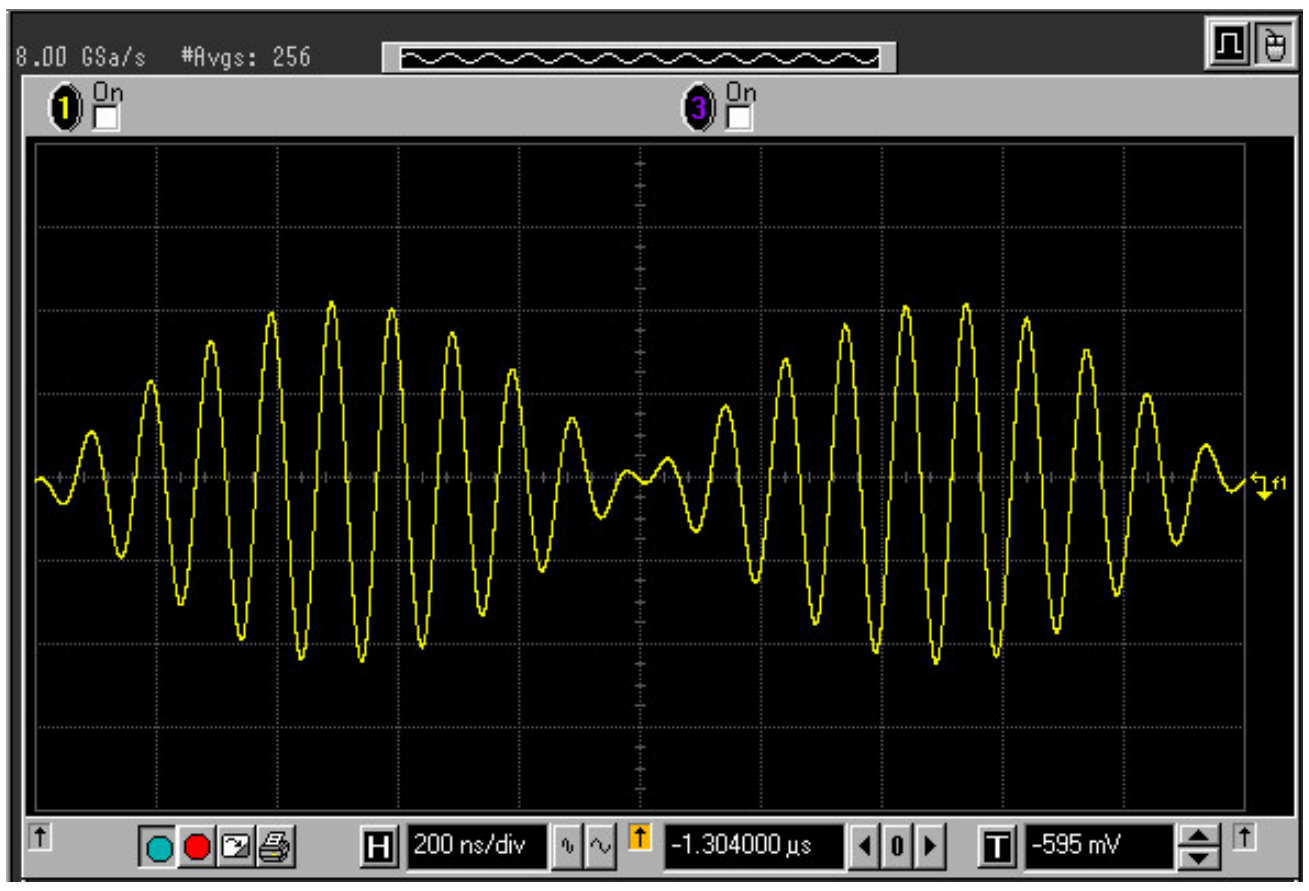

**Figure 6.17:** Differential driver output. Showing two 4-QAM symbols on a 10MHz carrier with input power 0dBm and symbol rate 1MS/s.

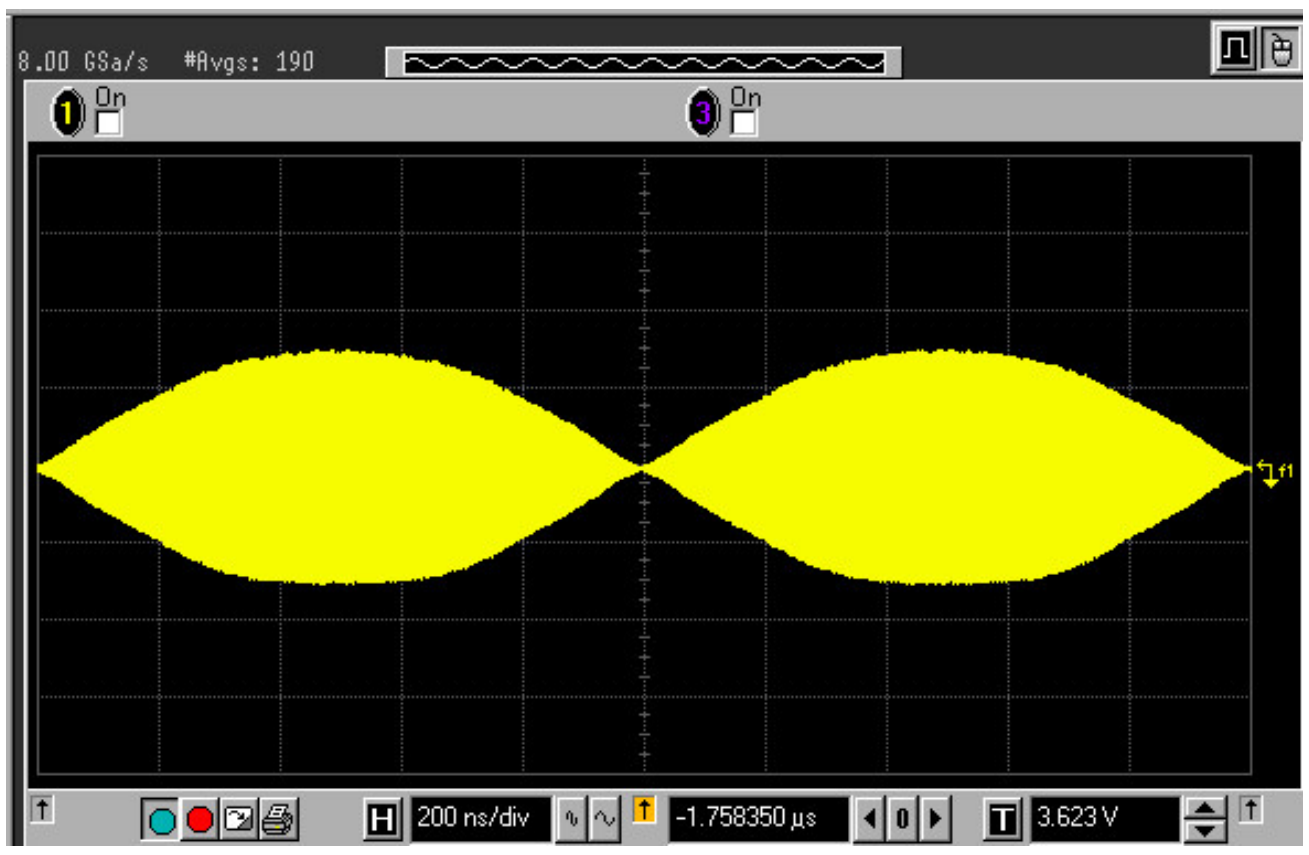

**Figure 6.18:** Differential driver output. Showing two 4-QAM symbols on a 850MHz carrier with input power 0dBm and symbol rate 1MS/s.

4 Root Nyquist filtering was used. Signal compression appears as 'flattening' at large amplitudes.

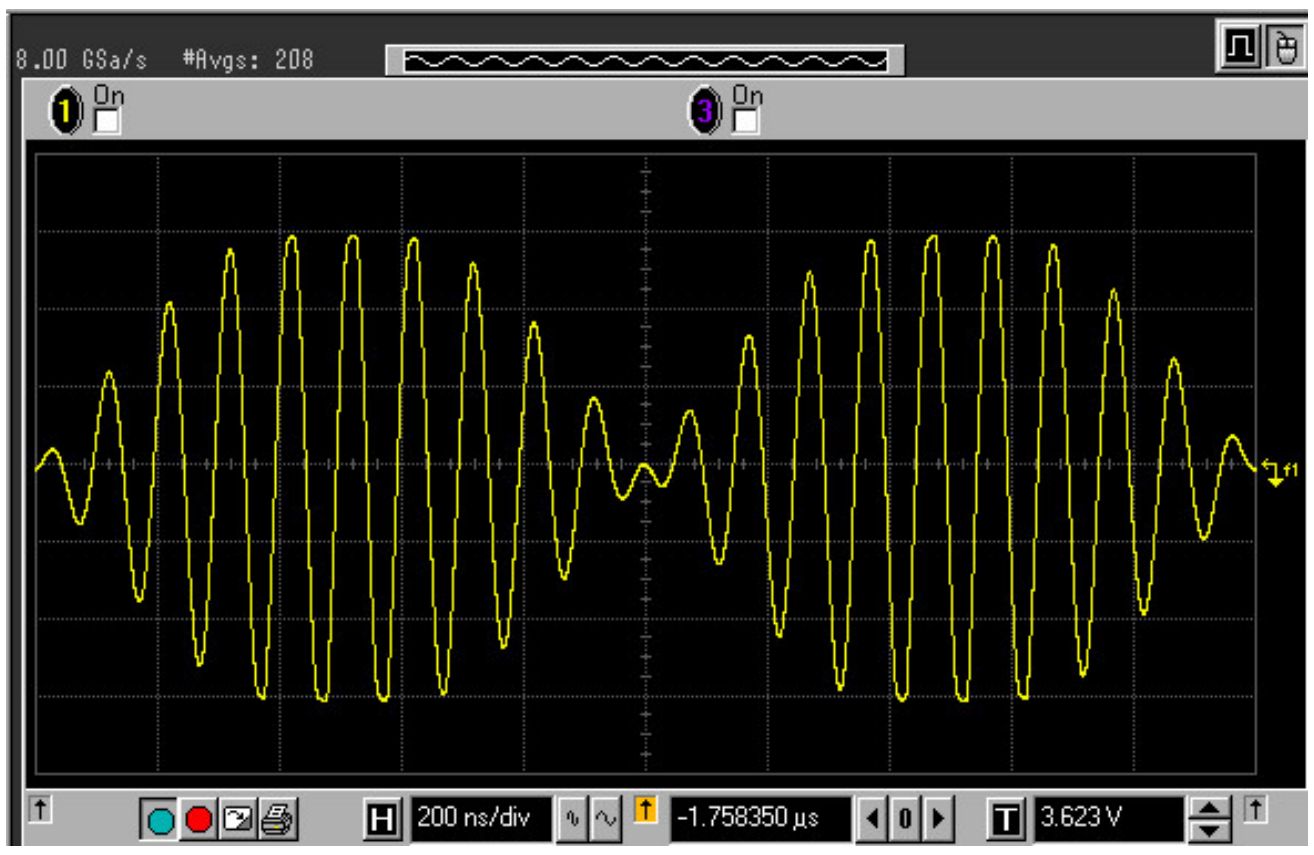

**Figure 6.19:** Differential driver output. Showing two 4-QAM symbols on a 10MHz carrier with input power 4dBm and symbol rate 1MS/s.

The significance of these results is that the test system imposes a limit on the power that can be used to transmit signals through the optical link. As will be seen in later sections, this is higher than the maximum power that can be tolerated by the optical link. Given the frequency response of the optical link, the maximum useable frequency should be ~600-700MHz (Figure 6.14). Figure 6.16 shows that there should be no performance degradation due to the test system for powers up to ~ -2dBm. Moreover, it has been shown that the maximum SNR that can be measured in this test system is ~52dB.

It is worth noting that in the presence of amplitude distortion on the QAM signals, the SNR-EVM relationship of Equation (6.3) is not necessarily accurate. The reason for this is discussed in section 6.4, where the errors and limitations of the QAM tests are detailed. Nevertheless, performance degradation due to clipping can still be identified with this test method, even if the SNR values may not be exact. For the purposes of this section's results this is satisfactory: It is the range of input powers that *do not* cause degradation, and not the *amount* of degradation, that is of interest.

## 6.3 Single-Carrier QAM Tests

### 6.3.1 Method

The test setup of Figure 6.5 was used for the single-carrier QAM tests. Unless stated otherwise, the method described in this section was followed for all SNR tests. Five different modulation schemes were used in the tests: 4, 16, 32, 64 and 256-QAM. The modulated RF carrier was swept in frequency (20MHz to 1000MHz, in 20MHz steps) and input power (0dBm to ~ -50dBm), for each modulation scheme (Figure 6.20). The symbol rate was kept at a constant 1MS/s, and root Nyquist filtering with a filter alpha of 1 was used at the transmitter and the receiver. The internal baseband generator of the vector signal generator was set to produce a pseudo-random bit pattern which was then used to modulate the carrier. For every point in frequency and power, the average EVM and SNR was calculated over 2 000 received symbols.

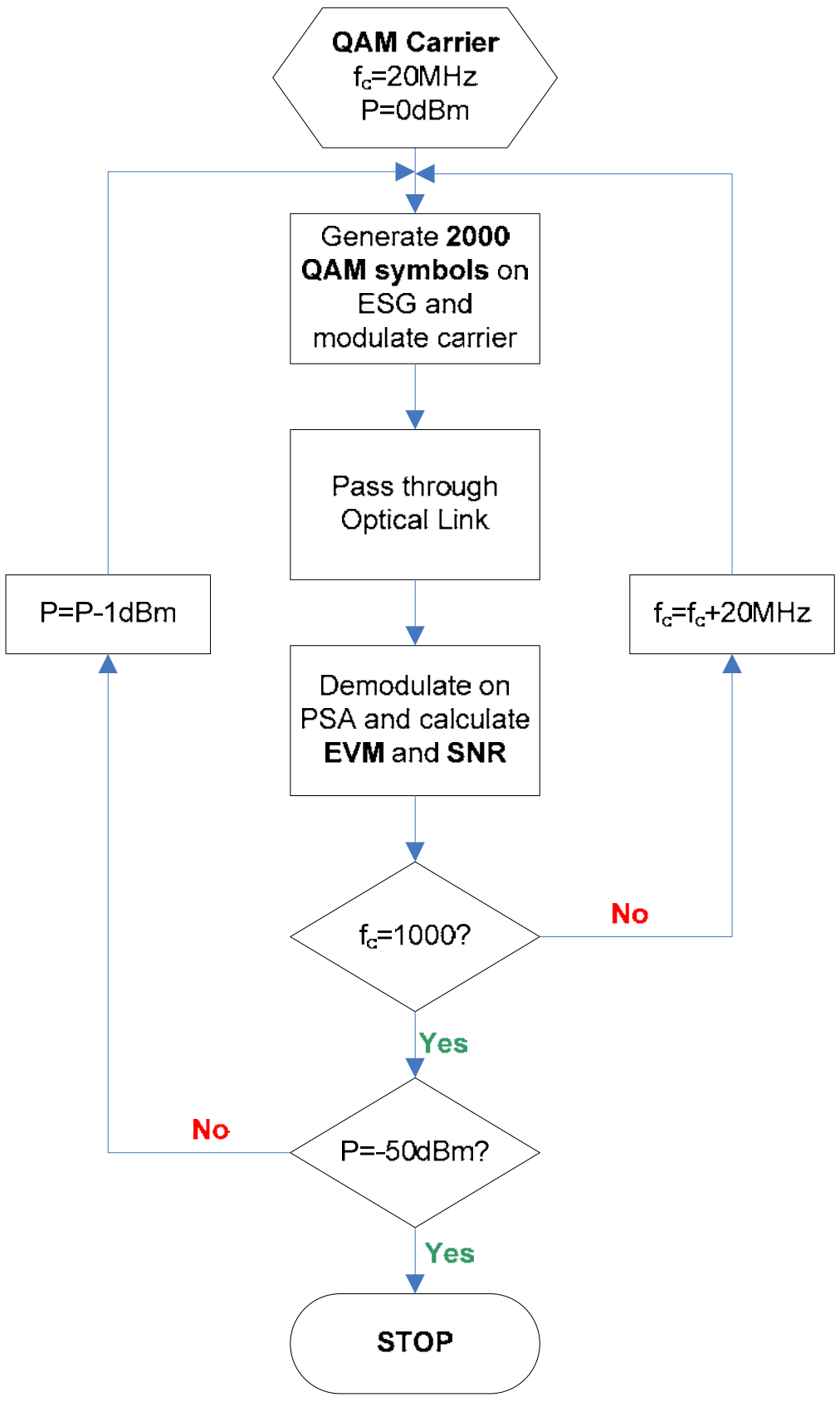

**Figure 6.20:** QAM test flow chart, showing the procedure followed for each modulation scheme (4-QAM, 16-QAM, 32-QAM, 64-QAM, 256-QAM).

It should be noted that the spectrum analyzer has no knowledge of the symbol pattern being sent by the modulator. This means that in order to calculate the EVM, it has to infer what symbol has been sent from the position of the received (noisy) IQ point, and the modulation scheme selected by the user. The constellation point closest to the received point is the one used for the EVM calculation. Clearly, if a transmitted IQ point is received beyond the decision boundary of its corresponding symbol, the demodulator will select the wrong constellation point as the reference for the EVM calculation (Figure 6.21). It follows that if the SNR is low enough to allow many errors in the transmission, the EVM calculation will fail. In the following measurements, the SNRs resulting from measurements with too many errors have been ignored. More specifically, if the SNR calculated corresponds to a symbol error rate above $10^{-3}$ for the particular modulation scheme, it is disregarded. This minimum allowed SNR is easily found by re-arranging Equation (6.5), and setting $P_e=10^{-3}$. Since the EVM calculation is made over 2 000 symbols, an error rate of $10^{-3}$ means that, on average, only 2 symbols will be received within the wrong decision boundary. Clearly, this will have a negligible effect on the calculation.

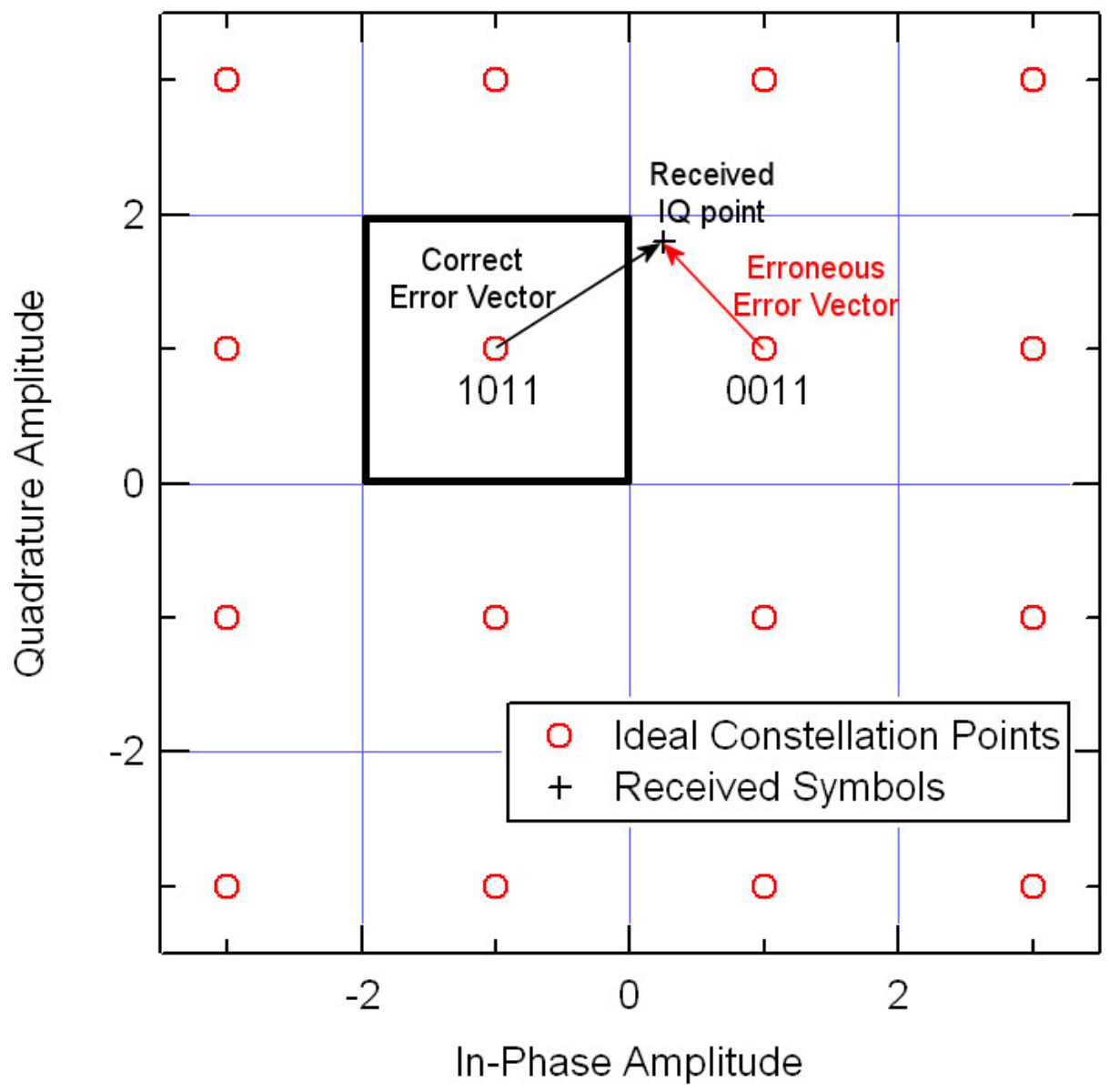

**Figure 6.21:** Illustration of EVM calculation error. The transmitted symbol is 1011, but has been corrupted by noise and received within the decision boundaries of symbol 0011. The spectrum analyzer assumed 0011 has been sent, and calculates the wrong error vector.

## 6.3.2 Results

SNR VS FREQUENCY EXAMPLE: 4-QAM

Figure 6.22 shows the SNR of a 4-QAM 1MS/s carrier, as a function of frequency. By sweeping the frequency and input power, the parameter space of the two quantities which are needed for the link's capacity calculation have been fully explored. From Equation (6.5), a symbol error rate of $10^{-3}$ is obtained when the SNR is 10.34dB. Hence this is the minimum SNR that can be measured with satisfactory accuracy by the test system (indicated by the dotted line in Figure 6.22).

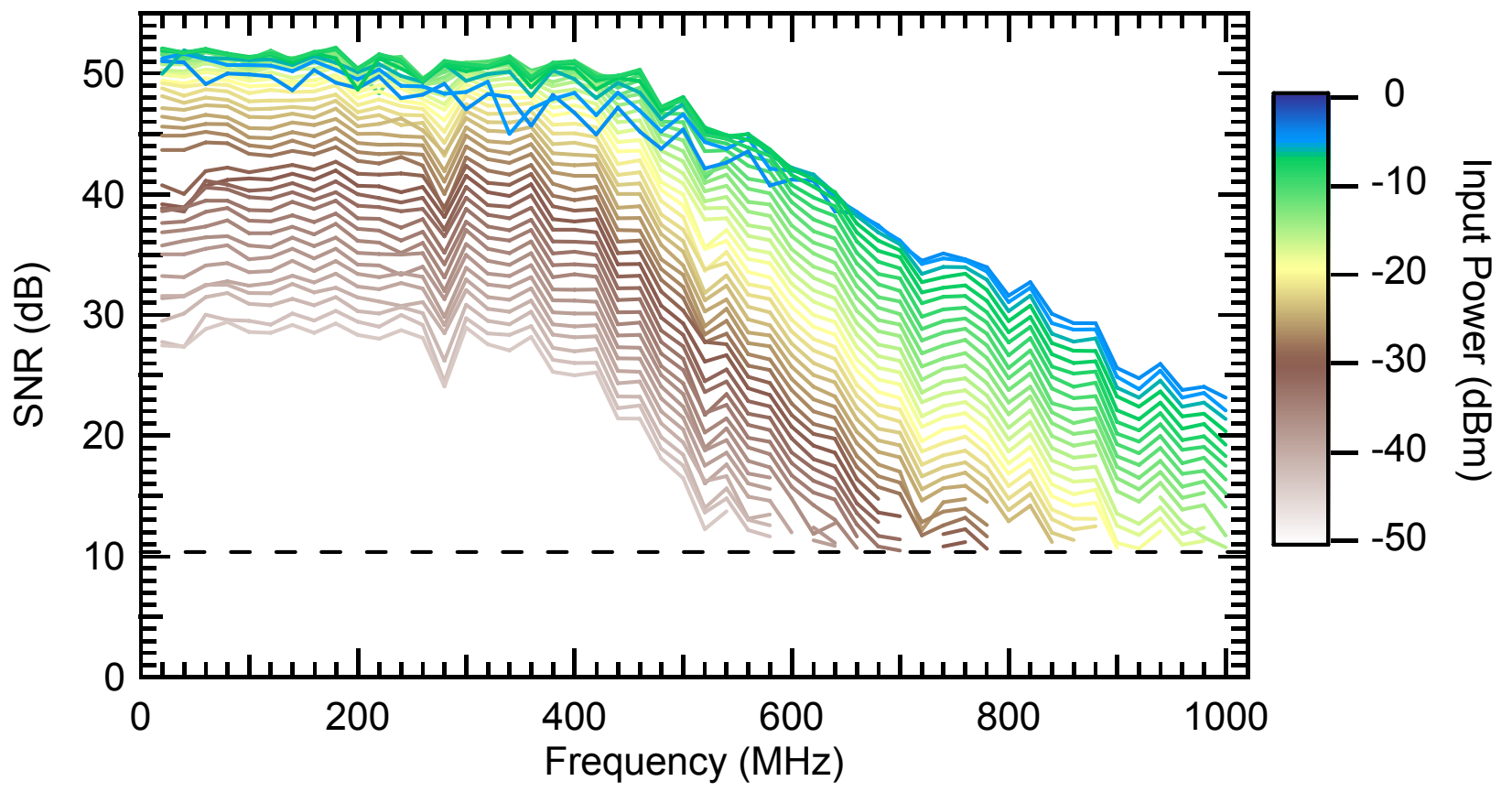

**Figure 6.22:** SNR as a function of frequency, estimated from a 4-QAM signal being transmitted through the optical link.

AVAILABLE TRANSMISSION POWER

Increased transmission power translates to higher SNR. However, at high enough power the SNR seems to degrade, indicating amplitude saturation somewhere in the optical link. In order to study this more closely, it is possible to plot the SNR as a function of power, for all frequency points. Figure 6.23 is obtained essentially by 'cutting' vertically –perpendicular to the frequency axis– across the plots of Figure 6.22.

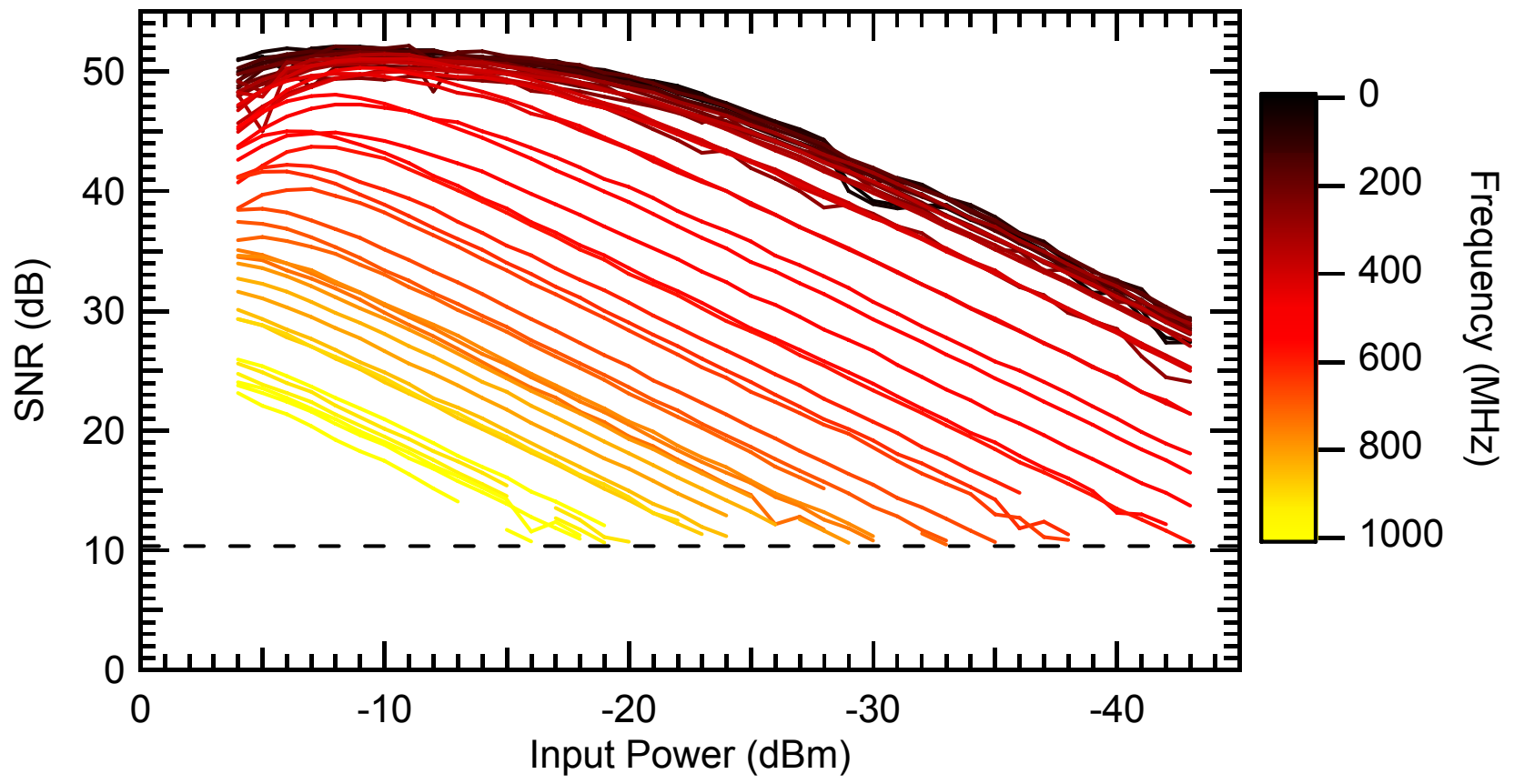

**Figure 6.23:** SNR as a function of input power, estimated from a 4-QAM signal being transmitted through the optical link.

Not surprisingly, at low powers (<-10dBm) there is a linear relationship between SNR and input power. Over -10dBm the SNR starts to roll-off, eventually decreasing with higher power after ~ -6dBm. This is due to the non-linear effects of amplitude compression somewhere in the link. Moreover, the peak of each plot tends to move toward higher power values with increasing frequency. This could be explained if the saturation is occurring at the receiver, rather than at the input of the link. The analog bandwidth of the LLD is specified at ~100MHz, and hence the signal is attenuated at higher frequencies. This explains why there is no saturation at higher frequencies, even at the highest power levels. The fact that the saturation observed in Figure 6.23 is occurring at the receiver is reinforced by the fact that SNR degradation occurs at just over -6dBm, which results in signal amplitudes that are well within the linear operating range of the AOH. As shown in section 6.2.1, a 4-QAM signal requires -3.5dBm to use the full linear range of the AOH, with this test system.

From Figure 6.23 the total power available to the system can be determined, given that in 4-QAM the average power is equal to the peak power (ignoring the effects of pulse shaping). The total transmission power available to the system which does not cause significant SNR degradation is ~ -6dBm.

### AVERAGE SNR VS FREQUENCY FROM ALL TESTS

The same procedure as that for the 4-QAM case was followed for all other modulation schemes used (16, 32, 64 and 256-QAM). The results from all schemes were combined into one graph, by averaging the SNRs obtained from all tests (Figure 6.24). The error bars denote the standard deviation of the SNR values. The results were quite consistent across all modulation schemes, giving confidence in the test method.

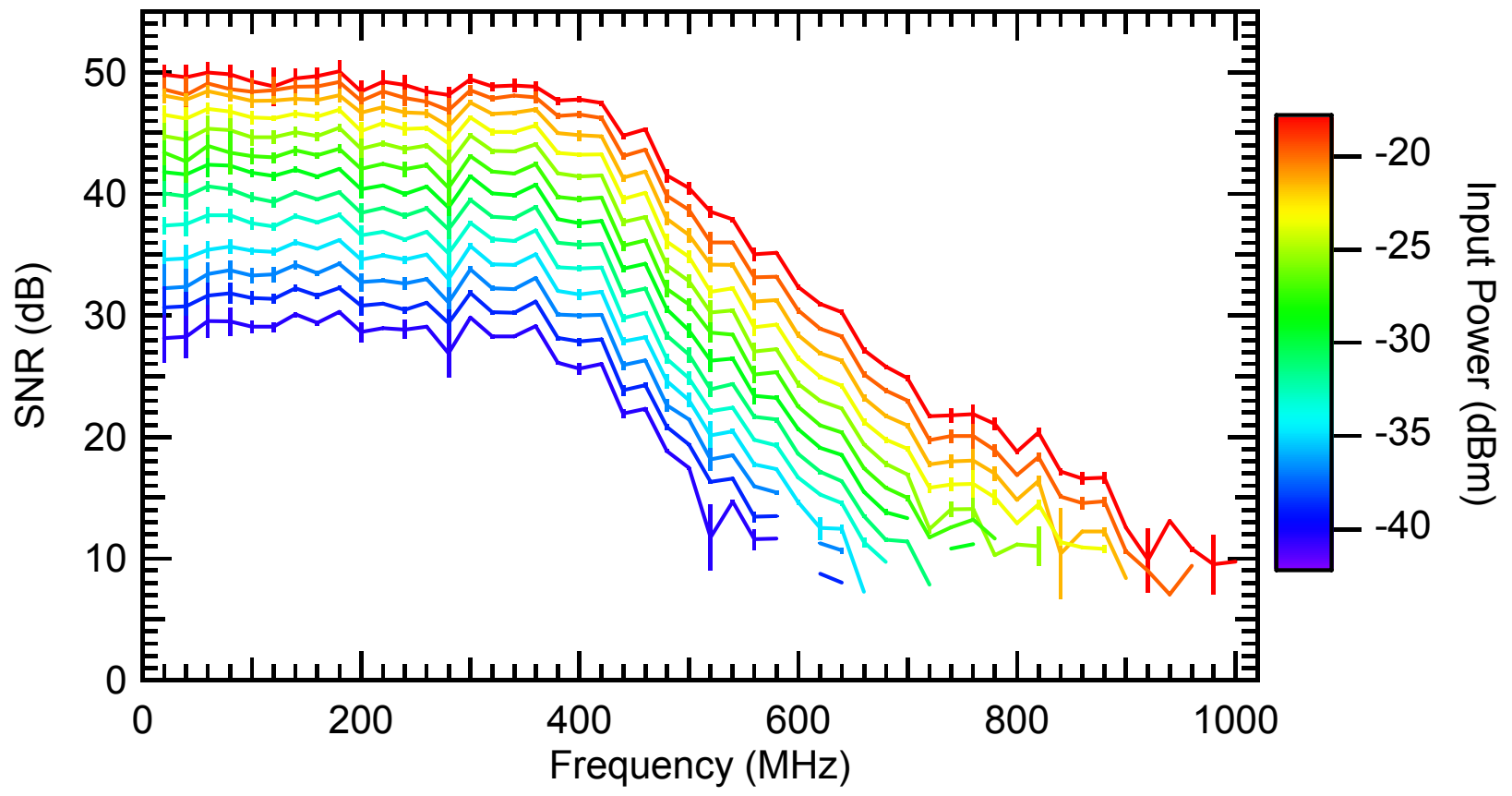

**Figure 6.24:** SNR Vs carrier frequency for a 1MS/s carrier, for several input powers. The plots were obtained by averaging the results from all tests (4, 16, 32, 64, 256-QAM).

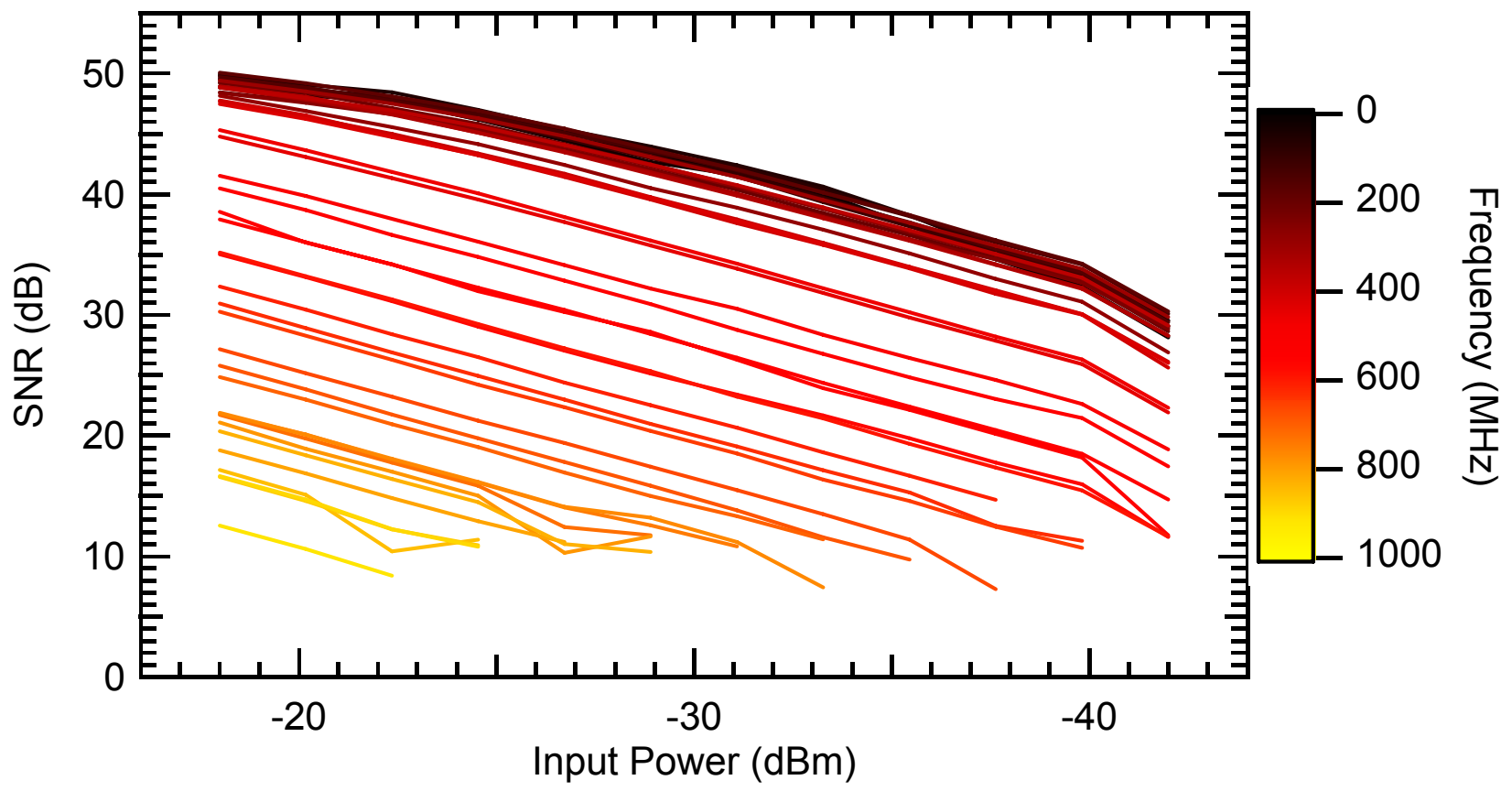

**Figure 6.25:** SNR Vs input power for a 1MS/s carrier, for several carrier frequencies.

Figure 6.25 shows the corresponding plots of SNR as a function of input power, for the range of tested frequencies. The SNR plots show linear dependence on the input power, having almost identical slopes (0.88dB/dBm) regardless of the carrier frequency. This suggests that the noise is higher for higher input powers (and therefore laser intensity).

Given the total available power determined previously, the SNR plots of Figure 6.24 and Figure 6.25 contain all the information needed to calculate the maximum achievable data rate over the Tracker optical links, using an RF digital communication system based on QAM.

### 6.3.3 Data Rate Calculation

POWER AND BIT ALLOCATION

As shown in Chapter 5, the data rate calculation relies on the knowledge of the SNR(s) of the QAM carrier(s) in the channel, which itself depends on the specific power and bit allocation for the particular system. In our analysis, we will use the results obtained in section 6.3.2 to emulate an OFDM system and determine its performance limits. The default loading algorithm is used, where the total power is distributed equally among all carriers (see section 5.2.6). Due to the uncertainty in the total power available to the system (section 6.3.2), the calculation will later be performed for various cases, from $P_{TOT}$=-6dBm to -10dBm. To illustrate the method followed, $P_{TOT}$=-6dBm will be considered in this section.

The SNR per carrier is obtained by extrapolation from the QAM test results. From Figure 6.22 it is known that the SNR curves show almost identical degradation with frequency, regardless of the input power. In fact, the change in SNR versus input power is a linear relationship with a slope of 0.88dB/dBm. The implication of this is that any of the SNR plots of Figure 6.22 can be used in the analysis, since all others can be obtained by simple extrapolation. It is preferable to use a higher input power (and therefore SNR) plot, since it covers a larger channel bandwidth.

Arbitrarily, the plot corresponding to a carrier of -22dBm can be chosen (this is just an example, and any other plot will produce almost identical results). This gives the SNR versus carrier frequency for the range 20-901MHz. The measurement points are in steps of 20MHz, while the analysis is based on using carriers spaced 1MHz apart. Hence, the SNRs of the carriers in between the measurement points are obtained by simple interpolation. The result gives the SNRs of 901 QAM carriers, with -22dBm power allocated to each and spaced at 1MHz. In OFDM this implies a symbol rate of 1MS/s for each carrier. The total power allocated is $-22+10\cdot\log_{10}(901) = 7.54$dBm, which is well above the link's limit.

The SNR versus frequency needs to be calculated first. For $P_{TOT}$ = -6dBm, the power allocated to each carrier will be:

$$-6+10\cdot\log_{10}(1/901) = -35.5\text{dBm} \quad (6.6)$$

The corresponding SNR versus frequency plot is then simply obtained by extrapolating from the -22dBm plot, and remembering that, from the slopes of Figure 6.25, a decrease of 1dBm in transmission power corresponds to a decrease of 0.88dB in SNR:

$$\begin{aligned} SNR_{-35.5dBm}(f) &= SNR_{-22dBm}(f) - 0.88\cdot\left[(-22)-(-35.5)\right] \\ &= SNR_{-22dBm}(f) - 11.9 \end{aligned} \quad (6.7)$$

The result is illustrated in Figure 6.26.

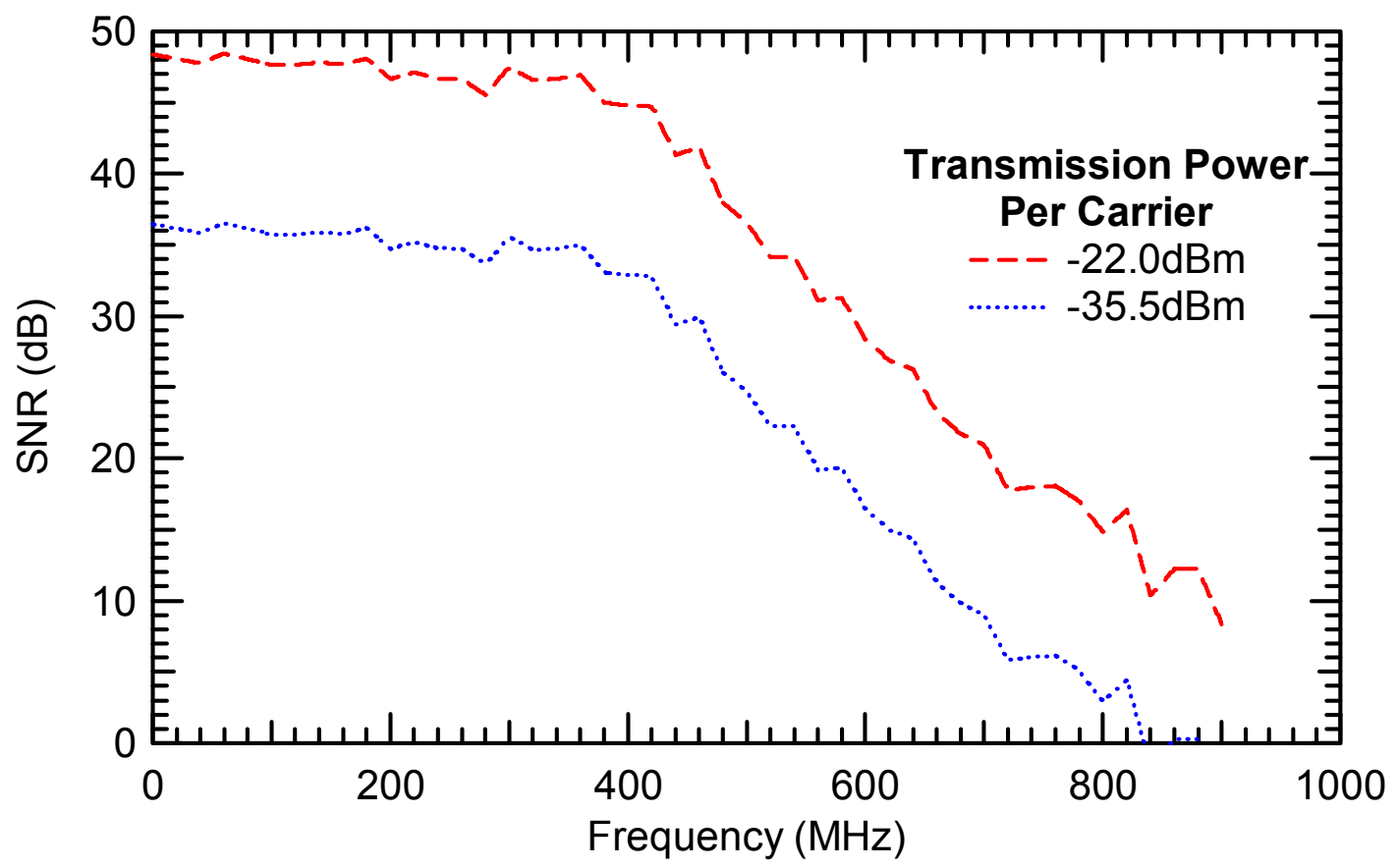

**Figure 6.26:** SNR versus carrier frequency for a 1MS/s QAM carrier, for a transmission power of -35.5dBm, obtained by extrapolation from the measured -22dBm SNR plot. The plot corresponds to a system with total allocated power of -6dBm, and 901 carriers.

### AVAILABLE BANDWIDTH

In the previous section, the allocated power (and SNR) were calculated for a system of 901 carriers in the channel. However, this does not mean that all of the 901 carriers will be able transmit information. Consider the example of the -35.5dBm SNR plot in Figure 6.26. A number of carriers in the higher frequencies of the channel have very low SNR, and may have to be 'turned off' (i.e. no power allocated). This will have an impact on the total number of carriers used, and therefore the total occupied channel bandwidth.

To calculate how many carriers out of the 901 will be used, the boundary condition is set that any carrier that cannot transmit a minimum of 1bit/symbol at the target error rate, will be turned off. As an example, for a target error rate of $10^{-9}$, the minimum SNR required for transmission of 1bit/symbol in QAM can be approximated from Equation (5.33):

$$SNR = \Gamma \cdot (2^C - 1) = 12 \cdot (2^1 - 1) = 12 \qquad (6.8)$$

In dB, the minimum SNR is 10·log10(12) = 10.8dB.

For the -35.5dBm SNR plot in Figure 6.26, it is obvious that several of the 901 carriers have SNRs below the 10.8dB limit. In order to find the maximum number of carriers (and channel bandwidth) available to an OFDM system, the carrier with the worst SNR is turned off. The carrier's allocated power is then evenly distributed over the remaining 900 carriers, and the SNRs re-computed. The process, illustrated in Figure 6.27, is repeated until no carrier has an SNR below 10.8dB. For a total power available of -6dBm, this results in 682 carriers of -34.3dBm each, covering 682MHz of channel bandwidth (Figure 6.28).

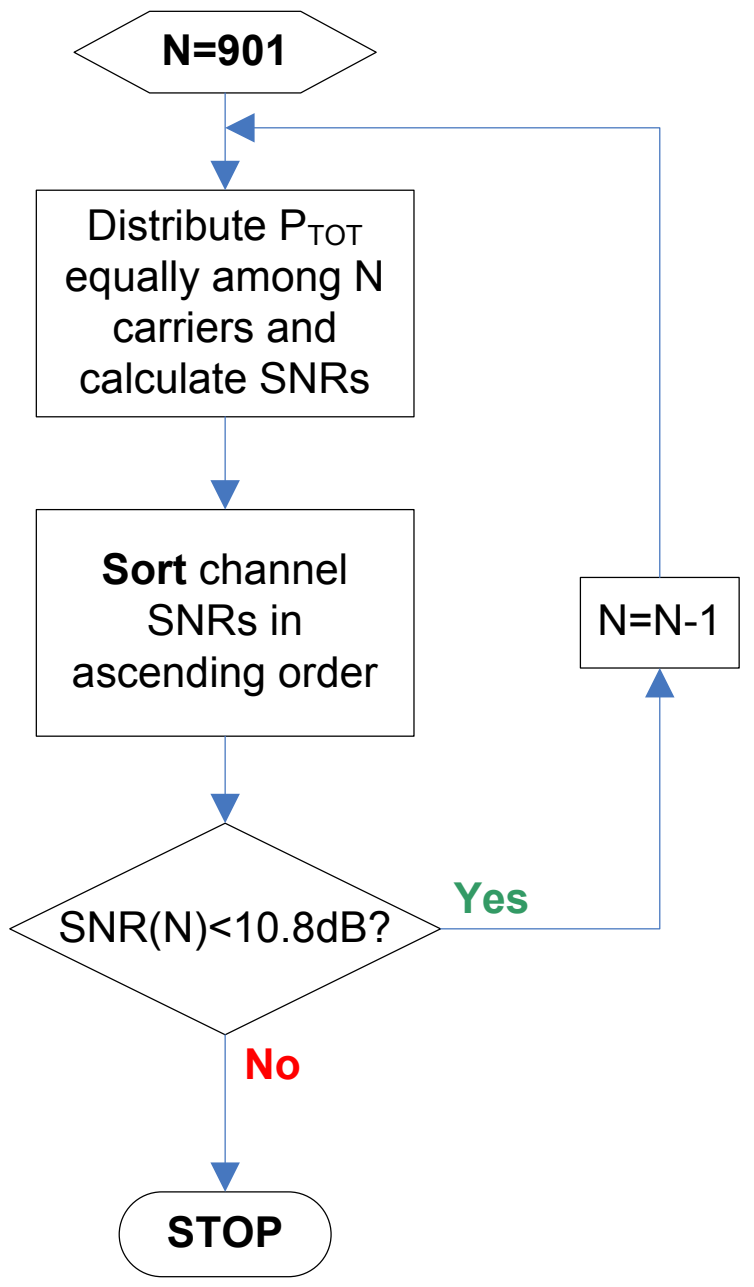

**Figure 6.27:** Flowchart of algorithm for determining the maximum bandwidth available to an OFDM system with uniform power distribution.

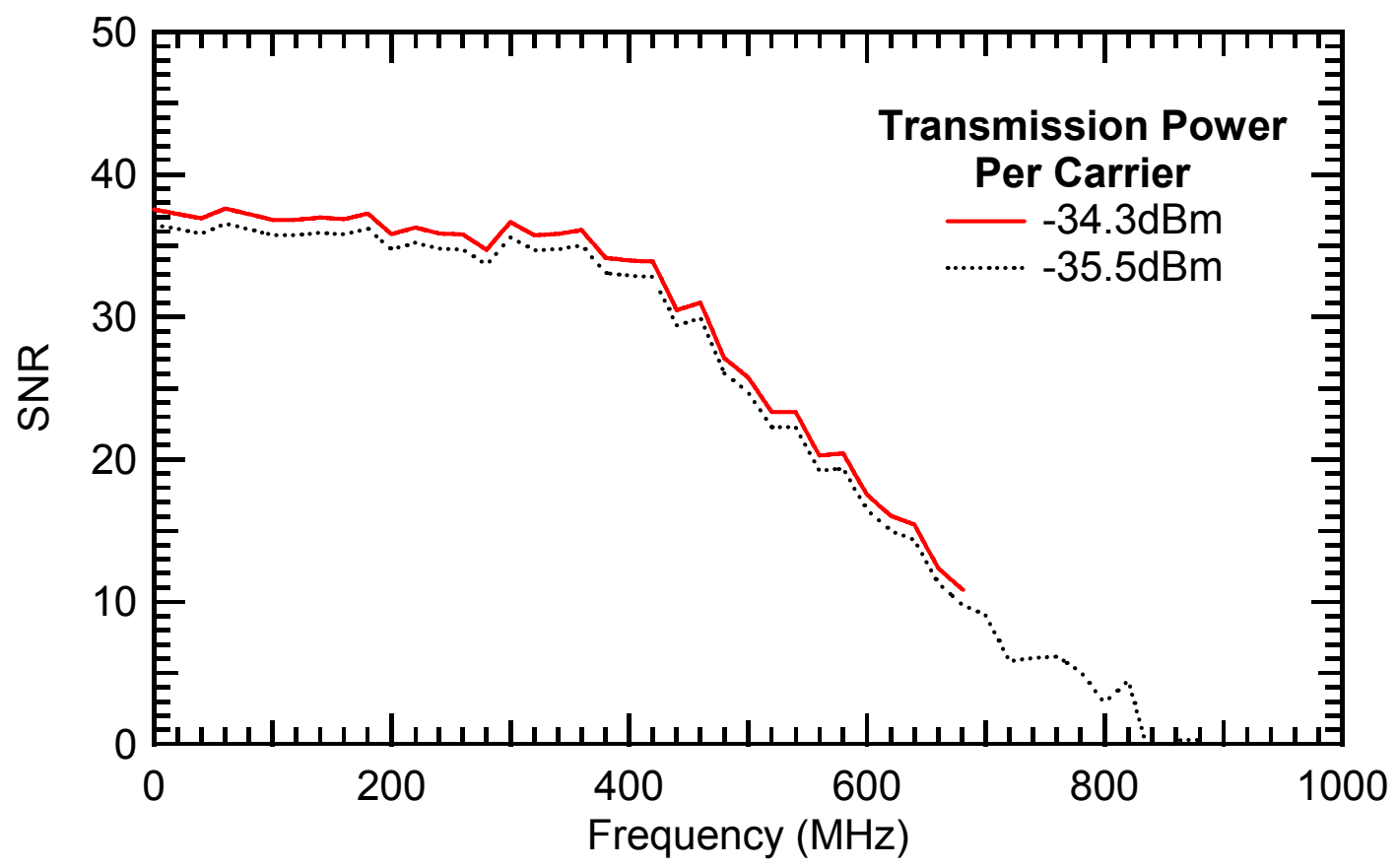

**Figure 6.28:** Showing the SNR versus carrier frequency plots for a 1MS/s QAM carrier, with -34.3dBm and -35.5dBm allocated power. The plots correspond to OFDM systems with 901 and 682 carriers respectively. $P_{TOT}$ = -6dBm and uniform carrier power distribution is assumed.

### OPTIMUM BANDWIDTH USE

Optimum rate-adaptive allocation algorithms attempt to maximize the data rate for a given target error rate, not only as a function of power and bits/symbol, but for the occupied bandwidth as well. Given a set of $N$ carriers of symbol rate SR, there exists a number of carriers, $N^* \leq N$, that maximizes the overall data rate, for a given carrier power distribution. Qualitatively, for a given channel, it is not always beneficial to use all the bandwidth available at the expense of power/carrier. A very straightforward approach to find $N^*$ is to calculate the data rate for all $N$, with the method used in Chapter 5. The optimum bandwidth will then be the one that results in the maximum data rate. To illustrate this, consider the example of the OFDM system with $P_{TOT}$ = -6dBm, power backoff = 0dB and a target error rate of $10^{-9}$ (Figure 6.29). The data rate increases faster at low $N$, and starts to roll off at $N$~400-500. Though not visible in the plot, the maximum data rate of 4.5739Gbit/s occurs at $N^*$=666, in this case. However, at $N$=682 the data rate is 4.5722Gbit/s, which is only 1.7Mbit/s lower. While the benefit of optimizing the use of bandwidth is negligible in this case, it will still be used in the subsequent calculations for completeness.

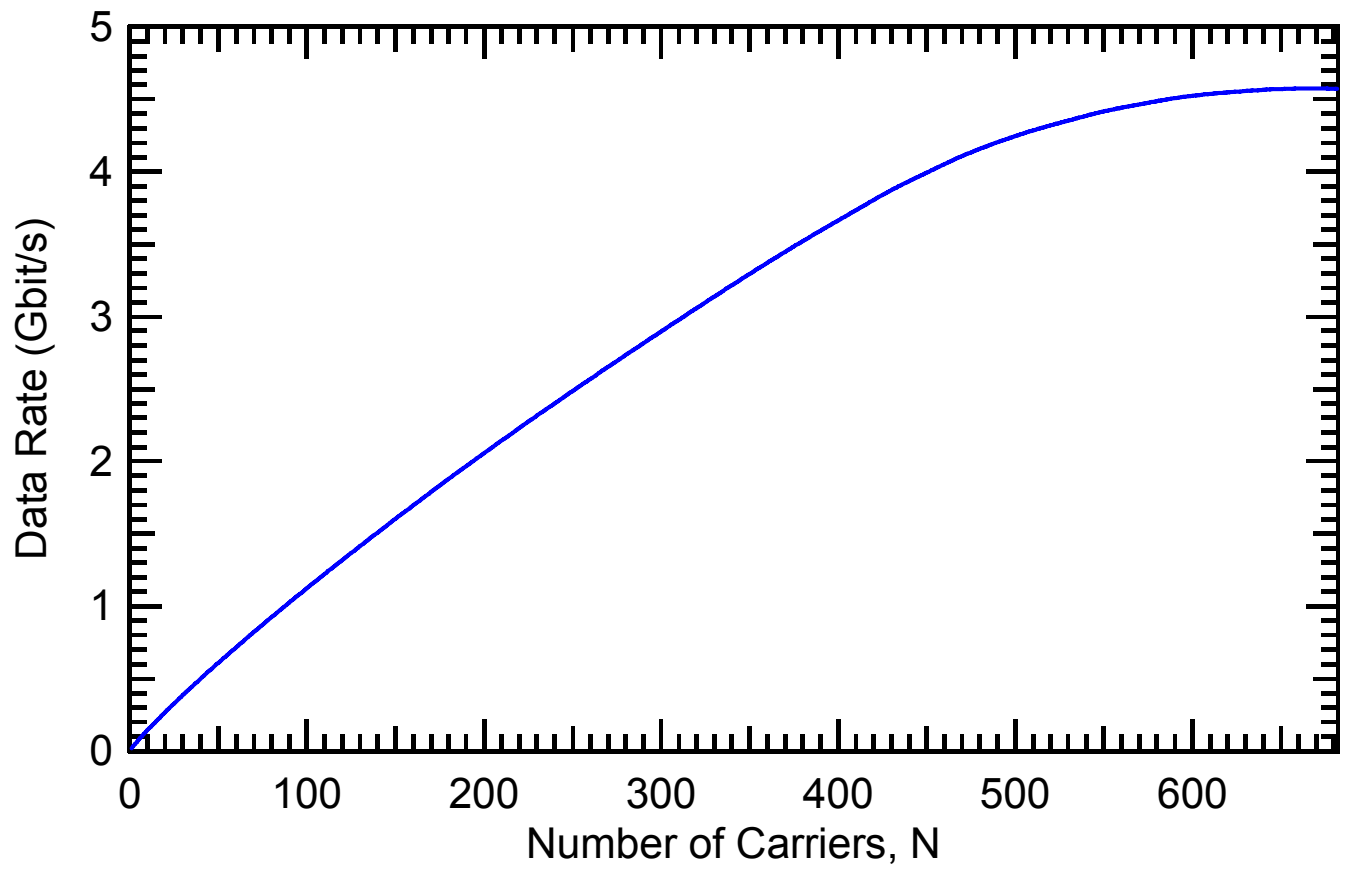

**Figure 6.29:** Data rate as a function of number of 1MS/s QAM carriers for an OFDM system with $P_{TOT}$=-6dBm, power backoff = 0dB and a target error rate of $10^{-9}$.

DATA RATE

The same procedure followed in Chapter 5 was used to calculate the data rate as a function of normalized transmit power (or, equivalently, power backoff) for various target error rates. The results for $P_{TOT}$ = -6dBm are shown in Figure 6.30. Due to the uncertainty in the total power available in the system, the data rate calculation was repeated for various values of $P_{TOT}$ and a target error rate of $10^{-9}$. The results are plotted in Figure 6.31.

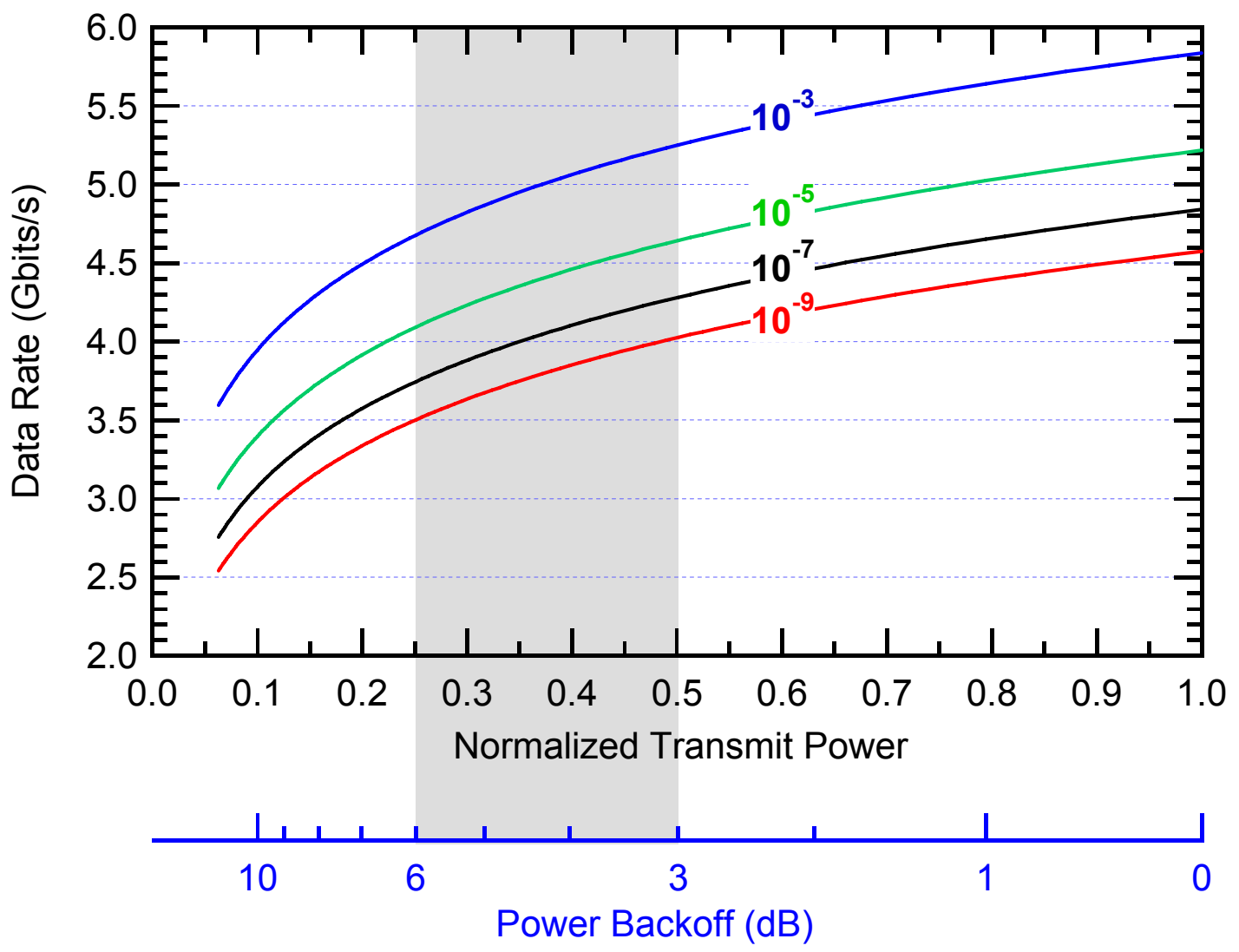

**Figure 6.30:** Data rate as a function of transmission power, for various target error rates.

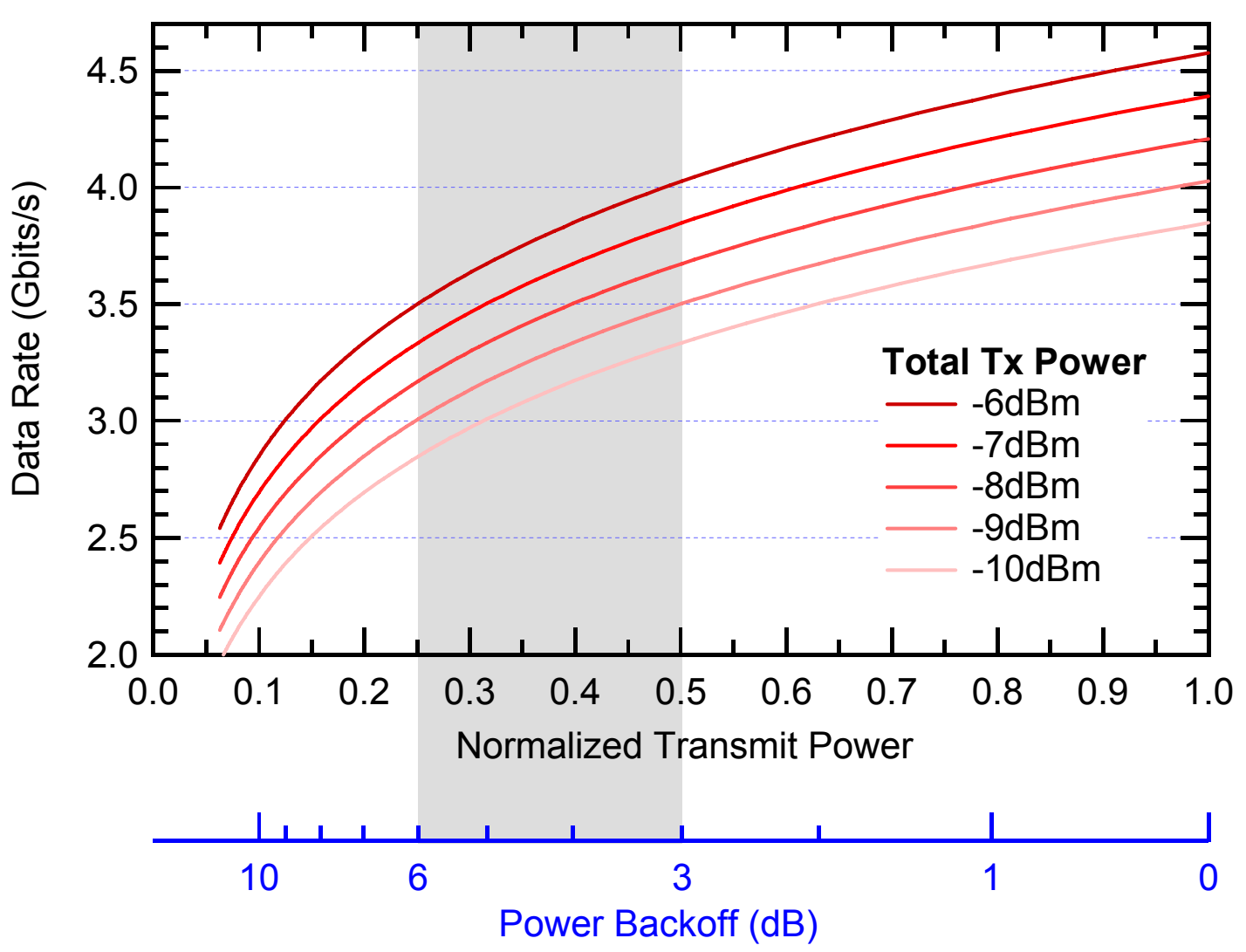

**Figure 6.31:** Data rate as a function of transmission power for various values of $P_{TOT}$, and a target error rate of $10^{-9}$.

The shaded areas in both plots indicate typical power backoff values. The calculation ignores the effects of amplifier compression, and hence one would expect SNR degradation (and therefore lower data rate for a given target error rate) in an OFDM system operating with little or no backoff. The exact amount of backoff required is a design choice, and depends on the modulation schemes used, as well as the techniques employed for mitigating the effects of compression. The results indicate that, for a target error rate of $10^{-9}$, data rates of at least 3-4Gbit/s are possible, depending on the total transmission power available, and the backoff employed in a real implementation.

It should be noted that the results shown are valid for an uncoded OFDM system. Techniques such as Forward Error Correction (FEC) and Trellis Code Modulation (TCM) can be used to lower error rate for a given transmission power, though a certain amount of data overhead would be incurred. Encoding is beyond the scope of this thesis, and will not be investigated further.

### 6.3.4 Example Plots: 256-QAM

An example of a received constellation diagram from the laboratory tests, as well as the associated eye diagrams, will be shown here for illustrative purposes. Eye diagrams are plotted for each component of the complex QAM signal, i.e. the in-phase amplitude ($I$) and the quadrature ($Q$) amplitude. From Chapter 5, a QAM signal is essentially made up of two pulse amplitude modulated (PAM) sinusoids that are orthogonal to each other. If the carriers are removed (as is the case in a QAM receiver), the resulting signals will be two baseband $M$-PAM waveforms ($M$ being the number of levels). Eye diagrams can then be displayed for each component separately.

Figure 6.32 shows 2 000 received symbols of a 256-QAM constellation transmitted through the optical link at a carrier frequency of 20MHz, with 1MS/s symbol rate and input power of -36dBm. The example was chosen to correspond to a hypothetical multi-carrier system based on the calculations made in previous sections. The plot essentially belongs to one of the carriers in an OFDM system with a total available power of -9dBm, and a power backoff of 6dB. The SNR is ~35dB (calculated from the EVM), and the corresponding symbol error rate is $10^{-9}$. The total bit rate in such a system would be ~3Gbit/s (see Figure 6.31).

Figure 6.33 and Figure 6.34 show the quadrature and in-phase eye-diagrams corresponding to the 256-QAM example. There are 16 distinct amplitude levels in each component, as would be expected in 256-QAM (16·16=256). Moreover, the effect of filtering can be seen in the smooth transitions between each symbol, as well as on the slight overshoot observed (root Nyquist filtering with alpha=1 was used).

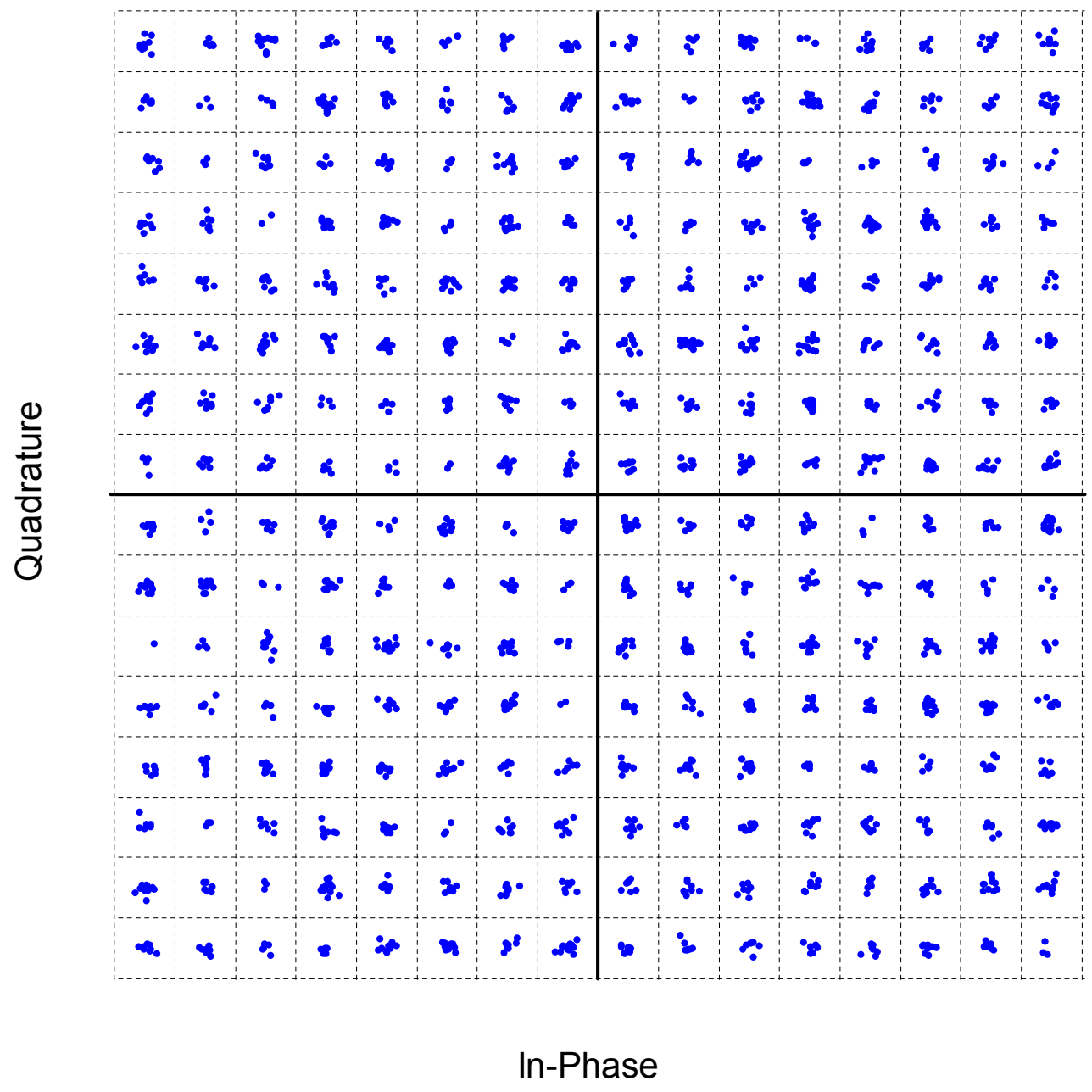

**Figure 6.32:** Example of a 256-QAM received constellation at an SNR~35dB, from the lab tests. Showing 2 000 decoded symbols. Carrier frequency = 20MHz, symbol rate = 1MS/s, input power = -36dBm.

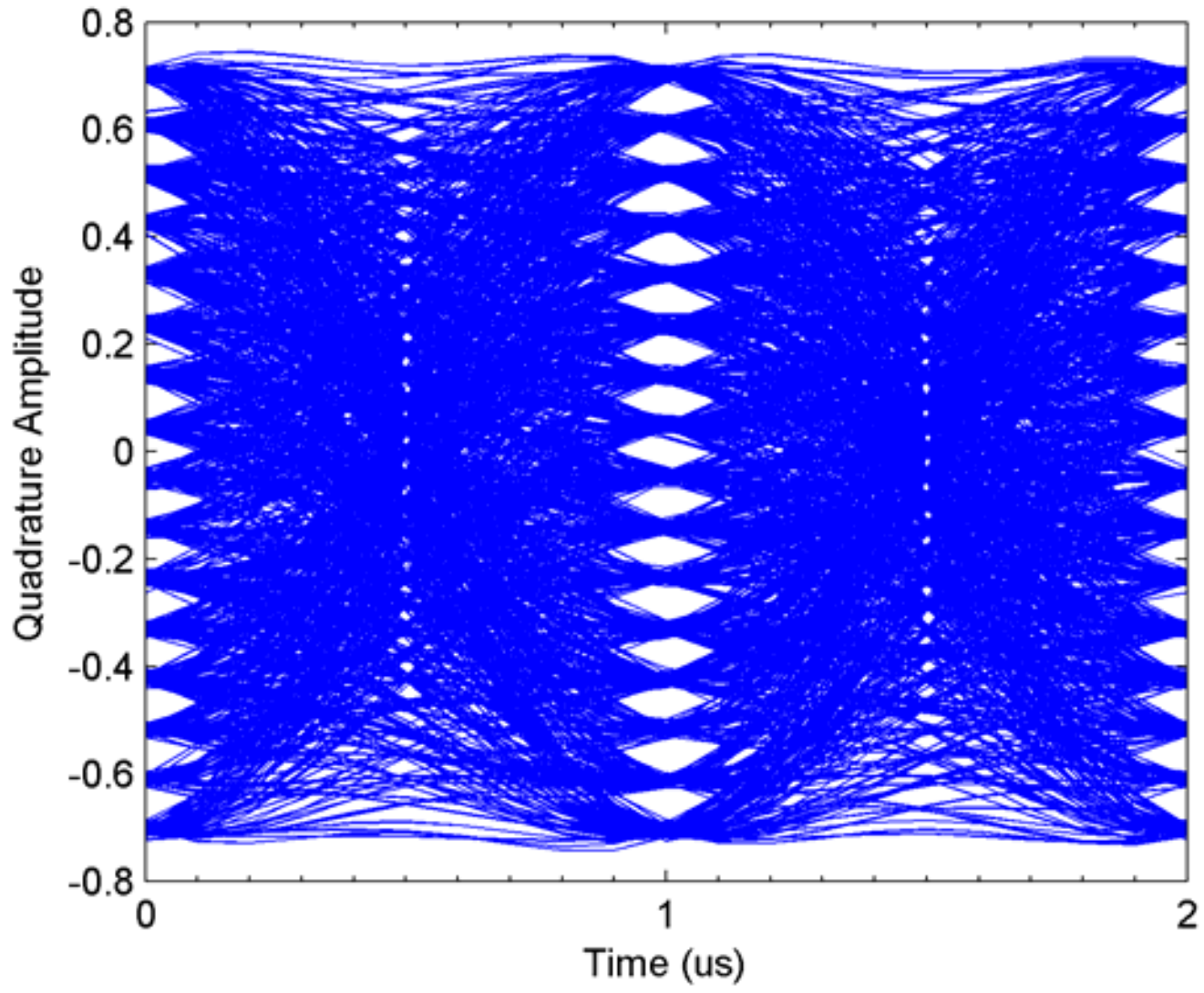

**Figure 6.33:** Quadrature amplitude eye diagram of the 256-QAM signal corresponding to the to received constellation in Figure 6.32.

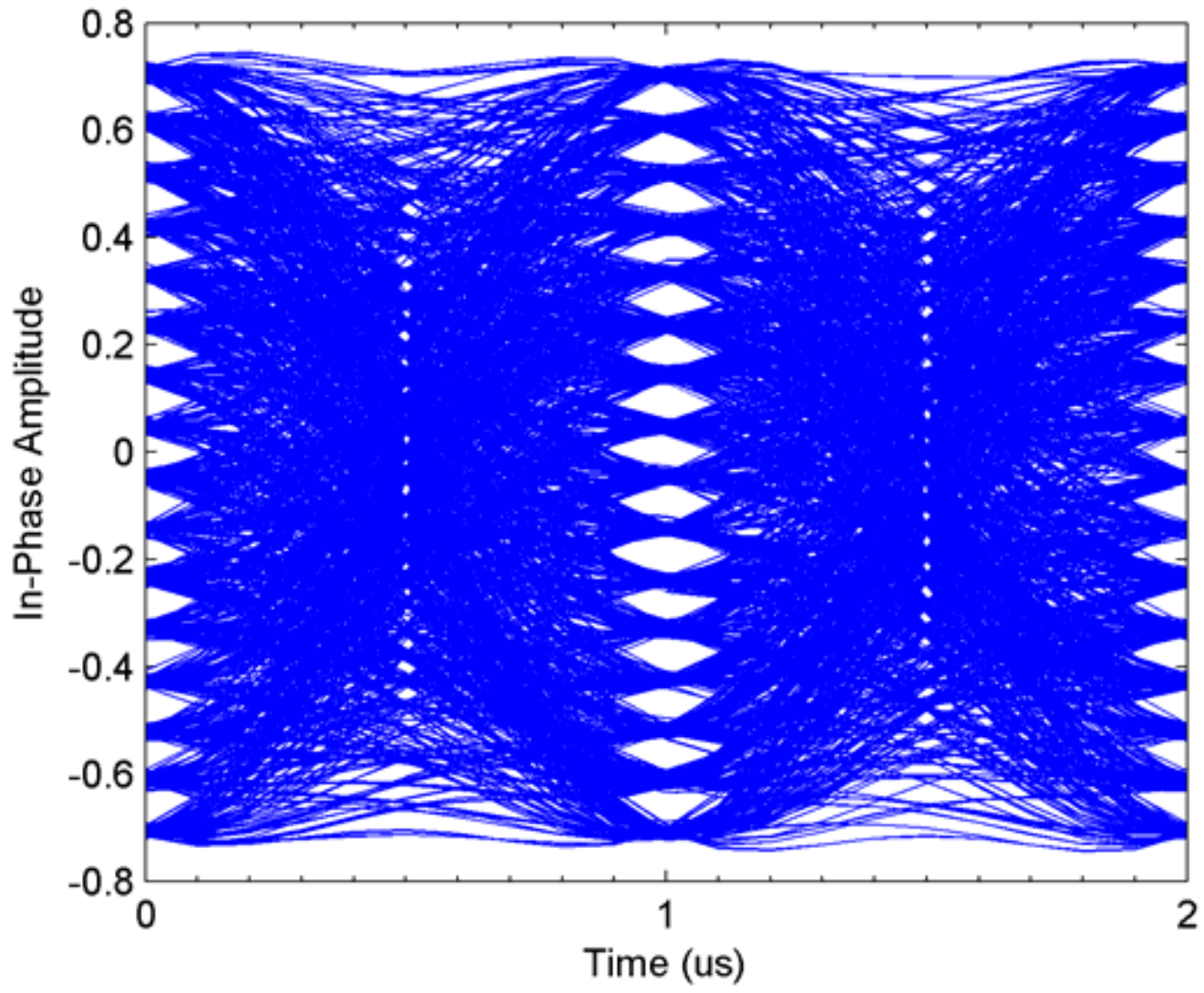

**Figure 6.34:** In-Phase amplitude eye diagram of the 256-QAM signal corresponding to the to received constellation in Figure 6.32.

## 6.4 Errors and Limitations

The main source of error is in the calculation of SNR from the EVM. EVM can only provide an estimate of the BER and SNR, and depends on the assumption that the noise is Gaussian. It has been shown (section 6.2.2) that this is the case for our test system, as long as the link is operating within its linear range. However, when operating at the limits of the link's linearity (usually at higher powers), the EVM calculation is less accurate, and can eventually fail. The effects of amplifier compression on OFDM are non-linear and are the subject of a lot of research [7-11].

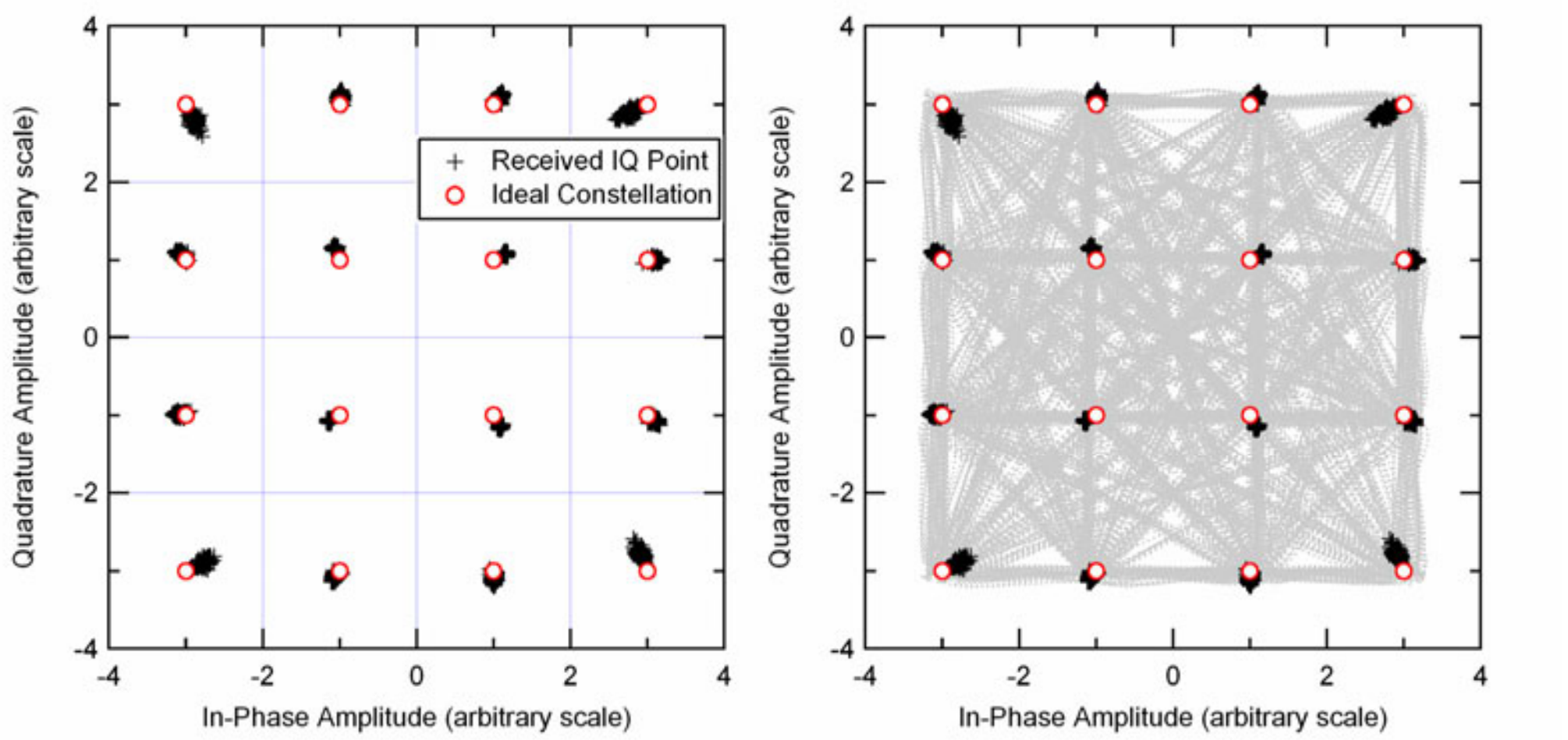

**Figure 6.35:** Example of amplitude compression at high transmission power. 16-QAM signal with carrier frequency = 420MHz, input power = 0dBm and symbol rate = 1MS/s. Decision boundaries are shown on the left, and symbol transitions included on the right (gray lines).

Figure 6.35 is an actual example of the compression observed during the QAM tests. It is immediately obvious that the noise is no longer Gaussian, and the EVM-SNR relationship used earlier no longer holds. The higher amplitude symbols are most severely affected, as can be seen by the 'smearing' of the received IQ points toward the center of the plot. This has a knock-on effect on the ability of the receiver to properly normalize the incoming waveform, causing severe offsets (with respect to the ideal constellation) even for the lower amplitude symbols. Moreover, as can be seen in Figure 6.35, right, amplitude overshoot due to the pulse shaping cannot be accommodated. Since equalization cannot account for non-linear distortion, the receiver has no way of compensating for this.

Including these non-linear effects in the determination of SNR from EVM clearly supersedes the scope of this work. The SNR results shown for higher powers should therefore be interpreted with care. It should be stressed that the error introduced due to signal compression only affects the determination of the peak transmission power available to the system. By providing data rate results for various assumed values of available transmission power, it has been possible to quantify (to a certain extent) the amount of uncertainty introduced to the final data rate calculation. Moreover, EVM will reveal performance degradation due to compression, and it is still a useful quantity on its own when operating beyond the link's linear range. It is only the link with the BER and SNR that is affected.

The results presented in this chapter are obviously directly relevant only to the test system used. The modulator and the demodulator were high-end desktop instruments developed specifically for testing telecom applications, and therefore one would expect them to provide exceptional performance (e.g. in terms of phase noise and dynamic range). There is no guarantee that it will be possible to produce viable dedicated hardware that can achieve this level of performance for a QAM-OFDM system. This is further compounded given the harsh environment of a HEP experiment like CMS, and the low power consumption requirement.

An attempt has been made to calculate the upper theoretical limit to the data rate of an OFDM system. To this end, fractional bit allocations in the bit-loading algorithm were allowed. While it is possible to use encoding to achieve this, the added complexity means that in many real implementations only integer bit allocations are allowed. This could cause a deviation from the optimal situation, leading to lower data rates in practice. On the other hand, this work serves as a reference to which the performance of a future implementation can be compared.

One of the limitations of using EVM to calculate the SNR is that it will fail when the error rate is too high. As explained in section 6.3.1, this is due to the receiver having no knowledge of the transmitted sequence, and using erroneous error vectors in the calculation. In the laboratory tests, the lowest detectable SNR was just over 10dB. Hence the linear relationship of SNR with transmission power was used to extrapolate to lower SNRs. Nevertheless, in the calculation the 'worst' carrier in the system was required to carry at least one bit of information (~10-11dB needed at SER=$10^{-9}$). Hence the limitation described here should have a negligible effect on the accuracy of the results.

## 6.5 Conclusions

It has been shown, experimentally, that RF digital modulation is a possible candidate for a future upgrade of the CMS Tracker analog optical links. 3-4Gbit/s at an error rate of $10^{-9}$ can be achieved using an uncoded OFDM-QAM type of implementation. Whether this will be enough for a future CMS Tracker operating in the SLHC environment remains to be seen. Factors such as implementation complexity, power consumption, and detector data volume will all drive any future decision.

Compared to the data rate calculated in the analytical approach followed in Chapter 5, the results are much more optimistic. The main reason for this is that the assumptions made previously turned to be highly conservative. The noise specification of the link was used in the calculation, and assumed to be additive white (i.e. with constant PSD over the entire frequency spectrum). Furthermore, the system was assumed to be a frequency selective channel with a constant noise source at the output. This is clearly an over-simplified approach, which greatly affected the SNR calculation. Finally, due to the better than expected SNR, the available channel bandwidth was larger by about 150-200MHz, with a corresponding increase in overall data rate.

# Chapter 7

## Conclusions and Future Work

## 7.1 Overview

The performance of the LHC will offer unprecedented potential for discovery and measurement of new physics processes by the CMS experiment. The Higgs Boson will almost certainly be discovered, should it exist as predicted by the Standard Model. The discovery potential of CMS will depend in part on the performance of its readout system, including the analog optical links. Moreover the next iteration of the experiment will operate in an upgraded accelerator (the Super LHC) in approximately ten years time. The CMS Tracker will be upgraded in order to take advantage of the machine's increased performance, and will include faster data readout.

This thesis deals with the characterization and performance of the CMS Tracker readout optical links, both at the component and system levels. Furthermore, armed with the knowledge gained from extensive testing on the optical links, the feasibility of a potential future upgrade using bandwidth efficient digital modulation with the current components has been explored.

## 7.2 The Current CMS Tracker Analog Readout Optical Links

The demanding environment of a high energy physics (HEP) experiment such as CMS has been described, with particular focus on the requirements this poses on the readout system's optoelectronic components. Radiation levels in the CMS Tracker will reach unprecedented levels for any application, prompting designers to either develop customized hardware or extensively qualify commercial components for use at the expected radiation levels. In addition, the physics performance goals require that as much of the sub-detector volume as possible be instrumented, while at the same time minimizing the material budget of the readout electronics that can degrade detector performance. Taking into account that the Tracker consists of over 10 million microstrip channels, this places significant constraints on the components' size and mass. Moreover, each readout link in the Tracker will transfer multiplexed data from 256 microstrip channels, thus requiring ~40 000 point-to-point links to be deployed.

This very unique optical link system has been developed with strict budgetary constraints. Hence commercial components have been used wherever possible, with minimum customization, in order to reduce developmental costs. The Tracker optical links have benefited from the driving trends in the telecommunications industry for devices with low threshold current (and hence low power consumption), uncooled operation and small packaging.

### COMPONENT PERFORMANCE FROM PRODUCTION TESTING

The main performance metrics of the readout system, together with the environmental constraints have been outlined. The optoelectronic components have been extensively tested during the qualification and production phases. Results from production tests on the Analog OptoHybrid (AOH), Linear Laser Driver (LLD), laser transmitter and 12-channel Analog Optoelectronic Receiver (ARx12) illustrate that the components will meet the specifications required of the readout system in terms of the main performance metrics relevant to an analog system operating in a HEP experiment.

The overall signal to noise ratio (SNR) specification of 48dB will be met, allowing the readout system to differentiate between 256 distinct levels (i.e. an amplitude resolution of 8 bits). All components are within their gain specifications and the system will be capable of delivering signal sizes of ~3.2MIPs within the 8 bit range of the back-end. All components have been shown to be very linear and the overall link Integrated Non Linearity (INL) will be better than 1%. The measured bandwidths of each component in the optical link comfortably exceed the 70MHz minimum specification, yielding sufficient dynamic performance for sampling at a rate of 40MS/s. Finally, all front-end electronics will survive the harsh radiation environment in CMS and will function adequately for the lifetime of the experiment.

Of particular interest is the ARx12 that includes a ‘compensation capacitance’ ($C_{ARx}$) setting which allows optimization of the dynamic behaviour of the device by matching the pin photodiode capacitance. Selecting the wrong compensation capacitance can result in extremely slow signal pulse settling times, exceeding the link’s specifications. Hence a method for selecting the correct setting was developed. It is, however, impractical to perform this algorithm on a module by module basis, given the number of receivers that will be installed in the final system. A potential solution is reported in this thesis: Analysis of production test results reveal a correlation between optimum capacitance setting and the intrinsic bandwidth of the device, as reported by the manufacturer. The significance of this is that the optimum $C_{ARx}$ value can be found a priori based on the bandwidth data supplied by the manufacturer, thus requiring no further testing in-situ. The results suggest that nearly all optical links should be set to $C_{ARx}$=800fF.

### SYSTEM PERFORMANCE – LABORATORY TESTS

The optoelectronic receivers are hosted on Front End Driver (FED) boards located in the back end (in the 'counting room'). The FED includes an analog electronics stage after the ARx12 for signal conditioning prior to digitization. The analog electronics are part of the analog portion of the readout chain and hence their performance should match that of the optical link components. Laboratory tests were conducted on the full analog readout chain, in a setup which involved connecting an optical link to a number of optoelectronic receivers mounted on FEDs. The results showed that dynamic performance of the readout links is not compromised by the FED's analog electronics and the specifications are met. In addition, the value of the output load resistor of the ARx12 was intended to provide an additional tuning handle for the gain of the optical links. Hence the value (100Ω) chosen for the first version FED (FEDv1) was not frozen. The tests verified that a potential change in the output load resistor of the ARx12 to lower values (down to 50Ω) would not adversely affect the dynamic performance of the system.

### SYSTEM PERFORMANCE – TEST BEAMS

In addition to experiments in the laboratory, the Tracker test beams have offered the opportunity to perform system tests on an unprecedented number of deployed optical links made up of final-version components in CMS-like conditions. Two important contributions to the performance assessment of the current links were made as a direct result of the data obtained from these large system tests: The development of robust optical link setup algorithms and the prediction of the link gain spread in the final system.

Setting up the optical links in the CMS Tracker is critical to the correct operation and performance of the readout system. The design of the optical link allows tuning of the gain as well as of the laser operating point, for which automated setup routines are required in the final system, given the immense number of deployed links in the final system. Previous routines failed if the link gain was high enough to cause saturation in the FED ADC. New algorithms for setting up the links have been developed that overcome this limitation. Additionally, the bias point selection algorithm has been optimized for operation at low noise and optimum dynamic range. Thus signals up to ~3.2MIPs can be measured reliably.

In the setup routine, the gain is derived from a known electrical quantity at the input of the link: The APV's digital header height (or APV 'tick'). This provides a fast and efficient gain estimation. However, this assumes that the tick height correlates well with the signals produced by particles traversing the detectors. The amplification of these signals is termed 'particle' or 'MIP' gain in this thesis and ultimately is the quantity which is of most importance for the readout system's performance goals. The accuracy of the setup routine's gain calculation has, for the first time, been reliably evaluated using physics data from test beams. The Tracker Outer Barrel (TOB) Cosmic Rack (CRack) was exposed to beams of muons and pions in several physics runs. Histograms of the signal sizes at the output of the optical links (produced by the particles hitting the detectors) were compiled and Landau fits to the data were performed to determine the particle gain of the readout links. The calculation method was presented in detail and the results used to show that the setup routines (based on measurement of the APV tick height) are accurate enough to be used in the setup of the readout system.

The Tracker has opted for an analog optoelectronic readout system. The choice of optical technology is an obvious one, given the massively parallel nature of the system. Electromagnetic interference (which would be a problem with twisted-pair copper links) is avoided, while mass is minimized. Analog transmission has the advantage of reducing complexity and power consumption in the front-end electronics, since there is no need for digitization. Moreover, compared to a binary digital system, the positional resolution of the Tracker can be improved by taking advantage of charge-sharing in the microstrip detectors.

The gain is, therefore, an important parameter in the readout system, since it affects the dynamic range of the data being read out. Production test data has been used to compile histograms of the gain of each optical link component. These were then used in a Monte Carlo simulation to predict the gain spread that will be observed throughout the optical links deployed inside CMS. With the tests taking place at room temperature, it was not possible to incorporate the effect of the Tracker's low temperature environment (-10°C) on the component gains based on production test data alone. However, test beams were conducted on cooled systems and data from these was used to derive the relationship between optical link gain and temperature.

This was incorporated in the simulation model, allowing determination of the gain distribution at any operating temperature.

The simulation results have been compared to real systems, and excellent agreement was found. Moreover, the results confirmed the suspicion that the optical link gains were too high. Corrective action was taken as a result of the simulation results, and the dynamic range of the Tracker readout system restored to within specifications (3.2MIPs/8bits for thin detectors). This was achieved by lowering the ARx12 load resistor from 100Ω to 62Ω (resulting in 38% gain reduction). The implication of this is an important one: It has been possible to use production test data to make useful and reliable predictions for a large-scale system that will be deployed in the future.

### 7.2.1 Future Prospects

Once the optical links are installed in the final system it will be possible to study their performance from a system point of view. Firstly, the accuracy of the gain spread prediction will be tested and any discrepancies analyzed. Future systems could benefit from the lessons learned in using production test data to predict performance of a system under development.

Monitoring of readout system performance metrics can be carried out during regular Tracker operation. For example, effects of radiation damage and temperature on link gain, laser slope efficiency and laser threshold can be extracted easily from the setup algorithms described in this thesis.

## 7.3 Upgrading the CMS Tracker Analog Readout Optical Links for Super LHC

The next iteration of the CMS Tracker will be operated in the Super LHC (SLHC) environment and will have to cope with significantly increased data rates due to the foreseen tenfold increase in luminosity. The cost of the optoelectronic components represents a large fraction of the CMS Tracker electronics budget. Hence, a digital system reusing the existing optoelectronic components while delivering sufficient performance for SLHC operation could potentially be a cost-effective alternative to a full replacement of the installed links. The first step in the feasibility study of such a conversion has been carried out, and an accurate estimate of the performance that can be achieved has been made.

The extensive testing and characterization carried out on the optoelectronic components described in the first part of the thesis has facilitated the selection of an appropriate bandwidth efficient modulation scheme for a future upgrade. The laser driver and the receiving amplifier ASICs present bottlenecks in the bandwidth of the Tracker optical link, with each component having a 3-dB bandwidth of ~100MHz. Given the excellent noise performance and linearity of the system, it is possible to employ a Quadrature Amplitude Modulation (QAM)-based scheme in order to reach Gbit/s data rates within the available bandwidth.

Established digital communication principles have been used to estimate the potential rate in a theoretical future upgrade, taking into account all relevant system parameters such as noise, bandwidth and available transmission power. The current analog link was treated as the communication channel over which digitally modulated signals are to be transmitted. The calculation method has been demonstrated for an uncoded multi-carrier QAM system, though the exact rate achieved in a future system will be implementation-dependent. The method and results are also readily applicable to a single-carrier QAM system. In fact, the single-carrier equivalent system has been used in the calculations (see Chapter 5).

Experimental evidence of the feasibility of QAM transmission in a Tracker optical link has been provided. Low-symbol rate QAM signals with carrier frequencies up to 1GHz were transmitted –one at a time– through an analog optical link with final-version components. This effectively allowed the complete characterization of the link, by evaluating the digital Signal to Noise Ratio (SNR) across the entire bandwidth of the optical link. The experimental data was fed into the QAM multi-carrier model, and this has been used to show that, for an uncoded system, 3-4Gbit/s for error rates of $10^{-9}$ are a realistic possibility.

### 7.3.1 QAM Upgrade: Feasibility of Implementation

It has been proven that QAM can be used to maximize the data rate in a future optical readout link based on the current CMS Tracker components. However, the applicability of such an upgrade (or any upgrade for that matter) also depends on the suitability of the hardware implementation to the unique requirements of a HEP experiment.

The fundamental trade-off in a digital communication system is that of complexity and bandwidth. In general, for a given data rate, a simple modulation scheme (such as binary NRZ) is relatively easy to implement, but requires significant bandwidth. Conversely, using bandwidth efficient modulation (e.g. QAM) one can achieve the same data rate and occupy less channel bandwidth, at the expense of more complex transmitters and receivers. The proposed QAM-based upgrade must therefore be evaluated on the basis of whether the increased data rate and complexity, as well as the cost of development of additional electronic components, can be justified by the savings in the development and purchase of new optoelectronic components.

Of primary importance is the transmitter, since it will be placed inside the detector volume (assuming the unidirectional point-to-point link model is retained for the future system). In particular, the following issues will need to be addressed for a future system:

- Power consumption
- Radiation hardness
- Low-mass and small size
- Hardware complexity and R&D effort required

SINGLE VS MULTI-CARRIER

A QAM-based system could employ one or multiple carriers. The two approaches offer different advantages and disadvantages and their suitability for a HEP readout link needs to be evaluated.

In the case of a multi-carrier system, an FFT will have to be performed by the front-end electronics at a very fast rate. With current electronic technology, this may be impossible to achieve while keeping the power consumption within Tracker specifications. For the single-carrier case there is no need to synthesize the carrier frequencies and thus reducing complexity in the front-end.

The SNR vs frequency measurements carried out (see Chapter 6) show that if low-symbol rate carriers are used, the SNR can reach over 35dB for many of the sub-channels. It follows that as many as 8-10 bits/symbol will be allocated to these high-SNR carriers, thus requiring an extremely high amplitude/phase resolution in the system. If only one high-symbol rate carrier is used, it will be susceptible to the full noise spectrum of the channel. The resulting (inferior) SNR will mean less bits/symbol (~5-6) will be allocated to the single carrier, hence requiring much less resolution than in the multi-carrier case.

Perhaps the most serious disadvantage of a multi-carrier system is that of the Peak to Average Power Ratio (PAPR). This can be an order of magnitude higher than in a single-carrier system and requires a significant amount of amplifier backoff (as much as 6-9dB) to avoid saturation that results in symbol errors. Backing-off the transmission power also results in decreased SNR, so a balance between the two needs to be found.

One of the disadvantages of the single-carrier system is that of InterSymbol Interference (ISI) due to the frequency response of the channel. A multi-carrier system is designed such that each carrier occupies a small slice of the available bandwidth that has a flat frequency response, effectively eliminating ISI. This is not the case for a carrier occupying the entire channel spectrum and equalization would be required to mitigate the effects of ISI. Hence more complexity is required in the receiver, where very fast equalization (with a sufficiently large number of coefficients) would have to be performed. Given that the frequency response of each analog optical link will not be known a priori, adaptive equalization would be required. This would be computationally expensive at Gbit/s data rates. On the other hand, this may not be a serious drawback if equalizer ‘training’ needs to be performed only once during the initial setup (i.e. when the link is first assembled). Since each analog optical link has a constant frequency response (i.e. it is not a fading environment as is the case in the wireless world), this may be feasible.

An upgrade based on one, high-symbol rate QAM carrier would be more demanding on the front-end electronics in terms of the requirement for higher clock speeds compared to multiple, low-symbol rate carriers. In the case of the Tracker optical link, a very high symbol rate (as high as 500-600MS/s) would be necessary to take full advantage of the available bandwidth of over ~600MHz. This would require a carefully designed high-speed mixed-signal ASIC, including very high-speed DACs (in the GS/s range, assuming oversampling is employed).

Finally, an advantage of the multi-carrier case is the ability to adjust the data rate with finer granularity due to the lower system symbol rate. The implication of this is that the data rate can be easily tailored to each optical link channel, depending on the individual frequency response and noise characteristics. A single QAM carrier operating at, say, 500MS/s would only be adjustable in steps of 500Mbits/s steps (ignoring any encoding).

In conclusion, a QAM system consisting of a single carrier may be more appropriate than a multi-carrier system for a HEP environment. Despite the drawback of the high symbol-rate, the single-carrier system avoids the need to perform a very fast FFT in the front-end. In general, multi-carrier systems are more appropriate in fading environments such as wireless networks. This is not the case with the Tracker optical links, where the frequency response is fixed. Hence, the added complexity, as well as the problem of PAPR, make a multi-carrier system less attractive for a future HEP readout link.

#### HARDWARE COMPLEXITY

An upgrade will require additional components in the front-end. Digitization will have to be performed on the front-end hybrids and components for single or multi-carrier QAM modulation would also be required. These must meet the same strict requirements as the current front-end components. In this section, a single-carrier implementation is examined.

Figure 7.1 shows the basic mixed-signal implementation of a single-carrier QAM modulator. Assuming the symbol rate is 600MS/s and 32-QAM is used (5bits/symbol), the total data rate would be 3Gbit/s. The 3Gbit/s bitstream from the detector ADCs is grouped in 5-bit words. The I/Q mapper translates these into in-phase and quadrature values (3 bits each) and must operate at the symbol rate of 600MS/s. If pulse shaping is used, a certain amount of oversampling is required. With x4 oversampling, the sample rate at the output of the digital filters would be 2.4GS/s, which requires very high-speed DACs to be used prior to the analog stage. Even if no pulse shaping is used, the DACs must operate at 600MS/s. The analog portion of the modulator would require RF components with a modulation bandwidth of over 600MHz.

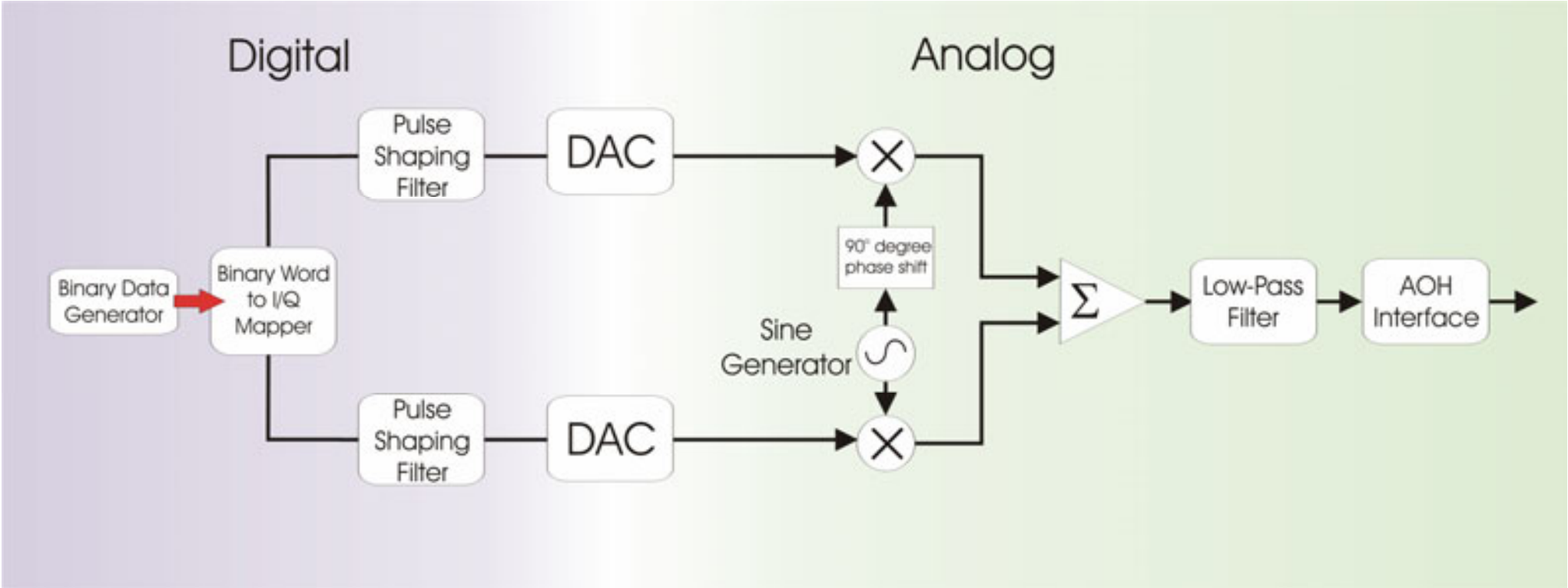

**Figure 7.1:** Block diagram of a QAM modulator.

It is clear that this upgrade path represents a huge challenge in terms of the effort required to overcome technological obstacles and develop a robust readout system. QAM modulators capable of Gbit/s data rates are not currently in use by any other standardized application. The implication of this is that a custom ASIC employing a proprietary modulation scheme would have to be built. This undoubtedly requires an enormous amount of research and development effort that cannot be carried out uniquely at CERN. Collaboration with an industrial partner (and possibly experienced academic departments) would therefore be essential. Moreover, this would represent a departure from the conventional wisdom adopted by the HEP community in the development of readout systems for the LHC era, which normally involves the use of commercial off-the-shelf (COTS) components with minimal customization. Hence the proposed upgrade cannot take advantage of the cost-effective approach of following industrial trends.

Given the modest data rate predicted in this thesis (3-4Gbit/s), coupled with the immense challenge of developing such a high data rate QAM system, this upgrade may not be the most sensible approach. While preservation of the optoelectronic components (transmitter, fiber and receiver) is an obvious advantage, this will most likely be offset by higher investments in the R&D effort for a proprietary and novel new system. 10 and 40Gbit/s optical transmission is a commercial reality, and may be a far easier and cheaper solution for the future of the CMS Tracker.

#### THE EXAMPLE OF THE AGERE SYSTEMS QAM TRANSCEIVER FOR OPTICAL FIBER

Currently there are no commercial applications employing QAM modulation at data rates that are sufficiently high for a future HEP readout link. Instead, high speed links use simple modulation schemes at the expense of bandwidth which is abundant in state-of-the-art optical fiber-based systems. Hence there are no Gbit/s QAM modems available commercially that could be used as an paradigm for the Tracker upgrade. However, bandwidth efficient RF modulation over fiber has recently attracted the attention of researchers as demonstrated by the example of this section. Azadet and Saibi of Agere Systems have proposed a QAM system for 40Gbit/s optical links (Figure 7.2) [1, 2]. Their approach is very similar to the upgrade considered in this thesis and is included here to give an idea of the complexity and hardware required for implementation.

The Agere proposal involves Sub-Carrier Multiplexing (SCM) of 16 16-QAM carriers, each with a symbol rate of 666MS/s. SCM differs from OFDM in that the carrier spacing is not equal to the symbol rate. Instead, the Agere proposal employs 833MHz spacing which, the authors claim, avoids the stringent requirements of OFDM systems for mitigating inter-carrier interference.

This system is designed for an optical link with a much higher bandwidth (the last carrier is placed at ~13GHz). However, the idea is relevant since the symbol rate of each individual carrier is compatible with the bandwidth available in the Tracker optical links. The motivation behind using bandwidth efficient modulation is to produce a more cost-effective system, by employing signalling rates that allow implementation in conventional IC technology such as SiGe or CMOS (transmission speeds of over 10Gbit/s normally require more expensive IC technologies).

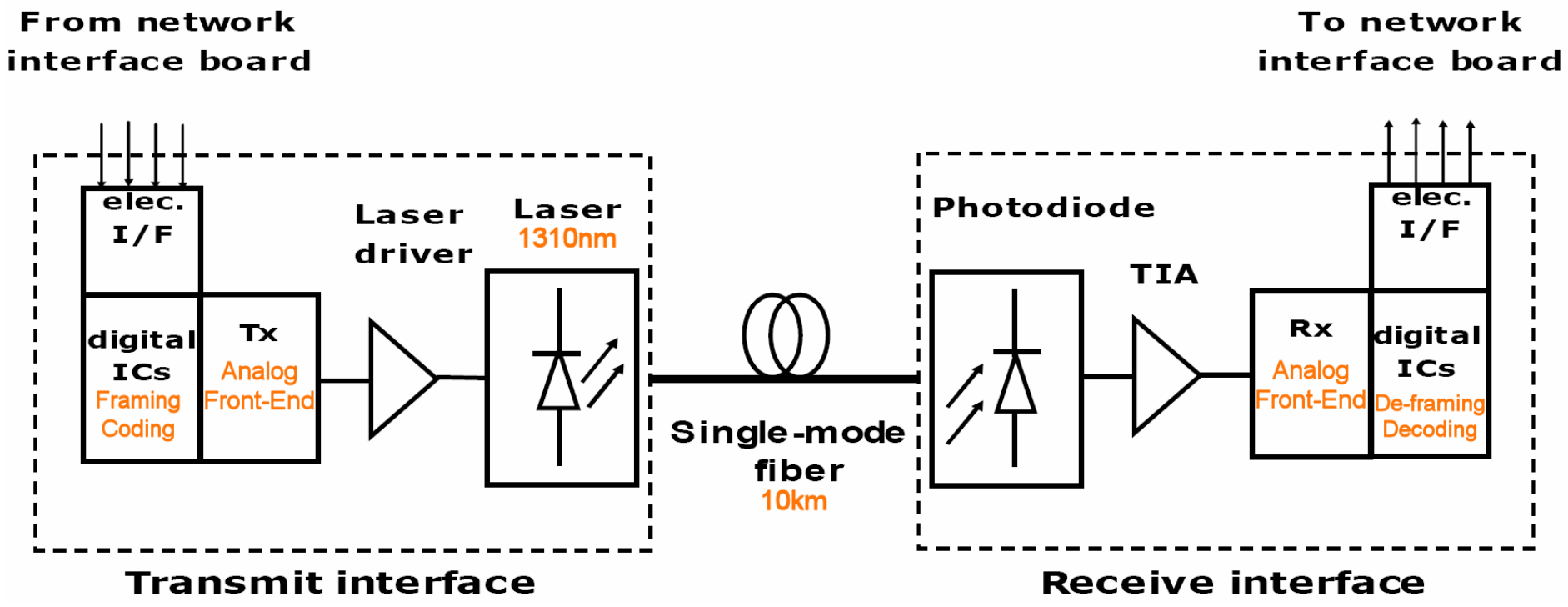

**Figure 7.2:** The Agere proposal: A QAM-based multi-carrier system [2].

To prove the viability of the concept, Agere have implemented a single-carrier 16-QAM transceiver operating at a symbol rate of 666MS/s [1], in a 1.5V, 0.14μm CMOS ASIC (Figure 7.3). The total area is 3.6mm$^2$, and the power consumption is 340mW. This is roughly six times higher than the consumption of the Tracker AOH transmitter, though one would expect that lower power consumption could be achieved in a future implementation. In addition, the Agere ASIC includes FEC in the form of Reed-Solomon (255,239), offering a coding gain of 5.5dB (BER is reduced to $10^{-15}$ from the input BER of $10^{-4}$) and burst error correction of up to 1024 bits (using interleaving). Hence, given the features and performance of this transceiver, the power consumption of the chip can be considered reasonable.

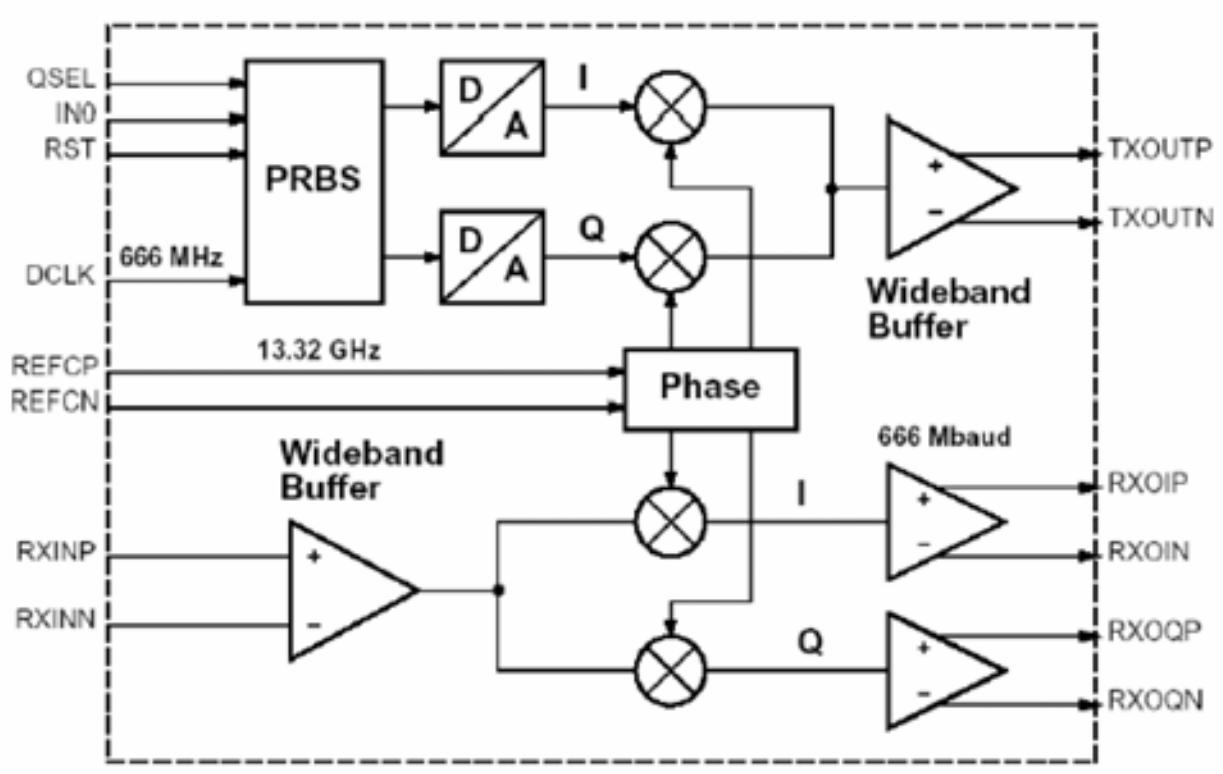

**Figure 7.3:** Block diagram of the Agere QAM transceiver ASIC [1].

Agere's implementation proves that the concept of QAM modulation over optical links at Gbit/s rates is viable, though not directly applicable to a HEP readout link yet. A future, lower-power ASIC with similar characteristics and performance would be necessary for the Tracker upgrade, but will only materialize if the telecoms industry finds sufficient motivation to adopt the concept (possibly for low-cost, multi-Gbit/s optical links).

### 7.3.2 Future Prospects

The future CMS Tracker sub-detector will have to be replaced for operation in the SLHC. Hence, it is not known how many channels will have to be read out, and at what data rate. Since the installed optical fibers will very likely remain regardless of the future implementation, it remains to be seen if 40 000 optical links transmitting data at 3-4Gbit/s will be sufficient for the readout system.

The future Tracker optical links will include some form of Forward Error Correction (FEC) if sufficient performance is to be achieved at the Gbit/s data rates envisaged. This is especially important for readout links operating in the HEP environment because of the radiation environment which can cause Single Event Upsets (SEUs) [3] as well as degradation in the SNR over time. For FEC to be beneficial, encoding needs to be tailored to the type of modulation chosen as well as the error characteristics. Hence there is research to be carried out in this area.

Another possibility for future research is the investigation of source compression prior to transmission over the optical links. Entropy-based coding is a possibility, since the sources' (i.e. the particle detectors) signal statistics are normally well known prior to system deployment (because of extensive testing during production and in test beams). Compression could potentially greatly reduce the amount of data to be transferred from the experiment. Future work would involve studying established codes and their applicability to the characteristics of the information source (in this case, the particle detectors). Of particular interest would be to examine if entropy-based coding can be employed, given that the statistics of the signals produced by the detectors are usually well-known from extensive testing prior to deployment of the system. Finally, the complexity of implementation also needs to be considered.